\documentstyle[preprint,aps]{revtex}

\begin{document}

\draft


\tighten

\draft

\title{Gravitational Correspondence of Two Types of Superconformal Anomaly}

\author{W.F. Chen\renewcommand{\thefootnote}{\ddagger}\footnote{E-mail:
wchen@wlu.ca, wfchen@phys.cts.nthu.edu.tw}}

\address{Department of Mathematics\\ Wilfrid Laurier University\\ Waterloo,
Ontario, Canada\\ and\\
Physics Division\\ National Center for Theoretical Sciences\\
Hsinchu, Taiwan, R.O.C. }


\maketitle

\begin{abstract}
\noindent In a classical conformal invariant supersymmetric gauge
theory, the chiral $R$-symmetry current $j^\mu$, the supersymmetry
current $s^\mu$ and the energy-momentum tensor $\theta_{\mu\nu}$
constitute a supercurrent multiplet. There are two different
superconformal anoamly multiplets in four-dimensional supersymmetric
gauge theories, one originating from the supersymmetric gauge
dynamics and the consequent nonvanishing $\beta$-function, and the
other one coming from the coupling of the supercurrent multiplet to
the external supergravity multiplet with non-trivial topology. We
emphasize that in the gauge/gravity dual correspondence these two
types of superconformal anomaly multiplets have distinct reflections
in the classical supergravity: the anomaly multiplet due to the
supersymmetry gauge dynamics is dual to the spontaneous symmetry
breaking and the consequent super-Higgs effect in $AdS_5$ bulk
supergravity, while the anomaly multiplet originating from the
non-trivial topology of external conformal supergravity mutiplet is
a boundary effects of the $AdS_5$ space.


\end{abstract}


\vspace{3ex}


\section{Introduction}

In Ref.\,\cite{kow}, Klebanov, Ouyang and Witten has investigated
the gravity dual of the chiral $U_R(1)$ symmetry anomaly in ${\cal
N}=1$ cascading $SU(N+M)\times  SU(N)$
  supersymmetric gauge theory
 using  the gauge theory/gravity duality
correspondence \cite{mald,gkp,witt1}. They showed that on the
gravity side the chiral $U_R(1)$-symmetry  anomaly of above
supersymmetric gauge theory  is a classical feature and behaves as a
spontaneous breaking of gauge symmetry in the ${\cal N}=2$ $U(1)$
gauged $AdS_5$ supergravity. This statement is supported by the
observation that the the vector field dual to the $R$-symmetry
current of the supersymmetric gauge theory has got a mass through a
Higgs mechanism. In the paper, the authors wrote ``{\it It is
interesting that the anomaly  appear as a bulk effect in AdS space,
in contrast to previously studied examples \cite{witt1,hesk} where
anomalies arose from boundary terms} ".  The aim of this paper is to
emphasize the origin for this interesting phenomenon in the
gauge/gravity dual.

In fact, in the gauge theory side the answer to this puzzle is very
straightforward. The anomaly discussed by Klebanov, Ouyang and
Witten \cite{kow} and those studied in Refs.\,\cite{witt1,hesk} have
completely different physical origin. The former comes from the
quantum dynamics ofa ${\cal N} <4$ supersymmetric gauge theory and
the anomaly coefficient is the beta function of the theory
\cite{crewther,abb,grisaru,afgj}, which is usually an internal
anomaly \cite{afgj}; While the later arises from the coupling  of a
suerpsymmetric gauge theory  with an external gauge field. If the
geometry and topology of the external gauge field configuration is
non-trivial, then this type of anomaly arises
\cite{abb,grisaru,afgj}. The former anomaly is actually a quantum
correction  to the latter one \cite{oweyl}.  This is the reason why
there exist two distinct gravitational correspondences to the
superconformal aonmaly.

Once the field theory origins for these two types of anomaly have
been understood, the next step is to reveal the distinct features of
their gravitational correspondence in terms of the gauge/gravity
dual. However, it is necessary to emphasize the specific feature of
the anomaly for space-time symmetry in a supersymmetric field theory
before we proceed. As it is well known,  the ${\cal N}=1$ space-time
supersymmetry is composed of the Poincar\'{e} symmetry,
supersymmetry and the chiral $R$-symmetry with generators
$M_{\mu\nu}$, $P_\mu$, $Q_\alpha$ ($\alpha=1,\cdots,4$) and  $R$,
respectively. The corresponding three N\"{o}ther currents, the
energy-momentum tensor $\theta^{\mu\nu}$, the supersymmetry current
$s^\mu$ and the chiral $R$-symmetry current $j^\mu$ must
 constitute a supercurrent multiplet \cite{ilio,anm}. Further,
 in a supersymmetric theory with classical scale
symmetry this supercurrent multiplet
($j^\mu$,$s^\mu$,$\theta^{\mu\nu}$) is actually a superconformal
current multiplet \cite{ilio,anm}. At quantum level, when  one
member in the supercurrent multiplet  suffers from anomaly, the
other two members should also become anomalous. These anomalies
should constitute an anomaly multiplet since the Poincar\'{e}
supersymmetry persists \cite{abb,chdu,grisaru,afgj}. Therefore, when
we investigate the gravitational dual description to the space-time
symmetry anomaly in a supersymmetric theory, we should consider the
whole anomaly multiplet. In this paper we consider the
superconformal anomaly $(\partial_\mu j^\mu, \gamma^\mu s_\mu,
\theta^\mu_{~\mu})$, which form an ${\cal N}=1$ chiral
supermultiplet.

The approaches of revealing the gravitational descriptions  to these
two types of anomaly in terms of the gauge/gravity correspondence
are completely different. The external superconformal anomaly arises
from the coupling of a supersymmetric gauge field theory  with an
external conformal supergravity in four dimensions \cite{abb,chdu}.
This anomaly multiplet can be directly calculated with the
holographic definition on the $AdS/CFT$  correspondence
\cite{gkp,witt1}. As it is well known, the original AdS/CFT
correspondence conjecture \cite{mald}
  states that  the type IIB string theory in the $AdS_5\times S^5$
  background with $N$ units of $R-R$ flux passing through $S^5$ should
  describes the same physics
  as ${\cal N}=4$ $SU(N)$ supersymmetric Yang-Mills theory.
Later,  Gubser et al \cite{gkp} and Witten \cite{witt1} made
  this conjecture concrete and proposed a holographic version on
  this correspondence. This concrete definition has  two
  formulations. One is based on the canonical quantization of a field
  theory system, which states
that there  exists a one-to-one correspondence between a quantum
state of type II superstring in the $AdS_{d+1}\times X^{9-d}$ target
space-time and a gauge invariant operator of  a $d$-dimensional
conformal field theory defined on the boundary of $AdS_{d+1}$,
 $X^{9-d}$ being a compact Einstein manifold \cite{witt1}; The other formulation
 is established through the path integral quantization on a field
 theory system. It states that the partition function of type II
superstring theory as a functional of the boundary value of the
$AdS_{d+1}$ bulk supergravity field  should be identical with the
generating functional for the correlation function of some gauge
invariant operators of a $d$-dimensional conformal field theory in
the external field furnished by the boundary value of the
$AdS_{d+1}$ bulk field \cite{witt1}. At low-energy level, the type
II superstring theory is approximated by its low-energy effective
theory, the type II supergravity \cite{gsw}.  Subsequently, the
partition function of type IIB superstring can be calculated with
the saddle point approximation, i.e., it can be evaluated as the
exponential of the supergravity action in a field configuration
which is determined by the classical equation of motion with initial
condition $AdS_{d+1}\times X^{9-d}$. The dual field theory must be
in its large-$N$ limit. Further, taking into account the spontaneous
compactification \cite{freu} on $X^{9-d}$ of  type II supergravity
\cite{marcus,kim} and assuming that there exists a consistent
nonlinear truncation of the massless modes from the massive
Kaluza-Klein spectrum \cite{dupo},  we should a gauged $AdS_{d+1}$
supergravity. The $AdS/CFT$ correspondence between a  on-shell
gauged $AdS_5$ supergravity \cite{gst} and a four-dimensional
  supersymmetric  gauge theory in an off-shell conformal supergravity
  background can thus be established. In particular, this off-shell
  four-dimensional  conformal supergravity \cite{kaku,fra} comes from the reduction of
  five-dimensional gauged supergravity on the boundary of $AdS_5$
  space \cite{nish,bala}. Consequently, the bulk supersymmetry in gauged $AdS_5$
  supergravity convert into the superconformal symmetry of four-dimensional
  conformal supergravity  \cite{chch1}. That is, the bulk gauge symmetry
  decomposes into a vector- and an axial
vector gauge symmetry \cite{witt1}; the bulk diffeormorphism
symmetry becomes a diffeormorphism symmetry  and a Weyl symmetry
\cite{imbi}; the bulk supersymmetry separates into a  Poincar\'{e}
supersymmetry and a super-Weyl symmetry \cite{nish,bala,wfchen}.
However, the on-shell action of the asymptotically gauged  $AdS_5$
supergravity cannot keep the symmetries in each pair simultaneously
\cite{witt1,hesk,wfchen}. Usually  the vector gauge symmetry, the
diffeormorphism symmetry and the Poincar\'{e} are required to be
satisfied. Therefore we can get the external superconformal anomaly
from the on-shell action of gauged $AdS_5$ supergravity near the
$AdS_5$ boundary. This is usually called the holographic
superconformal anomaly \cite{witt1,hesk,wfchen,chch1}.

For the internal superconformal anomaly in an ${\cal N}<4$
 four-dimensional supersymmetric gauge theory, the case becomes
 complicated. Since this anomaly stem from the quantum dynamics
 of a supersymmetric gauge theory itself \cite{abb,grisaru,afgj,anm}
 and has nothing  to do with the external
 conformal supergravity background, we cannot look for its
 gravitational correspondence using the holographic version on
 gauge/grvaity correspondence.  In fact, the existence of this type
 of superconformal anomaly means that the field theory at
 quantum level has no conformal symmetry. This is different
 from the external superconformal anomaly, its existence does not
 affect the superconformal symmetry of a supersymmetry gauge theory
since it arises from the coupling of field theory with the external
background field and originates in the non-trivial topology of
external field configuration. One typical example is ${\cal N}=4$
supersymmetric Yang-Mills theory in four dimensions. Its
energy-momentum tensor, supersymmetry current and $SU_R(4)$
$R$-symmetry all become anomalous when the theory couples with an
${\cal N}=4$ external conformal supergravity. But the theory itself
is still a superconformal theory since its beta function still
vanishes. According to the symmetry identification in the $AdS/CFT$
correspondence \cite{agmo}, the global superconformal symmetry on
the field theory has the exactly  counterpart with the local
supersymmetry in gauged $AdS_5$ supergravity. That is, the
$R$-symmetry in  the gauge field theory corresponds to the gauge
symmetry of the $AdS_5$ gauged supergravity;  the space-time
conformal symmetry $SO(2,4)$ is exactly the isometry symmetry of the
$AdS_5$ space; the conformal supersymmetry of the field theory goes
to the local supersymmetry of the $AdS_5$ supergravity.  Therefore,
we must go beyond the $AdS/CFT$ correspondence to to find the
gravitational dual description to the internal superconformal
anomaly.

How can we carry on to get the dual of the internal superconformal
anomaly? Let us recall how the gauge/gravity correspondence is
established from the non-perturbative type II superstring theory.
It is based on the fact that a $Dp$-brane has two distinct
features \cite{polch}. on one hand, in weakly coupled type-II
superstring theory, it behaves as a dynamical and geometrical
object with open string ending on it. Thus a stack of coincided
$Dp$-branes at low-energy will yield a $p+1$-dimensional
supersymmetric gauge theory on the world-volume of $Dp$-branes. We
can use various $D$-branes and solitonic branes as well as  other
possible geometric objects like orientifold planes to construct
various possible brane configurations. In this way any gauge
theory in principle can be obtained by such a so-called ``brane
engineering" \cite{giku,direv}; On the other hand, a $Dp$-brane,
as  string soliton, carries R-R charge and couples with $p+1$-rank
antisymmetric field originating from the R-R boundary condition.
This physical property of a $D$-brane determines that it provides
source  to the low-energy effective theory of strongly coupled
type-II  superstring theory --- type-II supergravity. Therefore,
it can modify space-time background of string theory and arises as
a $p$-brane solution to  type-II supergravity \cite{stel,bcpz}.
This is the physical reason behind the  gauge/gravity duality
conjecture.

This suggests  that we should resort to  a $D$-brane configuration
to look for the gravity dual of the internal superconformal anomaly.
Since a supersymmetric gauge theory stems from a $D$-brane
configuration,  all its quantum behaviors including the quantum
anomaly  should be sensible from the corresponding $D$-brane
dynamics \cite{giku}. Thus we first find  how the superconformal
anomaly is manifested  in the $D$-brane configuration, then we use
the gravitational perspective of  $D$-branes  to observe how it
emerge in the gravity side. In this way, we can work out the gravity
dual of the internal superconformal anomaly  \cite{wfchen2}.

In Sect.\,II we review the field theory  results on both the
external and internal superconformal anomalies of a supersymmetric
gauge theory and demonstrate their physical difference. In
introducing the external superconformal anomaly, we show how the the
conservation of  the superconformal current multiplet is equivalent
to the super-Weyl symmetry of the external conformal supergravity
background. So the arising of the external supercomformal anomaly is
converted into an violation of the super-Weyl symmetry in the
external conformal supergravity. This paves the way for the
derivation on superconformal anomaly in terms of the $AdS/CFT$
correspondence. In discussing the internal superconformal anomaly,
we stress that its anomaly coefficient is proportional to the beta
function of the theory.  Further, we show that  in the quantum
effective action it is manifested in the running of gauge coupling
and the shift of the strong CP violation angle. To uncover how the
$\gamma$-trace anomaly of is concealed in the quantum effective
action, we write the quantum effective action  in  the superspace.
This facilitate us to find how the superconformal anomaly is
reflected in the $D$-brane configuration. We also give a specific
introduction to ${\cal N}=1$ $SU(N+M)\times SU(N)$ supersymmetric
gauge theory since its gravity dual description is well studied.
Later we shall use this theory to illustrate the dual of internal
superconformal anomaly. Sect.\,III is about the holographic version
on $AdS/CFT$ correspondence and its low-energy approximation. We
start from Witten's definition \cite{witt1} on the $AdS/CFT$
correspondence between type II superstring in $AdS_{d+1}\times
X^{9-d}$ background and a $d$-dimensional $SU(N)$ superconformal
gauge theory living on the boundary of $AdS_{d+1}$ space. Then using
the well-known fact that the low-energy effective theory of type II
superstring is the type II supergravity,  we get the approximation
correspondence between $d$-dimensional $SU(N)$ superconformal gauge
theory at large-$N$ limit and on-shell type II supergravity whose
classical solution asymptotically approaches $AdS_{d+1}\times
X^{9-d}$. Further, considering the spontaneous compactification  on
$X^{9-d}$ of type II supergravity and assuming that there exists a
consistent truncation of the massive Kaluza-Klein modes \cite{dupo},
we finally establish a correspondence between the gauged $AdS_{d+1}$
supregravity and a $d$-dimensional superconformal gauge theory in a
conformal supergravity  background furnished by the $AdS_{d+1}$
boundary data of the bulk supergravity multiplet. To
straightforwardly verify this conclusion acquired from string theory
level,  we start directly from  five-dimensional  gauged
supergravity and check its solution to classical equations of motion
which asymptotically approaches  the $AdS_5$ geometry. It is shown
that  at the leading order of the expansion in the radial coordinate
the on-shell fields of gauged ${\cal N}=2,4$ $AdS_5$ supergravity
indeed constitute the off-shell multiplet for ${\cal N}=1,2$
conformal supergravity in four dimensions. This reveals the essence
that we can get the external superconformal anomaly in four
dimensions from gauged $AdS_5$ supergravity. Based on this
conclusion, in Sect.\,IV we derive the external supercoformal
anomaly from gauged supergravity in five dimensions. The holographic
chiral $R$-symmetry anomaly comes from  the Chern-Simons five-form
term in gauged $AdS_5$ supergravity; The holographic Weyl symmetry
anomaly lies in the IR divergence of on-shell action of gauged
$AdS_5$ supergravity near the $AdS_5$ boundary  due to the infinite
volume of $AdS_5$ space. This requires a so-called holographic
renormalization procedure to implement  and thus leads to the
holographic Weyl anomaly; Finally the arising of super-Weyl anomaly
is revealed through the supersymmetry variation of gauged $AdS_5$
supergravity. As a supersymmetric field theory, its variation under
supersymmetric transformation  should be a total derivative. When we
take this total derivative to the $AdS_5$ boundary and consider the
on-shell fields, it is found that these total derivatives cannot
preserve the four-dimensional supersymmetry and supr-Weyl symmetry
simultaneously. In this way the super-Weyl anomaly is uncovered.
From Sect.\,V we turn to the search for the dual description to
internal superconformal anomaly. First, a  general review is given
on how a $D$-brane configuration can on one hand yield a
supersymmetric gauge theory in a weakly coupled type II superstring
theory and on the other hand  behave as a classical solution to type
II supergravity in a strongly coupled type II superstring.  We
emphatically point out  the $D$-brane configuration comprised of $N$
$D3$-branes and $M$ fractional $D3$-branes in the traget space-time
$M^4\times {\cal C}^6$, here ${\cal C}^6$ is the six-dimensional
conifold with the base $T^{1,1}=\left[SU(2)\times
SU(2)\right]/U(1)$. This $D$-brane configuration in the field theory
side gives the ${\cal N}=1$ $SU(N+M)\times SU(N)$ supersymmetric
gauge theory with four matter fields in the bi-fundamental
representation $(N+M,\overline{N})$ and $(\overline{N+M},N)$,
respectively, of gauge groups, and in the gravity side produces the
celebrated Klebanov--Strassler solution to type IIB supergravity. In
Sect.\,VI, we consider the Dirac-Born-Infeld action and the
Wess-Zumino term of describing the low-energy dynamics of $D$-branes
and show that the internal superconformal anomaly originates from
the physical effects of fractional $D$-branes in the brane
configuration. Then we use the feature of $D$-brane in strongly
coupled type II superstring theory to argue that these fractional
branes deform the three-brane solution to type IIB supergravity
whose near horizon limit is $AdS_5\times T^{1,1}$ and convert this
solution into the K-S solution. Further, by comparing  the
symmetries reflected in $AdS_5\times T^{1,1}$ and K-S solution
background geometries, we find how space-time supersymmetries
symmetry decrease due to fractional branes. Sect.\,VII shows how the
dual of internal superconformal anomaly can be considered as a
spontaneous breaking of local suersymmetry and the consequent
super-Higgs mechanism in gauged asymptotical $AdS_5$ supergravity.
We choose the K-S solution as a classical vacuum configuration   and
observe the dynamical behavior of type IIB supergravity. We realize
that the spontaneous compactification of type IIB supergravity on
the deformed $T^{1,1}$ should occur and the resultant theory should
be certain gauged supergravity in five dimensions. In contrast to
the case of type IIB supergravity in the $AdS_5\times T^{1,1}$
space-time background, which gives rise to $N=2$ gauged $AdS_5$
supergravity coupled with $SU(2)\times SU(2)$ supersymmetric
Yang-Mills theory and some Betti multiplets, we find that this
gauged five-dimensional supergravity should be the gauged $AdS_5$
supergravity but with spontaneous breaking of local ${\cal N}=2$
supersymmetry to ${\cal N}=1$. We further exhibit the consequent
super-Higgs mechanism through which the ${\cal N}=2$ graviton
multiplet acquires mass. Finally Sect.\,VIII is a summary and
emphasis on the significance of distinguishing the gravitational
correspondences of these two types of superconformal anomalies.

\section{Two types superconformal anomalies of
supersymmetric gauge theory}

\subsection{Supercurrent and Superconformal Anomaly}

In a supersymmetric field theory, the Poincar\'{e} supersymmetry
algebra makes the energy-momentum tensor $\theta^{\mu\nu}$,
supersymmetry current $s^{\mu}$ and  chiral R-current $j^{\mu}$
constitute a supermultiplet \footnote{Here and in the following, the
possible spinorial and chiral $R$-symmetry group indices in
$s^{\mu}$ and $j^{\mu}$ are suppressed  in the supersymmetry
currents in the supercurrent.}. Further, if the theory contains no
parameter with mass dimension, these currents at classical level are
not only conserved,
\begin{eqnarray}
\partial_\mu \theta^{\mu\nu}=\partial_\mu s^\mu=\partial_\mu
j^{\mu}=0,
\end{eqnarray}
but also satisfy further algebraic relations,
\begin{eqnarray}
\theta^{\mu}_{~\mu}=\gamma_\mu s^\mu=0. \label{superconf}
\end{eqnarray}
 We can use these relations to construct three more conserved
currents,
\begin{eqnarray}
&& d^\mu {\equiv} x_\nu \theta^{\nu\mu}, ~ k_{\mu\nu}{\equiv} 2
x_\nu x^\rho\theta_{\rho\mu}-x^2\theta_{\mu\nu}, ~ l_\mu{\equiv}
ix^\nu \gamma_\nu s_\mu; \nonumber\\
&& \partial_\mu d^\mu= \partial_\mu k^{\mu\nu}=\partial_\mu l^\mu=0.
\end{eqnarray}
These three new conserved currents gives rise to the generators for
dilatation, conformal boosts and special supersymmetry,
respectively. Consequently, the Poincar\'{e} supersymmetry is
promoted to the superconformal symmetry.

 However,  the superconformal symmetry at quantum level may become
 anomalous. In the case that all of them, the trace $\theta^\mu_{~\mu}$ of
 energy-momentum tensor, the $\gamma$-trace $\gamma^\mu j_\mu$
 of supersymmetry current  and the divergence $\partial_\mu j^{\mu}$
 of chiral $R$-current receive
 contribution from quantum effects,
\begin{eqnarray}
\left(\partial_\mu j^{\mu},\gamma^\mu s_\mu,
\theta^{\mu}_{~\mu}\right) \label{cas}
\end{eqnarray}
  shall form a (on-shell) chiral supermultiplet with the $\partial_\mu
j^{\mu}$ playing the role of the lowest
componentt\cite{ilio}\footnote{In case that
  $\theta^{\mu}_{~\mu}{\neq}0,~~\gamma^\mu j_\mu{\neq}0$,
whereas  the chiral $R$-currents keeps conserved, $\partial_\mu
j^{(5)\mu}=0$, there will arise a linear anomaly multiplet
$(\theta^{\mu}_{~\mu},\gamma^\mu s_\mu, t_{\mu\nu})$. In this
multiplet, the antisymmetric field $t_{\mu\nu}$ satisfies
$\partial^\nu t_{\nu\mu}=0$.  It appears in both multiplets of
currents and anolies.  But $t_{\mu\nu}$  does not  yield a
conservative charge for supersymmetry algera. One can get other
types of anomaly multiplets by considering other possibilities.}.

There are usually two possible sources for above
 chiral supermultiplet anomaly. One is that the supersymmetric gauge theory
 couples with an external conformal supergravity background composed of
 the gravitational supermultiplet ($g_{\mu\nu}$,$\psi_\mu$,$A_\mu$).
  Note that for a supersymmetric gauge theory in four-dimensional Minkowski
 space-time, the space-time symmetries which include
 the Poincar\'{e} symmetry, supersymmetry and chiral $R$-symmetry are all
 global ones and there are no gauge fields within the theory itself to
 couple with the supercurrent multiplet
 ($\theta_{\mu\nu}$, $s_\mu$, $j_\mu$). Once we consider
 the theory in a curved space-time background,  the
  superconformal anomaly (\ref{cas}) will arise if the field
  configuration of the
  external conformal supergravity multiplet
  ($g_{\mu\nu}$,$\psi_\mu$,$A_\mu$) has non-trivial topology.

 The other type of superconformal anomaly supermultiplet
 originates  from the dynamics of a less supersymmetric (${\cal N}
 <4$) gauge theory itself and has nothing to do with external conformal
 supergravity background.
 Eq.\,(\ref{cas}) in this case is actually a scale anomaly
  supermultiplet. In an ${\cal N}<4$ supersymmetric gauge theory,
    an energy scale is generated dynamically due to the renormalization
 effect. Consequently,  the scale anomaly and the corresponding multiplet
required by supersymmetry must arise. To distinguish this scale
anomaly supermutiplet with the one orginating from  external
supergravity fields, we call this anomaly chiral supermultiplet
 as the internal superconformal anomaly, while the former one as
external superconformal anomaly. It should be emphasized that the
internal
 anomaly has distinct features with the external one. The
 coefficient of
 internal superconformal anomaly depends
 on the beta function of the theory. For
 a superconformal quantum gauge theory such as ${\cal N}=4$
 supersymmetric Yang-Mills theory, its beta function vanishes
  and there is no internal superconfromal anomaly, but the
  external one persists.

 In the following we list the external and internal superconformal
 anomalies in several typical supersymmetric gauge theories.

\subsection{Superconformal Anomaly and Current Supermutiplet in
  ${\cal N}=1$ Supersymmetric Yang-Mills Theory}

A four-dimensional supersymmetric $SU(N)$ Yang-Mills theory consists
of a  vector field $V_\mu$, a Majorana spinor $\lambda$ and  an
auxiliary $D$. All of them are in the adjoint representation of
$SU(N)$. The classical Lagrangian density reads
\begin{eqnarray}
{\cal L}=-\frac{1}{4}G_{\mu\nu}^a G^{\mu\nu a}+\frac{1}{2}
i\overline{\lambda}^a\gamma^\mu (\nabla_\mu \lambda)^a+\frac{1}{2}
(D^a)^2.
\end{eqnarray}
The current multiplet $(j_\mu,s_\mu,\theta_{\mu\nu})$ can be easily
derived from the the chiral $U_R(1)$ symmetry, supersymmetry and
Poincar\'{e} symmetry with the N\"{o}ther theorem. The internal
anomaly chiral multiplet was obtained long time ago with an explicit
perturbative calculation \cite{salam,abb,grisaru},
\begin{eqnarray}
\partial_\mu j^{\mu} &=& 2\frac{\beta
(g)}{g}
\left[\frac{1}{12}G^a_{\mu\nu}\widetilde{G}^{a\mu\nu}+\frac{1}{6}
\partial^\mu \left( \overline{\lambda}^a\gamma_\mu\gamma_5\lambda^a\right)
\right],\nonumber\\
 \gamma^\mu s_\mu &=& 2\frac{\beta
(g)}{g}\left(-\frac{1}{2} \sigma^{\mu\nu}G^a_{\mu\nu}+\gamma_5
D^a\right)\lambda^a,
\nonumber\\
\theta^\mu_{~\mu} &=& 2\frac{\beta
(g)}{g}\left[-\frac{1}{4}G_{\mu\nu}^a G^{\mu\nu a}+\frac{1}{2}
i\overline{\lambda}^a\gamma^\mu (\nabla_\mu \lambda)^a+\frac{1}{2}
(D^a)^2\right], \nonumber\\
P &=& -\beta (g)\overline{\lambda}^a\lambda^a, ~~Q=-\beta (g)
\overline{\lambda}^a\gamma_5\lambda^a. \label{intanoma}
\end{eqnarray}
It is obvious that the internal superconformal anomaly coefficient
is proportional to the beta function of ${\cal N}=1$ supersymmetric
Yang-Mills theory. The origins of scale anomaly $\theta^\mu_{~\mu}$
and chiral anomaly are the same as in a usual field theory: The UV
divergence requires  the process of  performing renormalization to
make the theory well defined and hence need to introduces a
dynamical generated energy scale, the scale anomaly thus arises; The
acquirement of chiral $U_R(1)$ anomaly $\partial_\mu j^\mu$ is
identical to the derivation on axial $U_A(1)$ anomaly in QCD. This
$U_R(1)$ symmetry is not compatible with the global vector $U_V(1)$
rotational symmetry among gauginoes at quantum level. When the
vector $U_V(1)$ symmetry is preserved, the $U_R(1)$ symmetry becomes
anomalous. The production of $\gamma$ trace anomaly $\gamma^\mu
s_\mu$ lies in the incompatibility of the conservation of
supersymmetry current and non-vanishing $\gamma$-trace of
supercurrent $s_\mu$ at quantum level. Since the supersymmetry
current should keep conserved, the $\gamma$-trace anomaly thus
emerges.

 In the following we turn to the external superconformal anomaly multiplet.
As stated above, the currents must couple with an external
supergravity multiplet in the following form,
\begin{eqnarray}
{\cal L}_{\rm ext}=\int d^4x \sqrt{-g} \left(\theta^{\mu\nu}
g_{\mu\nu}+ B_\mu j^{\mu}+\overline{\psi}_\mu s^{\mu}\right).
\label{excoupl}
\end{eqnarray}
The above Lagrangian density for a supersymmetric Yang-Mills theory
in an supergravity  background means that the covariant
conservations of energy-momentum tensor, supersymmetry current and
chiral $R$-symmetry current, $\nabla_\mu \theta^{\mu\nu}=\nabla_\mu
s^\mu=0$, are equivalent to the diffeormorphism tansformation
invariance, local supersymmetry and gauge symmetry of the external
supergravity, respectively,
\begin{eqnarray}
\delta g_{\mu\nu} (x) &=& \nabla_\mu \xi_\nu (x)+\nabla_\nu
\xi_\mu (x),\nonumber\\
\delta\psi_\mu (x)&=& \nabla_\mu \chi (x), \nonumber\\
\delta B_\mu (x) &=& \nabla_\mu \Lambda (x).
 \label{egt1}
\end{eqnarray}
Further, the vanishing of both $\gamma$-trace of supercurrent and
trace of energy-momentum tensor, $\gamma^\mu s_\mu
=\theta^{\mu}_{~\mu}=0$, implies  the Weyl- and super-Weyl
symmetries in the external supergravity,
\begin{eqnarray}
\delta g_{\mu\nu}&=&g_{\mu\nu} \sigma (x),\nonumber\\
\delta \psi_\mu &=& \gamma_\mu \eta (x), \label{busy}
\end{eqnarray}
These symmetries show that the external supergravity must be ${\cal
N}=1$ conformal supergravity in four dimensions \cite{kaku}. Later
we will see explicitly that in the context of $AdS/CFT$
correspondence or generally gauge/gravity dual, this ${\cal N}=1$
external conformal supergravity in four dimensions comes from the
$AdS_5$ boundary value of ${\cal N}=2$ gauged $AdS_5$ supergravity
multiplet and the ${\cal N}=1$ local superconformal symmetry in this
conformal supergravity originates from the $AdS_5$ boundary
reduction of the ${\cal N}=2$ bulk supergravity.
 The external superconformal anomaly in ${\cal N}=1$ supersymmetric
Yang-Mills theory will be reflected in the explicit violations of
the bulk symmetries of ${\cal N}=2$ gauged $AdS_5$ supergravity on
the boundary of $AdS_5$ space.

With no consideration the quantum correction from the dynamics of
supersymmetric gauge theory, the external anomaly is exhausted at
one-loop level. The external superconformal anomaly multiplet for
${\cal N}=1$ pure supersymmetric Yang-Mills theory is the following
\cite{grisaru},
\begin{eqnarray}
\nabla_\mu j^{\mu} &=&\frac{N^2-1}{16\pi^2}
\left(K_{\mu\nu}\widetilde{K}^{\mu\nu} -\frac{1}{24}
\widetilde{R}_{\mu\nu\lambda\rho}R^{\mu\nu\lambda\rho}\right)
\nonumber\\
\gamma^\mu s_\mu &=&\frac{N^2-1}{16\pi^2} \left(\frac{1}{16}
 R^{\mu\nu\lambda\rho}\sigma_{\lambda\rho}+\frac{1}{16}K^{\mu\nu}\right)
 \left(\overline{\nabla}_\mu \psi_\nu-
 \overline{\nabla}_\nu \psi_\mu\right),
\nonumber\\
\theta^\mu_{~\mu} &=&
\frac{N^2-1}{16\pi^2}\left(\frac{1}{8}C_{\mu\nu\lambda\rho}
C^{\mu\nu\lambda\rho}- \frac{3}{16}
\widetilde{R}_{\mu\nu\lambda\rho}\widetilde{R}^{\mu\nu\lambda\rho}
+\frac{1}{3}K^{\mu\nu}K_{\mu\nu}\right), \label{exone}
\end{eqnarray}
In above equation, $K_{\mu\nu}=\partial_\mu B_\nu-\partial_\nu
B_\mu$ is the field strength of external $U_R(1)$ vector field
$B_\mu$; $R_{\mu\nu\lambda\rho}$ and $C_{\mu\nu\lambda\rho}$  are
Riemannian and Weyl tensors corresponding to  gravitational
background $g_{\mu\nu}$; $\overline{\nabla}_\mu$  is the covariant
derivative with respect to both gravitational and $U_R(1)$ fields.
The common factor $N^2-1$ in the anomaly coefficients is due to the
fact that the gaugino is in the adjoint representation of $SU(N)$
gauge group.

In contrast to the internal superconformal anomaly, the external one
listed in (\ref{exone}) comes only from one-loop contribution of the
theory and the anomaly coefficient is proportional to the number of
gauge particles in supersymmetric Yang-Mills theory \cite{duff1}.
When we consider quantum dynamics of ${\cal N}=1$ supersymmetric
Yang-Mills theory, the anomaly coefficient receives quantum
correction \cite{duff2}. In particular, the internal superconformal
anomaly arises as a correction to the external one.

Let us analyze the manifestation of internal superconformal anomaly
(\ref{intanoma}) in the quantum effective action of ${\cal N}=1$
supersymmetric Yang-Mills theory. First, we re-scale the fields
 $(A_\mu,\lambda,D)\rightarrow (A_\mu/g, \lambda/g,D/g)$
 and consider the strong CP violation term. Consequently,
 the classical action of ${\cal N}=1$ supersymmetric $SU(N)$
 Yang-Mills theory can be re-written as the following,
\begin{eqnarray}
 S_{\rm cl}= \frac{1}{g^2}\int d^4x\left[-\frac{1}{4}G_{\mu\nu}^a G^{\mu\nu
a}+\frac{1}{2} i\overline{\lambda}^a\gamma^\mu (\nabla_\mu
\lambda)^a + \frac{1}{2} \left(D^a\right)^2\right]
 +\frac{\theta^{\rm (CP)}}{32\pi^2}\int d^4x
 G_{\mu\nu}^a\widetilde{G}^{\mu\nu a}.
 \label{claco}
\end{eqnarray}
The local part of the gauge invariant quantum effective action takes
the form,
\begin{eqnarray}
\Gamma_{\rm eff}=\frac{1}{g^2_{\rm eff}}\int
d^4x\left[-\frac{1}{4}G_{\mu\nu}^a G^{\mu\nu a}+\frac{1}{2}
i\overline{\lambda}^a\gamma^\mu (\nabla_\mu \lambda)^a + \frac{1}{2}
\left(D^a\right)^2\right]
 +\frac{\theta^{\rm (CP)}_{\rm eff}}{32\pi^2}\int d^4x
 G_{\mu\nu}^a\widetilde{G}^{\mu\nu a},
 \label{quanac}
\end{eqnarray}
where all the fields are renormalized quantities. The scale- and
chiral anomalies are reflected in the running of gauge coupling and
the shift of $\theta$-angle due to the non-vanishing $\beta (g)$
\cite{pequ},
\begin{eqnarray}
\frac{1}{g^2_{\rm
eff}(q^2)}&=&\frac{1}{g^2(q_0^2)}+\frac{3N}{8\pi^2}\ln \frac{q_0}{q}
\equiv
\frac{ 3N}{8\pi^2}\ln \frac{q}{\Lambda}, \nonumber\\
\theta^{\rm (CP)}_{\rm eff} &=& \theta^{\rm (CP)}+3N. \label{effcou}
\end{eqnarray}

To recognize  the $\gamma$-trace anomaly of supersymmetry current in
above gauge invariant quantum effective action, we must resort to
the superfield form of quantum effective action (\ref{quanac})
expressed in the superspace,
\begin{eqnarray}
{\cal L}=\frac{1}{32\pi} \int d^2\theta\, \mbox{Im}\left[\tau
\mbox{Tr} \left(W^\alpha W_\alpha\right) \right] +\mbox{h.c.}.
\end{eqnarray}
In above equation, the parameter $\tau$ is a complex combination of
gauge coupling and  CP-violated $\theta$-angle,
\begin{eqnarray}
\tau=\frac{\theta^{\rm (CP)}}{2\pi} +i\frac{4\pi}{g^2}.
\end{eqnarray}
$W_\alpha$ is the field strength corresponding to
  vector superfield  $V$,
\begin{eqnarray}
 W_\alpha =\frac{1}{g}\overline{D}^2\left[e^{-gV}
 \left( D_\alpha e^{gV}\right)\right].
\end{eqnarray}
In the Wess-Zumino gauge,
\begin{eqnarray}
V&=&-\theta\sigma^\mu \overline{\theta} A_\mu +i
\overline{\theta}^2\theta\lambda-i\theta^2\overline{\theta}
\overline{\lambda}-\frac{1}{2}\theta^2\overline{\theta}^2
D,\nonumber\\
W_\alpha  &=&\overline{D}^2\left( D_\alpha V-\frac{g}{2} [V,
D_\alpha V] \right)\nonumber\\
 &=&\lambda_\alpha-\frac{i}{2}\left(\sigma^{\mu\nu}
\right)_{\alpha\beta}\theta^\beta F_{\mu\nu}+i\theta_\alpha
D-i\theta^2\left(\sigma^\mu\right)_{\alpha\dot{\alpha}}
 \nabla \overline{\lambda}^{\dot{\alpha}},
\end{eqnarray}

We assume that $\tau$ is the lowest component of certain constant
chiral superfield,
\begin{eqnarray}
\Sigma \equiv \tau+\theta \chi +\theta^2 d
\end{eqnarray}
Further, we define that classically  $\chi_{\rm cl}=d_{\rm cl}=0$
and only at quantum level $\chi$ and $d$ get non-vanishing
expectation values,
\begin{eqnarray}
\Sigma_{\rm eff} &\equiv& \tau_{\rm eff}+\theta \chi_{\rm eff}
+\theta^2 d_{\rm eff}, \label{supercou}
\end{eqnarray}
where $\theta^{\rm (CP)}_{\rm eff}$ and $g^2_{\rm eff}$ are listed
in (\ref{effcou}).
 In the Wess-Zumino gauge, the generalized quantum effective action
 in superspace takes the form,
\begin{eqnarray}
{\Gamma}_{\rm eff}&=& \frac{1}{32\pi}\int d^2\theta\,
\mbox{Im}\left[\Sigma_{\rm eff} \mbox{Tr} \left(W^\alpha
W_\alpha\right) \right]+\mbox{h.c.}
 \nonumber\\
&=&\frac{1}{32\pi}\int d^2\theta\,
\mbox{Im}\left\{\left(\frac{\theta^{\rm (CP)}_{\rm eff}}{2\pi}
+\frac{4i\pi}{g^2_{\rm eff}} +\theta \chi_{\rm eff} +\theta^2 d_{\rm
eff} \right) \left[ \lambda\lambda +i\theta \left(2\lambda
D+\sigma^{\mu\nu}\lambda G_{\mu\nu}\right)
\right.\right.\nonumber\\
&&\left.\left. +\theta^2 \left(-\frac{1}{2}G_{\mu\nu}G^{\mu\nu}+
\frac{i}{2}G_{\mu\nu}\widetilde{G}^{\mu\nu}-2i \lambda \sigma^\mu
\nabla_\mu \overline{\lambda} -D^2 \right)\right]
\right\}+\mbox{h.c.}. \label{superac}
\end{eqnarray}
 A comparison between  Eq.\,(\ref{superac}) and the superconformal
anomaly equation (\ref{intanoma}) reveals the presence of fermionic
anomaly in the quantum effective action.
 The $\gamma$-trace anomaly of supersymmetry current due to the
non-vanishing $\beta$-function is  reflected in the shift of the
superpartner  parameter $\chi$, exactly like trace- and chiral
anomalies are represented by the shifts of the gauge couplings and
the $\theta$-angle. This is the advantage of promoting $\tau$
parameter to a chiral superfield. Of course, this should be a
natural conclusion since the superconformal anomaly constitutes a
supermultiplet. However, later we will see  that from the viewpoint
of brane dynamics, the running of gauge coupling and the shift of
$\theta$ angle originate from fractional branes and further get down
to the Goldstone fields. Therefore, introducing an artificial
superpartner for the gauge coupling and $\theta$-angle
 is very helpful for us to identify the Goldstone supermultiplet on
 supergravity side and find the super-Higgs mechanism due to the
 internal
 superconformal anomaly.

\subsection{Supercurrent in ${\cal N}=2$ Supersymmetric $SU(N)$
Yang-Mills Theory}

The field content of ${\cal N}=2$ pure supersymmetric Yang-Mills
theory consists of a vector field $V_\mu$, a complex scalar
$\phi{\equiv} (A+iB)/2$ and a $SU_I(2)$ symplectic Majorana spinor
$\lambda^i=\epsilon^{ij}i\gamma_5 C\overline{\lambda}^T_j$,
\begin{eqnarray}
\lambda^i=\left(\begin{array}{c}-i\epsilon^{ij}\lambda_{\alpha j}\nonumber\\
\overline{\lambda}^{\dot{\alpha} i}\end{array}\right), ~~~
\overline{\lambda}_i=\left(\lambda_i^\alpha, i\epsilon_{ij}
\overline{\lambda}^j_{\dot{\alpha}}\right)
\end{eqnarray}
$i,j=1,2$ being the $SU_I(2)$ indices and $C$ being the charge
conjugation matrix. All of these fields are in the adjoint
representation of gauge group $SU(N)$. The Lagrangian density of
this theory reads
\begin{eqnarray}
{\cal L}&=& -\frac{1}{4} F_{\mu\nu}^a F^{a\mu\nu}
+\frac{1}{2}i\overline{\lambda}_i^a\gamma^\mu D_\mu \lambda^{ai}
+\frac{1}{2}\left(D_\mu A\right)^a\left(D^\mu
A\right)^a+\frac{1}{2}\left(D_\mu B\right)^a\left(D^\mu
B\right)^a\nonumber\\
&& +\frac{i}{2}gf^{abc} \overline{\lambda}_i^a (A^b+i\gamma_5
B^b)\lambda^{ci}+\frac{1}{2}\left([A,B]^a\right)^2.
\end{eqnarray}
Note that now the $R$-symmetry is $U_R(2)=SU_I(2)\times U_R(1)$. The
$SU_I(2)$ transformation rotates only the $\lambda$ doublet,
\begin{eqnarray}
\delta \lambda^i=i\theta^A \left(t^A\right)^i_{~j} \lambda^j, ~~~
\delta \overline{\lambda}_i =-i\overline{\lambda}_j
\left(t^A\right)^{j}_{~i} \theta^A, \label{rota1}
\end{eqnarray}
while the chiral $U_R(1)$ acts both of the spinor and the complex
scalar fields,
\begin{eqnarray}
&& \delta \phi (x)= -2i\alpha \phi (x), ~~~\delta \phi^\star
=2i\alpha \phi^\star, \nonumber\\
&& \delta \lambda^i (x) =i\alpha\gamma_5 \lambda^i (x), ~~~
\delta\overline{\lambda}_i=i\alpha \overline{\lambda}_i\gamma_5.
\label{rota2}
\end{eqnarray}

Let us write down the current supermultiplet for ${\cal N}=2$
supersymmetric Yang-Mills theory, which was actually derived
long-time ago in the form of Weyl spinor \cite{fisher,dewit}. First,
the improved energy-momentum tensor reads \cite{fisher,dewit},
\begin{eqnarray}
\theta_{\mu\nu}&=&-F_{\mu\lambda}^aF_{\nu}^{a\lambda}
+\frac{1}{4}\eta_{\mu\nu} F_{\lambda\rho}^aF^{a\lambda\rho}\nonumber\\
&+&\frac{1}{2}i\left[\overline{\lambda}_i^a (\gamma_\mu D_\nu
+\gamma_\nu D_\mu)\lambda^{ia}- (D_\mu
\overline{\lambda}_i^a\gamma_\nu +D_\nu
\overline{\lambda}_i^a\gamma_\mu) \lambda^{ai}\right]-\frac{1}{4}i
\eta_{\mu\nu} \left(\overline{\lambda}_i^a \gamma^\lambda
\nabla_\lambda^{ai}
-\nabla_\lambda\overline{\lambda}_i^a\gamma^\lambda
\lambda^{ai}\right)
\nonumber\\
&+&\frac{2}{3}\left(D_\mu A^{a} D_\nu A^a +D_\mu B^{a} D_\nu
B^a\right) -\frac{1}{6}\eta_{\mu\nu} \left(D_\lambda A^{a} D^\lambda
A^a + D_\lambda B^{a} D^\lambda B^a\right)
\nonumber\\
&-&\frac{1}{3}\left(A^{a} D_\mu D_\nu A^a+ B^a D_\mu D_\nu
B^{a}\right)
 +\frac{1}{12} \eta_{\mu\nu} \left(A^{a} D_\lambda D^\lambda A^a+
B^a D_\lambda D^\lambda B^{a}\right).
\end{eqnarray}
The improved supersymmetry current $s_{\mu \alpha}^i$ can be
obtained from ${\cal N}=2$ supersymmetry transformation but with the
modification of adding some total derivative terms,
\begin{eqnarray}
s_\mu^{i} &=& -\sigma_{\nu\rho} \gamma_\mu \lambda^{ia}F^{a\nu\rho}
+2i(A^a-i\gamma_5 B^a) D_\mu  \lambda^{ia}
\nonumber\\
&&+\frac{i}{2}(A^a-i\gamma_5 B^a)\gamma_\mu \gamma^\nu D_\nu
\lambda^{ai} -\frac{2}{3}\sigma^{\mu\nu}\partial_\nu\left[
(A^a+i\gamma_5 B^a)\lambda^{ia}\right],
\end{eqnarray}
which is a $SU(2)_I$ doublet.  Both the energy-momentum tensor and
supersymmetry current satisfy the superconformal condition listed in
Eq.\,(\ref{superconf}).

The  $SU_I(2)$ current $j_{\mu}^A$ and chiral $U_R(1)$ current
$j_\mu$, can be read out from rotations (\ref{rota1}) and
(\ref{rota2}), respectively,
\begin{eqnarray}
j_\mu^A &=& \frac{1}{2}\overline{\lambda}_i^a\gamma_\mu
\left(t^A\right)^i_{~j}
\lambda^{aj},~~~~A=1,2,3\nonumber\\
j_\mu
&=&\frac{1}{2}\overline{\lambda}^a_i\gamma_\mu\gamma_5\lambda^{ia}
+2i \left[\phi^{\star a} \left( D_\mu\phi\right)^a -\left(D_\mu
\phi^\star\right)^a\phi^a\right]
\nonumber\\
&=& \frac{1}{2}\overline{\lambda}^a_i\gamma_\mu\gamma_5\lambda^{ia}
+A^a D_\mu B^a - B^a D_\mu A^a.
\end{eqnarray}
So we have following $24+24$-component current supermultiplet,
\begin{eqnarray}
(\theta_{\mu\nu}, s_{\mu\alpha}^i, j_\mu^A,j_\mu, t_{\mu\nu},
\varsigma^i,D), \label{ntwoc}
\end{eqnarray}
where
\begin{eqnarray}
\varsigma^i &=& \frac{i}{2}\left(A^a+i\gamma_5 B^a\right)\lambda^{ia},
\nonumber\\
t_{\mu\nu} &=&\frac{1}{2}\overline{\lambda}^{a}_i
\sigma_{\mu\nu}{\lambda}^{ai} +i\phi^a
F_{\mu\nu}^{a-},~~~~F_{\mu\nu}^{-}=\frac{1}{2}
\left(F_{\mu\nu}-\widetilde{F}_{\mu\nu}\right),\nonumber\\
D &=&
\phi^{a\star}\phi^a =\frac{1}{4} \left(A^{a2}+B^{a2}\right).
\end{eqnarray}

Similar to ${\cal N}=1$  case, one can discuss the superconformal
anomaly of above current supermultiplet. It is well known that the
${\cal N}=2$ supersymmetric Yang-Mills theory has non-vanishing beta
function at only one-loop \cite{howe}, so the internal
superconformal anomaly arises only at the first order of the
perturbative expansion. The trace $\theta^\mu_{~\mu}$ of the
energy-momentum tensor $\theta_{\mu\nu}$, the $\gamma$-trace of
supersymmetry current $s^i_\mu$ and the divergence of $U(1)_R$
current $j^\mu$ form an anomaly supermutiplet dominated by $N=2$
Poincar\'{e} supersymmetry. The structure of this anomaly is similar
to the ${\cal N}=1$ supersymmetric Yang-Mills case \cite{marcu}. It
should be mentioned that $SU(2)_R$ is non-anomalous since it is a
vector-like current and is responsible for the global $SU(2)_R$
rotation symmetry of ${\cal N}=2$ Poincar\'{e} supersymmetry after
the superconformal symmetry breaks down.

The external superconformal anomaly multiplet arises if the external
${\cal N}=2$ conformal supergravity fields ($g_{\mu\nu}$,
$\psi_\mu^i$, $V^A_\mu$, $B_\mu$, $B_{\mu\nu}$, $X$, $\chi^i$ )
couple with the off-shell supercurrent (\ref{ntwoc}). Like the
${\cal N}=1$ case, the classical superconformal symmetry of ${\cal
N}=2$ supersymmetric Yang-Mills theory correspond to the ${\cal
N}=2$ superconformal transformation invariance of the external
conformal supergravity.  The structure of this ${\cal N}=2$ external
superconformal anomaly is similar to Eq.\,(\ref{exone}) of the
${\cal N}=1$ case.



\subsection{Current Supermultiplet in
${\cal N}=4$ Supersymmetric Yang-Mills Theory}

  ${\cal N}=4$  $SU(N)$ supersymmetric
 Yang-Mills theory in four dimensions is a non-trivial
 superconformal quantum
 gauge theory since its $\beta$-function vanishes identically
 \cite{sohn}.
Its R-symmetry is the $SU_R(4)$  rather than
 $U(4)$. The field content consists of $SU(N)$ gauge fields
 $A_\mu$, a Majorana spinor $\lambda_i$ in the
 fundamental representation of $SU_R(4)$, and  complex scalar fields
 $\phi_{ij}{\equiv}1/2 (A_{ij}+iB_{ij})$ in the six-dimensional
 representation  of $SU_R(4)$
 and subject to an $SU_R(4)$ covariant reality constraint
 $(\phi_{ij})^*=1/2
 \epsilon^{ijkl}\phi_{kl}{\equiv}\phi^{ij}$.
 To manifest the $SU_R(4)$ symmetry in the classical
 action, it is necessary
 not to insisting on the symplectic-Majorana feature of $\lambda_i$. Instead
 we choose the chiral component $\lambda_{\alpha i}$ and define
 $\overline{\lambda}^i_{\dot{\alpha}}
 \equiv \left(\lambda_{\alpha i}\right)^\dagger$ so
 that $\lambda_{\alpha i}$
 and $\overline{\lambda}^i_{\dot{\alpha}}$ constitute a four-component
 Dirac spinor $\lambda_i$. Correspondingly, The classical Lagrangian density
 of ${\cal N}=4$ supersymmetric
 Yang-Mills theory is
 \cite{sohn}
 \begin{eqnarray}
{\cal L} &=& -\frac{1}{4}F_{\mu\nu}^a
F^{a\mu\nu}+i\overline{\lambda}^{ai}\gamma^{\mu} D_\mu\lambda^{a}_i
+ \frac{1}{2}D_\mu \phi^{ij a} D^\mu \phi^{a}_{ij}
\nonumber\\
&& + \frac{1}{2}f^{abc} \left[\overline{\lambda}^{ia}(1+\gamma_5)
\lambda_{j}^{b} \phi_{ij}^c-\overline{\lambda}_i^{a}(1-\gamma_5)
{\lambda}_j^b \phi^{cij}\right] +\frac{1}{4}[\phi_{ij},
\phi_{kl}][\phi^{ij}, \phi^{kl}]
 \end{eqnarray}
 The members of current multiplet contains not only  energy-momentum
 tensor $\theta_{\mu\nu}$, supersymmetry current $s_\mu^i$
 and chiral $SU_R(4)$
 current $j_{\mu~j}^{~i}$ \cite{bergshoe},
\begin{eqnarray}
\theta_{\mu\nu} &=& \frac{1}{4}\eta_{\mu\nu} F_{\lambda\rho}^a
F^{a\lambda\rho}- F^a_{\mu\lambda}F_{\nu}^{a\lambda}\nonumber\\
&+&\frac{1}{2}i\left[\overline{\lambda}^{ia}(\gamma_\mu D_\nu
+\gamma_\nu D_\mu)\lambda_{i}^{a}-
\left(D_\mu\overline{\lambda}^{ia}\gamma_\nu +
D_\nu\overline{\lambda}^{ia}\gamma_\mu
\right)\lambda_{i}^a\right]-\frac{1}{4}\eta_{\mu\nu}
\left(\overline{\lambda}^{ia}\gamma_\rho D_\mu-D_\lambda
\overline{\lambda}^{ia}\gamma^\rho\right)\lambda_{i}^a
\nonumber\\
& & -\eta_{\mu\nu} D_\rho\phi^{aij} D_\rho\phi^a_{ij} +2
D_\mu\phi^{aij} D_\nu\phi^a_{ij} -\frac{1}{3}\left(\partial^2
\eta_{\mu\nu}-\partial_\mu\partial_\nu\right)\left( \phi^{ij}
\phi_{ij}\right),\nonumber\\
s_{\mu i} &=&-\frac{1}{2}\sigma^{\nu\rho} F_{\nu\rho}^a\gamma_\mu
\left(1+\gamma_5\right)
\lambda_i^a\nonumber\\
&&+i\left(1-\gamma_5\right)\left(\phi_{ij}^a D_\mu\lambda^{aj}-
D_\mu\phi_{ij}^a\lambda^{aj}\right)+\frac{2}{3}i
\left(1-\gamma_5\right) \sigma_{\mu\nu}
\partial^\nu \left(\phi_{ij}^a\lambda^{aj}\right),\nonumber\\
j_{\mu~j}^{~i} &=& j_\mu^A \left(t^A\right)^i_{~j}= \phi^{aik} D_\mu
\phi^a_{kj}-\left(D_\mu \phi^{aik}\right) \phi_{kj}
\nonumber\\
&& +\frac{1}{2}\overline{\lambda}^{ai}\gamma_\mu
(1-\gamma_5)\lambda_j^a -\frac{1}{8}\delta^i_{~j}
\overline{\lambda}^{ak}\gamma_\mu (1-\gamma_5)\lambda^a_k,
\end{eqnarray}
but also the following compensating quantities,
\begin{eqnarray}
{\cal C}&=&F_{\mu\nu}^{a-}F^{\mu\nu{a-}},
~~~~F^{a\pm}_{\mu\nu}=\frac{1}{2} \left(F_{\mu\nu}^a\pm
\frac{1}{2}\epsilon_{\mu\nu\lambda\rho}
F^{a\lambda\rho}\right),\nonumber\\
\chi_i &=& \frac{1}{2}\sigma_{\mu\nu} (1-\gamma_5)\lambda^{Ca}_i
F^{a\mu\nu},
\nonumber\\
 t_{ij} &=& \frac{1}{2}
  \overline{\lambda}_i^a \left(1-\gamma_5\right)\lambda^{a}_j,\nonumber\\
  t_{\mu\nu}^{~~ij}&=&\frac{1}{2}\overline{\lambda}^{ai}
  \left(1-\gamma_5\right) \sigma_{\mu\nu}\lambda^{a^j}
  +2i \phi^{aij}F_{\mu\nu}^{a-},\nonumber\\
  \tau^{ij}_{~~k} &=& \frac{1}{4}(1-\gamma_5)\epsilon^{ijmn}\left(
  \phi_{mn}^a \lambda^{a}_k+\phi_{kn}^a\lambda^{a}_m\right),\nonumber\\
  {\cal D}^{ij}_{~~kl} &=& \phi^{aij}\phi^a_{kl}-\frac{1}{12}
  \left(\delta^i_{~k}\delta^j_{~l}-\delta^i_{~l}\delta^j_{~k}
   \right)\phi_{mn}^a\phi^{amn},
\end{eqnarray}
where $A=1, \cdots, 15$ denote  $SU_R(4)$ group generators and
$i,j,k,m,n=1,\cdots,4$ are its fundamental representation indices.

${\cal N}=4 $ supersymmetric Yang-Mills theory has no internal
superconformal anomaly since it has the vanishing beta function.
However, the external superconformal anomaly can arise if above
currents couple with external $N=4$ conformal supergavity fields
($g_{\mu\nu}$, $\psi_\mu$, $A_{\mu i}^j$, $\varphi$, $\psi_i$,
$E^{ij}$, $B_{\mu\nu}^{~~ij}$, $\psi^k_{ij}$, $D_{ij}^{~~kl}$). The
structure of the external superconformal anomaly is similar to
${\cal N}=1$ case.

\subsection{Supercurrent and Internal Superconformal Anomaly in
 ${\cal N}=1$ Supersymmetric $SU(N+M)\times SU(N)$
Gauge Theory with Two Flavors in Bifundamental Representations}

The dual description  for ${\cal N}=1$ supersymmetric $SU(N+M)\times
SU(N)$ gauge theory with two flavors in the representations
$(N+M,\overline{N})$ and $( \overline{N+M},N)$ on string theory side
is the type IIB superstring in the geometric background furnished
by Klebanov-Strassler solution \cite{klst}.  The classical action of
this supersymmetric gauge theory in superspace takes the following
form \cite{wfchen2},
\begin{eqnarray}
S&=& \int d^4x\left\{\frac{1}{32\pi}\int  d^2\theta
\sum_{i=1}^2\mbox{Im}\left[\tau_{(i)} \mbox{Tr}\left(W^{(i)\alpha}
W^{(i)}_\alpha \right)\right]+\mbox{h.c.}\right. \nonumber\\
&+&\left. \int d^2\theta d^2\overline{\theta}\sum_{j=1}^2\left(
\overline{A}^j e^{ \left[g_1V^{(1)} (N+M)+g_2
V_2(\overline{N})\right]} A_j +\overline{B}_j e^{\left[g_1V^{(1)}
(\overline{N+M})+g_2 V_2(N)\right]}B^j\right)\right\},
\end{eqnarray}
where the two flavors $A_j$ and $B^j$ are chiral superfields in the
bifundamental representations $(N+M,\overline{N})$ and $(
\overline{N+M},N)$, respectively. This model also admits a quartic
superpotential,
$W=\lambda\epsilon^{ik}\epsilon^{jl}\mbox{Tr}\left(A_iB_jA_k B_l
\right)$, which normalizes the R-charge of chiral superfields $A_i$,
$B^j$ to be $1/2$.
 This model is a vector supersymmetric gauge theory and
 $A_i$ and $B^j$  can be considered as  quark- and anti-quark chiral
 superfields, respectively.
 The two-component form of above action in the Wess-Zumino gauge
is
\begin{eqnarray}
{\cal
L}&=&\sum_{i=1}^2\frac{1}{g_{(i)}^2}\left[-\frac{1}{4}G^{(i)a_i\mu\nu}
G^{(i)a_i}_{\mu\nu} +i\lambda^{(i)a_i}\sigma^{\mu}
\left(\nabla_{\mu}\overline{\lambda}^{(i)}\right)^{a_i}
+\frac{1}{2}D^{(i)a_i}D^{(i)a_i} \right] +\frac{\theta_{(i)}^{\rm
(CP)}}{32\pi^2}G^{(i)a_i\mu\nu} \widetilde{G}^{(i)a_i}_{\mu\nu}
\nonumber\\
 &+& \left(D^\mu \phi_A\right)^{\dagger j} D_\mu \phi_{Aj}+
\widetilde{D}^\mu \phi_B^j \left(\widetilde{D}_\mu
\phi_B\right)^\dagger_j
 + i\overline{\psi}^j_A \overline{\sigma}^\mu D_{\mu} \psi_{Aj}
+ i{\psi}_{B}^j{\sigma}^\mu\widetilde{D}_\mu \overline{\psi}_{Bj}
\nonumber\\
&+&  \sqrt{2}i \sum_{i=1}^2 (-1)^{i+1} g_{(i)}\left[ \phi^{\dagger
j}_A
 T^{a_i} \lambda^{a_i}_{(i)}\psi_{Aj}
 -\overline{\lambda}^{a_i}_{(i)}\overline{\psi}^j_A
 T^{a_i}\phi_{Aj}-
 \psi^j_B\lambda^{a_i}_{(i)} T^{a_i}{\phi}^\dagger_{Bj}+
 {\phi}^j_BT^{a_i}\overline{\psi}_{Bj}
 \overline{\lambda}_{(i)}^{a_i}\right] \nonumber\\
 &+& \sum_{i=1}^2 g_{(i)} (-1)^{i+1} D_{(i)}^{a_i}
 \left[\phi^{\dagger j}_A T^{a_i}\phi_{Aj}-{\phi}^j_BT^{a_i}
 {\phi}^\dagger_{Bj}\right]\nonumber\\
 &+&F^{\dagger j}_A F_{Aj}+{F}^j_B{F}^\dagger_{Bj}
 +\left[ \left(F_{Aj}\frac{\partial W}{\partial \phi_{Aj}}
 +F_{B}^j\frac{\partial W}{\partial \phi_{B}^j}\right)\right.
 \nonumber\\
 &+&\left.
 \frac{1}{2}\left(\frac{\partial^2 W}{\partial \phi_{Ai}
\partial \phi_{Aj}}\psi_{Ai} \psi_{Aj}
+\frac{\partial^2 W}{\partial \phi_{B}^i
\partial \phi_{B}^j}\psi_{B}^i \psi_{B}^j
 +2\frac{\partial^2 W}{\partial \phi_{Ai}
\partial \phi_B^j}\psi_{Ai} \psi_{B}^j
  \right)+\mbox{h.c.} \right],
  \label{lag}
\end{eqnarray}
where $\nabla_\mu$, $D_\mu$ and $\widetilde{D}_\mu$ are gauge
covariant derivatives in adjoint-, fundamental and anti-fundamental
representations.

The theory has global flavor symmetry $SU_L(2)\times SU_R(2) \times
U_B(1)\times U_A(1)$  at classical level and the $U_A(1)$ symmetry
becomes anomalous at quantum level. In addition, the theory has
R-symmetry chiral $U_R(1)$. It is this chiral $U_R(1)$ symmetry
anomaly that enters the internal superconformal anomaly multiplet.

 The supercurrent multiplet (in four-component form) is similar to
those in pure supersymmetric Yang-Mills theory  and the  difference
is the involvement of the matter fields,
\begin{eqnarray}
\theta_{\mu\nu}&=&\sum_{i=1}^2\left\{-G_{\mu\rho}^{(i)a_i}
G_{~~\nu}^{(i)a_i\rho} +\frac{1}{2}i\left[\overline{\lambda}^{a_i}
(\gamma_\mu \nabla_\nu +\gamma_\nu \nabla_\mu)\lambda^{a_i}-
(\nabla_\mu \overline{\lambda}^{a_i}\gamma_\nu +\nabla_\nu
\overline{\lambda}^{a_i}\gamma_\mu)
\lambda^{a_i}\right]\right.\nonumber\\
&&\left.+\frac{1}{4}g_{\mu\nu} G_{\lambda\rho}^{(i)a_i}
G^{(i)a_i\lambda\rho}-\frac{1}{4}i g_{\mu\nu}
\left[\overline{\lambda}^{a_i} \gamma^\rho\left( \nabla_\rho
\lambda^{a_i}\right) -\left(\nabla_\rho\overline{\lambda}^{a_i}
\right) \gamma^\rho \lambda^{a_i}\right]\right\}
\nonumber\\
&&+\left[\left(D_\mu\phi_A\right)^{\dagger j}D_\nu \phi_{Aj}+
\left(D_\nu\phi_A\right)^{\dagger j}D_\mu \phi_{Aj}\right]+
\left[\left(\widetilde{D}_\mu\phi_B\right)^{\dagger}_j
\widetilde{D}_\nu \phi_{B}^j +
\left(\widetilde{D}_\nu\phi_B\right)^{\dagger}_j
\widetilde{D}_\mu\phi_{B}^j\right]
\nonumber\\
&&-\frac{1}{3}\left(\partial_\mu\partial_\nu-g_{\mu\nu}
\partial^2
\right)\left(\phi_A^{\dagger j}\phi_{Aj}\right)
-\frac{1}{3}\left(\partial_\mu\partial_\nu-g_{\mu\nu}\partial^2
\right)\left(\phi_{Bj}^{\dagger}\phi_{B}^j\right) \nonumber\\
&&+i\overline{\psi}^j\left(\gamma_\mu D_\nu+\gamma_\nu D_\mu \right)
\psi_j -g_{\mu\nu}\left\{i\overline{\psi}^j\gamma^\lambda
D_\lambda\psi_j +\left(D^\lambda \phi_A\right)^{\dagger j} D_\mu
\phi_{Aj}+ \widetilde{D}^\lambda \phi_B^j
\left(\widetilde{D}_\lambda \phi_B\right)^\dagger_j
\right.\nonumber\\
&&  + \frac{i} {\sqrt{2}} \sum_{i=1}^2 (-1)^{i+1} g_{(i)} \left[
\phi^{\dagger j}_A
 T^{a_i} \overline{\lambda}^{a_i}(1-\gamma_5)\psi_{j}
 -\overline{\psi}^j(1+\gamma_5){\lambda}^{a_i}
 T^{a_i}\phi_{Aj}\right.\nonumber\\
 && \left.-
 \overline{\psi}^j(1-\gamma_5)\lambda^{a_i}
 T^{a_i}{\phi}^\dagger_{Bj}
 +{\phi}^j_BT^{a_i}\overline{\lambda}^{a_i}(1+\gamma_5)
 {\psi}_{j}
 \right]-\frac{1}{2}\sum_{i=1}^2 g_{(i)}^2
 \left(\phi^{\dagger j}_A T^{a_i}\phi_{Aj}-{\phi}^j_BT^{a_i}
 {\phi}^\dagger_{Bj}\right)^2\nonumber\\
 &&\left.+\mbox{superpotential terms}
\right\}
 \nonumber\\
 s_\mu &=& -\sum_{i=1}^2\sigma_{\nu\rho}
 \gamma_\mu \lambda^{a_i} G^{(i)a_i\nu\rho}
+\frac{1}{2}\left(D^\nu \phi_A\right)^{\dagger j}\gamma_\nu
(1+\gamma_5)\gamma_\mu \psi_j -\frac{1}{2}\widetilde{D}^\nu\phi_B^j
\gamma_\nu
(1-\gamma_5)\gamma_\mu \psi_j \nonumber\\
&& +\frac{1}{2} \gamma_\nu (1-\gamma_5)\gamma_\mu
\left(C\overline{\psi}^{Tj}\right) D_\nu\phi_{Aj} -\frac{1}{2}
\gamma_\nu (1-\gamma_5)\gamma_\mu
\left(C\overline{\psi}^{Tj}\right)\left(\widetilde{D}_\nu\phi_B
\right)^\dagger
 \nonumber\\
&& -\frac{1}{3}i\sigma_{\mu\nu}\partial^\nu \left[
\phi_{Aj}(1+\gamma_5)\left(C\overline{\psi}^{Tj}\right)
+\phi_A^{j\dagger} (1-\gamma_5)\psi_j\right]\nonumber\\
&& +\frac{1}{3}i\sigma_{\mu\nu}\partial^\nu \left[
\phi_{Bj}^\dagger(1-\gamma_5) \left(C\overline{\psi}^{Tj}\right)
+\phi_B^j (1+\gamma_5)\psi_j
  \right]
 \nonumber\\
 &&+2\sqrt{2}\sum_{i=1}^2 (-1)^{i+1}g_{(i)}\gamma_5\gamma_\mu
 \lambda^{a_i}\left[\phi_A^{j\dagger}T^{a_i}\phi_{Aj}
 -\phi_{Bj}T^a\phi_B^{\dagger j}  \right]
 + \mbox{superpotential terms};\nonumber\\
j_\mu &=&\frac{1}{2}\sum_{i=1}^2\overline{\lambda}^{a_i}
\gamma_\mu\gamma_5 \lambda^{a_i}-\frac{1}{3}\overline{\psi}^j
\gamma_\mu\gamma_5 \psi_j -\frac{2i}{3}\left[\phi_A^{\dagger j}
D_\mu \phi_{Aj}-
\left(D_\mu\phi_A\right)^{\dagger j} \phi_{Aj} \right]\nonumber\\
&&-\frac{2i}{3}\left[\phi_B^{j} \left(\widetilde{D}_\mu
\phi_{Bj}\right)^\dagger- \left(\widetilde{D}_\mu\phi_B\right)^j
\phi_{Aj}^\dagger \right]+\mbox{superpotential terms}. \label{cus2}
\end{eqnarray}
In above equation, $\lambda^a$ and $\psi$ are Majorana and Dirac
spinors, respectively,
\begin{eqnarray}
\lambda^a=\left(\begin{array}{c}\lambda_\alpha^a\\
\overline{\lambda}^{a\dot{\alpha}} \end{array}\right),~~
\psi_j=\left(\begin{array}{c}\psi_{Aj\alpha}\\
\overline{\psi}^{\dot{\alpha}}_{Bj} \end{array}\right),
~~\left(C\overline{\psi}^T\right)^j
=\left(\begin{array}{c}\psi_{B\alpha}^j\\
\overline{\psi}^{j\dot{\alpha}}_A \end{array}\right). \label{curr1}
\end{eqnarray}
Classically the above conservative currents satisfy
$\theta^\mu_{~\mu}=\gamma^\mu s_\mu=0$ up to the classical
superpotential terms. At quantum level  the superconformal anomaly
arises due to the non-vanishing $\beta$-functions of two gauge
couplings. Let us first observe $U_R(1)$ symmetry anomaly.
Eq.\,(\ref{curr1}) shows that the chiral current $j^\mu$ is composed
of gluinoes $\lambda^{a_i}$, $\left(\psi_{Aj}\right)_{~r}^m$ and
$\left(\psi_B^j\right)_{~m}^{r}$. With respect to the first gauge
group $SU(N+M)$, there are $2N$ flavor matters and $\lambda^{a_1}$
in its adjoint represenation. Hence they contribute $2N\times
(-1/2)=-N$ and $C_2[SU(N+M)]=N+M$ to the anomaly coefficient. So for
the first gauge field background, the chiral $R$-anomaly coefficient
is $N+M-N=M$. A similar analysis for the second gauge group $SU(N)$
gives the chiral anomaly coefficient $-M$.  Therefore, the chiral
$U_R(1)$ $R$-symmetry anomaly reads
\begin{eqnarray}
\partial_\mu j^\mu &=&\frac{M}{16\pi^2}\left(g_1^2 G_{\mu\nu}^{a_1}
\widetilde{G}^{a_1\mu\nu}-g_2^2 G_{\mu\nu}^{a_2}
\widetilde{G}^{a_2\mu\nu}\right) +\mbox{classical superpotential
contribution}. \label{chira2}
\end{eqnarray}

 To show clearly the
 origin of the scale anomaly, we first consider the
 $SU(N)\times SU(N)$ gauge theory with chiral superfields
 $A_j$ in $(N,\overline{N})$ and $B^j$ in $(\overline{N},N)$
 representations. The exact
 NSVZ $\beta$-functions for those two gauge couplings,
\begin{eqnarray}
\beta (g_1^2)&=&-\frac{g^3_1}{16\pi^2}\left[ 3N-2N (1-\gamma
(g))\right],\nonumber\\
\beta (g_2^2)&=& \frac{g^3_2}{16\pi^2}\left[ 3N-2 N(1-\gamma
(g))\right],
\end{eqnarray}
shows the $\beta$-functions have zero-points with the anomalous
dimension $\gamma(g)=-1/2$, at which the theory is a superconformal
field theory. For the $SU(N+M)\times SU(N)$ gauge theory with chiral
superfields
 $A_j$ in $(N+M,\overline{N})$ and $B^j$ in $(N,\overline{N+M})$
 representations, the above IR fixed points are removed since now
 the
 above $\beta$-functions become
\begin{eqnarray}
\beta (g_1^2)&=&-\frac{3Mg^3_1}{16\pi^2},~~~~ \beta (g_2^2)=
\frac{3Mg^3_2}{16\pi^2}. \label{bfunction}
\end{eqnarray}
Consequently, the scale anomaly arises,
\begin{eqnarray}
\theta^{\mu}_{~\mu}&=&\frac{3M}{8\pi^2} \sum_{i=1}^2
(-1)^{i+1}{g_{i}^2}\left[-\frac{1}{4}G^{(i)a_i\mu\nu}
G^{(i)a_i}_{\mu\nu}
+\frac{1}{2}i\overline{\lambda}^{a_i}\gamma^{\mu}
\left(\nabla_{\mu}{\lambda}\right)^{a_i} \right]\nonumber\\
&& +\mbox{classical superpotential contribution}. \label{scal2}
\end{eqnarray}
Further, the beta functions (\ref{bfunction}) also implies the
$\gamma$-trace anomaly of supersymmetry current,
\begin{eqnarray}
\gamma^\mu s_\mu &=& \frac{3M}{8\pi^2}
\sum_{i=1}^2(-1)^{i+1}g_i^2\left(-\sigma^{\mu\nu}G^{a_i}_{\mu\nu}
+\gamma_5 D^{a_i}\right)\lambda^{a_i}\nonumber\\
&+& \mbox{classical superpotential contribution}
\label{gammatrace2}.
\end{eqnarray}

Similar to the ${\cal N}=1$ pure supersymmetric Yang-Mills theory
case, the above superconformal anomaly manifest itself in the
quantum effective action as gauge coupling in superspace. The
 local part of the explicit gauge invariant quantum effective
action composed only of gauge fields takes following form in
superspace,
\begin{eqnarray}
{\Gamma}_{\rm eff}&=& \frac{1}{32\pi}\int d^2\theta\,\sum_{i=1}^2
\mbox{Im}\left[\Sigma_{(i)\,\rm eff} \mbox{Tr} \left(W^{(i)\alpha}
W^{(i)}_\alpha\right) \right]+\mbox{h.c.},
\end{eqnarray}
where $\Sigma_{(i)\,\rm eff}$ is given  in Eq.\,(\ref{supercou}).
 The effect of superconformal anomaly reflects as
\begin{eqnarray}
\frac{1}{g^2_{(i)\,{\rm eff}}(q^2)}&=&\frac{1}{g^2_{(i)}(q_0^2)}
+(-1)^{i+1}\frac{3M}{8\pi^2}\ln
\frac{q_0}{q},  \nonumber\\
\theta^{\rm (CP)}_{(i)\,{\rm eff}} &=& \theta^{\rm
(CP)}_{(i)}+(-1)^{i+1}M,\nonumber\\
\chi_{(i)\,{\rm eff}} &=& 0+(-1)^{i+1}M. \label{effcou1}
\end{eqnarray}
This ends our discussions on the superconformal anomaly in ${\cal
N}=1$ $SU(N+M)\times SU(N)$ supersymmetric gauge theory.

\section{Holographic version on AdS/CFT Correspondence, On-Shell gauged
$AdS_5$ Supergravity and off-shell four-dimensional conformal
supergravity}

\subsection{AdS/CFT Correspondence and  Gauged Supergravity}

The original AdS/CFT correspondence conjecture \cite{mald}
  states that type IIB superstring theory in $AdS_5\times S^5$
  geometrical background
 with $N$ units of $R-R$ flux passing through $S^5$
 describes the same physics
  as ${\cal N}=4$ $SU(N)$ supersymmetric Yang-Mills theory
   in four dimensions does.
  A concrete  holographic definition on the AdS/CFT correspondence
  was  given
  in Refs.\,\cite{gkp} and \cite{witt1}, which has two versions.
  The first one is based on canonical quantization of a field theory,
  it states that quantum states of  type IIB string in $AdS_5\times S^5$
  space-time background should have one-to-one correspondence with
gauge invariant operators of ${\cal N}=4$ supersymmetric Yang-Mills
theory in four dimensions. Due to the difficulty in quantizing
string theory in curved space-time, one can can only establish
correspondence between supergravity modes on string theory side and
certain operators of ${\cal N}=4$ supersymmetric Yang-Mills theory.
The other version is expressed in terms of the path integral
quantization, which states that the partition function of type II
superstring   should equal to the generating functional for the
correlation functions of  composite operators  in a  superconformal
field theory defined on the boundary of $AdS_5$ space. Concrete
speaking, given type II superstring theory in target space-time
background $AdS_{d+1}\times X^{9-d}$ with $X^{9-d}$ being a compact
Einstein manifold,  its partition function $Z_{\rm string}$ with
bulk fields having non-trivial $AdS_{d+1}$ boundary value  takes the
following form,
\begin{eqnarray}
\left.Z_{\rm
string}\left[\phi\right]\right|_{\phi\rightarrow\phi_{0}}
=\int_{\phi (x,0)=\phi_{0}(x)}{\cal D}\phi (x,r) \exp\left(-S[\phi
(x,r)]\right), \label{spf}
\end{eqnarray}
where $\phi_{0}(x)$ is the boundary value of the field function
$\phi (x,r)$ in $AdS_{d+1}$ bulk  such as the graviton, gravitino,
NS-NS and R-R antisymmetric tensor fields etc. On the other hand,
the generating functional for correlation functions of gauge
invariant operators ${\cal O}(x)$ in  conformal field theory on the
$AdS_{d+1}$ boundary  is
\begin{eqnarray}
Z_{\rm CFT}\left[\phi_{0}\right]&=& \left\langle \exp\int_{M^d} d^dx
{\cal O} (x)  \phi_{0}(x)\right\rangle \nonumber\\
&=&\sum_n\frac{1}{n!}\int \prod_{i=1}^n d^d x_i \left\langle {\cal
O}_1 (x_1)\cdots {\cal O}_n (x_n) \right\rangle \phi_{0} (x_1)
\cdots \phi_{0} (x_n)\nonumber\\
&{\equiv}& \exp\left(-\widetilde{\Gamma}_{\rm CFT}
[\phi_{0}]\right), \label{acc1}
\end{eqnarray}
where $\Gamma [\phi_{0}]$ denotes the quantum effective action
describing  composite operators  interacting with background field
$\phi_{0}(x)$. The AdS/CFT correspondence means
 \begin{eqnarray}
   \left.Z_{\rm
string}\left[\phi\right]\right|_{\phi\rightarrow\phi_{0}}
   =Z_{\rm CFT}\left[\phi_{0}\right] \,.
   \label{acc2}
 \end{eqnarray}
 Since the string correction to type II
supergravity is proportional to  $1/\sqrt{g_sN}$, $g_s$ being the
string coupling, thus in the large-$N$ limit, one can neglect string
effect and just consider its low-energy effective theory,  i.e., the
type IIB supergravity. In this case, the  partition function
(\ref{spf}) of type IIB superstring can be evaluated with
saddle-point approximation. That is, it is approximately equal to
the exponential of  supergravity action evaluated at the field
configuration $\phi^{\rm cl}[\phi^{0}]$ that satisfies the classical
equation of motion of type II supergravity with the boundary value
$\phi_{0}$,
\begin{eqnarray}
\left.Z_{\rm
string}\left[\phi\right]\right|_{\phi\rightarrow\phi_{0}}
 =\exp\left(-S_{\rm SUGRAII}[\phi^{\rm cl}[\phi_{0}]]\right) \,.
\label{acc3}
\end{eqnarray}
Comparing \,(\ref{acc3}) with (\ref{acc1}) and (\ref{acc2}), we
conclude that the background effective action of
 $d$-dimensional conformal field theory living on $AdS_{d+1}$ boundary
  at large-$N$ limit is
approximately equal to on-shell classical action of $AdS_{d+1}$
supergravity,
\begin{eqnarray}
\left.\widetilde{\Gamma}_{\rm
CFT}[\phi_{0}]\right|_{N\to\infty}=S_{\rm SUGRAII}[\phi^{\rm
cl}[\phi_{0}]]
\label{acc4}
\end{eqnarray}

Let us specialize to the $d=4$ case and  analyze the role of
five-dimensional gauged supergravities \cite{gst,awada,guna2} played
in the AdS/CFT correspondence \cite{ferr1}. The $AdS_5\times S^5$
space-time background arises from the near-horizon limit of
$D3$-brane solution of type IIB supergravity \cite{agmo}, so in
$AdS_5\times S^5$ background the spontaneous compactification on
$S^5$ of type IIB supergravity must occur \cite{freu}. Based on the
assumption that there exists a consistent nonlinear truncation of
massless modes from the whole Kaluza-Klein (K-K) spectrum of type
IIB supergravity compactified on $S^5$ \cite{dupo,marcus,kim}, the
resultant theory should be $SO(6)(\cong SU(4))$ gauged ${\cal N}=8$
$AdS_5$ supergravity since the isometry group $SO(6)$ of internal
space $S^5$ becomes gauge group of the compactified theory and
specifically, the $AdS_5\times S^5$ background preserves all
supersymmetries in type IIB supergravity \cite{guna2}. In
particular, the massive K-K modes correspond to irrelevant gauge
invariant operators on field theory side \cite{witt1}. The
truncation of massive K-K modes in compactified type IIB
supergravity means ignoring irrelevant operators in the quantum
effective action $\widetilde{\Gamma}_{\rm CFT}[\phi_{0}]$ for gauge
invariant field operators. So the AdS/CFT correspondence finally
reduces to the equivalence between on-shell ${\cal N}=8$ $SO(6)$
gauged $AdS_5$ supergravity and the large-$N$ limit of ${\cal N}=4$
supersymmetric Yang-Mills theory in ${\cal N}=4$ conformal
supergravity background furnished by the boundary data of gauged
$AdS_5$ supergravity \cite{witt1,liu},
\begin{eqnarray}
\left.{\Gamma}_{\rm CFT}[\phi_{0}]\right|_{N\to\infty}=S_{\rm
AdS-SUGRA}[\phi^{\rm cl}[\phi_{0}]] \label{acc50}
\end{eqnarray}
 Further, there is
a straightforward generalization. If the background for type IIB
supergravity is $AdS_5 \times X^5$ with $X^5$ being an Einstein
manifold less symmetric than $S^5$ such as $T^{1,1}=(SU(2)\times
SU(2))/U(1)$, then due to the singularity in the internal manifold,
the number of preserved supersymmetries in gauged $AdS_5$
supergravity is reduced and the gauge group of
 becomes smaller \cite{roman2,kw2,kach,law}.  One can thus obtain
the gauged ${\cal N}=2,4$ $AdS_5$ supergravities and their dual
field theories  should be $N=1,2$ supersymmetric gauge theories
\cite{ferr1,kw2,kach,law}. Strictly speaking, the ${\cal N}=1,2$
quantum supersymmetric gauge theories  are not conformal invariant
since they have non-vanishing beta functions. However, it was shown
that renormalization group flow of these supersymmetric gauge
theories have the fixed point, at which the conformal invariance
arises \cite{kach,law,shif,seib,lei2,early1,early2}. Therefore, the
$AdS/CFT$ correspondence between ${\cal N}=2,4$ gauged $AdS_5$
supergravities and  the four-dimensional ${\cal N}=1,2$
supersymmertric gauge theories at fixed-points of their
renormalization group flow be established \cite{ferr1,kw2,kach,law}.

 Eq.\,(\ref{acc50}) means that the on-shell action of gauged $AdS_5$
 supergravity equals to the quantum effective action
of superconformal gauge theory in an external conformal supergravity
background \cite{witt1,liu}, only now the background field comes
from the boundary value of supergravity in  one-dimension-higher
space-time. As introduced in Sect.\,II, a classical superconformal
gauge theory in an external supergravity background suffers from
superconformal anomaly, which should be contained in the background
effective action.  Therefore, the external superconformal anomaly
can be extracted out from the on-shell action of gauged $AdS_5$
supergravity in terms of the AdS/CFT correspondence.

In following subsections we shall recall some typical features
gauged ${\cal N}=2,4,8$ supergravity in five dimensions and  see how
 ${\cal N}=1,2,4$ off-shell conformal supergravity in four dimensions
 arise from gauged
 supergravities.

\subsection{${\cal N}=2$ $U(1)$ Gauged $AdS_5$ Supergravity
and ${\cal N}=1$ Conformal Supergravity in Four Dimensions}

 The ungauged ${\cal N}=2$ supergravity in five dimensions
 contains a graviton
$\widehat{e}_\alpha^{~m}$, two gravitini $\widehat{\psi}_\alpha^a$
and an Abelian vector field $\widehat{A}_\alpha$
\cite{crem2,gst,guna}, $a=1,2$ being $SU(2)$ doublet index. The
theory has a global $USp(2)\cong SU(2)$ symmetry. The gravitini are
$USp(2)\cong SU(2)$ doublets and symplectic Majorana spinors. The
classical Lagrangian density is \cite{crem2,gst,guna}
\begin{eqnarray}
\widehat{e}^{-1} {\cal L}_{\rm ungauged} &=&
-\frac{1}{2}\widehat{R}[\widehat{\omega}]-\frac{1}{2}
\overline{\widehat{\psi}}_\alpha^a
\widehat{\gamma}^{\alpha\beta\gamma}\widehat{\nabla}_\beta
\widehat{\psi}_{\gamma a}-\frac{1}{4}a_{00}
\widehat{F}^{\alpha\beta} \widehat{F}_{\alpha\beta}\nonumber\\
&&+\frac{1}{6\sqrt{6}}\widehat{e}^{-1}C_{000}
\epsilon^{\alpha\beta\gamma\sigma\delta}
\widehat{F}_{\alpha\beta}\widehat{F}_{\gamma\sigma}
\widehat{A}_\delta
\nonumber\\
&& -\frac{3i}{8\sqrt{6}}h_0\left(
\overline{\widehat{\psi}}_\alpha^a\widehat{\gamma}^{\alpha\beta\gamma\delta}
\widehat{\psi}_{\beta
a}\widehat{F}_{\gamma\delta}+2\overline{\widehat{\psi}}^{\alpha
a}\widehat{\psi}^\beta_a \widehat{F}_{\alpha\beta}\right)
+\mbox{four-fermion terms}, \label{ungaugedf}
 \end{eqnarray}
 The supersymmetry transformation laws are \cite{gst}
\begin{eqnarray}
\delta \widehat{e}_\alpha^{~m}&=&
\frac{1}{2}\overline{\widehat{\epsilon}}^a\widehat{\gamma}^m
\widehat{\psi}_{\alpha a},
\nonumber\\
\delta \widehat{\psi}_{\alpha }^a &=&
\widehat{\nabla}_{\alpha}{\epsilon}^a+ \frac{1}{4\sqrt{6}} i h_0
\left(\widehat{\gamma}_{\alpha}^{~\beta\gamma}-4\delta_{\alpha}^{~\beta}
\widehat{\gamma}^\gamma \right)
\widehat{F}_{\beta\gamma}\widehat{\epsilon}^a,
\nonumber\\
\delta \widehat{A}_\alpha &=& \frac{\sqrt{6}}{2}i
h^0\overline{\widehat{\psi}}_\alpha^a\widehat{\epsilon}_a.
\label{ugtwost}
\end{eqnarray}

 The gauged ${\cal N}=2$ supergravity can be obtained from above
 ungauged supergravity by converting $U(1)$ subgroup of
 the global
 $SU(2)$ group into a local symmetry group and considering the vector
 field $\widehat{A}_\alpha$ as $U(1)$  gauge field. The space-time
 covariant derivative on  gravitini
 should be enlarged to include $U(1)$ gauge covariance,
\begin{eqnarray}
\widehat{D}_\alpha \widehat{\psi}_\beta^a=\widehat{\nabla}_\alpha
\widehat{\psi}_\beta^a+g V_0
\widehat{A}_{\alpha}\delta^{ab}\widehat{\psi}_{\beta b},
\end{eqnarray}
where $g$ is the $U(1)$ gauge coupling.
 The gauged ${\cal N}=2$ supergravity action is \cite{gst}
\begin{eqnarray}
\widehat{e}^{-1} {\cal L} =\widehat{e}^{-1} {\cal L}_{\rm un-gauged}
+g^2 P_0^2-\frac{i\sqrt{6}}{8}g
\overline{\widehat{\psi}}_\alpha^a\widehat{\gamma}^{\alpha\beta}
\widehat{\psi}^{b}_\beta \delta_{ab}P_0. \label{gaugedf0}
\end{eqnarray}
In Eqs.\,(\ref{ungaugedf}) and (\ref{gaugedf0}),  $a_{00}$, $h_0$
and  $C_{000}$ are parameters that can be determined by ${\cal N}=2$
supersymmetry; $P_0$ is the scalar potential when this  ${\cal N}=2$
supergravity couples with matter fields, while in the  case at hand,
it is a pure supergravity, $P_0$ is just a parameter.

We require that the above $U(1)$ gauged ${\cal N}=2$ supergravity
(\ref{gaugedf0}) has domain wall solution which should
asymptotically approach  $AdS_5$ vacuum configuration. The solution
thus presents the following standard form,
\begin{eqnarray}
   ds^2 &=& \frac{l^2}{r^2}\left[\widehat{g}_{\mu\nu}(x,r)dx^\mu dx^\nu
   +dr^2\right],
 \nonumber\\
   \widehat{A}_\alpha &=&\widehat{\psi}_\alpha^a=0.
   \label{metrican1}
   \end{eqnarray}
This requirement fixes the parameters in the Lagrangian
(\ref{gaugedf0}) as the following \cite{gst,bala},
 \begin{eqnarray}
g &=& \frac{3}{4},
~~h_0=\frac{l}{2}\sqrt{\frac{3}{2}},~~h^0=\frac{1}{h_0},~~V_0=1,
~~P_0=2 h^0 V_0=\frac{4}{l}\sqrt{\frac{2}{3}}, \nonumber\\
a_{00} &=& \left(h_0\right)^2=\frac{3l^2}{8},~~ C_{000}
=\frac{5}{2}h_0^3-\frac{3}{2}a_{00} h_0=\frac{3\sqrt{6}l^3}{32}.
\end{eqnarray}
It should be emphasized that the space-time background described by
$AdS_5$ solution (\ref{metrican1}) preserves the full ${\cal N}=2$
supersymmetry. Consequently, the  Lagrangian density (\ref{gaugedf})
  up to the quadratic terms of fermionic field becomes
\begin{eqnarray}
\widehat{e}^{-1} {\cal L} &=&
-\frac{1}{2}\widehat{R}[\widehat{\omega}
]-\frac{1}{2}\overline{\widehat{\psi}}_\alpha^a
\widehat{\gamma}^{\alpha\beta\gamma}\widehat{D}_\beta
\widehat{\psi}_{\gamma a}-\frac{3l^2}{32} \widehat{F}^{\alpha\beta}
\widehat{F}_{\alpha\beta}
+\frac{l^3}{64}\widehat{e}^{-1}\epsilon^{\alpha\beta\gamma\sigma\delta}
\widehat{F}_{\alpha\beta}\widehat{F}_{\gamma\sigma}
\widehat{A}_\delta
\nonumber\\
&&-\frac{3}{4l}i\overline{\widehat{\psi}}_\alpha^a
\widehat{\gamma}^{\alpha\beta}\psi_{\beta}^b \delta_{ab}
 -\frac{3l}{32}i\left(\overline{\widehat{\psi}}_\alpha^a
 \widehat{\gamma}^{\alpha\beta\gamma\delta}
\widehat{\psi}_{\beta
a}\widehat{F}_{\gamma\delta}+2\overline{\widehat{\psi}}^{\alpha
a}\widehat{\psi}^\beta_a
\widehat{F}_{\alpha\beta}\right)-\frac{6}{l^2}. \label{gaugedf}
\end{eqnarray}
The supersymmetry transformation at the leading order of fermionic
field reads
\begin{eqnarray}
\delta \widehat{e}_\alpha^{~m}&=&
\frac{1}{2}\overline{\widehat{\epsilon}}^a\widehat{\gamma}^m
\widehat{\psi}_{\alpha a},
\nonumber\\
\delta \widehat{\psi}_{\alpha }^a &=&
\widehat{D}_{\alpha}\widehat{\epsilon}^a+ \frac{il}{16}
\left(\widehat{\gamma}_{\alpha}^{~\beta\delta}-4\delta_{\alpha}^{~\beta}
\widehat{\gamma}^\delta \right)
\widehat{F}_{\beta\delta}\epsilon^a+\frac{i}{2l}g\widehat{\gamma}_\alpha
\delta^{ab}\widehat{\epsilon}_b,
\nonumber\\
\delta \widehat{A}_\alpha &=&
\frac{i}{l}\overline{\widehat{\psi}}_\alpha^a\widehat{\epsilon}_a.
\label{twost1}
\end{eqnarray}

In the following we choose the above $AdS_5$ solution
(\ref{metrican1}) as a vacuum configuration for this ${\cal N}=2$
$U(1)$ gauged supergravity. Then we  expand the theory around this
vacuum configuration and derive classical equations of motion for
$\widehat{g}_{\alpha\beta}$, $\widehat{\psi}_\alpha^i$ and
$\widehat{A}_\alpha$.  The dynamical behavior of gauged supergravity
near $AdS_5$ vacuum configuration can be observed by solving
linearized equations of motion. Geometrically, this is actually a
process of revealing the asymptotic behavior of the bulk fields near
the boundary of $AdS_5$ space.

The linearized classical equations of motion for graviton
$\widehat{g}_{\alpha\beta}$, gravi-photon $\widehat{A}_\alpha$ and
gravitini $\widehat{\psi}_\alpha^i$ are (linearized) Einstein-,
Maxwell- and Rarita-Schwinger equations, respectively,
\begin{eqnarray}
\widehat{R}_{\alpha\beta}-\frac{1}{2}
\widehat{g}_{\alpha\beta}\widehat{R}-\frac{6}{l^2}
\widehat{g}_{\alpha\beta}=0 \label{eequation1}
\end{eqnarray}
\begin{eqnarray}
\widehat{e}^{-1}\partial_\alpha
\left[\widehat{g}^{\alpha\delta}\partial_\delta \widehat{A}_\beta
(x,r) \right]=0, \label{mequation1}
\end{eqnarray}
\begin{eqnarray}
\widehat{\gamma}^{\alpha\beta\gamma}\widehat{D}_\beta \psi_{\gamma
a}+\frac{3i}{2}\delta_{ab}\widehat{\gamma}^{\alpha\beta}
\widehat{\psi}_{\beta}^b=0. \label{rsequation1}
\end{eqnarray}
The solutions to these equations near $AdS_5$ boundary (given by
$r\to 0$) can be expressed as a series expansion  in terms of $r/l$.
Here we consider only the leading order in the expansion given in
Ref.\,\cite{bala}. First, the $AdS_5$ vacuum configuration
   (\ref{metrican1}) and the partial gauge-fixing choice on diffeomorphism
   symmetry in $r$-direction
   determine that the solution to Einstein
   equation  (\ref{eequation1}) at the leading order  of $r/l$ expansions
   should be the following  \cite{bala},
   \begin{eqnarray}
   \widehat{e}_\mu^{~s} (x,r)&=&\frac{l}{r} {e}_\mu^{~s}
    (x)+{\cal O}(r/l), ~~
   \widehat{e}_4^{~r}=\widehat{e}_{\mu}^{~4}=0,
   ~~~\widehat{e}_4^{~4}=\frac{l}{r}+{\cal O}(r/l).
   \label{rgf1}
   \end{eqnarray}
The background solution and torsion-free condition
$d\widehat{e}^r+\widehat{\omega}^r_{~s}\wedge \widehat{e}^s=0$ yield
spin connections,
\begin{eqnarray}
\widehat{\omega}^{r}_{~4}(x,r)&=& -\widehat{e}^{r}(x,r)
=-\frac{l}{r}{e}^{a}(x),\nonumber\\
\widehat{\omega}^{r}_{~s}(x,r) &=&{\omega}^{r}_{~s}(x).
\label{redsc1}
\end{eqnarray}
To solve the Maxwell equation (\ref{mequation1}), one should first
fix $U(1)$ gauge symmetry in $r$-direction. The most convenient
choice is
\begin{eqnarray}
\widehat{A}_4 (x,r)=0.
   \label{rgf2v}
   \end{eqnarray}
Subsequently, Eq.\,(\ref{mequation1}) leads to
\begin{eqnarray}
\widehat{A}_\mu (x,r) ={A}_\mu (x)+{\cal O}(r/l). \label{redg1}
\end{eqnarray}
For gravitino equation (\ref{rsequation1}),  the gauge choice of
fixing local supersymmetry in radial direction that is consistent
with bulk supersymmetry transformation is
\begin{eqnarray}
  \widehat{\psi}_4^a(x,r)=0.
   \label{rgf2a}
   \end{eqnarray}

The spin connection (\ref{redsc1}) and the gauge field (\ref{redg1})
as well as the gauge choices (\ref{rgf2v}) and (\ref{rgf2a}) lead to
the near-$AdS_5$ boundary reduction of bulk covariant derivatives,
 \begin{eqnarray}
 \widehat{D}_4&=&\partial_4, ~~~
 \widehat{D}_\mu
 =\widetilde{D}_\mu (x)-\frac{1}{2r}{\gamma}_\mu \gamma_{5},
 \nonumber\\
 \widetilde{D}_\mu (x) &{\equiv}& \nabla_\mu
 +\frac{1}{4}{\omega}_\mu^{~rs}
 {\gamma}_{~rs}
 +\frac{3}{4}{A}_\mu,
 \label{rederi}
 \end{eqnarray}
 where the convention for $\widehat{\gamma}$-matrix is chosen
 as the following,
 \begin{eqnarray}
 {\gamma}_{r}=\widehat{\gamma}_{r},~~ \widehat{\gamma}_\mu=
 \widehat{e}_\mu^{~r}\widehat{\gamma}_{r}=\frac{l}{r}
 {\gamma}_{\mu},
 ~~\widehat{\gamma}^\mu = \widehat{e}^\mu_{~r}\widehat{\gamma}^{r}
 =\frac{r}{l} {\gamma}^{\mu},
 ~~ \gamma_5=\widehat{\gamma}^{4}
 =\widehat{\gamma}_{4},~~\gamma_5^2=1.
 \label{rega}
 \end{eqnarray}
The linearized gravitino equation (\ref{mequation1}) reduces to
 \begin{eqnarray}
 \widehat{\gamma}^{\mu\nu}\left(\delta_{ab}\partial_r
 -\frac{1}{r}\delta_{ab}
 -\frac{3}{2r}\gamma_5 \epsilon_{ab}\right)\widehat{\psi}_{\nu}^b (x,r)
 -\widehat{\gamma}^{\mu\nu\rho}\widetilde{D}_\nu (x)\gamma_5
 \widehat{\psi}_{\rho a} (x,r)=0.
 \label{redugraeq}
 \end{eqnarray}
The above equation can be diagonalized  by combining those two
components of symplectic Majorana spinor, $\widehat{\psi}_\mu\equiv
\widehat{\psi}_{\mu 1}+i\widehat{\psi}_{\mu 2}$, and performing
chiral decomposition $\widehat{\psi}^R_\mu\equiv
(1-\gamma_5)/2\,\widehat{\psi}_\mu$,
 $\widehat{\psi}^L_\mu \equiv (1+\gamma_5)/2\,\widehat{\psi}_\mu$.
 Near $AdS_5$ boundary ($r\to 0$), Eq.\,(\ref{redugraeq}) gives
 the equation for $\widehat{\psi}^R_\mu$ \cite{bala}
\begin{eqnarray}
\left(\partial_r+\frac{1}{2r}\right)\widehat{\psi}_\mu^R(x,r)=0,
\end{eqnarray}
and hence the radial dependence of $\widehat{\psi}_\mu^R$ is
\begin{eqnarray}
\widehat{\psi}_\mu^R(x,r)=\left(\frac{2l}{r}\right)^{1/2}{\psi}_\mu^R(x).
\label{redsl}
\end{eqnarray}
The  left-handed  component is not independent, and its radial
coordinate dependence turned out to be \cite{bala}
\begin{eqnarray}
\widehat{\psi}_\mu^L &= & \left(2lr\right)^{1/2}{\chi}_\mu^L(x), \nonumber\\
\chi_\mu^L &=&
\frac{1}{3}\gamma^\nu\left(\widetilde{D}_\mu\psi_\nu^R
-\widetilde{D}_\nu\psi_\mu^R\right)-\frac{i}{12}
\epsilon_{\mu\nu}^{~~\lambda\rho} \gamma_5\gamma^\nu
\left(\widetilde{D}_\lambda\psi_\rho^R
-\widetilde{D}_\rho\psi_\lambda^R\right). \label{redsr1}
\end{eqnarray}

The near-$AdS_5$ boundary solutions listed in Eqs.\,(\ref{rgf1}),
(\ref{redg1}), (\ref{redsl}) and (\ref{redsr1}) show that on-shell
${\cal N}=2$ $U(1)$ gauged  supergravity multiplet
($\widehat{e}_\alpha^{~m}$, $\widehat{\psi}_\alpha^a$,
 $\widehat{A}_\alpha$) reduces to the boundary field (${e}_\mu^{~a}$,
${\psi}_\mu^{R}$, ${A}_\mu$). To reveal the physical features
described by these boundary fields, we check the boundary reduction
of bulk supersymmetry transformation law realized on
$({e}_\mu^{~a}$, ${\psi}_\mu^{R}$, ${A}_\mu$). First, the same as
that done on bulk gravitino fields, we combine the bulk
supersymmetric transformation parameter, $\widehat{\epsilon}
(x,r)=\widehat{\epsilon}_1(x,r)+i\widehat{\epsilon}_2(x,r)$ and
decompose it into chiral components. Then we choose the radial
coordinate dependence of $\widehat{\epsilon}^{L,R}$ in the same way
as bulk gravitino,
\begin{eqnarray}
\widehat{\epsilon}^R(x,r)=\left(\frac{2l}{r}\right)^{1/2}{\epsilon}^R(x),
~~\widehat{\epsilon}^L(x,r)=
\left(2lr\right)^{1/2} {\eta}^L(x). \label{strp}
\end{eqnarray}
At the leading order of $r/l$ expansion, the bulk supersymmetry
transformation (\ref{twost1}) becomes the following \cite{bala},
\begin{eqnarray}
\delta {e}_\mu^{~r} &=&-\frac{1}{2}
\overline{\psi}_\mu\gamma^{r}{\epsilon},
\nonumber\\
 \delta {\psi}_\mu
&=&\widetilde{D}_\mu {\epsilon} -\gamma_\mu {\eta} ={\nabla}_\mu
{\epsilon} -\frac{3i}{4}A_\mu \gamma_5 {\epsilon}-\gamma_\mu {\eta},
\nonumber\\
\delta {A}_\mu  &=& i\left(\overline{\psi}_\mu\gamma_5{\eta}
-\overline{\chi}_\mu \gamma_5\epsilon \right), \label{desut}
\end{eqnarray}
where all the spinorial quantities, $\psi_\mu (x)$, ${\chi}_\mu (x)$
${\epsilon} (x)$ and ${\eta}(x)$
 are Majorana spinors constructed from
their two-component chiral spinors $\psi^R_\mu (x)$, $\chi^L_\mu
(x)$, ${\epsilon}^R (x)$ and ${\eta}^L(x)$. One can immdietely
recognize that Eq.\,(\ref{desut}) is the supersymmetric
transformation law for the four-dimensional ${\cal N}=1$ conformal
supergravity with $\epsilon$ and $\eta$ being  Poincar\'{e}- and
conformal supersymmetry transformation parameters, respectively
\cite{kaku,fra}.

Further, it can be easily checked that other symmetries in ${\cal
N}=2$ $U(1)$ gauged supergravity near $AdS_5$ vacuum configuration
 also convert into those in the four-dimensional ${\cal N}=1$ conformal
supergravity. For examples, the bulk diffeormorphism invariance
decomposes into the four-dimensional diffeomorphism symmetry and the
Weyl symmetry \cite{imbi}, and the  bulk supersymmetry converts into
the Poincar\'{e} supersymmetry and the conformal supersymmetry  in
four dimensions \cite{bala}. Therefore, the $AdS_5$ boundary
reduction of on-shell ${\cal N}=2$ $U(1)$ gauged supergravity
constitutes
 the 8+8 off-shell multiplet of ${\cal N}=1$ conformal supergravity
in four dimensions \cite{bala}. The matching of degrees of freedom
between the on-shell multiplet ($\widehat{e}_\alpha^{~m}$,
$\widehat{\psi}_\alpha^a$,
 $\widehat{A}_\alpha$) and off-shell multiplet (${e}_\mu^{~r}$,
${\psi}_\mu^{R}$, ${A}_\mu$) are listed in Table I.

\begin{center}
\begin{table}

  \begin{tabular}{|c|c|}
\hline
  On-shell 5-D   ${\cal N}=2$ $U(1)$
    & Off-shell 4-D ${\cal N}=1$ \\
   Gauged Supergravity  &   Conformal Supergravity \\
\hline
  $\widehat{e}_\alpha^{~m}$:~~ $1/2 \times (5-2) (5-2+1)-1=5$   & $e_\mu^{~r}$:~~
   $4\times 4-4-6-1=5$
     \\
  \hline
 $\widehat{A}_\alpha$:~~ $5-2=3$    & $A_\mu$:~~ $4-1=3$  \\
 \hline
$\widehat{\psi}_\alpha^a$:~~ $1/2 \times (5-3) \times 4\times 2=8$ &
 $\psi_\mu$:~~ $4\times 4-4-4=8$
  \\
 \hline
  $8+8$  & $8+8$ \\
  \hline
     \end{tabular}

\vspace{4mm}

 \caption{Field correspondence and matching of degrees
of freedom between  on-shell five-dimensional ${\cal N}=2$
  $U(1)$ gauged supergravity and off-shell
 ${\cal N}=1$ conformal conformal in four dimensions.}

\label{tablen2}
 \end{table}
\end{center}

\begin{table}

  \begin{center}

   \begin{tabular}{|c|c|}
   \hline
  On-shell 5-D  ${\cal N}=4$ $SU(2)\times U(1)$
    & Off-shell 4-D ${\cal N}=2$ \\
   Gauged Supergravity  &   Conformal Supergravity \\
\hline
  $\widehat{e}_\alpha^{~m}$:~~ $1/2\times(5-2) (5-2+1)-1=5$  & $e_\mu^{~r}$:~~
  $4\times 4-4-6-1=5$
    \\
  \hline
 $\widehat{W}_\alpha^I$:~~$(5-2)\times 3=9$  &  $A_{\mu j}^i$:~~$(4-1)\times 3=9$   \\
 \hline
 $\widehat{A}_\alpha$:~~$5-2=3$    & $a_\mu$:~~ $4-1=3$  \\
 \hline
 $ \widehat{B}_{\alpha\beta}^p$:~~$1/2\times 3\times 2\times 2=6$ & $B_{\mu\nu}$:~~ $1/2 \times
 4\times 3=6$  \\
 \hline
 $\widehat{\phi}$: ~~ 1    & $\phi$:~~ 1    \\
  \hline
  $\widehat{\psi}_\alpha^a$:~~$1/2 \times (5-3) \times 4\times 4=16$  &
 $\psi_\mu^i$:~~
 $(4\times 4-4-4)\times 2=16$
  \\
 \hline
 $\widehat{\chi}^a$:~~$1/2 \times 4 \times 4=8$  & $\chi^i$:~~ $4\times 2=8$  \\
 \hline
  24+24  & 24+24 \\
  \hline
 \end{tabular}

\vspace{4mm}

\caption{Field correspondence and degree of freedom matching
    between  five-dimensional on-shell  ${\cal N}=4$
    $SU (2)\times U(1)$ gauged supergravity and off-shell
   ${\cal N}=2$ conformal supergravity in four dimensions.}
\end{center}
\label{tablen4}
 \end{table}

\begin{table}
  \begin{center}

  \begin{tabular}{|c|c|}
\hline
  On-shell 5-D  ${\cal N}=8$ $SO(6)$
    & Off-shell 4-D ${\cal N}=4$ \\
   Gauged Supergravity  &   Conformal Supergravity \\
\hline
  $\widehat{e}_\alpha^{~~m}$:~~$1/2\times(5-2) (5-2+1)-1=5$
   & $e_\mu^{~r}$:~~$4\times 4-4-6-1=5$
   \\
  \hline
 $\widehat{W}_\alpha^{IJ}$:~~$(5-2)\times 15=45$   &$V_{\mu j}^i$:~~$(4-1)\times (4^2-1)=45$
   \\
 \hline
 $ \widehat{B}_{\alpha\beta}^{Ip}$:~~$1/2\times 3\times 2\times 6\times 2=36$
  &$B_{\mu\nu}^{-[ij]}$:~~ $1/2 \times
 4\times 3 \times 6=36$  \\
 \hline
  $V^{IJab}$,~ $V_{Ip}^{ab}$:  & $\phi$:~~ 2  \\
   & $E_{ij}$:~~$1/2\times 4\times (4+1)\times 2=20$   \\
    42   & $D^{ij}_{~~kl}$:~~$6\times 6-4\times 4=20$ \\
           \hline
 $\widehat{\psi}_\alpha^a$:~~ $1/2 \times (5-3) \times 4\times 8=32$  &
 $\psi_\mu^i$:~~
 $(4\times 4-4-4)\times 4=32$
   \\
 \hline
$\widehat{\chi}^{abc}$:~~  $ 1/2\times (56-8)\times 4=96$ &
$\chi^i$:~~ $4\times 4=16$   \\
  &$\lambda^{ij}_{~~k}$:~~ $ (6\times 4-4)\times 4=80$ \\
  \hline
  128+128  & 128+128 \\
  \hline
     \end{tabular}

\vspace{4mm} \caption{Field correspondence and degree of freedom
matching between  on-shell five-dimensional ${\cal N}=8$ $SO(6)$
gauged supergravity and off-shell
    ${\cal N}=4$ conformal supergravity in four dimensions.}

  \end{center}
\label{table3}
 \end{table}

\subsection{Off-shell ${\cal N}=2$ Conformal Supergravity in Four Dimensions from
On-shell ${\cal N}=4$ $SU(2)\times U(1)$ Gauged $AdS_5$ Supergravity
}

All  ${\cal N}=2k<8$ (k=1,2,4) supergravities in five dimensions can
be obtained through consistent truncations on ${\cal N}=8$
supergravity. This procedure is actually the dimensional reduction
in the representation spaces of global $E_{6(6)}$ group and local
$USp(8)$ group in ${\cal N}=8$ supergravity \cite{crem2}. In the
following we introduce ${\cal N}=4$ $SU(2)\times U(1)$ gauged
supergravity in five dimensions \cite{crem2,awada,roma,agat} and
show how its classical dynamics near $AdS_5$ vacuum configuration
leads to the off-shell conformal supergravity in four dimensions.
The ungauged ${\cal N}=4$ supergravity in five dimensions has both
global $USp(4)$ and local $USp(4)$ symmetries. The field content
consists of a graviton $\widehat{e}_\alpha^{~m}$, four gravitini
$\widehat{\psi}_{\alpha a}$ and four spin-1/2 fields
$\widehat{\chi}_a$
 in the fundamental representation \textbf{4} of the
global and local $USp(4)$, five vector fields
$\widehat{W}_\mu^{[ab]}$ in  representation \textbf{5} of $USp(4)$,
 one vector field $\widehat{A}_\mu$ and one scalar field $\widehat{\phi}$ in
the trivial representation of $USp(4)$. All the fermionic fields are
$USp(4)$ symplectic Majorana spinors, i.e., they satisfy
$\overline{\widehat{\lambda}}^a
=\widehat{\lambda}_a^\star\widehat{\gamma}_0=\widehat{\lambda}^{iT}
\widehat{C}$. The indices $a,b=1,\cdots,4$ label the fundamental
representation of $USp(4)$. The fundamental representation of
$USp(4)$ is actually a $SO(5)$ spinor representation  since there
exists the isomorphism $USp(4)\cong \mbox{Spin}(5)$. Thus, the
various representations of $USp(4)$ can be obtained with
 the representation of  five Euclidean dimensional Clifford algebra
 generated by $(\Gamma_i)_a^{~b}$,
\begin{eqnarray}
\left(\Gamma_i\right)_a^{~c}\left(\Gamma_j\right)_c^{~b}
+\left(\Gamma_j\right)_a^{~c}\left(\Gamma_i\right)_c^{~b}
=2\delta_{ij}\delta_a^{~b} \label{5dclifa}
\end{eqnarray}
where $i=1,\cdots,5$ are the vector indices of Spin(5). In
particular,
 the adjoint
representation of $USp(4)\cong \mbox{Spin}(5)$ is given by
$(\Gamma_{ij})_a^{~b}={i}/{2}\,[\Gamma_i,\Gamma_j]_a^{~b}$.

The Lagrangian density for the ungauged case is \cite{awada}
\begin{eqnarray}
\widehat{e}^{-1}{\cal L}_{\rm ungauged} &=& -\frac{1}{2}
\widehat{R}[\widehat{\omega}]-\frac{1}{2}\overline{\widehat{\psi}}^a_\alpha
\widehat{\gamma}^{\alpha\beta\delta}
\widehat{\nabla}_\beta\widehat{\psi}_{\delta a}-\frac{1}{4}\xi^2
\widehat{W}_{\alpha\beta}^{ab}
\widehat{W}_{ab}^{\alpha\beta}-\frac{1}{4}\xi^{-4}
\widehat{F}_{\alpha\beta}\widehat{F}^{\alpha\beta}
-\frac{1}{2}\overline{\widehat{\chi}}^a\widehat{\gamma}^\alpha
\widehat{\nabla}_\alpha \widehat{\chi}_{a}\nonumber\\
&&-\frac{1}{2}\partial^\alpha\widehat{\phi}\partial_\alpha\widehat{\phi}
-\frac{1}{2}i\overline{\widehat{\chi}}^a\widehat{\gamma}^\alpha
\widehat{\gamma}^\beta\widehat{\psi}_{\alpha
a}\partial_\beta\widehat{\phi} +\frac{\sqrt{3}}{6}\xi
\overline{\widehat{\chi}}_a\widehat{\gamma}^\alpha
\widehat{\gamma}^{\beta\delta}\widehat{\psi}_{\alpha b}
\widehat{W}_{\beta\delta}^{ab}-\frac{1}{2\sqrt{6}}\xi^{-2}
\overline{\widehat{\chi}}^a
\widehat{\gamma}^\alpha\gamma^{\beta\delta}
\widehat{\psi}_{\alpha a}\widehat{F}_{\beta\delta}\nonumber\\
&&-\frac{i}{12}\xi
\overline{\widehat{\chi}}_a\widehat{\gamma}^{\beta\delta}
\widehat{\psi}_{\alpha b} \widehat{W}_{\beta\delta}^{ab}
+\frac{5i}{24\sqrt{2}}\xi^{-2}\overline{\widehat{\chi}}^a
\widehat{\gamma}^{\alpha\beta}\widehat{\chi}_a
\widehat{F}_{\alpha\beta}- \frac{1}{4}i\xi
\left(\overline{\widehat{\psi}}^a_\alpha
\widehat{\gamma}^{\alpha\beta\delta\sigma}\widehat{\psi}_\beta^b
+2\overline{\widehat{\psi}}^{\delta a}\widehat{\psi}^{\sigma
b}\right) \widehat{W}_{\delta\sigma ab}
\nonumber\\
&&-\frac{i}{8\sqrt{2}}\xi^{-2}\left(\overline{\widehat{\psi}}^a
\widehat{\gamma}^{\alpha\beta\delta\sigma}\widehat{\psi}_{\beta a}
+2\overline{\widehat{\psi}}^{\delta
a}\widehat{\psi}^{\sigma}_a\right) \widehat{F}_{\delta\sigma}
+\frac{\sqrt{2}}{8}\widehat{e}^{-1}\epsilon^{\alpha\beta\gamma\delta\sigma}
\widehat{W}_{\alpha\beta}^{ab}
\widehat{W}_{\gamma\delta ab} \widehat{A}_\sigma\nonumber\\
&&+\mbox{four-fermion terms}, \label{fgauge}
\end{eqnarray}
where
\begin{eqnarray}
 \xi &=& \exp (1/\sqrt{3}\widehat{\phi}),\nonumber\\
\widehat{F}_{\alpha\beta}&=& \partial_\alpha \widehat{A}_\beta
-\partial_\beta \widehat{A}_\alpha,\nonumber\\
\widehat{W}_{\alpha\beta}^{ab} &=& \partial_\alpha
\widehat{W}_\beta^{[ab]}-\partial_\beta \widehat{W}_\alpha^{[ab]}
+[\widehat{W}_\alpha,\widehat{W}_\beta]^{[ab]}.
\end{eqnarray}

The gauge group is the subgroup $SU(2)\times U (1)$ of  global
$USp(4)$ group \cite{roma}. The reason for choosing such a gauge
group is based on the existence of  $AdS_5$ solution to the ${\cal
N}=4$ gauged supergravity. The ${\cal N}=4$ anti-de Sitter
supergroup in five dimensions is $SU(2,2|2)$ and its maximal bosonic
subgroup is $SU(2,2)\times SU(2)\times U(1)$, while $SU(2,2)\cong
SO(4,2)$  is the space-time symmetry group of $AdS_5$, thus one can
gauge the $SU(2)\times U(1)$ group to get an ${\cal N}=4$ gauged
supergravity. To complete this gauge procedure one should first
branch  field representations of $USp(4)$ into the representations
$SU(2)\times U(1)$.In this way the transformation behaviors of
fields with respect to gauge group $SU(2)\times U(1)$ can be
naturally obtained from the $USp(4)$  representations in the
ungauged case. First, the adjoint representation of $USp(4)$
decomposes into
\begin{eqnarray}
10 &=& 3\oplus 1\oplus 3\oplus 3 \nonumber\\
J_{ij}&= & J_{IJ}\oplus J_{45}\oplus J_{I4}\oplus J_{I5},
\end{eqnarray}
where $I,J,K=1,2,3$ denote $SU(2)$ group indices. Note that above
generators have  representations in terms of Clifford algebra
generators , one has the representation
\begin{eqnarray}
  J_{ij}&=& \frac{i}{2}[\Gamma_i,\Gamma_j],\nonumber\\
  J_{IJ}&=&\frac{i}{2}[\Gamma_I,\Gamma_J]\sim
  \epsilon_{IJK}\Gamma^{K45},~~\nonumber\\
  J_{45} &=& i\Gamma_4\Gamma_5,~~J_{4I}=i\Gamma_4\Gamma_I,~~
  J_{5I}=i\Gamma_5\Gamma_I.
\end{eqnarray}
 The spinor and vector representations  of  the global $USp(4)$
 are branched with respect to $SU(2)\times U(1)$ as the following,
  \begin{eqnarray}
  4&=&2_{1/2}\oplus2_{-1/2};\nonumber\\
 5 &=& 3_0\oplus 1_1\oplus 1_{-1}.
 \label{groupd}
  \end{eqnarray}
That is,  the five vector fields $\widehat{W}_\alpha^{[ab]}$ convert
into the $SU(2)$ gauge fields $\widehat{W}_\alpha^I \Gamma_{I45}$
and two vector fields $\widehat{W}_\alpha^p$ ($p=1,2$) with
$SO(2)\cong U(1)$-charged. Further,
 $\widehat{W}_\alpha^p$ are replaced their Hodge dual
 $\widehat{B}_{\alpha\beta}^p$ and specifically the tensor
field $\widehat{B}_{\alpha\beta}^p$  should satisfy self-dual
condition, $\widehat{B}^p\propto \epsilon^{pq} \widehat{B}_q$, which
keeps $\widehat{B}^p$ (equivalently $\widehat{W}_\alpha^p$
massless). The vector $\widehat{W}_\alpha^I$ and
$\widehat{A}_\alpha$ shall act as $SU(2)\times U(1)$ gauge fields
with the coupling constant $g_1$ and $g_2$. Therefore, the
$SU(2)\times U(1)$ covariant derivatives acting on a spinor
$\widehat{\psi}_a$, a vector $\widehat{X}^{I p}$ and scalar
$\widehat{\phi}$ should take the form,
\begin{eqnarray}
\widehat{D}_\alpha\psi_a &=&
\widehat{\nabla}_\alpha\widehat{\psi}_a+\frac{1}{2}g_1
\widehat{A}_\alpha (\Gamma_{45})_a^{~b}
\widehat{\psi}_b+\frac{1}{2}g_2 \widehat{W}_\alpha^I(\Gamma_{I45})_a^{~b}
\widehat{\psi}_b, \nonumber\\
\widehat{D}_\alpha X^{I p}&=&\widehat{\nabla}_\alpha \widehat{X}^{I
p}+g_1 \widehat{A}_\alpha\epsilon^{pq}
\widehat{X}^{I q}+g_2\epsilon^{IJK} \widehat{W}_\alpha^J \widehat{X}^{K p},
\nonumber\\
\widehat{D}_\alpha \widehat{\phi} &=&
\partial_\alpha\widehat{\phi}-ig_1 \widehat{A}_\alpha\widehat{\phi}. \label{gcderi}
\end{eqnarray}

The Lagrangian of gauged ${\cal N}=4$ supergravity with the gauge
group $SU(2)\times U(1)$ is \cite{roma}
\begin{eqnarray}
\widehat{e}^{-1}{\cal L}
&=&-\frac{1}{4}\widehat{R}[\widehat{\omega}]
-\frac{1}{2}i\overline{\widehat{\psi}}_\alpha^a
\widehat{\gamma}^{\alpha\beta\delta}\widehat{D}_\beta
\widehat{\psi}_{\delta a}+\frac{1}{2}i\overline{\widehat{\chi}}^a
\widehat{\gamma}^\alpha \widehat{D}_\alpha
\widehat{\chi}_a+\frac{1}{2}\widehat{D}^\alpha\widehat{\phi}
\widehat{D}_\alpha\widehat{\phi}-\frac{1}{4}\xi^{-4}
\widehat{F}^{\alpha\beta}\widehat{F}_{\alpha\beta}\nonumber\\
&&-\frac{1}{4}\xi^2\left(\widehat{W}^{I\alpha\beta
}\widehat{W}^{I}_{\alpha\beta} +\widehat{B}^{\alpha\beta
p}\widehat{B}_{\alpha\beta}^p \right)
+\frac{1}{4}\widehat{e}^{-1}\epsilon^{\alpha\beta\gamma\delta\sigma}
\left(\frac{1}{g_1} \epsilon_{pq}\widehat{B}_{\alpha\beta}^p
\widehat{D}_\gamma \widehat{B}_{\delta\sigma}^q-
\widehat{W}_{\alpha\beta}^I \widehat{W}_{\gamma\delta}^I
\widehat{A}_\sigma \right)\nonumber\\
&&+\frac{i}{4\sqrt{2}}\left(H_{\alpha\beta}^{ab}+\frac{1}{\sqrt{2}}
h_{\alpha\beta}^{ab} \right)
\overline{\widehat{\psi}}_a^\delta\widehat{\gamma}_{[\delta}
\widehat{\gamma}^{\alpha\beta}\widehat{\gamma}_{\sigma
]}\widehat{\psi}_b^\sigma
+\frac{i}{2\sqrt{6}}\left(H_{\alpha\beta}^{ab}
-\sqrt{2}h_{\alpha\beta}^{ab}\right)
\overline{\widehat{\psi}}_a^\delta
\widehat{\gamma}^{\alpha\beta}\widehat{\gamma}_{\delta]}\chi_b\nonumber\\
&&-\frac{i}{12\sqrt{2}}\left(H_{\alpha\beta}^{ab}
-\frac{5}{\sqrt{2}}h_{\alpha\beta}^{ab}\right)
\overline{\widehat{\chi}}_a
\widehat{\gamma}^{\alpha\beta}\widehat{\chi}_b
+\frac{i}{\sqrt{2}}\widehat{D}_\beta\widehat{\phi}
\overline{\widehat{\psi}}^a_\alpha\widehat{\gamma}^\beta
\widehat{\gamma}^\alpha \widehat{\chi}_a+\frac{3}{2}iT_{ab}
\overline{\widehat{\psi}}^a_\alpha\widehat{\gamma}^{\alpha\beta}
\widehat{\psi}_\beta^b\nonumber\\
&&-iA_{ab}\widehat{\psi}_\alpha^a\widehat{\gamma}^\alpha
\widehat{\chi}^b+ i\left(\frac{1}{2}T_{ab}-\frac{1}{\sqrt{3}}A_{ab}
\right)\overline{\widehat{\chi}}^a \widehat{\chi}^b
+\left(3T_{ab} T^{ab}-\frac{1}{4}A^{ab}A_{ab}\right)\nonumber\\
&&+P[\widehat{\phi}]+\mbox{four-fermion terms}, \label{nadsgauge4}
\end{eqnarray}
where
\begin{eqnarray}
H_{\alpha\beta}^{ab} &=&\xi\left[\widehat{W}_{\alpha\beta}^I
(\Gamma_I)^{ab}
+\widehat{B}_{\alpha\beta}^p (\Gamma_p)^{ab} \right),\nonumber\\
h_{\alpha\beta}^{ab}&=&\xi^{-2}\Omega^{ab} \widehat{F}_{\alpha\beta},\nonumber\\
T^{ab} &=&
\left(\frac{1}{6\sqrt{2}}g_2\xi^{-1}+\frac{1}{12}g_1\xi^2\right)
(\Gamma_{45})^{ab},\nonumber\\
A^{ab} &=& \left(\frac{1}{6\sqrt{2}}g_2\xi^{-1}-\frac{1}{2\sqrt{3}}
g_1\xi^2\right)(\Gamma_{45})^{ab},
\end{eqnarray}
and the scalar potential $P[\widehat{\phi}]$ is
\begin{eqnarray}
P[\widehat{\phi}] &=& 3T_{ab} T^{ab}-\frac{1}{4}A^{ab}A_{ab}\nonumber\\
&=& \frac{1}{8}g_2 \left( g_2\xi^{-2}+2\sqrt{2}g_1\xi\right).
\label{spot}
\end{eqnarray}
 The corresponding supersymmetry transformations to the
leading order of fermionic terms is listed as following \cite{roma},
\begin{eqnarray}
\delta \widehat{e}_\alpha^{~m} &=&
i\overline{\widehat{\psi}}^a_\alpha
\widehat{\gamma}^m \widehat{\epsilon}_a,\nonumber\\
\delta \widehat{W}_\alpha^I &=&
\frac{i}{\sqrt{2}}\xi^{-1}\left(-\overline{\widehat{\psi}}^a
\widehat{\epsilon}^b
+\frac{1}{\sqrt{3}}\overline{\widehat{\chi}}^a\widehat{\gamma}_\alpha
\widehat{\epsilon}^b\right)
(\Gamma^I)_{ab},\nonumber\\
\delta \widehat{B}_{\alpha\beta}^p &=& 2
\widehat{D}_{[\alpha}\left[\frac{i}{\sqrt{2}}\xi^{-1}
\left(-\overline{\widehat{\psi}}^a_{\beta
]}\widehat{\epsilon}^b+\frac{1}{\sqrt{3}}\overline{\widehat{\chi}}
\widehat{\gamma}_{\beta]}\widehat{\epsilon}^b\right)
(\Gamma^p)_{ab} \right]\nonumber\\
&&-\frac{1}{\sqrt{2}}ig_1\epsilon^{pq} \left(\Gamma_q\right)_{ab}
\xi\left(
\overline{\widehat{\psi}}^a_{[\alpha}\gamma_{\beta]}\widehat{\epsilon}^b+
\frac{1}{2\sqrt{3}}\overline{\widehat{\chi}}^a
\widehat{\gamma}_{\alpha\beta}\widehat{\epsilon}^b\right),
\nonumber\\
\delta \widehat{A}_\alpha &=& \frac{i}{2}\xi^2 \left(
\overline{\widehat{\psi}}^a_\alpha\widehat{\epsilon}_a
+\frac{2}{\sqrt{3}}\overline{\widehat{\chi}}^a\widehat{\gamma}_\alpha
\widehat{\epsilon}_a\right),\nonumber\\
\delta \widehat{\phi} &=& \frac{i}{\sqrt{2}}
\overline{\widehat{\chi}}^a\widehat{\epsilon}_a,\nonumber\\
\delta \widehat{\psi}_{\alpha}^a &=&
\widehat{D}_\alpha\widehat{\epsilon}^a+\widehat{\gamma}_\alpha
T^{ab}\widehat{\epsilon}_b
-\frac{1}{6\sqrt{2}}\left(\widehat{\gamma}_\alpha^{~\beta\delta}
-4\delta_\alpha^{~\beta}\gamma^\delta
 \right)\left(H_{\beta\delta}^{ab}+\frac{1}{\sqrt{2}}h_{\beta\delta}^{ab}
  \right)\widehat{\epsilon}_b,
\nonumber\\
\delta \widehat{\chi}_a
&=&\frac{1}{\sqrt{2}}\widehat{D}_\alpha\widehat{\phi}
\widehat{\epsilon}_a+A_{ab}\widehat{\epsilon}^b
-\frac{1}{2\sqrt{6}}\widehat{\gamma}^{\alpha\beta}\left(
H_{\alpha\beta ab}-\sqrt{2}h_{\alpha\beta ab}
\right)\widehat{\epsilon}^b. \label{nfsupert01}
\end{eqnarray}

It was found in Ref.\,\cite{roma} that when $g_2=\sqrt{2}g_1$ there
exists a classical $AdS_5$ solution to this gauged supergravity
which preserves the full ${\cal N}=4$ supersymmetry
\cite{roma,vn,war}. The explicit form of $AdS_5$ solution is the
same as that given in  (\ref{metrican1}) and all other fields
vanish. The cosmological constant $\Lambda$ comes from the scalar
potential at its extremum $\phi=0$
 \footnote{ Note that this convention is
different from that in Ref.\,\cite{roma}.}
\begin{eqnarray}
\Lambda=4 P[\widehat{\phi}_0=0]=\frac{3}{2}g^2=\frac{12}{l^2}.
\end{eqnarray}
and subsequently
\begin{eqnarray}
g=\frac{2\sqrt{2}}{l}. \label{rcar}
\end{eqnarray}

We expand above ${\cal N}=4$ gauged supergravity around the $AdS_5$
vacuum configuration and rescale on bulk gauge fields and gauge
coupling,
\begin{eqnarray}
\widehat{W}_\alpha^I \longrightarrow l \widehat{W}_\alpha^I, ~~
\widehat{A}_\alpha \longrightarrow l\widehat{A}_\alpha,~~~ g
\longrightarrow l\,g.
\end{eqnarray}
 The linearized classical action  near above $AdS_5$ vacuum configuration
 is
\begin{eqnarray}
\widehat{e}^{-1}{\cal L} &=&-\frac{1}{4}
\widehat{R}[\widehat{\omega}]
-\frac{1}{2}i\overline{\widehat{\psi}}_\alpha^a
\widehat{\gamma}^{\alpha\beta\delta}\widehat{D}_\beta\widehat{\psi}_{\delta
a}+\frac{1}{2}i\overline{\widehat{\chi}}^a
\widehat{\gamma}^\alpha \widehat{D}_\alpha\chi_a
+\frac{1}{2}\partial^\alpha\widehat{\phi}
\partial_\alpha\widehat{\phi}\nonumber\\
&&-\frac{l^2}{4}  \widehat{F}^{\alpha\beta}
\widehat{F}_{\alpha\beta}-\frac{1}{4}\left(l^2 \widehat{W}^{I
\alpha\beta} \widehat{W}^{I}_{\alpha\beta}
+\widehat{B}^{\alpha\beta p}\widehat{B}_{\alpha\beta}^p \right)\nonumber\\
&&+\frac{1}{4}\widehat{e}^{-1}\epsilon^{\alpha\beta\gamma\delta\sigma}
\left(\frac{l}{2} \epsilon_{pq} \widehat{B}_{\alpha\beta}^p
\widehat{D}_\gamma \widehat{B}_{\delta\sigma}^q-
l^3 \widehat{W}_{\alpha\beta}^I \widehat{W}_{\gamma\delta}^I
 \widehat{A}_\sigma \right)\nonumber\\
&&+\frac{il}{4\sqrt{2}}\left[\widehat{W}_{\alpha\beta}^I
\left({\Gamma}_I\right)^{ab} +\frac{1}{l}
\widehat{B}_{\alpha\beta}^p \left({\Gamma}_p \right)^{ab}
+\frac{1}{\sqrt{2}}\Omega^{ab}\widehat{F}_{\alpha\beta}\right]
\overline{\widehat{\psi}}_a^\delta\widehat{\gamma}_{[\delta}
\widehat{\gamma}^{\alpha\beta}\widehat{\gamma}_{\sigma ]}
\widehat{\psi}_b^\sigma\nonumber\\
&& +\frac{il}{2\sqrt{6}} \left[ \widehat{W}_{\alpha\beta}^I
\left({\Gamma}_I\right)^{ab} +\frac{1}{l}\widehat{B}_{\alpha\beta}^p
\left({\Gamma}_p \right)^{ab} -{\sqrt{2}}\Omega^{ab}
\widehat{F}_{\alpha\beta}\right] \overline{\widehat{\psi}}_a^\delta
\widehat{\gamma}^{\alpha\beta}\widehat{\gamma}_{\delta ]}
\widehat{\chi}_b\nonumber\\
&&-\frac{il}{12\sqrt{2}} \left[ \widehat{W}_{\alpha\beta}^I
\left({\Gamma}_I\right)^{ab} +\frac{1}{l}\widehat{B}_{\alpha\beta}^p
\left({\Gamma}_p \right)^{ab} -\frac{5}{\sqrt{2}}\Omega^{ab}
\widehat{F}_{\alpha\beta}\right] \overline{\widehat{\chi}}_a
\gamma^{\alpha\beta}\widehat{\chi}_b\nonumber\\
&& +\frac{i}{\sqrt{2}}\partial_\alpha\widehat{\phi}
\overline{\widehat{\psi}}^a_\beta\widehat{\gamma}^\alpha
\widehat{\gamma}^\beta\widehat{\chi}_a
+\frac{3i}{4l}\left({\Gamma}_{45}\right)_{ab}
\overline{\widehat{\psi}}^a_\alpha\widehat{\gamma}^{\alpha\beta}
\widehat{\psi}_\beta^b +\frac{i}{4l}\left({\Gamma}_{45}\right)_{ab}
\overline{\widehat{\chi}}^a\widehat{\chi}^b\nonumber\\
&& +\frac{3}{l^2}+\mbox{four-fermion terms}. \label{gauge4}
\end{eqnarray}
The supersymmetry transformation law (\ref{nfsupert01}) at the
leading order of fermionic terms becomes
\begin{eqnarray}
\delta \widehat{e}_\alpha^{~m} &=&
i\overline{\widehat{\psi}}^a_\alpha\widehat{\gamma}^m
\widehat{\epsilon}_a,\nonumber\\
\delta  \widehat{W}_\alpha^I &=& \frac{i}{\sqrt{2}l}
\left(-\overline{\widehat{\psi}}^a_\alpha \widehat{\epsilon}^b
+\frac{1}{\sqrt{3}}\overline{\widehat{\chi}}^a\widehat{\gamma}_\alpha
\widehat{\epsilon}^b\right)
({\Gamma}^I)_{ab},\nonumber\\
\delta  \widehat{A}_\alpha &=& \frac{i}{2l}\left(
\overline{\widehat{\psi}}^a_\alpha \widehat{\epsilon}_a
+\frac{2}{\sqrt{3}}\overline{\widehat{\chi}}^a
\widehat{\gamma}_\alpha \widehat{\epsilon}_a\right),\nonumber\\
\delta \widehat{B}_{\alpha\beta}^p &=& \frac{i}{\sqrt{2}}
\widehat{D}_\alpha \left(-\overline{\widehat{\psi}}^a_{\beta}
\widehat{\epsilon}^b+\frac{1}{\sqrt{3}} \overline{\widehat{\chi}}^a
\widehat{\gamma}_{\beta}{\widehat{\epsilon}}^b\right)
({\Gamma}^p)_{ab}
\nonumber\\
&& -\frac{i}{\sqrt{2}} \widehat{D}_\beta
\left(-\overline{\widehat{\psi}}^a_{\alpha}
\widehat{\epsilon}^b+\frac{1}{\sqrt{3}} \overline{\widehat{\chi}}^a
\widehat{\gamma}_{\alpha} \widehat{\epsilon}^b\right)
({\Gamma}^p)_{ab}\nonumber\\
&&-\frac{i}{\sqrt{2}l}\epsilon^{pq} \left({\Gamma}_q\right)_{ab}
\left( \overline{\widehat{\psi}}^a_{\alpha}\widehat{\gamma}_{\beta}
\widehat{\epsilon}^b
-\overline{\widehat{\psi}}^a_{\beta}\widehat{\gamma}_{\alpha}
\widehat{\epsilon}^b+ \frac{1}{\sqrt{3}}\overline{\widehat{\chi}}^a
\widehat{\gamma}_{\alpha\beta} \widehat{\epsilon}^b\right),
\nonumber\\
\delta \widehat{\phi} &=& \frac{i}{\sqrt{2}}
\overline{\widehat{\chi}}^a \widehat{\epsilon}_a,\nonumber\\
\delta\widehat{\psi}_{\alpha a} &=& \widehat{D}_\alpha
\widehat{\epsilon}_a+\frac{1}{2l}\widehat{\gamma}_\alpha
\left({\Gamma}_{45}\right)_{ab} \widehat{\epsilon}^b \nonumber\\
&& -\frac{l}{6\sqrt{2}}\left(\widehat{\gamma}_\alpha^{~\beta\delta}
-4\delta_\alpha^{~\beta}\widehat{\gamma}^\delta
 \right)\left[ \widehat{W}_{\beta\delta}^I \left({\Gamma}_I\right)_{ab}
+\frac{1}{l}\widehat{B}_{\beta\delta}^p\left({\Gamma}_p\right)_{ab}
+\frac{1}{\sqrt{2}}\Omega_{ab} \widehat{F}_{\beta\delta}\right]
\widehat{\epsilon}^b,
\nonumber\\
\delta \widehat{\chi}_a
&=&\frac{1}{\sqrt{2}}\widehat{\gamma}^\alpha\partial_\alpha\widehat{\phi}
\widehat{\epsilon}_a
-\frac{l}{2\sqrt{6}}\widehat{\gamma}^{\alpha\beta}\left[
\widehat{W}_{\alpha\beta}^I \left({\Gamma}_I\right)_{ab}
+\frac{1}{l} \widehat{B}_{\alpha\beta}^p \left({\Gamma}_p
\right)_{ab} -{\sqrt{2}}\Omega_{ab} \widehat{F}_{\alpha\beta}
\right] \widehat{\epsilon}^b. \label{nfsupert}
\end{eqnarray}

The linearized equations of motion for ${\cal N}=4$ $SU(2)\times
U(1)$ gauged supergravity multiplet $\Phi=$
($\widehat{e}_{\alpha}^{~m}$, $\widehat{W}_\alpha^I$,
 $\widehat{A}_\alpha$, $\widehat{B}_{\alpha\beta}^p$, ${\phi}$,
 $\widehat{\psi}_{\alpha a}$, $\widehat{\chi}_a$) are the
Einstein equation and the following equations,
\begin{eqnarray}
\widehat{e}^{-1}\partial_\alpha (\widehat{g}^{\alpha\delta}
\partial_\delta \widehat{W}^I_\beta)=0;
\end{eqnarray}
\begin{eqnarray}
\widehat{e}^{-1}\partial_\alpha (\widehat{g}^{\alpha\delta}
\partial_\delta \widehat{A}_\beta)=0;
\end{eqnarray}
\begin{eqnarray}
\widehat{e}^{-1}\partial_\alpha
\left(\widehat{e}\widehat{g}^{\alpha\beta}\partial_\beta
\widehat{\phi}\right) -\frac{\partial P (\widehat{\phi})} {\partial
\widehat{\phi}}=0; \label{scalar01}
\end{eqnarray}
\begin{eqnarray}
 \widehat{B}_{~\alpha\beta}^{p} &=& \frac{3l}{2} \widehat{e}\epsilon^{pq}
\epsilon_{\alpha\beta\gamma\delta\sigma} \widehat{D}^{[\gamma}
\widehat{B}^{\delta\sigma]}_{~~~q} = \frac{3l}{2}\epsilon^{pq}
\widehat{e} {\epsilon}_{\alpha\beta\gamma\delta\sigma}
\widehat{g}^{\gamma\gamma^\prime}  \widehat{g}^{\delta\delta^\prime}
 \widehat{g}^{\sigma\sigma^\prime} \widehat{D}_{[\gamma^\prime}
 \widehat{B}_{\delta^\prime\sigma^\prime]q}; \label{bequation}
\end{eqnarray}
\begin{eqnarray}
\widehat{\gamma}^\alpha \widehat{D}_\alpha
\widehat{\chi}_a+\frac{1}{2}\left({\Gamma}_{45}\right)_{ab}
\widehat{\chi}^b=0; \label{se1}
\end{eqnarray}
\begin{eqnarray}
\widehat{\gamma}^{\alpha\beta\delta}\widehat{D}_\beta
\widehat{\psi}_{\delta a}-\frac{3}{2l} \left(
{\Gamma}_{45}\right)_{ab}\Omega^{bc}
\widehat{\gamma}^{\alpha\beta}\widehat{\psi}_{\beta c}=0.
\label{se20}
 \end{eqnarray}
The following task is to find the solutions  to these equations
which asymptotically approach $AdS_5$ vacuum  at $r\to 0$.  A
detailed process of solving these linearized equations is displayed
in Ref.\,\cite{chch2}. As in ${\cal N}=2$ case,  the gauge choices
in radial direction and  the solutions to above equations \emph{near
$AdS_5$ boundary} take the following forms \cite{chch2},
\begin{eqnarray}
&& \widehat{e}_\mu^{~s} (x,r)\sim \frac{l}{r} {e}_\mu^{~s}(x), ~~
   \widehat{e}_4^{~r}=\widehat{e}_{\mu}^{~4}= 0,
   ~~\widehat{e}_4^{~4}= \frac{l}{r};
 \nonumber\\
 && \widehat{A}_\mu (x,r) \sim {A}_\mu (x), ~~ A_4 (x,r)=0; \nonumber\\
 && \widehat{W}_\mu^I (x,r) \sim {W}_\mu (x), ~~ W_4^I (x,r)=0; \nonumber\\
&&  \widehat{\phi} (x,r) \sim \left(\frac{r}{l}\right)^2 \phi (x);\nonumber\\
&& \widehat{B}_{\mu\nu}^p (x,r)\sim \left(\frac{r}{l}\right)^{-1/3}
B_{\mu\nu}^-(x); \nonumber\\
&& \widehat{\chi}^a (x,r)\sim \left(\frac{r}{l}\right)^{3/2}\chi^i
(x);
\nonumber\\
&& \widehat{\psi}_\mu^a (x,r) \sim
\left(\frac{r}{l}\right)^{-1/2}\psi^i_\mu (x), ~~\widehat{\psi}_4^a
(x,r)=0; \label{nere}
\end{eqnarray}
In above equation we need to make some explanations on near-$AdS_5$
boundary behaviors of antisymmetric field
$\widehat{B}_{\alpha\beta}^p$ and fermionic fields $\chi^a$,
$\psi_\alpha^a$ \cite{chch2}. First, we define
\begin{eqnarray}
\widehat{B}_{\alpha\beta}=\frac{1}{\sqrt{2}}
\left(\widehat{B}_{\alpha\beta}^4
-i \widehat{B}_{\alpha\beta}^5\right).
\end{eqnarray}
This is actually converting the $SO(2)$ vector representation on
$\widehat{B}_{\mu\nu}^p$ into $U(1)$ representation on
$\widehat{B}_{\mu\nu}$. Further, we decompose $\widehat{B}_{\mu\nu}$
into self-dual and anti-self-dual sectors,
\begin{eqnarray}
\widehat{B}_{\mu\nu}^{\pm}&=& \frac{1}{2}\left( \widehat{B}_{\mu\nu}
\pm \widetilde{\widehat{B}}_{\mu\nu} \right);\nonumber\\
\widetilde{\widehat{B}}_{\mu\nu}&=& \frac{1}{2}\widehat{e}
\epsilon_{\mu\nu\lambda\rho}
\widehat{g}^{\lambda\sigma}\widehat{g}^{\rho\delta}
\widehat{B}^{\lambda\delta}.
 \label{santd}
\end{eqnarray}
 $\widehat{B}_{4\mu}^p$ is  not an independent quantity and its
 relation with $\widehat{B}_{\mu\nu}^p$ is given by
  $(\tau\mu)$ component of Eq.\,(\ref{bequation}).
\begin{eqnarray}
 \widehat{B}_{~\tau\mu}^p=\frac{3r}{2l}e\epsilon^{pq}
\epsilon_{\mu\nu\lambda\rho} g^{\nu\sigma}
g^{\lambda\delta}g^{\rho\xi} D_{[\sigma} \widehat{B}_{\delta\xi]q}.
\label{homb}
\end{eqnarray}
This equation also shows that the radial coordinate dependence
$\widehat{B}_{~\tau\mu}^p$ is one power higher than $\widehat{
B}_{~\mu\nu}^p$.

As for the near-$AdS_5$ boundary behavior of on-shell fermionic
fields $\psi_\mu^a$ and $\chi^a$, they are in the fundamental
representation of $USp(4)$. According to Eq.\,(\ref{groupd}), these
fermions should decompose into two $SU(2)$ doublets with opposite
$U(1)$ charges. We choose following explicit representations of
${\Gamma}$-matrices for the five-dimensional Euclidean Clifford
algebra (\ref{5dclifa}) \cite{chch2},
\begin{eqnarray}
{\Gamma}^{1} =1 \otimes \sigma_1,~~
 {\Gamma}^{2}= 1 \otimes \sigma_2,~~
{\Gamma}^{3} =\sigma_3 \otimes \sigma_3,~~ {\Gamma}^{4}= \sigma_1
\otimes \sigma_3~~  {\Gamma}^{5} =  \sigma_2\otimes \sigma_3,
\label{rsgamma}
\end{eqnarray}
and hence obtain the representation for $SU(2)\times U(1)$
generators,
\begin{eqnarray}
Q &=& -\frac{i}{2}{\Gamma}^{45} =\frac{1}{2}\sigma_3 \otimes 1
=\left(\begin{array}{cc} 1/2 & 0 \\
                           0 & -1/2 \end{array}\right),
\nonumber\\
 T^{I} &=& -\frac{i}{2}{\Gamma}^{I45},\nonumber\\
 T^1 &=& \frac{1}{2}\sigma_3\otimes \sigma_1, ~~ T^2
=\frac{1}{2}\sigma_3\otimes \sigma_2  ~~ T^3 =\frac{1}{2}
1 \otimes \sigma_3 \nonumber\\[2mm]
 [Q,T^I]&=& 0, ~~ [T^I,T^J]=i
{\epsilon}^{IJK} T^K. \label{epre}
\end{eqnarray}
These explicit representations for $SU(2)\times U(1)$ generators
determines that the first two components of $\Psi_a$ in the
fundamental representation of $USp(4)$ should be chosen as the
$SU(2)$ doublet with $U(1)$ charge $+1/2$ and the last two
components as the doublet of $U(1)$ charge $-1/2$ \cite{chch2},
\begin{eqnarray}
\left(\Psi_a\right)=\left(\begin{array} {c} \Psi_1 \\ \Psi_2 \\
\Psi_3 \\
\Psi_4 \end{array}\right)=\left(\begin{array} {c} \Psi_i \\
\Psi^i
\end{array}\right),~~
\left(\Psi_i\right)=\left(\begin{array}{c} \Psi_1 \\ \Psi_2
\end{array}\right), ~~~ \left(\Psi^i\right)
= \left(\Psi_i^\star\right) =\left(\begin{array}{c}
 \Psi_3 \\ \Psi_4
\end{array}\right),~~i=1,2.
\label{uspr1}
\end{eqnarray}
In above equation, $\Psi_a$ represents either $\widehat{\psi}_\mu^a$
or $\widehat{\chi}^a$. Therefore, the equation of motion (\ref{se1})
((\ref{se20}) for $\widehat{\chi}^a$ ($\widehat{\psi}_\mu^a$)
reduces to that for $\widehat{\psi}^i_\mu$ ($\widehat{\chi}^i$).
Further, like in the ${\cal N}=2$ case, we  perform chiral
decomposition on the four-dimensional spinors
 $\widehat{\psi}_\mu^i$ and $\widehat{\chi}^i$ and  the
 corresponding reductions of equations of motion on chiral
 spinors give the
 asymptotical behaviors of fermionic fields shown in
 Eq.\,(\ref{nere}).
 Note that $\psi_\mu^i (x)$ ($\chi^i (x)$) is
  $SU(2)$ symplectic  Majorana spinors constructed from the right-handed
 spinor $\psi_{\mu}^{Ri} (x)$ ($\chi^{Ri} (x)$. The near-$AdS_5$
 boundary behavior of the
   left-handed spinor $\widehat{\psi}_{\mu}^{Li} (x)$
 ($\widehat{\chi}^{Li} (x)$ is not independent, and it can be expressed  in terms
 of the right-handed spinor $\psi_{\mu}^{Ri} (x)$ ($\chi^{Ri} (x)$)
 \cite{chch2}. Substituting Eq.\,(\ref{nere}) into the supersymmetry
 transformation law (\ref{nfsupert})
 and making the same operation on the supersymmetry transformation
 parameter $\widehat{\epsilon}^a$ as $\widehat{\chi}^a$, we obtain
 the reduction of bulk supersymmetry transformation on $AdS_5$
 boundary up to the leading order of fermionic fields \cite{chch2},
\begin{eqnarray}
 \delta {e}_\mu^{~s}&=&\frac{1}{2}
\overline{{\epsilon}}^{i}
{\gamma}^{s}{\psi}_{\mu i},\nonumber\\
\delta {\psi}_\mu^i &=& {D}_\mu {\epsilon}^i
-\frac{1}{2}{\gamma}^{\nu\rho} {B}_{\nu\rho}^-{\gamma}_\mu
{\epsilon}^i -{\gamma}_\mu {\eta}^i,
 \nonumber\\
  \delta {W}_{\mu~j}^{~i}
 &=& \left(-\overline{\zeta}_\mu^i\gamma_5\epsilon_j
 +\frac{1}{2}\delta^i_{~j}\overline{\xi}_\mu^k\gamma_5
 {\epsilon}_k
 -\overline{{\psi}}^i\gamma_5{\eta}_j
 +\frac{1}{2}\delta^i_{~j}\overline{{\psi}}_\mu^k\gamma_5\eta_k
 \right),\nonumber\\
 \delta A_\mu &=& -i\overline{\zeta}_\mu^i\gamma_5
 {\epsilon}_i
 +i\overline{{\psi}}_\mu^i \gamma_5{\eta}_i,\nonumber\\
 \delta \phi &=& \frac{1}{2}\overline{\epsilon}^i {\chi}_i, \nonumber\\
 \delta {B}_{\mu\nu}^- &=&
\overline{{\epsilon}}^i
 R_{\mu\nu i}[Q]
+\overline{\epsilon}^i
{\gamma}_{\mu\nu}{\chi}_i,\nonumber\\
 \delta {\chi}_i &=&
 \gamma^\mu\partial_\mu \phi {\epsilon}_i
 +\frac{1}{2}\gamma^{\mu\nu} B_{\mu\nu}^-\epsilon_i.
 \label{final1}
 \end{eqnarray}
where ${\xi}_\mu^i$ and  the curvature $R_{\mu\nu}^i[Q]$
corresponding to local supersymmetry are defined as the following,
\begin{eqnarray}
{R}_{\mu\nu\,i}[Q]&=& {D}_\mu{\psi}_{\nu i} -{D}_\mu{\psi}_{\nu i}
-\left({\gamma}_\mu{\psi}_{\nu i} -{\gamma}_\nu {\psi}_{\mu i}
\right),\nonumber\\
{\xi}_\mu^i &=& \frac{1}{3}{\gamma}^\nu \left( {D}_\nu {\psi}_\mu^i
-{D}_\mu {\psi}_\nu^i+\frac{1}{2}
\gamma_5{e}\epsilon_{\mu\nu\lambda\rho} {D}^\lambda
{\psi}^{i\rho}\right). \label{losucur}
\end{eqnarray}
Eq.\,(\ref{final1}) is exactly the supersymmetry transformation law
${\cal N}=2$ conformal supergravity in four dimensions. Further, we
find that the bulk diffeomorphism transformation law of the on-shell
field on $AdS_5$ boundary reduces to the four-dimensional
deffeomorphism- and Weyl transformations and that the bulk
$SU(2)\times U(1)$ gauge transformations on $AdS_5$ boundary convert
into the axial $SU(2)\times U(1)$ gauge transformation. These
reductions of symmetry transformations determine that the ${\cal
N}=4$ five-dimensional $SU(2)\times U(1)$ gauged supergravity near
its $AdS_5$ vaccum configuration behaves like the ${\cal N}=2$
off-shell conformal supergravity in four dimensions. The matching of
degrees of freedom  between on-shell multiplet
($\widehat{e}_\alpha^{~m}$, $\widehat{W}_\alpha^I$,
$\widehat{B}^p_{\alpha\beta}$, $\widehat{A}_\alpha$,
$\widehat{\phi}$, $\widehat{\psi}_\alpha^a$, $\widehat{\chi}^a$)
 of ${\cal N}=4$ $SU(2)\times U(1)$ gauged $AdS_5$ supergravity
  and off-shell
multiplet ($e_\mu^{~r}$, $W_\mu^I$, $A_\mu$, $\phi$, $B_{\mu\nu}^-$,
$\psi_\mu^i$, $\chi^i$) of ${\cal N}=2$ conformal supergravity in
four dimensions is listed in Table II.

\subsection{On-shell ${\cal N}=8$ $SO(6)$ Gauged $AdS_5$ Supergravity and
  Off-shell ${\cal N}=4$ Conformal Supergravity in Four Dimensions}

In this subsection, we review ${\cal N}=8$  $SO(6)$ gauged
supergravity in five dimensions and argue that its classical
dynamical behavior near $AdS_5$ vacuum configuration leads to ${\cal
N}=4$ conformal supergravity in four dimensions.

The ungauged ${\cal N}=8$ supergravity in five dimensions can be
obtained from the dimensional reduction of ${\cal N}=1$ supergravity
in eleven dimensions \cite{crem1,crem2}. It has global $E_{6(6)}$
symmetry and local $USp(8)$ $R$-symmetry. If counted in terms of the
representation of $USp(8)$ $R$-symmetry group, the field content
consists of 1 graviton $\widehat{e}_{\alpha}^{~m}$, 8 gravitini
$\widehat{\psi}_{\alpha}^a$, 27 vector fields
$\widehat{A}_{\alpha}^{ab}$, 48 spin-1/2 fields
$\widehat{\chi}^{abc}$ and 42 scalar fields $\widehat{\phi}^{abcd}$,
${\alpha}=0,\cdots,4$ and ${m}=0,\cdots,4$ denote the
five-dimensional space-time and local $SO(1,5)$ Lorentz indices,
respectively, and $a,b,c,d=1,\cdots,8$ label fundamental
representation indices of $USp(8)$. The above particle spectrum can
also be viewed from the representation of the global $E_{6(6)}$
group by introducing a 27-bein $V_{AB}^{~~~ab}$, $A,B=1,\cdots,8$,
which is an element of the coset group $E_{6(6)}/USp(8)$ and
furnishes both the 27-dimensional vector representation of $USp(8)$
and the fundamental representation of $E_{6(6)}$. According to this
viewpoint the spinor fields $\widehat{\psi}_{\alpha}^a$ and
$\widehat{\chi}^{abc}$ still behave as the $USp(8)$ tensors, but
those 27 $USp(8)$ vector fields $\widehat{A}_\alpha^{ab}$ should be
considered as in the fundamental representation of $E_{6(6)}$ with
the identification
$A_\mu^{AB}=\widetilde{V}^{AB}_{~~~ab}A_\mu^{ab}$. $
\widetilde{V}^{AB}_{~~~ab}$ is the inverse of $V_{AB}^{~~~ab}$, it
furnishes the conjugate representation $\overline{27}$ of $E_{6(6)}$
 and satisfies the following relation,
\begin{eqnarray}
\widetilde{V}^{AB}_{~~~ab}V_{AB}^{~~~cd}=
\frac{1}{2}\left(\delta^c_a\delta^d_b -\delta^d_a\delta^c_b
\right)+\frac{1}{8}\Omega_{ab}\Omega^{cd},
\end{eqnarray}
where $\Omega_{ab}$ and $\Omega^{ab}$ are the symplectic metric of
$USp(8)$ and its inverse, respectively. The 42 scalar fields
$\widehat{\phi}^{abcd}$  parameterize the non-compact coset space
$E_{6(6)}/USp(8)$ since $E_{6}$ and $USp(8)$ have 78 and 36
generators, respectively.  Based on this observation, one can
replace these  scalar fields $\widehat{\phi}^{abcd}$ by $
\widetilde{V}^{AB}_{~~~ab}$ or equivalently by  vector fields
$P_{\alpha abcd}$ defined by
\begin{eqnarray}
\widetilde{V}^{AB}_{~~~cd}\partial_\alpha V_{AB}^{~~~ab}= 2
Q_{\alpha [c}^{~~~[a}\delta^{~b]}_{d]}+P_{\alpha cd}^{~~~ab}.
\label{27b}
\end{eqnarray}
$Q_{\alpha a}^{~~~b}$  are   $USp(8)$ Lie algebra valued and
transform as gauge fields of $USp(8)$, and $P_{\alpha abcd}$ are in
the 42-dimensional representation $USp(8)$ exactly like  scalar
fields $\widehat{\phi}^{abc}$. The above process is actually
furnishing  field function representation to $E_{6}$ group through
its maximal subgroup $USp(8)$.

The Lagrangian density for the ungauged ${\cal N}=8$ Poincar\'{e}
supergravity in five dimensions can thus be constructed based on the
global $E_{6(6)}$ and local $USp(8)$ symmetries as well as the
${\cal N}=8$ supersymmetry \cite{crem1,crem2},
\begin{eqnarray}
\widehat{e}^{-1}{\cal L}_{\rm
ungauged}&=&-\frac{1}{4\kappa^2}\widehat{R}[\widehat{\omega}]
-\frac{1}{2}i\overline{\widehat{\psi}}^a_{\alpha}
\widehat{\gamma}^{\alpha\beta\delta}\widehat{\nabla}_\alpha
\widehat{\psi}_{\delta
a}-\frac{1}{8}G_{AB,CD}\widehat{F}^{AB}_{\alpha\beta}
\widehat{F}^{\alpha\beta
CD}+\frac{1}{12}i\overline{\widehat{\chi}}^{abc}
\widehat{\gamma}^\alpha\widehat{\nabla}_\alpha \chi_{abc}
\nonumber\\
&& -\frac{1}{24\kappa^2}P_{\alpha abcd}P^{\alpha abcd}
-\frac{1}{12}\widehat{e}^{-1}\epsilon^{\alpha\beta\gamma\delta\sigma}
\widehat{F}_{\alpha\beta~B}^{~~A}
\widehat{F}_{\gamma\delta~C}^{~~B}\widehat{A}_{\sigma~A}^{~C}
+\frac{i}{3\sqrt{2}}P_{\alpha
abcd}\overline{\widehat{\psi}}^a_\beta\widehat{\gamma}^\alpha
\widehat{\gamma}^\beta \chi^{bcd}\nonumber\\
&&+\frac{1}{4}i\kappa V_{AB}^{~~~ab}\widehat{F}_{\alpha\beta}^{~~AB}
\left( \overline{\widehat{\psi}}_a^\delta
\widehat{\gamma}_{[\delta}\widehat{\gamma}^{\alpha\beta}
\widehat{\gamma}_{\sigma]}\widehat{\psi}^\sigma_b
+\frac{i}{\sqrt{2}}\overline{\widehat{\psi}}^c_\delta
\widehat{\gamma}^{\alpha\beta}\widehat{\gamma}^\delta
\widehat{\chi}_{abc} +\frac{1}{2}\overline{\widehat{\chi}}_{acd}
\widehat{\gamma}^{\alpha\beta}\widehat{\chi}_b^{~cd}
 \right)\nonumber\\
 &&+\mbox{four-fermion terms},
\end{eqnarray}
where $\widehat{F}_{\alpha\beta}^{~~AB}$ is the field strength
corresponding to the vector field $A_\alpha^{~AB}$,
\begin{eqnarray}
\widehat{F}_{\alpha\beta}^{~~AB}=\partial_\alpha
\widehat{A}_\beta^{~AB}-\partial_\beta \widehat{A}_\alpha^{~AB}
+[\widehat{A}_\alpha,\widehat{A}_\beta]^{AB};
\end{eqnarray}
$G_{AB,CD}$ is $E_{6(6)}$ covariant and $USp(8)$ invariant metric,
 \begin{eqnarray}
 G_{AB,CD}=V_{AB}^{~~~ab}\Omega_{ac}\Omega_{bd} V_{CD}^{~~~cd};
 \end{eqnarray}
 The covariant derivative $\widehat{\nabla}_\alpha$ acting on spinor fields
 is defined with respect to both  spin connection $\widehat{\omega}_\alpha$
  and $USp(8)$ gauge field $Q_{\alpha~b}^{~a}$,
 \begin{eqnarray}
 \widehat{\nabla}_\alpha \widehat{\psi}_\beta^a
 &=&\left( \partial_\alpha \delta^a_{~b}
 -Q_{\alpha~b}^{~a}+
 \frac{1}{4}\widehat{\omega}_{\alpha\,mn}\widehat{\gamma}^{mn}\delta^a_{~b}
 \right)\widehat{\psi}^b_\beta, \nonumber\\
 \widehat{\nabla}_\alpha\widehat{\chi}^{abc} &=&\left( \partial_\alpha
 \delta^{[a}_{~~d}-3Q_{\mu~~d}^{~[a}+
 \frac{1}{4}\widehat{\omega}_{\alpha mn}\widehat{\gamma}^{mn}
 \delta^{[a}_{~~d}
 \right)\widehat{\chi}^{dbc]}.
 \end{eqnarray}

The gauging of above supergravity is just turning some of the 27
vector
 fields $A_\alpha^{~AB}$ into gauge fields of certain simple subgroup
 of  $E_{6(6)}$ \cite{guna2}. That is, the global symmetry group
 of $E_{(6)}$ in the ungauged ${\cal N}=8$ supergravity should break to
 the gauge group. It is obvious that one cannot convert
 all of those 27 vector
 fields into gauge fields since there exists no  simple group with
27 generators. Thus the gauge group should be a subgroup of
$E_{6(6)}$ with dimensionality less than 27. This determines that
the possible gauge group should be $SO(6)$ and its noncompact real
forms $SO(n, 6-n)$, $1<n\leq 3$. This is because $SO(n,6-n)$,
($0\leq n \leq 3$) are all  subgroups of $SL(6,R)$, while
$SL(6,R)\times SL(2,R)$ is the maximal subgroup of $E_{6(6)}$. To
complete this gauging procedure, one should first rearrange  above
field representations of $E_{6(6)}$ in terms of the representations
realized on its maximal subgroup $SL(6,R)\times SL(2,R)$. That is,
we branch $E_{6(6)}$ with respect to $SL(6,R)\times SL(2,R)$. As a
consequence, the vector fields $\widehat{A}_\alpha^{~AB}$, the
27-bein $V_{AB}^{~~~ab}$ and its inverse $V^{AB}_{~~~ab}$, all of
which are in the 27- or $\overline{27}$-dimensional fundamental
representation $E_{6(6)}$, decompose as
\begin{eqnarray}
27&=&(\overline{15},1)\oplus(6,2), ~~~
\overline{27}=(15,1)\oplus(\overline{6},\overline{2}),
\nonumber\\
\widehat{A}_\alpha^{AB}&=&\left(\begin{array}{c} \widehat{A}_{\alpha IJ}\\
\widehat{A}^{K p}_\alpha\end{array}\right), ~~~\widehat{A}_{\alpha
IJ}=-\widehat{A}_{\alpha JI}; \nonumber\\
V_{AB}^{~~~ab}&=& \left(V^{IJab}, V_{Ip}^{~~ab} \right), ~~~
\widetilde{V}^{AB}_{~~~ab}=\left(\widetilde{V}_{IJab},
\widetilde{V}^{Ip}_{~~ab} \right),
 \label{decom}
\end{eqnarray}
and there exist that
\begin{eqnarray}
V^{IJab}\widetilde{V}_{abKL}=\delta^{IJ}_{~~~KL},~~
V_{Jq}^{~~ab}\widetilde{V}_{ab}^{~~Ip}=\delta^{I}_{~J}\delta^p_{~q},~~
V_{KL}^{~~~ab}\widetilde{V}_{ab}^{~~Ip}
={V}_{Ipab}\widetilde{V}_{abIJ}=0.
\end{eqnarray}
In above equations, $I,J,K=1,\cdots,6$ and $p=1,2$ are the
fundamental representation indices of $SL(6,R)$ and  $SL(2,R)$,
respectively. The vector fields $\widehat{A}_{\alpha IJ}$ naturally
become  gauge fields for
 $SO(n,6-n)$ group. For any quantity $X_{aI}$ which used to
transform with respect to certain representation of the global
$SL(6,R)$ and local $USp(8)$ in the ungauged theory, the covariant
derivative is defined as
\begin{eqnarray}
\widehat{D}_\alpha \widehat{X}_{aI}{\equiv}\partial_\alpha
\widehat{X}_{aI}+Q_{\alpha
a}^{~~~b}\widehat{X}_{bI}-g\widehat{A}_{\alpha
IJ}\eta^{JK}\widehat{X}_{aK}.
\end{eqnarray}
However,  the remained 12 vector fields $\widehat{A}_\alpha^{I p}$
transform nontrivially under the gauge group $SO(n,6-n)$. To keep
these 12 vector fields massless, one must replace them by its Hodge
dual, the second-rank antisymmetric field
$\widehat{B}_{\alpha\beta}^{I p}$ and specifically
$\widehat{B}_{\alpha\beta}$ should satisfy the following self-dual
field equation,
\begin{eqnarray}
\widehat{B}^{I p}&=& c\, \star \widehat{G}^{Ip},\nonumber\\
\widehat{B}^{I p} &=& \frac{1}{2}\widehat{B}_{\alpha\beta}^{Ip}dx^\alpha \wedge  dx^\beta,\nonumber\\
\widehat{G}&=& d\widehat{B},
\end{eqnarray}
where $c$ is certain constant. The Abelian symmetry $\widehat{B}\to
\widehat{B}+d \widehat{\Lambda}$ prevents  $\widehat{B}_\alpha^{I
p}$ from becoming massive.

 The gravitational field $\widehat{e}_\alpha^{~m}$
 keeps intact since it is  an $E_{6(6)}$ singlet. In the following, let us
 turn to the conversion of the 42 scalar fields  in the gauging
 process. As stated above, these scalar fields parametrize the
 non-compact space $E_{(6)}/USp(8)$ and are described by
 27-bein $V_{AB}^{~~~cd}$ or its inverse
 $\widetilde{V}_{ab}^{~~AB}$.
Since  $V_{AB}^{~~~cd}$ (or $\widetilde{V}_{ab}^{~~AB}$) transforms
as $\overline{27}$ (or $27$) under $E_{(6)}$ and $\overline{27}$
($27$) under $Usp(8)$, it should be branched with respect to
$SL(6,R)\times SL(2,R)$ as shown in Eq.\,(\ref{decom}). To keep
$E_{(6)}/USp(8)$ coset space structure  for scalar fields as that in
the ungauged case, one just simply generalizes Eq.\,(\ref{27b}) in a
gauge covariant way and defines $P_\alpha^{~abcd}$ as the covariant
derivative of the 27-bein with respect to $SO(n,6-n)$ gauge fields
$\widehat{A}^{IJ}_\alpha$ and $USp(8)$ connection,
\begin{eqnarray}
\widetilde{V}^{abAB}D_\alpha V_{AB}^{~~~cd}=P_\alpha^{~abcd} \equiv
P_\alpha^{~[abcd]}. \label{covd1}
\end{eqnarray}
These definitions determine that in the gauged case
 the $USp(8)$ connection $Q_{\alpha a}^{~~b}$ can
 be expressed  in terms of the 27-bein and $SO(n,6-n)$ gauge fields,
\begin{eqnarray}
Q_{\alpha\,a}^{~~~b} = -\frac{1}{3}\left[
\widetilde{V}^{bcAB}\partial_\alpha V_{ABac}+gA_{\alpha IK}\eta^{JK}
\left(2V_{ac}^{~~IL}\widetilde{V}^{bc}_{~~JL}-V_{Jpac}
\widetilde{V}^{bcIp} \right) \right]
 \label{covd}
\end{eqnarray}

The Lagrangian density of five-dimensional $SO(n,6-n)$ gauged ${\cal
N}=8$ supergravity  takes the following form,
\begin{eqnarray}
\widehat{e}^{-1}{\cal
L}&=&-\frac{1}{4\kappa^2}\widehat{R}[\widehat{\omega}]
-\frac{1}{2}i\overline{\widehat{\psi}}^a_{\alpha}
\widehat{\gamma}^{\alpha\beta\delta}\widehat{D}_\beta
\widehat{\psi}_{\delta a}
+\frac{1}{12}i\overline{\widehat{\chi}}^{abc}
\widehat{\gamma}^\alpha\widehat{\nabla}_\alpha \widehat{\chi}_{abc}
-\frac{1}{24}P_{\alpha abcd}P^{\alpha abcd}
\nonumber\\
&& -\frac{1}{8} \left(\widehat{F}_{\alpha\beta
ab}+\widehat{B}_{\alpha\beta ab}\right)
\left(\widehat{F}^{\alpha\beta ab}+\widehat{B}^{\alpha\beta
ab}\right) +\frac{i}{3\sqrt{2}}P_{\beta
abcd}\overline{\widehat{\psi}}^a_\alpha\widehat{\gamma}^\beta
\widehat{\gamma}^\alpha \widehat{\chi}^{bcd}\nonumber\\
&&+\frac{1}{4}i\kappa \left(\widehat{F}_{\alpha\beta}^{~~ab}
+\widehat{B}_{\alpha\beta}^{~~ab}\right) \left(
\overline{\widehat{\psi}}_a^\delta
\widehat{\gamma}_{[\delta}\widehat{\gamma}^{\alpha\beta}
\widehat{\gamma}_{\sigma]}\widehat{\psi}^\sigma_b
+\frac{i}{\sqrt{2}}\overline{\widehat{\psi}}^c_\delta
\widehat{\gamma}^{\alpha\beta}\widehat{\gamma}^\delta
\widehat{\chi}_{abc} +\frac{1}{2}\overline{\widehat{\chi}}_{acd}
\widehat{\gamma}^{\alpha\beta}\widehat{\chi}_b^{~cd}
 \right)\nonumber\\
 &&-\frac{1}{15}ig T_{ab}\overline{\widehat{\psi}}^a_\alpha
 \widehat{\gamma}^{\alpha\beta}
 \widehat{\psi}^b_\beta
 +\frac{i}{6\sqrt{2}}g A_{abcd}\overline{\widehat{\chi}}^{abc}
 \widehat{\gamma}^\alpha\widehat{\psi}_\alpha^d
 +\frac{1}{2}ig\overline{\widehat{\chi}}^{abc}\left(
 \frac{1}{2}A_{bcde}-\frac{1}{45}
 \Omega_{bd}T_{ce}\right)\widehat{\chi}_a^{~de}\nonumber\\
 && +\frac{1}{96} g^2 \left[ \left(\frac{8}{15}\right)^2
 (T_{ab})^2-(A_{abcd})^2 \right]
 +\frac{1}{8g} \widehat{e}^{-1}\epsilon^{\alpha\beta\gamma\delta\sigma}
 \eta_{IJ}
 \epsilon_{pq}
 \widehat{B}_{\alpha\beta}^{~~I p}\widehat{D}_\gamma
 \widehat{B}_{\delta\sigma}^{~~Jq}
 \nonumber\\
 &&
 -\frac{1}{96}e^{-1}\epsilon^{\alpha\beta\gamma\delta\sigma}
 \epsilon^{IJKLMN}
 \left(\widehat{F}_{IJ\alpha\beta}\widehat{F}_{KL\gamma\delta}
 \widehat{A}_{MN\sigma}
 +g\eta^{PQ}\widehat{F}_{IJ\alpha\beta}\widehat{A}_{KL\gamma}
 \widehat{A}_{MP\delta}\widehat{A}_{QN\sigma}
 \right.\nonumber\\
 &&\left.+\frac{2}{5}g^2\eta^{PQ}\eta^{RS}\widehat{A}_{IJ\alpha}
 \widehat{A}_{KP\beta}\widehat{A}_{QL\gamma}\widehat{A}_{MR\delta}
 \widehat{A}_{QN\sigma}
  \right)
 +\mbox{four-fermion terms}.
 \label{neight}
\end{eqnarray}
In above Lagrangian, the $SO(n,6-n)$ gauge field strength
$\widehat{F}_{\alpha\beta IJ}$ and antisymmetric field
$\widehat{B}_{\alpha\beta}^{~~ab}$ are defined as the following,
\begin{eqnarray}
\widehat{F}_{\alpha\beta IJ}&=&\partial_\alpha \widehat{A}_{\beta
IJ}-\partial_\beta
\widehat{A}_{\alpha IJ}-g[\widehat{A}_\alpha,\widehat{A}_\beta]_{IJ},
\nonumber\\
\widehat{F}_{\alpha\beta}^{~~ab}&=&\widehat{F}_{\alpha\beta IJ}V^{IJ
ab}, ~~\widehat{B}_{\alpha\beta}^{~~ab}
=\widehat{B}_{\alpha\beta}^{~~~I p} V_{I p}^{~~ab}.
\end{eqnarray}
The $USp(8)$ tensors $T$, $A$ and the relevant $W$ are all
constructed from the $SL(6,R)\times SL(2,R)$ components of 27-bein
$V_{AB}^{~~~ab}$ and its inverse $\widetilde{V}^{AB}_{~~~ab}$,
\begin{eqnarray}
 T^a_{~bcd}& \equiv &
\left( 2V^{IKae}\widetilde{V}_{beJK}-V_{J
p}^{~~ae}\widetilde{V}_{be}^{~~Ip} \right)
\eta^{JL}\widetilde{V}_{cdIL},\nonumber\\
W_{abcd} &\equiv
&\epsilon^{pq}\eta^{IJ}V_{Ipab}V_{Jqcd}=-W_{cdab},~~~
A_{abcd} \equiv T_{a[bcd]}\nonumber\\
T_{abcd} &=& \Omega_{ae}T^{e}_{~bcd}=T_{bacd},~~~ T_{ab} \equiv
T^c_{~abc}.
 \label{scalar}
\end{eqnarray}
The above  $USp(8)$ tensors satisfy the following relations and
identities originating from their definitions,
\begin{eqnarray}
&& T_{ab}= T_{ba} = \frac{15}{4}W^c_{~acb},\nonumber\\
&& W_{abcd} = \frac{1}{6}\left(T_{acbd}-T_{bcad}+T_{bdac}-T_{adbc}
\right)+\frac{1}{5}\left(\Omega_{a[c}T_{d]b}-
\Omega_{b[c}T_{d]a}\right),\nonumber\\
&& A_{abcd} = -3W_{a[bcd]},~~~A_{[abcd]}=0,\nonumber\\
&& T_{abcd}W^{cd}_{~~ef}=0, ~~~A^e_{~bcd}W_{a]efg}=0, \nonumber\\
&& 9A^{cde}_{~~~a}A_{cdeb}-\frac{8}{5}T^{cd}A_{cdab}-\left(
A_a^{~cde}A_{bcde} +\Omega_{ab}A^{cdef}A_{cdef}\right)=0,\nonumber\\
&& \frac{6}{(45)^2} T_{ac}T_b^{~c}-\frac{1}{96}A_{cde}A_b^{~cde}=
 -\frac{1}{8}\Omega_{ab} \left[
 \frac{6}{(45)^2} T_{cd}T^{cd}-\frac{1}{96}A_{cdef}A^{cdef}
  \right].
 \label{identities}
\end{eqnarray}
These identities play some roles in proving the supersummetric
invariance of gauged supergravity.

 The supersymmetry transformation laws at the leading order of
fermionic fields are
\begin{eqnarray}
\delta \widehat{e}_\alpha^{~m}&=&
-i\overline{\widehat{\epsilon}}^a\widehat{\gamma}^m
\widehat{\psi}_{\alpha a}, \nonumber\\
\delta \widehat{A}_{\alpha IJ} &=&
2i\left(\overline{\epsilon}^a\widehat{\psi}^b_\alpha
+\frac{1}{2\sqrt{2}} \overline{\epsilon}_c\widehat{\gamma}_\alpha
\widehat{\chi}^{abc}\right)
\widetilde{V}_{abIJ},\nonumber\\
\delta \widehat{B}_{\alpha\beta}^{~~Ip} &=& 2D_{[\alpha } \left[2i
\left(\overline{\widehat{\epsilon}}^a\widehat{\psi}^b_\beta
+\frac{1}{2\sqrt{2}}
\overline{\widehat{\epsilon}}_c\widehat{\gamma}_\beta
\widehat{\chi}^{abc}\right)\widetilde{V}_{ab}^{~~Ip} \right]
\nonumber\\
&& -2ig\eta^{IJ}\epsilon^{pq}V_{J q ab}
\left(\overline{\widehat{\psi}}^a_{[\alpha}
\widehat{\gamma}_{\beta]}\widehat{\epsilon}^{b} +\frac{1}{4\sqrt{2}}
\overline{\widehat{\chi}}^{abc}\widehat{\gamma}_{\alpha\beta}
\widehat{\epsilon}_c\right),\nonumber\\
\delta \widehat{\psi}_{\alpha a} &=&
\widehat{D}_\alpha\widehat{\epsilon}_a-\frac{2}{45}gT_{ab}
\widehat{\gamma}_\alpha \widehat{\epsilon}^b-\frac{1}{6}\left(
\widehat{F}_{\beta\delta ab}+\widehat{B}_{\beta\delta ab}\right)
\left(\widehat{\gamma}^{\beta\sigma}\widehat{\gamma}_\alpha +2
\widehat{\gamma}^\beta \delta^\sigma_{~\alpha} \right)
\widehat{\epsilon}^b,
\nonumber\\
\delta\widehat{\chi}_{abc} &=& \sqrt{2}\widehat{\gamma}^\alpha
P_{\alpha abcd}\widehat{\epsilon}^d -\frac{1}{\sqrt{2}}g
\widehat{\epsilon}^d
A_{dabc}-\frac{3}{2\sqrt{2}}\widehat{\gamma}^{\alpha\beta} \left(
\widehat{F}_{\alpha\beta [ab}+\widehat{B}_{\alpha\beta [ab}
\right)\widehat{\epsilon}_{c]},
\nonumber\\
\left(\begin{array}{c} \delta V^{IJab} \\ \delta
V_{I\alpha}^{~~ab}\end{array}\right)
&=&-2\sqrt{2}i\left(\begin{array}{c} V^{IJ}_{~~~cd} \\ V_{I\alpha
cd}\end{array}\right) \overline{\widehat{\epsilon}}^{[a}
\widehat{\chi}^{bcd]} \label{steight}
\end{eqnarray}

It was found in Ref.\,\cite{guna2} that when the gauge group is
$SO(6)$ this gauged supergravity admits $AdS_5$ classical solution
which preserves the full ${\cal N}=8$ supersymmetry (or equivalent
speaking,  a classical solution exhibiting ${\cal N}=4$ anti-de
Sitter supersymmetry $SU(2,2|4)$) \cite{guna2,vn,war}. The existence
of this solution can be explained as the following
\cite{guna2,vn,war}. Considering a configuration of above gauged
supergravity with only nonvanishing metric and scalar field, one has
the Einstein  equation from the action (\ref{neight}),
\begin{eqnarray}
\widehat{R}_{\alpha\beta}-\frac{1}{2} \widehat{g}_{\alpha\beta}
\widehat{R}=2 P \widehat{g}_{\alpha\beta}. \label{ee}
\end{eqnarray}
In above equation, $P$ is the scalar potential  in (\ref{neight}),
which be rewritten as the following form  with the $T$-tensor
identities,
\begin{eqnarray}
P &=&
-g^2\left[\frac{6}{(45)^2}T_{ab}T^{ab}-\frac{1}{96}A_{abcd}A^{abcd}\right]
\nonumber\\
&=&-\frac{1}{32}g^2\left[2 W_{ab}W^{ab}-W_{abcd}W^{abcd} \right].
\label{tidenty1}
\end{eqnarray}
 Eq.\,(\ref{ee}) shows that if the the
scalar potential has a critical point at certain expectation value
of the scalar field, and the corresponding critical  value of the
scalar potential $P_0<0$,  the $AdS_5$ solution exists with the
cosmological constant $\Lambda=-4/3 P_0$.

To verify explicitly the existence of  such a critical point one has
to re-express the scalar
 potential in an $SO(n,6-n)$ gauge invariant form. So
 $W_{ab}$ and $W_{abcd}$ in the scalar potential should be rewritten
  in terms of the representations of $SL(6,R)\times SL(2,R)$ because
  it is the subgroup of $SL(6,R)$ that is gauged.
  Since $W_{ab}$ and $W_{abcd}$
  are tensors of $USp(8)$ group, the maximal compact subgroup of
  of $E_{6(6)}$, while $SL(6,R)\times SL(2,R)$ is the maximal
  subgroup of $E_{6(6)}$, one should establish an explicit connection
between the representations of $USp(8)$ and of $SL(6,R)\times
SL(2,R)$. This can be done by working out the explicit forms for the
$SL(6,R)\times SL(2,R)$ components of the 27-bein, $V_{AB}^{~~~ab}
=\left(V^{IJab}, V_{Ip}^{~~~ab}\right)$. The method is employing the
representation of $USp(8)$ generators constructed from $SO(7)$
Clifford algebra and the transformation law of the $SL(6,R)\times
SL(2,R)$ components of 27-dimensional basis
$Z^{AB}=(Z_{IJ},Z^{Ip}/\sqrt{2})$ under $E_{6(6)}$ action. This can
be proceeded as the following.  First, with $SO(7)$ gamma matrices
$\Gamma_i$, $i=0,I$, $I=1,\cdots,6$,
$\{\Gamma_i,\Gamma_j\}=2\delta_{ij}$,
 the representation of
generators of $USp(8)$ expressed in terms of its $SU(4)\times U(1)$
subalgebra is
\begin{eqnarray}
\left(A_{IJ} \right)_a^{~b}=\left( \Gamma_{IJ}\right)^{ab},~~
\left(A \right)_a^{~b}=\left( \Gamma_{0}\right)^{ab},~~ \left(
S_{IJK}\right)^{ab}=\left( \Gamma_{IJK}\right)^{ab}=\left(
\Gamma_{IJK}\right)^{ab},
\end{eqnarray}
$A_{IJ}$ and $A$ are generators of $SU(4)\times U(1)$. Using the
following defined $\Gamma$-matrices,
\begin{eqnarray}
\Gamma_{Ip}=\left(\Gamma_I,i\Gamma_I\Gamma_0\right)
\end{eqnarray}
and above $SU(4)$ generators, one can obatin the $USp(8)$ components
$Z^{ab}$ of  the  27-dimensional bases $Z^{AB}$ of $E_{(6)6}$ with
the projection coefficients  being the $SL(6,R)\times SL(2,R)$
components $(Z_{IJ},Z^{Ip})$,
\begin{eqnarray}
Z^{ab}=\frac{1}{4} \left(\Gamma^{IJ} \right)^{ab} Z_{IJ}
+\frac{1}{2\sqrt{2}}\left(\Gamma_{Ip} \right)^{ab} Z^{Ip} =Z^{AB}
V_{AB}^{~~~ab}. \label{rsubgroup}
\end{eqnarray}
The invariant inner product
\begin{eqnarray}
\widetilde{Z}_{ab}Z^{ab}=\widetilde{Z}_{AB}Z^{AB}=
\widetilde{Z}^{IJ}Z_{IJ}+\widetilde{Z}_{Ip}Z^{Ip}
\end{eqnarray}
determines the projection $\widetilde{Z}_{ab}$ in the $USp(8)$
representation space  of dual bases $\widetilde{Z}_{AB}$,
\begin{eqnarray}
\widetilde{Z}_{ab}&=& \frac{1}{4} \left(\Gamma_{IJ} \right)^{ab}
\widetilde{Z}^{IJ} -\frac{1}{2\sqrt{2}}\left(\Gamma^{Ip}
\right)^{AB} \widetilde{Z}_{Ip}= V^{~~AB}_{ab}\widetilde{Z}_{AB}.
\end{eqnarray}
where $\widetilde{Z}_{IJ}$ and $\widetilde{Z}^{Ip}$ are dual bases
in $SL(6,R)\times SL(2,R)$ conjugate representation spaces.

On the other hand, as a fundamental representation of  $E_{6(6)}$,
the infinitesimal transformations of $Z^{AB}$ under the action of
$E_{6(6)}$ read
\begin{eqnarray}
\delta \left(\begin{array}{c} Z_{IJ} \\ Z^{K\alpha}/\sqrt{2}
\end{array}\right) =X \left(\begin{array}{c} Z_{MN} \\ Z^{P\beta}/\sqrt{2}
\end{array}\right).
 \label{infe6tr}
\end{eqnarray}
The  $X$ in above equation is the $27\times 27$ matrix,
\begin{eqnarray}
X=\left(\begin{array}{cc} -4\lambda^{[M}_{~~[I}\delta^N_{~~J]} &
\sqrt{2}\Sigma_{IJP\beta}\\
\sqrt{2}\Sigma^{MNK\alpha} & \Lambda^K_{~~P}\delta^\alpha_{~\beta}+
\delta^K_{~~P}\Lambda^\alpha_{~\beta}
 \end{array}\right)\, ;
\end{eqnarray}
$\Lambda^K_{~~P}$ and $\Lambda^\alpha_{~\beta}$ are $SL(6,R)$ and
$SL(2,R)$ Lie algebra valued infinitesimal matrices, respectively,
and $\Sigma_{MNKp}$ is a real and antisymmetric self-dual tensor of
$SL(6,R)$, $\Sigma^{MNK p}=\Sigma^{[MNK] p}=1/3!
\epsilon^{pq}\epsilon^{IJKLMN} \Sigma_{LMNq}$. Eq.\,(\ref{infe6tr})
implies the following finite transformation of $Z^{AB}$ under the
action of $E_{6(6)}$,
\begin{eqnarray}
Z^\prime_{IJ} &=&\frac{1}{2} U^{~~~MN}_{IJ}Z_{MN}+\frac{1}{\sqrt{2}}
Z^{Lq}U_{LqIJ} , \nonumber\\
Z^{^\prime Kp} &=& Z^{Lq}
U_{Lq}^{~~~Kp}+\frac{1}{\sqrt{2}}U^{KpIJ}Z_{IJ},
\nonumber\\
U &\equiv & \exp \left( X\right). \label{estr}
\end{eqnarray}
Eqs.\,(\ref{rsubgroup}) and (\ref{estr}) give the action of
$E_{6(6)}$ on the $USp(8)$ components $Z^{ab}$ of $Z^{AB}$,
\begin{eqnarray}
Z^{\prime ab}&=&\left[ \frac{1}{8}\left( \Gamma^{IJ}\right)^{ab}
U^{~~~KL}_{IJ}+\frac{1}{4}\left(\Gamma_{Ip}
\right)^{ab}U^{IpKL}\right]Z_{KL}\nonumber\\
&& + Z^{Kq}\left[ \frac{1}{4\sqrt{2}}\left( \Gamma^{IJ}\right)^{ab}
U_{KqIJ}+\frac{1}{2\sqrt{2}}\left(\Gamma_{Ip}
\right)^{ab}U^{~~~Ip}_{Kq}\right]
\end{eqnarray}
and hence yield the 27-bein expressed  in terms of the fundamental
representation ($27\times 27$) matrix of $E_{6(6)}$,
\begin{eqnarray}
V^{KLab} &=& \frac{1}{8}\left( \Gamma^{IJ}\right)^{ab}
U^{~~~KL}_{IJ}+\frac{1}{4}\left(\Gamma_{Ip}
\right)^{ab}U^{IpKL},\nonumber\\
V_{Ip}^{~~~ab} &=& \frac{1}{4\sqrt{2}}\left( \Gamma^{MN}\right)^{ab}
U_{IpMN}+\frac{1}{2\sqrt{2}}\left(\Gamma_{Kq}
\right)^{ab}U^{~~~Kq}_{Ip}. \label{tstrm}
\end{eqnarray}
A substitution (\ref{tstrm}) into (\ref{scalar}) and
(\ref{tidenty1}) converts the $USp(8)$ tensors in the scalar
potential into the $SL(6,R)\times SL(2,R)$ tensors,
\begin{eqnarray}
W^{ab}&=&\frac{1}{8}i\epsilon^{pq}\eta^{IJ}\left\{ \left(
\Gamma^{KL}\Gamma_0\right)^{ab}U_{IpKM}U_{JqLM}-i\delta_{r}^{~s}
\left( \Gamma_{[KL}\Gamma_{M]s}\right)^{ab}U_{Ip}^{~~Mr}U_{JqKL}
\right.
\nonumber\\
&& \left.+\epsilon_{rs}U_{Ip}^{~~Kr}U_{Jq}^{~~Ls}\left[\left(
\Gamma_{KL}\Gamma_0\right)^{ab}+i\delta^{ab}\right]\right\};\nonumber\\
W_{abcd} W^{abcd} &=& \eta^{IK}\eta^{JL}\epsilon^{pr}\epsilon^{qs}
M_{IpJq}M_{KrLs};\nonumber\\
M_{IpJq} &=& (UU^T)_{IpJq}=\frac{1}{2}U_{Ip}^{~~KL}U_{JqKL}
+U_{Ip}^{~~Kr}U_{JqKr}.
\end{eqnarray}
Further, the scalar potential can be reduced to $SL(6,R)$ invariant
form by choosing the cross components $\sigma_{IJKp}$ to vanish.
This gives
\begin{eqnarray}
U^{IJ}_{~~~KL}&=&
2S_{[K}^{~~[I}S_{L]}^{~~J]},~~~U^{IJkp}=U_{KpIJ}=0,
~~~U_{Ip}^{~~Jq}=S_I^{~J}S^{\prime~\beta}_{\alpha},\nonumber\\
&& \left(S_J^{~I}\right)\in
SL(6,R),~~~\left(S^{\prime~\beta}_{\alpha}\right)\in SL(2,R)
\end{eqnarray}
and $USp(8)$ tensor $W_{ab}$  simplifies to
\begin{eqnarray}
W_{ab}=-\frac{1}{4}\mbox{Tr}(\eta M) \delta_{ab}, \label{uspetensor}
\end{eqnarray}
where $M_{IJ}=\left(SS^T\right)_{IJ}=S_I^{~K}S_{JK}$ is the symmetry
SL(6,R) metric. Consequently, the anticipated form of scalar
potential (\ref{tidenty1}) arises,
\begin{eqnarray}
P=-\frac{1}{32}g^2\left\{ \left[\mbox{Tr}(\eta M) \right]^2 -2
\left[\mbox{Tr}(\eta M \eta M )\right] \right\},
\end{eqnarray}
Since $M$ is a symmetric matrix of $SL(6,R)$, one can  always employ
$SO(6)$ gauge symmetry to make $M$ diagonal by its subgroup $SO(6)$
($\det M=1$) ,
\begin{eqnarray}
M=\mbox{diag}\left\{ e^{2\lambda_1}, e^{2\lambda_2}, e^{2\lambda_3},
e^{2\lambda_4}, e^{2\lambda_5},e^{2\lambda_6} \right\},~~~
\sum_{\lambda_i} \lambda_i=0. \label{diagform}
\end{eqnarray}
It was shown that only when all $\lambda_i=0$, i.e., the gauge group
is $SO(6)$, the  $AdS_5$ vacuum configuration exists. According to
Eq.\,(\ref{uspetensor}), the critical points of scalar fields are
\begin{eqnarray}
W_{ab}=-\frac{3}{2}\delta_{ab}=-\left(-\frac{3P_0}{g^2}
\right)^{1/2}\delta_{ab} \label{critensor}
\end{eqnarray}
 and the cosmological constant comes from the critical value of
the scalar potential,
\begin{eqnarray}
\Lambda=-\frac{4}{3}P_0=g^2. \label{ccpo}
\end{eqnarray}

 The preservation on ${\cal N}=8$ supersymmetry in this
 $AdS_5$ space-time background can be revealed by observing
the Killing spinor equation obtained from the supersymmetric
transformations for the fermionic fields $\widehat{\psi}_\alpha^a$
and $\widehat{\chi}_{abc}$
\begin{eqnarray}
\delta \widehat{\psi}_{\alpha}^a &=& \widehat{\nabla}_\alpha
\widehat{\epsilon}^a-\frac{1}{6}
g\Omega^{ab} W_{bc}\widehat{\gamma}_\alpha \widehat{\epsilon}^c=0,
 \label{killeq1}\\
 \delta \widehat{\chi}^{abc} &=& -\frac{1}{\sqrt{2}}gA_{abcd}
 \widehat{\epsilon}^d=0.
 \label{killeq2}
\end{eqnarray}
It was verified that these two Killing equations are equivalent and
trivially satisfied in above $AdS_5$ vacuum configuration. This can
be shown as the following. First, Eq.\,(\ref{killeq1}) leads to the
integrability condition,
\begin{eqnarray}
\left[\left(\widehat{R}_{\alpha\beta\delta\sigma}
\widehat{\gamma}^{\delta\sigma}\right) \delta^a_{~b} -\frac{2}{9}g^2
W^{ac}W_{bc}\widehat{\gamma}_{\alpha\beta}\right]\widehat{\epsilon}^b=0.
\label{cond}
\end{eqnarray}
This equation and the Riemannian curvature tensor of $AdS_5$
space-time
\begin{eqnarray}
 \widehat{R}_{\alpha\beta\delta\sigma} &=&
\frac{1}{4}\Lambda
\left(\widehat{g}_{\alpha\delta}\widehat{g}_{\beta\sigma}-
\widehat{g}_{\alpha\sigma}\widehat{g}_{\beta\delta} \right)\nonumber\\
&=& -\frac{1}{3}P_0
\left(\widehat{g}_{\alpha\delta}\widehat{g}_{\beta\sigma}-
\widehat{g}_{\alpha\sigma}\widehat{g}_{\beta\delta} \right)
\end{eqnarray}
imply that $\epsilon^a$ should be an eigenvector of the real
symmetric matrix $W_{ab}$,
\begin{eqnarray}
(W^2)^a_{~b}\widehat{\gamma}_{\alpha\beta}\widehat{\epsilon}^b=
-\frac{3}{g^2}P_0\widehat{\gamma}_{\alpha\beta}\widehat{\epsilon}^a.
\label{eigeneq}
\end{eqnarray}
The cosmological constant $\Lambda$ and the critical point $W_{ab}$
 given in Eqs.\,(\ref{critensor}) and (\ref{ccpo})
commit Eq.\,(\ref{cond}) to stand identically. Further, the
contraction of $\epsilon^a$ with  the last identity in
Eq.\,(\ref{identities}) yields
\begin{eqnarray}
\widehat{\epsilon}^a A_{acde} A_b^{~cde} =4\widehat{\epsilon}^a
\left[ (W^2)_{ab}-\frac{3}{g^2} P_0\Omega_{ab}
 \right].
\label{contraeq}
\end{eqnarray}
Eqs.\,(\ref{eigeneq}) and (\ref{contraeq}) determine that those two
Killing equations (\ref{killeq1}) and (\ref{killeq2}) are
equivalent. Since there is no constraint on the Killing spinor
$\widehat{\epsilon}^a$,  the $AdS_5$ vacuum configuration thus
preserves the full ${\cal N}=8$ supersymmetry.

 We see from above discussion that in the gauging process the global
$E_{6(6)}$ symmetry first breaks to its maximal subgroup
$SL(6,R)\times SL(2,R)$ and then $SL(6,R)$ breaks to its subgroup
$SO(6)$. The $SL(2,R)$ remains as a global symmetry in the
five-dimensional ${\cal N}=8$  gauged supergravity, which can be
considered as the $SL(2,R)$ symmetry  of type IIB supergravity if
the gauged supergravity is obtained from the compactification of
type IIB supergravity on $S^5$. Originally in the ungauged case
there are 42 scalar fields, two of which should be interpreted as
the dilaton and axion.  The quantities $W_{ab}$ and $W_{abcd}$
consisting of the scalar potential  are the complicated
$SL(6,R)\times SL(2,R)$ invariant combination of scalar fields. It
turned out that $W_{ab}$ and $W_{abcd}$ depends only on 20 scalar
fields $M_{IJ}$, which is obtained from the combination of those 40
scalar fields. $(M_{IJ})$ is symmetric matrix and parametrizes the
coset space $SL(6,R)/SO(6)$. Further, the $SO(6)$ symmetry can take
$M$ to be the diagonal form (\ref{diagform}). The shows that finally
there are only 5 scalar fields in the theory. Further, using an
orthonormal parametrization \cite{fgpw1},
$\lambda_I=\alpha_{Ii}\widehat{\phi}_i$, $i=1,\cdots,5$,
 one can replace $\alpha_I$ of Eq.\,(\ref{diagform})
  by five independent fields $\widehat{\phi}_i$. The explicit form
  if $\left(\alpha_{Ii}\right)$ is given in Ref.\,\cite{fgpw1}.
 The classical
 Lagrangian consisting of only gravitational and scalar fields
 is
 \begin{eqnarray}
 \widehat{e}^{-1}{\cal L}=-\frac{1}{4}\widehat{R}+\frac{1}{2}\partial_\alpha
 \widehat{\phi}_i\partial^\alpha\widehat{\phi}^i-P[\widehat{\phi}],
 \label{effect1}
 \end{eqnarray}
and the scalar potential can be written as
\begin{eqnarray}
P[\widehat{\phi}]=\frac{g^2}{8}\sum_{i=1}^5 \left(\frac{\partial
W}{\partial \widehat{\phi}_i} \right)^2-\frac{g^2}{3} W^2,.
\label{effect2}
\end{eqnarray}
Near the critical point, $\widehat{\phi}_i=0$, one has the
approximation expansions
\begin{eqnarray}
W[\widehat{\phi}] &= & -\left[\frac{3}{2}+\frac{1}{4}
\sum_{i=1}^5\left(\widehat{\phi}_i\right)^2+{\cal O}(\widehat{\phi})
 \right]
\nonumber\\
P[\widehat{\phi}] &= &
-\frac{3}{4}g^2-\frac{3}{32}g^2\sum_{i=1}^5\left(\widehat{\phi}_i\right)^2
+{\cal O}(\widehat{\phi}).
\end{eqnarray}

mmmmmmm

We can in principle  do the same thing on  ${\cal N}=8$ $SO(6)$
gauged supergravity: expanding the theory around the $AdS_5$ vacuum
and observing the solution to linearized equation of motion whether
they constitute an off-shell multiplet for ${\cal N}=4$ conformal
supergravity in four-dimensions. However, in practice, this
procedure  is much more complicated to implement than the ${\cal
N}=2,4$ cases. Despite of the complication of this procedure,  there
are quite a number of evidences supporting this identification
between on-shell five-dimensional ${\cal N}=8$ gauged $AdS_5$
supergravity and off-shell ${\cal N}=4$ conformal supergravity in
four dimensions.  The ${\cal N}=8$ supersymmetry group in $AdS_5$
space is $SU(2,2|4)$ and it is exactly the $N=4$ superconformal
group in four dimensions. In particular,  the on-shell degrees of
freedom of the gauged ${\cal N}=8$ $SO(6)$ gauged $AdS_5$
supergravity counted with $SO(3)$ little group matches identically
with the ${\cal N}=4$ off-shell conformal supergravity in four
dimensions.

The field content of ${\cal N}=4$ conformal supergravity in four
dimensions consists 1 graviton ${e}_\mu^{~r}$, 4 gravitini
${\psi}_\mu^i$, 15 $SU(4)$ (or $SO(6)$) gauge fields
${V}_{\mu\,j}^{i}$, 1 complex field ${\varphi}$, 4 spinors
${\chi}^i$, 10 complex scalars ${E}_{(ij)}$, 20 spinor fields
${\lambda}_{~[ij]}^{k}$, 6 antisymmetric tensors
${B}_{\mu\nu}^{-[ij]}$. The transformations under supersymmetry and
special supersymmetry generators for the ${\cal N}=4$ conformal
supergravity multiplets should be reproduced from
Eq.\,(\ref{steight}) near the $AdS_5$ vacuum configuration
\cite{bergshoe,fra,chch2},
\begin{eqnarray}
 \delta {e}_\mu^{~r}&=&\frac{1}{2}
\overline{\epsilon}^i
\gamma^r{\psi}_{\mu\,i }+\mbox{h.c},\nonumber\\
\delta{\psi}_\mu^i &=& {D}_\mu {\epsilon}^i
-\frac{1}{2}{\gamma}^{\nu\lambda} {B}_{\nu\lambda}^{-ij}{\gamma}_\mu
{\epsilon}_j -{\gamma}_\mu{\eta}^i,
 \nonumber\\
  \delta {V}_{\mu\,j}^{i}
 &=& \left(-\overline{\xi}_\mu^i\epsilon_j
 +\frac{1}{4}\delta^i_{~j}\overline{\xi}_\mu^k
 {\epsilon}_k
 +\frac{1}{2}\overline{\epsilon}^k{\gamma}_\mu
 {\chi}_{kj}^{~~i}-\overline{\psi}^i{\eta}_j
 +\frac{1}{4}\delta^i_{~j}\overline{\psi}_\mu^k\epsilon_k
 \right)+\mbox{h.c.},\nonumber\\
 \delta \phi &=&
 \frac{1}{2}\overline{\epsilon}^i {\chi}_i,\nonumber\\
 \delta {\chi}_i &=&
 {\gamma}^\mu {D}_\mu {\phi}
 {\epsilon}_i+\frac{1}{2}{E}_{ij}{\chi}^j
 +\frac{1}{2}\epsilon_{ijkl}{\gamma}^{\mu\nu}B_{\mu\nu}^{-kl}
 {\epsilon}^j, \nonumber\\
 \delta {E}_{ij} &=&
 \frac{1}{2}\left(\overline{\epsilon}_i
 {\gamma}^\mu {D}_\mu {\chi}_j
 +\overline{\epsilon}_j
 {\gamma}^\mu{D}_\mu {\chi}_i
 \right)-\frac{1}{2}\overline{\epsilon}^k\left(
 \overline{\lambda}^{lm}_{~~i}\epsilon_{jklm}
 +\overline{\lambda}^{lm}_{~~j}\epsilon_{iklm}\right)\nonumber\\
 &&-\left(\overline{\eta}_i{\chi}_j+
  \overline{\eta}_j {\chi}_i\right),\nonumber\\
  \delta {D}^{ij}_{~~kl}
  &=& -2 \left(\overline{\epsilon}^i
 {\gamma}^\mu {D}_\mu {\lambda}^j_{~kl}
 -\overline{\epsilon}^j
 {\gamma}^\mu {D}_\mu {\lambda}^i_{~kl}
 \right)
 +\delta^i_{k}\overline{\epsilon}^m
{\gamma}^\mu {D}_\mu {\lambda}^j_{~ml} \nonumber\\
&& -\delta^i_{~l}\overline{\epsilon}^m
 {\gamma}^\mu {D}_\mu {\lambda}^j_{~mk}
-\delta^j_{~k}\overline{\epsilon}^m
 {\gamma}^\mu {D}_\mu {\lambda}^i_{~ml}
 +\delta^j_{~l}\overline{\epsilon}^m
 {\gamma}^\mu {D}_\mu {\lambda}^i_{~mk},
\nonumber\\
\delta {B}^{-ij}_{\mu\nu} &=& \frac{1}{2}
\left(\overline{\epsilon}^i
 R^j_{~\mu\nu}[Q] -\overline{\epsilon}^j
 R^i_{~\mu\nu}[Q]\right)+\frac{1}{2}\overline{\epsilon}^k
 {\gamma}_{\mu\nu}{\lambda}^{ij}_{~~k}\nonumber\\
 && +
 \frac{1}{4}\epsilon^{ijkl}\overline{\epsilon}_k
 {\gamma}^\rho {D}_\rho {\gamma}_{\mu\nu}
 \widetilde{\chi}_l-\frac{1}{4}\epsilon^{ijkl}
 \overline{\epsilon}_k {\gamma}_{\mu\nu}
 {\chi}_l,\nonumber\\
 \delta {\lambda}^{ij}_{~~k} &=&
 -\frac{1}{2}{\gamma}^{\mu\nu}{\gamma}^\rho {D}_\rho
 \left[{B}^{-ij}_{\mu\nu}{\epsilon}_k
 +\frac{1}{3}\left(\delta^i_{~k}{B}^{-jl}_{\mu\nu}
 - \delta^j_{~k}{B}^{-il}_{\mu\nu} \right) {\epsilon}_l\right]
 \nonumber\\
&& -\frac{1}{2}{\gamma}_{\mu\nu} \left[\left(
R_{\mu\nu\,k}^i[V]{\epsilon}^j -R_{\mu\nu\,k}^j [V]
{\epsilon}^i\right)+\frac{1}{3} \left(\delta^i_{~k} R_{\mu\nu\,l}^j
[V]
-\delta^j_{~k}R_{\mu\nu\,l}^i[{V}]\right){\epsilon}^l \right]\nonumber\\
&&-\frac{1}{4}\epsilon^{ijlm}{\gamma}^\mu {D}_\mu
{E}_{kl}{\epsilon}_m +\frac{1}{2}{D}^{ij}_{~~kl} {\epsilon}^l
-\frac{1}{2}\epsilon^{ijlm}{E}_{kl}{\eta}_m\nonumber\\
&&+{\gamma}^{\mu\nu}\left[{B}^{-ij}_{\mu\nu}
{\eta}_k+\frac{1}{3}\left(\delta^i_{~k}
 {B}^{-jl}_{\mu\nu} -\delta^j_{~k}
 {B}^{-il}_{\mu\nu} \right) {\eta}_l \right],
\end{eqnarray}
where $\xi_\mu^i$ and $R^i_{\mu\nu}[Q]$ are listed in
(\ref{losucur}); The curvature $R_{\mu\nu\,j}^i[V]$ relevant to
local $SU_R(4)$ symmetry are the corresponding gauge field strength
but with fermionic terms required by supersymmetry,
\begin{eqnarray}
R_{\mu\nu\,j}^i[V]=\partial_\mu V_{\nu\,j}^i-\partial_\nu
V_{\mu\,j}^i+[V_\mu.V_\nu]^i_{~j}-2\left[\left(\overline{\psi}^i_{[\mu}
\xi_{\nu]j}-\frac{1}{4}\delta^i_{~j} \overline{\psi}^k_{[\mu}
\xi_{\nu]k}\right)+\mbox{h.c.}\right].
\end{eqnarray}

\section{External Superconforml Anomaly from AdS/CFT correspondence}

\subsection{Generality}

In this section we show how the external superconformal anomaly
(\ref{exone}) can be obtained from the gauged supergravity in the
$AdS_5$ vacuum configuration. First, the relation
Eq.\,(\ref{acc50}), which is established  based on $AdS_5/CFT_4$
correspondence, tells that the quantum effective action of the
supersymmetric $SU(N)$ gauge theory at large-$N$ limit in a
conformal supergravity background is equal to the on-shell action of
five-dimensional gauged  supergravity evaluated with the classical
solutions which asymptotically approach to the $AdS_5$ vacuum
configuration. In last section it was verified explicitly (in ${\cal
N}=2,4$ cases) that the $AdS_5$ boundary data actually constitute
certain off-shell conformal supergravity multiplet in four
dimensions. Correspondingly, on the $AdS_5$ boundary, the bulk
diffeomorphism symmetry decomposes the four-dimensional
diffeomorphism symmetry and the Weyl symmetry; the bulk
supersymmetry converts into the Poincar\'{e} supersymmetry and the
super-Weyl symmetry in four dimensions, and the bulk gauge symmetry
separates into four-dimensional vector- and axial vector gauge
symmetries. Eq.\,(\ref{acc50}) means that the external conformal
supergravity background in the context of $AdS/CFT$ correspondence
is furnished by the $AdS_5$ boundary data of on-shell gauged
supergravity in five dimensions. On the other hand,
Eqs.\,(\ref{excoupl}), (\ref{egt1}) and (\ref{busy}) imply that the
(covariant) conservation  of superconformal current multiplet of a
supersymmetric gauge theory is completely equivalent to the local
conformal supersymmetry of external  conformal supergravity
background. Thus the violation of local conformal supersymmetry in
the background effective action $\Gamma_{\rm CFT}[\phi_0]$ means the
arising of superconformal anomaly. Therefore, we calculate the above
on-shell action of five-dimensional gauged supergravity and check it
supersymmetry variations near $AdS_5$ boundary.  Due to the large
volume of $AdS_5$ boundary, the on-shell action of gauged
supergravity near $AdS_5$ boundary usually suffers from IR
divergence. One should employ a so-called holographic
renormalization procedure to make it well defined. This process
cannot preserve above bulk symmetry reduction in each pair
simultaneously. The physical consideration requires that the
four-dimensional diffeomorphism symmetry, the Poincar\'{e}
supersymmetry and the vector gauge symmetry should be preserved,
hence the external superconformal conformal anomaly arises. It is
usually called the holographic superconformal anomaly since it is
reproduced from the five-dimensional bulk theory.

In the following we review how the holographic superconformal
anomaly can be produced from the on-shell five-dimensional gauged
supergravity.

\subsection{Holographic Chiral $R$-symmetry Anomaly}

The reproduction of the holographic chiral $R$-symmetry anomaly lies
in the Chern-Simons five-form term in five-dimensional gauged
supergravity. As is well known, the Chern-Simons term has a
remarkable feature; its variation under gauge transformation is a
total derivative. According to differential geometry construction on
chiral anomaly \cite{zwz}, this total derivative term is just the
(consistent) chiral anomaly since it satisfies the Wess-Zumino
condition \cite{wezucons}. Therefore, based on Eq.\,(\ref{acc50}),
we perform gauge transformation on classical action of gauged
supergravity. All other terms are gauge invariant except that the
Chern-Simons five-form yields a total derivative term under gauge
transformation. We then take this total derivative to the $AdS_5$
boundary and use the boundary reductions of bulk gauge field and
gauge symmetry to extract out the chiral anomaly.
 The fact that the holographic chiral $R$-symmetry
anomaly of supersymmetric gauge theory can be obtained from the
Chern-Simons term  in five-dimensional  gauged supergravity side
was
 first pointed by Witten \cite{witt1} and later it was explored
 beyond the leading order of large-$N$ expansion \cite{apty,chu}.

\subsubsection{Holographic $SU(4)$ Bardeen Anomaly
from ${\cal N}=8$ Gauged  $AdS_5$ Supergravity}

In ${\cal N}=4$ supersymmetric Yang-Mills theory, the chiral
$R$-symmetry is $SU_R(4)$, which corresponds to $SO(6)\cong SU(4)$
gauge symmetry in  ${\cal N}=8$ five-dimensional supergravity. It
should be emphasized that the $SU_R(4)$ symmetry in  supersymmetric
Yang-Mills theory is a global symmetry. An external vector $A_\mu^a$
must couple with the chiral $SU_R(4)$ current $j_\mu^a$ to yield the
anomaly, here $a=1,\cdots, 15$ denoting $SU(4)$ group indices (or
$a=[IJ]$ SO(6) indices). The $AdS_5$ boundary value $A_\mu^a (x)$ of
the bulk $SU(4)$ gauge field $\widehat{A}_\alpha^a (x,r)$ provides
such an external field in terms of the holographic version on
$AdS/CFT$ correspondence. According to Eq.\,(\ref{acc4}) we have
\begin{eqnarray}
S_{\rm SUGRA}[\widehat{A}_\alpha^a[A^a_\mu(x)],\cdots]=\Gamma_{\rm
SYM}[A_\mu^a,\cdots] \label{acc5}
\end{eqnarray}
Recall that there exists the Chern-Simons term in five-dimensional
${\cal N}=8$ gauged supergravity,
\begin{eqnarray}
S_{\rm CS}[\widehat{A}]&=&\frac{l^3}{48 \pi
G^{(5)}}\int\mbox{Tr}\left(
\widehat{A}\widehat{F}^2-\frac{1}{2}\widehat{A}^3\widehat{F}
+\frac{1}{10} \widehat{A}^5\right)\nonumber\\
&=&\frac{l^3}{48 \pi G^{(5)}}\int \mbox{Tr}\left( \widehat{A}
(d\widehat{A})^2+\frac{3}{2}\widehat{A}^3d\widehat{A}
+\frac{3}{5}\widehat{A}^5\right)\nonumber\\
 &=&\frac{l^3}{192 \pi G^{(5)}}\int d^5x
 \epsilon^{\alpha\beta\gamma\delta\sigma}
d^{abc}\left(\widehat{A}_\alpha^a \widehat{F}_{\beta\gamma}^b
\widehat{F}_{\delta\sigma}^c- f^{ade}\widehat{A}_\alpha^d
\widehat{A}_\beta^e \widehat{A}_\gamma^b
\widehat{F}_{\delta\sigma}^c
\right.\nonumber\\
&& \left. +\frac{2}{5} f^{ade}f^{efg} \widehat{A}_\alpha^d
\widehat{A}_\beta^f \widehat{A}_\gamma^g \widehat{A}_\delta^b
\widehat{A}_\sigma^c \right),
\end{eqnarray}
where $d^{abc}=\mbox{Tr}\left(t^a t^b t^c\right)$, $[t^a,t^b]
=if^{abc} t^c$.
 Making the following $SU(4)$ gauge transformation on
five-dimensional gauged supergravity near $AdS_5$ vacuum
configuration,
\begin{eqnarray}
\delta \widehat{A}_\alpha^a (x,r)=(\widehat{D}_\alpha V(x,r))^a,
\end{eqnarray}
with $V(x,r)=V^a(x,r) t^a$ being the gauge transformation parameter,
we obtain
\begin{eqnarray}
\delta_V S_{\rm SUGRA}[\widehat{A}[A],\cdots] &=& \delta_V S_{\rm
CS}[\widehat{A}]=\int d\omega_4^1 (V,\widehat{A})
=\int \omega_4^1 (v,A)\nonumber\\
&=& \frac{l^3}{48 \pi G^{(5)}}\int
\mbox{Tr}\left[{v}\,d\left(AdA+\frac{1}{2}
A^3\right)\right]\nonumber\\
&=& \delta_v \Gamma[A_\mu^a,\cdots]= \int d^4x \frac{\delta \Gamma}
{\delta A_\mu^a (x) }\delta A_\mu^a (x)\nonumber\\
&=& \int d^4x \left\langle j^{a\mu}\right\rangle \delta A_\mu^a (x)
=\int d^4x
j^{a\mu}(x)[D_\mu v(x)]^a \nonumber\\
&=& -\int d^4x v^a (x)[D_\mu j^\mu (x)]^a, \label{gtr}
\end{eqnarray}
where we have used Eq,\,(\ref{acc5}) and the $AdS_5$ boundary
reduction of bulk gauge fields as well as the Stora-Zumino chain of
descent equations,
\begin{eqnarray}
&& \frac{i^{n+2}}{(2\pi)^n (n+1)!}\mbox{Tr}F^{n+1}
=d\omega_{2n+1}[A],\nonumber\\
&& \delta_v\omega_{2n+1}=d\omega_{2n}^1 (v,A).
\end{eqnarray}
If we consider the string origin of the AdS/CFT correspondence and
the fact that the ${\cal N}=8$ gauged $AdS_5$ supergravity can be
obtained from the type IIB supergravity around $AdS_5\times S^5$
background, there should exist following relations among the $AdS_5$
radius $l$, string coupling $g_s$, the flux $N$ passing through
$S^5$, and also a connection between the five- and ten-dimensional
gravitational constants due to the compactification of the type IIB
supergravity on $S^5$ of radius $l$ \cite{agmo},
\begin{eqnarray}
G^{(5)}=\frac{G^{(10)}}{\mbox{Volume}\,
(S^5)}=\frac{G^{(10)}}{l^5\pi^3}, ~~~ G^{(10)}=8\pi^6g^2_s,
~~~l=\left(4\pi N g_s\right)^{1/4}.
\end{eqnarray}
Hence we  immediately recognize from Eq.\,(\ref{gtr})  the Bardeen
anomaly up to the normalization constant \cite{bard,bertl}.
\begin{eqnarray}
[D_\mu j^{\mu} (x)]^a =-\frac{N^2}{24\pi^2}e^{-1}
\epsilon^{\mu\nu\lambda\rho}
\partial_\mu \mbox{Tr}t^a\left(A_\nu \partial_\lambda A_\rho+
\frac{1}{2}A_\nu A_\lambda A_\rho \right). \label{nfour1}
\end{eqnarray}

\subsubsection{Holographic Chiral $U_R(1)$ Anomaly from
$SU(2)\times U(1)$ Gauged ${\cal N}=4$ $AdS_5$ Supergravity}

In four-dimensional  ${\cal N}=2$ supersymmetric Yang-Mills theory,
the $R$-symmetry is the $U_R(2){\cong} SU_I(2)\times U_R(1)$. It is
the $U_R(1)$ that becomes anomalous. Observing the classical action
(\ref{gauge4}) of $SU(2)\times U(1)$ gauged ${\cal N}=4$
supergravity, we see that there exists a $SU(2)\times U(1)$-mixed
Chern-Simons term,
\begin{eqnarray}
S_{\rm CS}[\widehat{W},\widehat{A}] &=& -\frac{l^3}{8\times 16\pi
G^{(5)}} \int d^5x \epsilon^{\alpha\beta\gamma\delta\sigma}
\widehat{W}_{\alpha\beta}^a\widehat{W}_{\gamma\delta}^a
\widehat{A}_\sigma\nonumber\\
&=&-\frac{l^3}{16\pi G^{(5)}} \int \mbox{Tr}\left(\widehat{W}\wedge
\widehat{W}\right)\wedge \widehat{A}.
\end{eqnarray}
Under the bulk $U(1)$ gauge transformation $\delta_V \widehat{A}=d
V$ on ${\cal N}=4$ gauged supergravity near $AdS_5$ vacuum
configuration, we have
\begin{eqnarray}
&& \delta_V S_{\rm SUGRA}[ \widehat{W}[W], \widehat{A}[A],\cdots]=
\delta_V S_{\rm CS}[\widehat{W}[W],
\widehat{A}[A]]\nonumber\\
&=&-\frac{l^3}{16\pi G^{(5)}} \int \mbox{Tr}\left(\widehat{W}\wedge
\widehat{W} \right) \wedge dV =-\frac{l^3}{16\pi G^{(5)}} \int
d\left[V\mbox{Tr}\left(\widehat{W}\wedge \widehat{W}\right)
\right]  \nonumber\\
&=&-\frac{l^3}{16\pi G^{(5)}}\int v \mbox{Tr}\left(W\wedge W\right).
\label{grasiva}
\end{eqnarray}
According to the AdS/CFT correspondence (\ref{acc4}) at gauged
supergravity level, Eq.\,({grasiva}) should  equal to
\begin{eqnarray}
\delta_v \Gamma_{\rm SYM}[W,A] =-\int d^4x {v} D_\mu j^{\mu}.
\label{fourano}
\end{eqnarray}
Thus we obtain the chiral $U_R(1)$ anomaly of the ${\cal N}=2$
supersymmetric Yang-Mills theory at the leading-order of large-$N$
expansion,
\begin{eqnarray}
D_\mu j^\mu=\frac{N^2}{32\pi^2}\epsilon^{\mu\nu\lambda\rho}
W_{\mu\nu}^a W_{\lambda\rho}^a.
\end{eqnarray}

Finally, we briefly mention that the holographic  chiral $U_R(1)$
anomaly of the ${\cal N}=1$ supersymmetric $SU(N)$ Yang-Mills theory
can be easily obtained from  the ${\cal N}=2$ $U(1)$ gauged
supergravity near $AdS_5$ vacuum configuration. According to
Eq.\,(\ref{gaugedf}), the $U(1)$ Chern-Simons term reads
\begin{eqnarray}
S_{\rm CS}[\widehat{A}]&=&\frac{l^3}{16\pi G^{(5)}}\int d\widehat{A}
\wedge d\widehat{A}\wedge \widehat{A} =\frac{l^3}{64\pi G^{(5)}}\int
d^5x\,\epsilon^{\alpha\beta\gamma\sigma\delta}
\widehat{F}_{\alpha\beta}\widehat{F}_{\gamma\sigma}
\widehat{A}_\delta .
\end{eqnarray}
The AdS/CFT correspondence and the variation of Chern-Simons term
under bulk $U(1)$ gauge transformation  yields the $U(1)$ chiral
anomaly of ${\cal N}=1$ supersymmetric gauge theory in the external
conformal supergravity background.

\subsection{Holographic Weyl Anomaly}

It is non-trivial to recognize the trace anomaly of a supersymmetric
gauge theory from the supergravity side. The origin of trace anomaly
lies in the near-$AdS_5$ boundary behavior of the classical solution
of the gauged supergravity. In the earlier works on gauged
supergravity \cite{guna2,awada,roma} the dynamical behavior of
five-dimensional gauged supergravity near $AdS_5$ vacuum
configuration was neglected. It was first found in Ref.\,\cite{hesk}
that due to the infinity of the $AdS_5$ boundary, the on-shell
action of gauged supergravity near $AdS_5$ vacuum configuration is
not well defined and the infrared divergence arises. This IR
divergence  is  dual to the the UV divergence of the supersymmetric
gauge theory defined on $AdS_5$ boundary.  One must perform a ``
holomorphic renormalization '' procedure \cite{bian1,pil} to
manipulate this divergence. It is very similar to the
renormalization method of dealing with the UV divergence in a
perturbative quantum field theory. That is, one must first introduce
a cut-off (IR regulator) to make the space-time have a finite volume
and calculate the on-shell bulk gravitational action, then
introduces the counterterms according to
 the renormalization condition to make the on-shell  action
 of gauged supergravity near the $AdS_5$
boundary well defined. Finally one removes the cut-off and hence
obtains a finite on-shell bulk action. However, there exists an
ambiguity when adding the finite counterterms.  One usually chooses
the diffeomorphism symmetry on the boundary  to fix this ambiguity.
As stated above, the diffeomorphism symmetry of the gauged
supergravity near the $AdS_5$ boundary decomposes into the
diffeomorphism symmetry \cite{imbi} and the Weyl symmetry.  These
two symmetries cannot be preserved simultaneously during the
holomorphic renormalization process. When requiring the
diffeomorphism symmetry to be preserved,  one thus obtains the Weyl
anomaly of a supersymmetric gauge theory in the curved space-time.
In the following we show how the holographic Weyl anomaly comes from
the gauged supergravity.

  We must look for the domain wall solution of the gauged ${\cal N}=8$
supergravity which asymptotically approaches to the $AdS_5$
geometry. Without losing generality, in ${\cal N}=4,8$ cases, we
consider the gauged supergravity consisting of only gravitational-
and scalar fields.  The classical action is listed in
Eq.\,(\ref{effect1}) and the cosmological constant originates from
the critical value of the scalar potential. It should be emphasized
that the presence of scalar fields is necessary for the existence of
domain wall solution, otherwise there will be no non-trivial vacuum
configurations and the domain wall solutions cannot arise. We
consider only the scalar field for simplification. The corresponding
classical field equations are the Einstein equation (\ref{ee}) and
and the scalar field equation
\begin{eqnarray}
\nabla_\mu\partial^\mu \phi=\frac{\partial P}{\partial \phi}.
\end{eqnarray}
 For ${\cal N}=2$ case,  we consider the pure gravitational
field with the cosmological constant term  and assume that the
constant term comes from the critical value of certain scalar
potential.

We use the results in Ref.\,\cite{bian1},
The domain wall solution preserving the 4-dimensional general
covariance takes the form given in Eq.\,(\ref{metrican1}). Further,
near the boundary $r\rightarrow 0$, there must exist
$\widehat{g}_{\mu\nu\nu}(x,r)\rightarrow g_{\mu\nu}(x)$,
$\widehat{\phi} (x,r) \rightarrow 0$ \footnote{Precisely speaking,
the bulk scalar field should satisfy the Dirichlet boundary
condition, $\phi (x,r)\rightarrow r^{(4-\Delta)/2}$ as $r\rightarrow
0$, $\Delta$ being the scale dimension of the composite operator in
supersymmetric gauge theory coupled with $\phi(x)$, the boundary
value of $\widehat{\phi}(x,r)$}. Based on these requirements,  the
gravitational  fields near the boundary  should have the following
expansion in the radial coordinate,
\begin{eqnarray}
\widehat{g}_{\mu\nu}(x,r)&=&
g_{(0)\mu\nu}(x)+g_{(2)\mu\nu}(x)\frac{r^2}{l^2}\nonumber\\
&& +\left(\frac{r^2}{l^2}\right)^2\left[g_{(4)\mu\nu}+
h_{1(4)\mu\nu}\ln \frac{r^2}{l^2}+h_{2(4)\mu\nu}\left(\ln
\frac{r^2}{l^2}\right)^2\right]+\cdots\, . \label{expa}
\end{eqnarray}
Substituting this expansion into the Einstein equation (\ref{ee})
with the cosmological constant furnished by the value of the scalar
potential at the critical point $\widehat{\phi}=0$, one can
determine the coefficients $g_{(2k)\mu\nu}$'s and $h_{\mu\nu}$'s in
terms of the leading coefficient $g_{(0)\mu\nu}(x)$,
\begin{eqnarray}
g_{(2)\mu\nu} &=&\frac{l^2}{2}\left( R_{\mu\nu}
-\frac{1}{6}R g_{(0)\mu\nu}\right),\nonumber\\
h_{1(4)\mu\nu}
&=&\frac{l^4}{8}\left(R_{\mu\lambda\nu\rho}R^{\lambda\rho}
+\frac{1}{6}\nabla_\mu\nabla_\rho R-
\frac{1}{2}\nabla^2R_{\mu\nu}-\frac{1}{3}R R_{\mu\nu}\right)\nonumber\\
&&+\frac{l^4}{32}g_{(0)\mu\nu}\left(\frac{1}{3}\nabla^2
R+\frac{1}{3} R^2
-R_{\lambda\rho}R^{\lambda\rho}\right), \nonumber\\
h_{2(4)\mu\nu} &=& 0, \nonumber\\
g_{(4)\mu\nu} &=&\frac{1}{4}\left(g_{(2)}\right)^2_{\mu\nu},\nonumber\\
\nabla^\nu g_{(4)\mu\nu} &=& \nabla^\nu\left\{
-\frac{1}{8}\left[\mbox{Tr}g_{(2)}^2-\left(\mbox{Tr}g_{(2)}\right)^2
\right] g_{(0)\mu\nu}+\frac{1}{2}\left(g_{(2)}^2\right)_{\mu\nu}
-\frac{1}{4}g_{(2)\mu\nu}\mbox{Tr} g_{(2)}\right\}.
\end{eqnarray}
In above equation, $R_{\mu\nu\lambda\rho}$, $R_{\mu\nu}$ and $R$ are
the Riemannian curvature, Ricci tensor and scalar curvature defined
with respect to $g_{(0)\mu\nu}$.

 However, when one substitutes above solution into the
classical action of the bulk gauged supergravity, the on-shell
action is divergent near the $AdS_5$ boundary. So one has to
introduce a cut-off $\epsilon$ to perform IR regularization, that
is,taking the integral in radial coordinate to a finite domain by
choosing $r=\epsilon
>0$,
\begin{eqnarray}
S_{\rm reg} &=& \frac{1}{8\pi G^{(5)}}\int_{r=\epsilon >0}  \int
d^5X
\sqrt{-\widehat{g}}\left(-\frac{1}{2}\widehat{R}-P[\widehat{\phi}_0=0]\right)
\nonumber\\
&=& \frac{l^5}{8\pi G^{(5)}}\int_{\epsilon} \frac{dr}{r^5}\int
 d^4x \sqrt{\widehat{g} (x,r)}\,
\frac{2}{3}P[\widehat{\phi}=0]\nonumber\\
&=& -\frac{l^5}{8\pi G^{(5)}}\int_{\epsilon} \frac{dr}{r^5}\int
 d^4x \sqrt{\widehat{g} (x,r)}\,
\frac{4}{3}\Lambda = \frac{l^3}{\pi G^{(5)}}\int_{\epsilon}
\frac{dr}{r^5}\int
 d^4x \sqrt{\widehat{g} (x,r)}\nonumber\\
 &=&\frac{l^3}{\pi G^{(5)}}\int
 d^4x \sqrt{g_0(x)} \int_{\epsilon} \frac{dr}{r^5}
 \left\{1+\frac{r^2}{2l^2}\mbox{Tr}g_{(2)}
 +\frac{r^4}{4l^4}\left[\left( \mbox{Tr}g_{(2)}\right)^2
 -\mbox{Tr}g^2_{(2)}\right]+\cdots\right\}\nonumber\\
&=&\frac{l^3}{\pi G^{(5)}}
 \int d^4x\sqrt{g_{(0)}}\left[-\frac{1}{4\epsilon^4}
 -\frac{1}{12\epsilon^2}R
-\frac{1}{16}\ln\frac{\epsilon}{l}\,\left(R_{\mu\nu}R^{\mu\nu}
-\frac{1}{3}R^2 \right) +{\cal L}_{\rm
finite}\right]\nonumber\\
&=& S_{\epsilon^{-4}}+S_{\epsilon^{-2}}+S_{\ln\epsilon}+S_{\rm
finite}. \label{regaction}
\end{eqnarray}
In deriving above result, we have used the Einstein equation
(\ref{ee}) at the critical point of the scalar potential
$\widehat{\phi}_0=0$, the identification
$\Lambda=-2P[\phi=0]=-{6}/{l^2}$ and the matrix operation
\begin{eqnarray}
\sqrt{\det (1+A)}&=&\exp\left[\frac{1}{2}\mbox{Tr} \ln
\left(1+A\right)\right] =\exp\left[\frac{1}{2}\mbox{Tr} \left(A-
\frac{1}{2}A^2+\cdots \right)\right]
\nonumber\\
&=& 1+\frac{1}{2}\mbox{Tr}A+\frac{1}{4} \left[
\left(\mbox{Tr}A\right)^2 -\mbox{Tr}A^2\right]+\cdots,
\end{eqnarray}
${\cal L}_{\rm finite}$ consists of the terms that survive
$\epsilon{\rightarrow}0$ limit.

To make the on-shell action action well defined, one must first
define a subtracted action  by introducing the counterterms to
cancel the IR divergence in the limit $\epsilon\rightarrow 0$,
\begin{eqnarray}
S_{\rm sub}[g_{(0)\mu\nu},\epsilon,\cdots]=S_{\rm reg}+S_{\rm
counter}
\end{eqnarray}
A holographically renormalized on-shell action for the
five-dimensional gauged supergravity is obtained after  the IR
regulator is removed,
\begin{eqnarray}
S_{\rm
ren}[g_{(0)\mu\nu},\cdots]=\lim_{\epsilon{\rightarrow}0}S_{\rm
sub}[g_{(0)\mu\nu},\epsilon,\cdots]
\end{eqnarray}
In adding counterterms there appears a finite ambiguity similar to
that when canceling the UV divergence in a perturbative quantum
field theory. This ambiguity can be fixed by the symmetry
requirement. As is explicitly shown  in Ref.\,\cite{imbi}, the bulk
diffeomorphism symmetry
\begin{eqnarray}
\delta \widehat{g}_{\alpha\beta}=\widehat{\nabla}_\alpha
\widehat{\xi}_\beta+\widehat{\nabla}_\beta \widehat{\xi}_\alpha
\end{eqnarray}
preserving the form of domain wall solution (\ref{metrican1}) should
take the following form,
\begin{eqnarray}
\delta \widehat{g}_{\mu\nu}(x,r)=2\sigma (x)
\left(1-\frac{1}{2}r\partial_r \right)
\widehat{g}_{\mu\nu}(x,r)+\left[\widehat{\nabla}_\mu
\widehat{\xi}_\nu (x,r)+
 \widehat{\nabla}_\nu \widehat{\xi}_\mu (x,r)\right],
 \label{dec}
\end{eqnarray}
where $\sigma (x)=l^2\xi_5(x)/(2r^2)$ and the covariant derivative
$\widehat{\nabla}_\mu$ is defined with respect to
$\widehat{g}_{\mu\nu}(x,r)$. Eq.\,(\ref{dec}) means that the bulk
diffeomorphism transformation decomposes into a Weyl transformation
and a diffeormorphism transformation on $\widehat{g}_{\mu\nu}(x,r)$.
 Near the $AdS_5$
boundary $r\to 0$,  from Eq.\,(\ref{expa}) and the series expansion
for the Killing spinor,
\begin{eqnarray}
\widehat{\xi}^\mu (x,r) &=& \sum_{n=1}^\infty \xi^\mu_{(n)}(x)
\left(\frac{r^2}{l^2}\right)^n,
\end{eqnarray}
we can see that the 5-dimensional bulk diffeomorphism symmetry
reduces to the Weyl- and the diffeomorphism symmetries on $AdS_5$
boundary,
 \begin{eqnarray}
&& \delta g_{(0)\mu\nu}=\nabla_\mu \xi_\nu (x)+\nabla_\nu \xi_\mu, \label{morp}\\
&& \delta g_{(0) \mu\nu} =2\sigma (x)g_{(0)\mu\nu}. \label{weyl}
\end{eqnarray}
Requiring the four-dimensional diffeomorphism symmetry (\ref{morp})
 preserved in the subtraction process, one can only introduce
 the following counterterm in the sense of ``minimal subtraction" \footnote{The
 geometric meaning of the counterterm becomes much clearer  if it is expressed
 in terms of the induced metric on the boundary
 ${g}_{\mu\nu}(x)=\lim_{\epsilon\to 0}
 \widehat{g}_{\mu\nu}(x,\epsilon)/\epsilon$ \cite{sken}.},
 \begin{eqnarray}
 S_{\rm count} &=& \frac{l^3}{\pi G^{(5)}}
 \int d^4x\sqrt{-g_{(0)}}\left[\frac{1}{4\epsilon^4}
 +\frac{1}{12\epsilon^2}R
+\frac{1}{16}\left(R_{\mu\nu}R^{\mu\nu}-\frac{1}{3}R^2
\right)\ln\frac{\epsilon}{l}\right].
 \end{eqnarray}
Consequently, we have
\begin{eqnarray}
S_{\rm ren} &=& \frac{l^3}{\pi G^{(5)}}\int d^4x
\sqrt{-g_{(0)}}{\cal L}_{\rm finite} =\Gamma_{\rm SYM}[g_{(0)\mu\nu}
\langle T^{\mu\nu}\rangle, \cdots ].
\end{eqnarray}

One can directly calculate the variation of the renormalized
on-shell action under the Weyl transformation (\ref{weyl}) to
extract out the Weyl anomaly. However, the most convenient way is
choosing the Weyl transformation parameter $\sigma (x)$ as a
constant $\sigma$ and using the scale symmetry of the regularized
action (\ref{regaction}). In this way, the Weyl anomaly in the
renormalized on-shell action  can be equivalently obtained from the
scale transformation of the counterterms.  The IR regularized action
(\ref{regaction}) is invariant under the combination of scale
transformations, $\delta g_{(0)\mu\nu}=2\sigma g_{(0)\mu\nu}$,
$\delta \epsilon=2 \sigma \epsilon$ \cite{hesk},
\begin{eqnarray}
\left(\delta_{g_0}+\delta_\epsilon\right) S_{\rm reg}
=\left(\delta_{g_0}+\delta_\epsilon\right) \left(
S_{\epsilon^{-4}}+S_{\epsilon^{-2}}+S_{\ln\epsilon}+S_{\rm
finite}\right) =0.
\end{eqnarray}
It can be directly checked that \cite{hesk}
\begin{eqnarray}
\delta_{g_0} \left(S_{\epsilon^{-4}}+S_{\epsilon^{-2}}\right) =0,~~~
\delta_\epsilon \left( S_{\epsilon^{-4}}+S_{\epsilon^{-2}}\right)
=0,~~~ \delta_{g_0} S_{\ln\epsilon}=0, ~~~ \delta_\epsilon S_{\rm
finite} =0.
\end{eqnarray}
This leads to
\begin{eqnarray}
\delta_{g_0} S_{\rm finite}&=&\int d^4x \sqrt{-g_{(0)}}\langle
T^{\mu}_{~\mu}\rangle\sigma
=-\delta_\epsilon S_{\ln\epsilon}\nonumber\\
&=&\frac{l^3}{8\pi G^{(5)}}\int d^4x \sqrt{-g_{(0)}}
\left(R_{\mu\nu}R^{\mu\nu}-\frac{1}{3}R^2 \right)\sigma,
\end{eqnarray}
and yields the Weyl anomaly \cite{hesk,bian1},
\begin{eqnarray}
\langle T^\mu_{~\mu}\rangle =
\frac{N^2}{4\pi^2}\left(R_{\mu\nu}R^{\mu\nu}-\frac{1}{3}R^2\right).
\label{trace01}
\end{eqnarray}
\begin{eqnarray}
\delta_{\sigma} S_{\rm ren}=-\int d^4x \sqrt{-g_{(0)}}\langle
T^{\mu}_{~\mu}\rangle\sigma .
\end{eqnarray}
gives the Weyl anomaly in the gravitational background,
\begin{eqnarray}
{\cal A}=\langle T^\mu_{~\mu}\rangle =
\frac{N^2}{4\pi^2}\left(R_{\mu\nu}R^{\mu\nu}-\frac{1}{3}R^2\right).
\label{trace}
\end{eqnarray}
It can be further rewritten as the combination of the $A$- and
$B$-type anomalies up to the coefficients, i.e, the sum of the Euler
number and the square of Weyl tensor \cite{hesk,deser},
\begin{eqnarray}
\langle T^\mu_{~\mu}\rangle &=& -\frac{N^2}{\pi^2} \left(E_4+W_4\right), \nonumber\\
E_4 &=&
\frac{1}{8}\widetilde{R}_{\mu\nu\lambda\rho}\widetilde{R}^{\mu\nu\lambda\rho}
=\frac{1}{8}\left( R^{\mu\nu\lambda\rho}R_{\mu\nu\lambda\rho}
-4R^{\mu\nu}R_{\nu\nu}+R^2 \right),\nonumber\\
W_{4} &=& -\frac{1}{8}
{C}_{\mu\nu\lambda\rho}{C}^{\mu\nu\lambda\rho} =-\frac{1}{8}\left(
R^{\mu\nu\lambda\rho}R_{\mu\nu\lambda\rho}
-2R^{\mu\nu}R_{\mu\nu}+\frac{1}{3}R^2 \right). \label{trace2}
\end{eqnarray}
It is exactly the trace anomaly  of ${\cal N}=4$ supersymmetric
Yang-Mills theory in the external gravitational field at the leading
order of the large-$N$ expansion \cite{hesk}.

There are also contribution to Weyl anomaly from scalar- and gauge
fields, which  can be obtained in the same way as the gravitational
field.

The holographic Weyl anomalies for ${\cal N}=1,2$ supersymmetric
Yang-Mills theory take the same forms as that in Eqs.\,(\ref{trace})
and (\ref{trace2}), only the anomaly coefficients are different due
to different particle contents in the corresponding supersymmetric
gauge theories.

\subsection{Holographic Super-Weyl Anomaly of Supersymmetry Current}

The arising of holographic super-Weyl anomaly  is due to the
universal feature of a supersymmetric field theory. The
supersymmetric variation of the Lagrangian of the gauged
supergravity should be composed of the total derivative terms. This
terms cannot be naively ignored due to the non-emptiness of $AdS_5$
boundary. As shown in Sect.\,III,  the bulk supersymmetry
transformations for the on-shell fields near the $AdS_5$ boundary
convert into the four-dimensional supersymmetric transformation  and
the super-Weyl transformation. When we take those total derivative
terms yielded from the supersymmetric variation of on-shell action
of gauged supergravity to the $AdS_5$ boundary, the surface terms
may be IR divergent since the on-shell action has IR divergence due
to the large $AdS_5$ boundary. Like in calculating the holographic
Weyl anomaly, the holographic renormalization procedure is needed to
make the surface term behave well. This will make either the
four-dimensional Poincar\'{e} supersymmetry or the super-Weyl
symmetry on $AdS_5$ boundary violated. Usually the Poincar\'{e}
supersymmetry should be preserved, the surface term thus gives the
super-Weyl anomaly (the $\gamma$-trace anomaly). In the following we
take ${\cal N}=2$ $U(1)$ gauged supergravity (\ref{gaugedf}) as
illustrating example and show how the holographic super-Weyl anomaly
of ${\cal N}=1$ supersymmetric gauge theory arises.

The concrete calculation on supersymmetry
 variation of the  ${\cal N}=2$ $U(1)$ gauged supergravity
 (\ref{gaugedf}) was given in Ref.\,\cite{wfchen}. Here we list the
 main results.

 First, the supersymmetric variation of the Einstein-Hilbert action
 and cosmological term gives
 \begin{eqnarray}
 \delta  S_{\rm GR}&=&\delta \int d^5x
  \widehat{e}\left(-\frac{1}{2}\widehat{R}-\frac{6}{l^2}\right)
  \nonumber\\
  &=& \int d^5x \widehat{e}\left[ -\frac{1}{2}
  \overline{\widehat{\epsilon}}^a\widehat{\gamma}_m
  \widehat{\psi}^{\alpha}_a
  \left(\widehat{R}_\alpha^{~m}-\frac{1}{2}
  \widehat{e}_\alpha^{~m}\widehat{R}
  -\frac{6}{l^2}\widehat{e}_\alpha^{~m}\right)
  -\widehat{\nabla}_\alpha \left(
  \widehat{e}_m^{~\alpha}\widehat{e}_n^{~\beta}\delta
  \widehat{\omega}_\beta^{~mn}
  \right)\right].
  \label{superv1}
 \end{eqnarray}
 In above equation the inertial coordinate
 system is usually chosen so that the surface term  is  quadratic
 in fermionic fields and the supersymmetry transformation parameter.

 The variation of the pure gauge field terms (including Chern-Simons term)
 gives
 \begin{eqnarray}
&& \delta S_{\rm GAU}= \int d^5x \delta \left[\widehat{e}
\left(-\frac{3 l^2}{32}\right)\widehat{F}_{\alpha\beta}
\widehat{F}^{\alpha\beta}
 -\frac{l^3}{64} i\epsilon^{\alpha\beta\gamma\delta\sigma}
  \widehat{F}_{\alpha\beta}
 \widehat{F}_{\gamma\delta} \widehat{A}_\sigma\right]\nonumber\\
 &=& \int d^5x \left\{\widehat{e}\left[-\frac{3l^2}{64}
 \overline{\widehat{\epsilon}}^a
 \widehat{\gamma}^\alpha\widehat{\psi}_{\alpha a} \widehat{F}_{\beta\delta}
 \widehat{F}^{\beta\delta}+
 \frac{3il}{8}\widehat{\nabla}_\alpha
 \left(\widehat{F}^{\alpha\beta}
 \overline{\widehat{\epsilon}}^a\widehat{\psi}_{\beta a}\right)
 -\frac{3il}{8}\left(\widehat{\nabla}_\alpha
 \widehat{F}^{\alpha\beta}\right)
 \overline{\widehat{\epsilon}}^a\widehat{\psi}_{\beta a}
 \right]\right.\nonumber\\
 &&\left. -\frac{3l^2}{32}\epsilon^{\alpha\beta\gamma\delta\sigma}
 \left[\frac{2}{3}\widehat{\nabla}_\alpha \left(\widehat{A}_\sigma
 \widehat{F}_{\gamma\delta}\overline{\widehat{\epsilon}}^a
 \widehat{\psi}_{\beta a}\right)
 +\frac{1}{2}\widehat{F}_{\alpha\beta}\widehat{F}_{\gamma\delta}
 \overline{\widehat{\epsilon}}^a\widehat{\psi}_{\sigma a}
   \right]\right\}.
   \label{superv2}
 \end{eqnarray}

The supersymmtric variations
 of the terms concerning gravitino is quite lengthy. We first calculate the
 variation of the kinetic terms for gravitino,
 \begin{eqnarray}
 && \delta S_{\rm KF} =\delta \int d^5 x \left[-\frac{1}{2}
 \widehat{e}
 \overline{\widehat{\psi}}^a_\alpha \widehat{e}_m^{~\alpha}
 \widehat{e}_n^{~\beta} \widehat{e}_p^{~\delta}
 \widehat{\gamma}^{mnp} \left(\widehat{\nabla}_\beta
 \widehat{\psi}_{\delta a}
 -\frac{3}{4}\widehat{A}_\beta \delta_{ab}
 \widehat{\psi}_\delta^b\right)\right]\nonumber\\
 &=& \int d^5 x \left(-\frac{1}{2}\right) \widehat{e} \left\{
\left[\widehat{\nabla}_\alpha \overline{\widehat{\epsilon}}^a
+\frac{3}{4}\widehat{A}_\alpha\delta^{ab}\overline{\widehat{\epsilon}}_b
-\frac{il}{16} \overline{\widehat{\epsilon}}^a
\left(\widehat{\gamma}_\alpha^{~\delta\sigma}+4\delta_\alpha^{~\delta}
\widehat{\gamma}^\sigma
 \right)\widehat{F}_{\delta\sigma}-
\frac{i}{2l}\delta^{ab}\overline{\widehat{\epsilon}}_b
\widehat{\gamma}_\alpha\right]\right.\nonumber\\
&&\times
\widehat{\gamma}^{\alpha\beta\gamma}\left(\widehat{\nabla}_\beta
\widehat{\psi}_{\gamma a}
-\frac{3}{4}\widehat{A}_\beta\delta_{ac}
\widehat{\psi}_\gamma^c\right)\nonumber\\
&& + \overline{\widehat{\psi}}^a_\alpha
\widehat{\gamma}^{\alpha\beta\delta}
 \widehat{\nabla}_\beta \left[\widehat{\nabla}_\delta
 \widehat{\epsilon}_a-\frac{3}{4}\widehat{A}_\delta
 \delta_{ab}\widehat{\epsilon}^j +\frac{il}{16}
 \left(\widehat{\gamma}_\delta^{~\gamma\sigma}
 -4\delta_\delta^{~\gamma}\widehat{\gamma}^\sigma \right)
 \widehat{\epsilon}_a \widehat{F}_{\gamma\sigma}
 +\frac{i}{2l}\widehat{\gamma}_\delta\delta_{ac}
 \widehat{\epsilon}^c\right]\nonumber\\
&&-\frac{3}{4}\delta_{ab}\overline{\widehat{\psi}}^a_\alpha
\widehat{\gamma}^{\alpha\beta\delta}\widehat{A}_\beta
\left[\widehat{\nabla}_\delta
\widehat{\epsilon}^b+\frac{3}{4}\widehat{A}_\delta
 \delta^{bc}\widehat{\epsilon}_c +\frac{il}{16}
 \left(\widehat{\gamma}_\delta^{~\gamma\sigma}
 -4\delta_\delta^{~\gamma}\widehat{\gamma}^\sigma \right)
 \widehat{\epsilon}^b \widehat{F}_{\gamma\sigma}
 -\frac{i}{2l}\widehat{\gamma}_\delta\delta^{bc}
 \widehat{\epsilon}_c\right]\nonumber\\
 &&\left.+\cdots \right\}
 \nonumber\\
&=& \int d^5x \widehat{e} \left[\widehat{\nabla}_\alpha
 \left(\overline{\widehat{\epsilon}}^a\widehat{\gamma}^{\alpha\beta\delta}
 \widehat{\nabla}_\beta \widehat{\psi}_{\delta a}\right)
 +\left(\widehat{R}_{\alpha}^{~\beta}-\frac{1}{2}
 \widehat{R}\delta_{\alpha}^{~\beta}\right)
 \overline{\widehat{\epsilon}}^a \widehat{\gamma}^{\alpha}
 \widehat{\psi}_{\beta a} \right.\nonumber \\
 && -\frac{3}{4}\widehat{\nabla}_\alpha \left(\delta^{ab}
 \widehat{A}_\beta
 \overline{\widehat{\epsilon}}_a\widehat{\gamma}^{\alpha\beta\delta}
 \widehat{\psi}_{\delta b}\right)
 +\frac{3}{4}\overline{\widehat{\epsilon}}^a
 \widehat{\gamma}^{\alpha\beta\delta}\widehat{\psi}_\alpha^b \delta_{ab}
 \widehat{F}_{\beta\delta} \nonumber \\
 && +\frac{il}{16}
 \widehat{\nabla}_\alpha\left(\widehat{F}_{\gamma\sigma}
 \overline{\widehat{\epsilon}}^a\widehat{\gamma}_\delta^{~\gamma\sigma}
 \widehat{\gamma}^{\alpha\beta\delta}\widehat{\psi}_{\beta a}\right)
 -\frac{il}{8}\widehat{F}_{\gamma\sigma}
 \overline{\widehat{\epsilon}}^a\widehat{\gamma}^{\beta\delta\gamma\sigma}
 \left(\widehat{\nabla}_\beta \widehat{\psi}_{\delta}\right)_a+
 \frac{3il}{4}\widehat{F}^{\alpha\beta}\overline{\widehat{\epsilon}}^a
 \left(\widehat{\nabla}_\alpha
\widehat{\psi}_{\beta}\right)_a\nonumber\\
&& +\frac{3il}{32}
    \widehat{F}_{\gamma\sigma}\widehat{A}_\alpha \delta_{ab}
    \overline{\widehat{\epsilon}}^a
    \widehat{\gamma}^{\alpha\beta\gamma\sigma}\widehat{\psi}_\beta^b
    -\frac{9il}{16}\delta_{ab}\widehat{F}^{\alpha\beta}
    \widehat{A}_\alpha \overline{\widehat{\epsilon}}^a\widehat{\psi}_\beta^b
     \nonumber \\
 && + \frac{il}{4}
 \widehat{\nabla}_\alpha \left( \overline{\widehat{\epsilon}}^a
 \widehat{\gamma}^\sigma\widehat{\gamma}^{\alpha\beta\delta}
 \widehat{\psi}_\beta^a \widehat{F}_{\rho\sigma}\right)
 +\frac{il}{2} \widehat{F}_{\alpha\gamma}
 \overline{\widehat{\epsilon}}^a\widehat{\gamma}^{\alpha\beta\delta\gamma}
 \left(\widehat{\nabla}_\beta\widehat{\psi}_\delta\right)_a
   -\frac{3il}{8}
\widehat{F}_{\alpha\gamma}A_\beta\delta_{ab}
\overline{\widehat{\epsilon}}^a\widehat{\gamma}^{\alpha\beta\delta\gamma}
\widehat{\psi}_\delta^b \nonumber \\
   && \left.+\frac{3i}{2l}\widehat{\nabla}_\alpha
   \left(\overline{\widehat{\epsilon}}^a
\widehat{\gamma}^{\alpha\beta}\widehat{\psi}_{\beta}^b\right)
 -\frac{3i}{l}\delta^{ab}\overline{\widehat{\epsilon}}^a
 \widehat{\gamma}^{\alpha\beta}
 \widehat{\nabla}_\alpha\widehat{\psi}_{\beta}^b-\frac{9i}{4l}
 \widehat{A}_\alpha \overline{\widehat{\epsilon}}^a
  \widehat{\gamma}^{\alpha\beta}\widehat{\psi}_{\beta a}\right].
  \label{superv4}
 \end{eqnarray}
Note that in above and the following calculations, we take into
account only the terms which are at most quadratic in terms of the
fermionic quantities.

 Further, the supersymmetric variation of the gravitino mass-like term is
 \begin{eqnarray}
 && \delta S_{\rm GM} = \delta \left[\frac{3i}{4l}
 \int d^4x \widehat{e} \overline{\widehat{\psi}}_\alpha^a
  \widehat{\gamma}^{\alpha\beta}
 \widehat{\psi}_\beta^b \delta_{ab} \right]
\nonumber\\
 &=&\frac{3i}{4l}\int d^5x
 \widehat{e}\left\{\left[\widehat{\nabla}_\alpha
 \overline{\widehat{\epsilon}}^a
+\frac{3}{4}\widehat{A}_\alpha\delta^{ac}\overline{\widehat{\epsilon}}_c
-\frac{il}{16} \overline{\widehat{\epsilon}}^a
\left(\widehat{\gamma}_\alpha^{~\tau\sigma}
+4\delta_\alpha^{~\tau}\widehat{\gamma}^\sigma
 \right)\widehat{F}_{\tau\sigma}-
\frac{i}{2l}\delta^{ac}\overline{\widehat{\epsilon}}_c
\widehat{\gamma}_\alpha\right]
\widehat{\gamma}^{\alpha\beta} \widehat{\psi}_\beta^b\right.\nonumber\\
&& \left.+\overline{\widehat{\psi}}_\alpha^a
\widehat{\gamma}^{\alpha\beta} \left[\widehat{\nabla}_\beta
\widehat{\epsilon}^b+\frac{3}{4}\widehat{A}_\beta
 \delta^{bd}\widehat{\epsilon}_d +\frac{il}{16}
 \left(\widehat{\gamma}_\beta^{~\tau\sigma}
 -4\delta_\nu^{~\tau}\widehat{\gamma}^\sigma \right)
 \widehat{\epsilon}^b \widehat{F}_{\tau\sigma}
 -\frac{i}{2l}\widehat{\gamma}_\beta\delta^{bd}
 \widehat{\epsilon}_d\right]\right\}\delta_{ab}
 \nonumber\\
&=&\int d^5x \widehat{e} \left[\frac{3i}{2l}
 \left(\widehat{\nabla}_\alpha \overline{\widehat{\epsilon}}^a\right)
 \widehat{\gamma}^{\alpha\beta}\widehat{\psi}_\beta^b \delta_{ab}
 -\frac{9i}{8}\widehat{A}_\alpha \overline{\widehat{\epsilon}}^a
 \widehat{\gamma}^{\alpha\beta}\widehat{\psi}_{\beta a}
 \right.\nonumber\\
 &&\left.
 +\frac{3}{16}\delta_{ab}\widehat{F}^{\alpha\beta}
 \overline{\epsilon}^a\widehat{\gamma}_\alpha \widehat{\psi}_\beta^b
 -\frac{3}{16}\delta_{ab}\widehat{F}_{\alpha\beta}
 \overline{\widehat{\epsilon}}^a\widehat{\gamma}^{\alpha\beta\delta}
 \widehat{\psi}_\delta^b
 -\frac{3}{l^2}\overline{\widehat{\epsilon}}^a\widehat{\gamma}^\alpha
 \widehat{\psi}_{\alpha a}
 \right]
 \label{superv5}
 \end{eqnarray}
 Finally, the supersymmetric variation of the interaction terms of
 the gravitino and graviphoton produces
 \begin{eqnarray}
&& \delta S_{\rm GG} =\delta \int d^5 x \widehat{e}
\left(-\frac{3il}{32}\right)\left(
 \overline{\widehat{\psi}}_\alpha^a
 \widehat{\gamma}^{\alpha\beta\gamma\delta}
 \widehat{\psi}_{\beta a} \widehat{F}_{\gamma\delta}
 +2 \overline{\widehat{\psi}}_\alpha^a \widehat{\psi}_a^\beta
 \widehat{F}^{\alpha\beta}\right)\nonumber\\
 &=& -\frac{3il}{32}\int d^5x \widehat{e}
 \left\{ \widehat{F}_{\gamma\delta} \left(
 \left[\widehat{\nabla}_\alpha \overline{\widehat{\epsilon}}^a
+\frac{3}{4}\widehat{A}_\alpha\delta^{ac}\overline{\widehat{\epsilon}}_c
-\frac{il}{16} \overline{\widehat{\epsilon}}^a
\left(\widehat{\gamma}_\alpha^{~\tau\sigma}+4\delta_\alpha^{~\tau}
\widehat{\gamma}^\sigma
 \right)\widehat{F}_{\tau\sigma}-
\frac{i}{2l}\delta^{ac}\overline{\widehat{\epsilon}}_c
\widehat{\gamma}_\alpha\right]
\widehat{\gamma}^{\alpha\beta\gamma\delta}
\widehat{\psi}_{\beta a}\right.\right.\nonumber\\
&&\left.+\overline{\widehat{\psi}}_\alpha^a
\widehat{\gamma}^{\alpha\beta\gamma\delta}
\left[\widehat{\nabla}_\delta
\widehat{\epsilon}_a-\frac{3}{4}\widehat{A}_\delta
 \delta_{ab}\widehat{\epsilon}^b +\frac{il}{16}
 \left(\widehat{\gamma}_\delta^{~\tau\sigma}
 -4\delta_\delta^{~\tau}\widehat{\gamma}^\sigma \right)
 \widehat{\epsilon}_a \widehat{F}_{\tau\sigma}
 +\frac{i}{2l}\widehat{\gamma}_\delta\delta_{ab}
 \widehat{\epsilon}^b  \right]\right)\nonumber\\
&& +2 \widehat{F}^{\alpha\beta}\left(\left[\widehat{\nabla}_\alpha
\overline{\widehat{\epsilon}}^a
+\frac{3}{4}\widehat{A}_\alpha\delta^{ac}\overline{\widehat{\epsilon}}_c
-\frac{il}{16} \overline{\widehat{\epsilon}}^a
\left(\widehat{\gamma}_\alpha^{~\gamma\sigma}+4\delta_\alpha^{~\gamma}
\widehat{\gamma}^\sigma
 \right)\widehat{F}_{\gamma\sigma}-
\frac{i}{2l}\delta^{ac}\overline{\widehat{\epsilon}}_c
\widehat{\gamma}_\alpha\right]
\widehat{\psi}_{\beta a}\right.\nonumber\\
&& \left.+\overline{\widehat{\psi}}_{\alpha}^a
\left[\widehat{\nabla}_\beta
\widehat{\epsilon}_a-\frac{3}{4}\widehat{A}_\beta
 \delta_{ab}\widehat{\epsilon}^b +\frac{il}{16}
 \left(\widehat{\gamma}_\beta^{~\gamma\sigma}
 -4\delta_\nu^{~\gamma}\widehat{\gamma}^\sigma \right)
 \widehat{\epsilon}_a \widehat{F}_{\gamma\sigma}
 +\frac{i}{2l}\widehat{\gamma}_\beta\delta_{ac}\widehat{\epsilon}^c  \right]
 \right)\nonumber\\
&&\left. +\cdots \right\}\nonumber\\
&=& \int d^5x \widehat{e}\left[-\frac{3il}{8}
\left(\widehat{\nabla}_\alpha
\overline{\widehat{\epsilon}}\right)^a\widehat{\psi}_{\beta
a}\widehat{F}^{\alpha\beta}
-\frac{9il}{32}\delta^{ab}\widehat{A}_\alpha
\widehat{F}^{\alpha\beta} \overline{\widehat{\epsilon}}_a
\widehat{\psi}_{\beta b} \right.\nonumber\\
&& -\frac{3il}{16} \left(\widehat{\nabla}_\alpha
\overline{\widehat{\epsilon}}\right)^a\widehat{\gamma}^{\alpha\beta\gamma\delta}
\widehat{\psi}_{\beta
a}\widehat{F}_{\gamma\delta}-\frac{9il}{64}\widehat{A}_\alpha
\widehat{F}_{\gamma\delta}\delta^{ab}
\overline{\widehat{\epsilon}}_a\widehat{\gamma}^{\alpha\beta\gamma\delta}
\widehat{\psi}_{\beta b}\nonumber\\
&&
+\frac{3l^2}{64}\widehat{F}^{\alpha\beta}\widehat{F}_{\alpha\beta}
\overline{\widehat{\epsilon}}^a\widehat{\gamma}^\delta
\widehat{\psi}_{\delta a}+\frac{3l^2}{64}\widehat{e}^{-1}
\epsilon^{\alpha\beta\gamma\delta\sigma}
\widehat{F}_{\alpha\beta}\widehat{F}_{\gamma\delta}
\overline{\widehat{\epsilon}}^a
\widehat{\psi}_{\sigma a}\nonumber\\
&&+\frac{3l^2}{32}\overline{\widehat{\epsilon}}^a\widehat{\gamma}^{\delta}
 \widehat{\psi}_{\beta a}\widehat{F}^{\beta\gamma}\widehat{F}_{\gamma\delta}
-\frac{3l^2}{32}\overline{\widehat{\epsilon}}^a
\widehat{\gamma}^{\delta}
 \widehat{\psi}_{\beta
 a}\widehat{F}_{\alpha\delta}\widehat{F}_{\alpha\beta}\nonumber\\
&&\left.
-\frac{3}{16}\delta^{ab}\overline{\widehat{\epsilon}}_a\widehat{\gamma}_\alpha
\widehat{\psi}_{\beta b}
\widehat{F}^{\alpha\beta}+\frac{3}{16}\delta^{ab}
\overline{\widehat{\epsilon}}_a\widehat{\gamma}^{\alpha\beta\delta}
\widehat{\psi}_{\alpha b} \widehat{F}_{\beta\delta}\right].
\label{superv6}
 \end{eqnarray}
 In above calculation, the following definitions and algebraic
 operations are repeatedly used,
 \begin{eqnarray}
 \widehat{\gamma}_{\alpha\beta}&=&\frac{1}{2}[\widehat{\gamma}_\alpha,
 \widehat{\gamma}_\beta], ~~
\widehat{\gamma}^{\alpha\beta\delta}=-\frac{1}{2!}\widehat{e}^{-1}\,
\epsilon^{\alpha\beta\delta\sigma\tau}\widehat{\gamma}_{\sigma\tau},
\nonumber\\
\widehat{\gamma}^{\alpha\beta\delta\sigma}&=& \widehat{e}^{-1}\,
\epsilon^{\alpha\beta\delta\sigma\tau}\widehat{\gamma}_{\tau},
~~\widehat{\gamma}_{\alpha\beta\delta\sigma\tau}=\widehat{e}\,
\epsilon_{\alpha\beta\delta\sigma\tau}.
\nonumber \\
 \widehat{\gamma}_{\alpha\beta}\widehat{\gamma}_{\delta\sigma}
&=&
\widehat{e}\,\epsilon_{\alpha\beta\gamma\delta\sigma}\widehat{\gamma}^\sigma-
\left(\widehat{g}_{\alpha\delta}
\widehat{g}_{\beta\sigma}-\widehat{g}_{\alpha\sigma}\widehat{g}_{\beta\delta}\right),
\nonumber\\
\widehat{\gamma}_{\alpha}\widehat{\gamma}_{\beta\delta}
&=&\widehat{\gamma}_{\alpha\beta\delta}
+\widehat{g}_{\alpha\beta}\widehat{\gamma}_\delta-\widehat{g}_{\alpha\delta}
\widehat{\gamma}_\beta.
\end{eqnarray}
\begin{eqnarray}
&&  \widehat{\psi}^a =
C^{-1}\Omega^{ab}\overline{\widehat{\psi}}_b^T=C^{-1}
\overline{\widehat{\psi}}^{aT},
 ~~~\overline{\widehat{\psi}}^a=-\widehat{\psi}^{aT}C,\nonumber\\
 && \overline{\widehat{\psi}}^a\widehat{\gamma}_{\alpha_1
 \cdots\alpha_n}\widehat{\chi}_a =
 -\widehat{\psi}^{aT}C\widehat{\gamma}_{\alpha_1\cdots\alpha_n}
 C^{-1}\overline{\widehat{\chi}}_a^T\nonumber\\
&=&\left\{\begin{array}{l}
 -\widehat{\psi}^{aT}\widehat{\gamma}_{\alpha_1\cdots\alpha_n}
 \overline{\widehat{\chi}}^T_a
=\overline{\widehat{\chi}}_a\widehat{\gamma}_{\alpha_1\cdots\alpha_n}
\widehat{\psi}^a
=-\overline{\widehat{\chi}}^a\widehat{\gamma}_{\alpha_1\cdots\alpha_n}
\widehat{\psi}_a,~~ n=0,1,4,5,\\
\widehat{\psi}^{aT}\widehat{\gamma}_{\alpha_1\cdots\alpha_n}
\overline{\widehat{\chi}}^T_a
=-\overline{\widehat{\chi}}_a\widehat{\gamma}_{\alpha_1\cdots\alpha_n}
\widehat{\psi}^a
=\overline{\widehat{\chi}}^a\widehat{\gamma}_{\alpha_1\cdots\alpha_n}
\widehat{\psi}_a,~~ n=2,3,
\end{array}
\right.\,.
\end{eqnarray}
\begin{eqnarray}
 \widehat{\gamma}^{\alpha\beta\delta}\widehat{\nabla}_\beta
 \widehat{\nabla}_\delta
 \widehat{\chi}_a &=&
\frac{1}{2}\widehat{\gamma}^{\alpha\beta\delta}[\widehat{\nabla}_\beta,
\widehat{\nabla}_\delta] \widehat{\chi}_a
=\frac{1}{8}\widehat{\gamma}^{\alpha\beta\delta}\widehat{R}_{\beta\delta
mn}\widehat{\gamma}^{mn}\widehat{\chi}_a
\nonumber\\
&=&
\frac{1}{8}\widehat{\gamma}^{\alpha\beta\delta}\widehat{R}_{\beta\delta\sigma\tau}
\widehat{\gamma}^{\sigma\tau} \widehat{\chi}_a=
\frac{1}{2}\widehat{\gamma}^{\beta}\left(-\widehat{R}_{\beta}^{~\alpha}
+\frac{1}{2}\delta_{\beta}^{~\alpha}\widehat{R}\right)\widehat{\chi}_a.
 \end{eqnarray}
 We have also employed the Ricci and Bianchi identities for
 the Riemannian curvature tensor and $U(1)$ gauge field strength in five-dimensional
  curved space-time,
 \begin{eqnarray}
 \epsilon^{\alpha\beta\gamma\delta\sigma} \widehat{R}_{\tau\gamma\delta\sigma}=0, ~~~
 \epsilon^{\alpha\beta\gamma\delta\sigma}\widehat{\nabla}_\beta
  \widehat{R}_{\delta\sigma\tau\kappa}=0,~~~
 \epsilon^{\alpha\beta\gamma\delta\sigma}\widehat{\nabla}_\beta
 \widehat{F}_{\delta\sigma}=0.
 \end{eqnarray}
Moreover,  due to the nocommutativity between
$\widehat{\nabla}_\alpha$ and
 $\widehat{\gamma}_{\alpha_1\cdots\alpha_n}$,
 we have repeatedly performed the following operations in the calculation,
 \begin{eqnarray}
 \widehat{\gamma}_{\alpha_1\cdots\alpha_n} \widehat{\nabla}_\alpha (\cdots)
 &=&\left[\widehat{\gamma}_{\alpha_1\cdots\alpha_n},\widehat{\nabla}_\alpha\right] (\cdots)
 +\widehat{\nabla}_\alpha \left[\widehat{\gamma}_{\alpha_1\cdots\alpha_n}
 (\cdots)\right].
 \end{eqnarray}
 Specifically, we work in the inertial coordinate system,
 i.e., the Christoffel symbol $\widehat{\Gamma}^{\alpha}_{\beta\delta}=0$.
 Consequently, the metricity condition gives $
 \partial_\alpha \widehat{e}_\beta^m=0$
 and further the modified spin connection
 \begin{eqnarray}
 \widehat{\omega}_{\alpha\, mn}&=&\frac{1}{2}\widehat{e}_m^{~\beta}
 \left(\partial_\alpha \widehat{e}_{\beta\,n}
 -\partial_\beta \widehat{e}_{\alpha\,n}\right)
 -\frac{1}{2}\widehat{e}_n^{~\beta}\left(\partial_\alpha \widehat{e}_{\beta\,m}
 -\partial_\beta \widehat{e}_{\alpha\,m}\right)-\frac{1}{2}
 \widehat{e}_m^{~\delta}
 \widehat{e}_n^{~\sigma}\left(\partial_\delta \widehat{e}_{\sigma\,p}
 -\partial_\sigma \widehat{e}_{\delta\,p}\right)\widehat{e}_{\alpha}^{~p}
 \nonumber\\
 && +\frac{1}{4}\left(\overline{\widehat{\psi}}^a_\alpha
 \widehat{\gamma}_m\widehat{\psi}_{a\,n}+
 \overline{\widehat{\psi}}^a_\alpha\widehat{\gamma}_m\widehat{\psi}_{a\,n}
 -\overline{\widehat{\psi}}^a_\alpha\widehat{\gamma}_n\widehat{\psi}_{a\,m}
  \right)
 \end{eqnarray}
 has only quadratic fermionic terms.

Putting the supersymmetric variations (\ref{superv1})
--- (\ref{superv6}) together, we obtain the total derivative terms,
\begin{eqnarray}
\delta S &=& \int d^5 x \widehat{e} \widehat{\nabla}_\alpha
\left(-\widehat{e}^\alpha_{~m}\widehat{e}^\beta_{~n} \delta
\widehat{\omega}_{\beta}^{~mn}-
 \frac{9il}{16}
 \overline{\widehat{\epsilon}}^a\widehat{\psi}_{\beta a}\widehat{F}^{\alpha\beta}
 -\frac{1}{2}\overline{\widehat{\epsilon}}^a
 \widehat{\gamma}^{\alpha\beta\delta}\widehat{\nabla}_\beta
 \widehat{\psi}_{\delta a}+ \frac{3}{8}
  \overline{\widehat{\epsilon}}^a\widehat{\gamma}^{\alpha\beta\delta}
 \widehat{\psi}_{\delta}^b\delta_{ab}\widehat{A}_\beta \right.\nonumber\\
 &&\left. -\frac{3il}{32}
 \overline{\widehat{\epsilon}}^a\widehat{\gamma}^{\alpha\beta\delta\sigma}
 \widehat{\psi}_{\beta a}\widehat{F}_{\delta\sigma}
 +\frac{9}{4}\overline{\widehat{\epsilon}}^a\widehat{\gamma}^{\alpha\beta}
 \widehat{\psi}_{\beta}^b\delta_{ab}
 +\frac{l^2}{32}\widehat{e}^{-1}
 \epsilon^{\alpha\beta\gamma\delta\sigma}
 \overline{\widehat{\epsilon}}^a
 \widehat{\psi}_{\sigma}\widehat{A}_\beta
 \widehat{F}_{\gamma\delta}\right)\nonumber\\
 &=& \int d^4x \int dr \partial_\alpha \left[\widehat{e}\left(
 -\widehat{e}^\alpha_{~m}\widehat{e}^\beta_{~n} \delta
\widehat{\omega}_{\beta}^{~mn}-
 \frac{9il}{16}
 \overline{\widehat{\epsilon}}^a\widehat{\psi}_{\beta a}\widehat{F}^{\alpha\beta}
 -\frac{1}{2}\overline{\widehat{\epsilon}}^a
 \widehat{\gamma}^{\alpha\beta\delta}\widehat{\nabla}_\beta
 \widehat{\psi}_{\delta a}+ \frac{3}{8}
  \overline{\widehat{\epsilon}}^a\widehat{\gamma}^{\alpha\beta\delta}
 \widehat{\psi}_{\delta}^b\delta_{ab}\widehat{A}_\beta \right.\right.\nonumber\\
 &&\left.\left. -\frac{3il}{32}
 \overline{\widehat{\epsilon}}^a\widehat{\gamma}^{\alpha\beta\delta\sigma}
 \widehat{\psi}_{\beta a}\widehat{F}_{\delta\sigma}
 +\frac{9}{4}\overline{\widehat{\epsilon}}^a\widehat{\gamma}^{\alpha\beta}
 \widehat{\psi}_{\beta}^b\delta_{ab}
 +\frac{l^2}{32}\widehat{e}^{-1}
 \epsilon^{\alpha\beta\gamma\delta\sigma}
 \overline{\widehat{\epsilon}}^a
 \widehat{\psi}_{\sigma}\widehat{A}_\beta
 \widehat{F}_{\gamma\delta}
\right) \right]. \label{totderiv}
\end{eqnarray}

In the following we take these total derivative terms near $AdS_5$
boundary. Substituting the asymptotical forms listed in (\ref{rgf1})
-- (\ref{strp}) of on-shell fields and supersymmetry transformation
parameters as well as $\widehat{\gamma}$-matrices into
(\ref{totderiv}), one can see that the above supersymmetry variation
is IR divergent due to the infinite $AdS_5$ boundary ($r\to 0$).
Therefore, as the case of evaluating Weyl anomaly,  we must perform
holographic renormalization: integrating over
 the radial coordinate to the cut-off $r=\epsilon>0$, introducing
the counterterm and then taking the limit $\epsilon\to 0$ to remove
the regulator. The finite ambiguity  in adding the counter term is
fixed by the four-dimensional Poincar\'{e} supersymmetry. During
this process the super-Weyl symmetry is violated and the
corresponding super-Weyl anomaly arises. But we can use a shortcut
to extract out this super-Weyl anomaly. We use the induced metric
 \begin{eqnarray}
 {g}_{\mu\nu}(x)=\left.\frac{l^2}{\epsilon^2}\widehat{g}_{\mu\nu}(x,\epsilon)
 \right|_{\epsilon\to 0}
\end{eqnarray}
on $AdS_5$ boundary rather than ${g}_{\mu\nu}(x,\epsilon)$
\cite{hesk,bian1}. In this way we can directly take the $\epsilon\to
0$ limit and find
  that the non-vanishing contribution comes only from the
  term $\widehat{e}^{-1}\epsilon^{\alpha\beta\delta\sigma\tau}
  \overline{\widehat{\epsilon}}^a\widehat{\gamma}_{\tau}
 \widehat{\psi}_{\beta\,a}\widehat{F}_{\delta\sigma}$.
 We obtain \cite{wfchen}
 \begin{eqnarray}
 \delta S &=& \frac{3il^3}{8\times 32\pi G^{(5)}}
 \int d^4 x \epsilon^{\mu\nu\lambda\rho}
 F_{\nu\lambda}\overline{\eta}\gamma_\rho\chi_\mu,
 \label{fvar}
  \end{eqnarray}
  where ${\chi}_\mu$ is the Majorana spinor constructed from the
  left-handed spinor $\chi_\mu^L$ given in (\ref{redsr1}).

The non-vanishing term in Eq.\,(\ref{fvar}) definitely gives the
holographic super-Weyl anomaly  since it is proportional to the
confromal supersymmetry transformation parameter $\eta$.
Substituting the explicit form (\ref{redsr1}) of $\chi_\mu$ into
(\ref{fvar}), we have
 \begin{eqnarray}
 && \delta S = \int d^4x \overline{\eta}\gamma^\mu s_\mu \nonumber\\
  &=& -\frac{l^3}{8\times 16\pi G^{(5)}}\int d^4x
  \left[F^{\mu\nu}D_\mu \psi_\nu
 +\epsilon^{\mu\nu\lambda\rho}\gamma_5 F_{\mu\nu} D_\lambda\psi_\rho
 +\frac{1}{2}\sigma^{\mu\nu} F_{\nu\lambda}
  \left(D_\mu\psi^\lambda-D^\lambda\psi_\mu\right)\right],
  \label{intermerela}
 \end{eqnarray}
 where the $\gamma$-matrix algebraic relations are used,
 $\gamma^\mu\gamma^\nu=g^{\mu\nu}-i\gamma^{\mu\nu}$,
 $\gamma_5\gamma^{\mu\nu}={i}\epsilon^{\mu\nu\lambda\rho}
 \gamma_{\lambda\rho}/{2}$.
Eq.\,({intermerela}) gives the gauge field part of the  super-Weyl
anomaly of ${\cal N}=1$ $SU(N)$ supersymmetric gauge theory at the
leading-order of large-$N$ expansion,
\begin{eqnarray}
\gamma^\mu s_\mu &=& \frac{N^2}{64\pi^2}\left[F^{\mu\nu}D_\mu
\psi_\nu
  +\epsilon^{\mu\nu\lambda\rho}\gamma_5 F_{\mu\nu} D_\lambda\psi_\rho
  +\frac{1}{2}\sigma^{\mu\nu} F_{\nu\lambda}
  \left(D_\mu\psi^\lambda-D^\lambda\psi_\mu\right)\right].
\label{gtra}
\end{eqnarray}

The super-Weyl anomaly of  ${\cal N}=2(4)$ supersymmetric gauge
field theory in external conformal superconformal supergravity
background should be derived from on-shell action of ${\cal N}=4
(8)$ $SU(2)\times U(1)$ ($SO(6)$) gauged supergravity near $AdS_5$
vacuum configuration in a similar way, but the concrete calculation
on the supersymmetry variation is extremely complicated.

\subsection{Concluding Remarks on Holographic Superconformal Anomaly}

We have shown how the superconformal anomaly multiplet of a
supersymmetric gauge theory in a classical conformal supergravity
background can be reproduced from on-shell  gauged supergravity via
the AdS/CFT correspondence. It is amazing that these three distinct
anomalies can be extracted out in the framework of a
five-dimensional gauged supergravity. However, only partial results
 listed in Eq.\,(\ref{exone}) are reproduced from gauged
supergravity. For example, Eqs.\,(\ref{fourano}) and (\ref{gtra})
show that only the gauged field part of chiral- and super-Weyl
anomalies  arise holographically from gauged supergravity
\footnote{The holographic chiral $R$-symmetry anomaly in ${\cal
N}=4$ SYM
 receives no contribution from gravitational background due to its its
 nice field content \cite{blau}. Thus the gauge field CS term is
 fully responsible for the holographic
 origin of the chiral R-symmetry anomaly. This
 also explains the absence of the
 $R_{\mu\nu\lambda\rho}R^{\mu\nu\lambda\rho}$ term
 in the holographic Weyl
 anomaly of ${\cal N}=4$ SYM found in Ref.\,\cite{hesk}.
 For ${\cal N}=1,2$ supersymmetric gauge theories the
 gravitational background part
  should holographically  arise from the supergravity side.}. The
reason for the failure of revealing this gravitational part is that
the five-dimensional gauged supergravity (or the type IIB
supergravity in $AdS_5\times X^5$ background) is only the lowest
approximation to the type IIB superstring theory in $AdS_5\times
X^5$ background. Thus it is possible that the gravitational part
cannot be revealed within the five-dimensional gauged supergravity
itself, and one must consider the higher-order gravitational action
 such as the Gauss-Bonnet term generated from the superstring theory
 \cite{berg2}. It was found that in Ref.\,\cite{ahar}
   that for an ${\cal N}=2$  $USp(2N)$ supersymmetric
  gauge theory coupled to two hypermultiplets in the fundamental and
  antisymmetric tensor representations of the gauge group,
  the gravitational background part of its holographic chiral R-symmetry
  anomaly does come from a mixed CS term in a gauged supergravity with
  higher order gravitational correction. Specifically, this CS term originates
  from the compactification on $S^3$ of the Wess-Zumino
  term describing the interaction of the R-R 4-form field with
  eight $D7$-branes and one orientifold 7-plane system
  in type IIB superstring. In particular,
  this gravitational background term is at the subleading $N$ order
  rather than the leading $N^2$ order in the large-$N$ expansion.

 Finally, it should be pointed out that the essence of
  the holographic anomaly
  is the anomaly inflow from the bulk theory
  to the $AdS_5$ boundary \cite{callan}.

\section{Internal Superconformal Anomaly from $D$-brane Dynamics
and Refection in Brane Solution of Type II Supergravity}

\subsection{Fractional Brane as the Origin of Internal
Superconformal Anomaly in Low-energy Brane Dynamics}

In this section, we shall try to understand the gravity dual of the
internal superconformal anomaly. As stated in Sect. II, this type of
superconformal anomaly  originates from the dynamics of
supersymmetric gauge theory itself.  It has nothing to do with the
external background field and and its coefficient is proportional to
the beta function of the theory. Therefore, the gravity dual of this
type of superconformal anomaly should have no holographic meaning
and we should go to the string origin of a supersymmetric gauge
theory to look for its gravity dual.

As stated in Introduction, a supersymmetric theory can be obtained
through $D$-brane engineering in weakly coupled type-II superstring,
i.e., placing a stack of $D3$-branes (maybe several types of
$D$-branes) in a background space-time \cite{giku} Thus any
dynamical phenomenon like the anomaly of a supersymmetric gauge
theory should originate from certain brane configuration setting-up.
We will see that the internal superconformal conformal anomaly
actually relates to the fractional D-brane fixed at the singularity
of a singular background space-time. On the other hand, in strong
coupled type II superstring theory, $Dp$-branes appear as a
$p$-brane solutions to type II supergravity, i.e., space-time
background to type II superstring or type II supergravity.
Therefore, we can observe how the existence of D-branes modify the
space-time background and further investigate  how this modification
affect the supergravity in this space-time background to find the
dual description to the internal superconformal anomaly. In the
following we shall reveal how the internal superconformal anomaly is
embodied in the construction of a $D$-brane configuration.

Since a $Dp$-brane is characterized by  open strings ending on it.
 The low-energy dynamics of a $Dp$-brane system
describes the dynamical behavior of   open string modes trapped on
world-volume of $Dp$-branes and their interaction with bulk
supergravity. The corresponding effective action consists of two
parts. One is the Dirac-Born-Infeld (DBI) action (in Einstein
framework) and it describes the interaction between $Dp$-branes and
NS-NS fields of type II superstring,
\begin{eqnarray}
S_{\rm DBI}&=&-\tau_{p} \int_{V_{p+1}} d^{p+1}x e^{(p-3)\phi/4}
\mbox{Tr}\sqrt{-\det\left[g_{\mu\nu}+ e^{-\phi/2}
\left(B_{\mu\nu}+2\pi\alpha^\prime F_{\mu\nu}\right)\right]}.
\nonumber\\
 \mu, \nu &=&0,\cdots, p;
 \label{dbiaction}
\end{eqnarray}
 The other one represents
 the interaction of $Dp$-branes with R-R fields described
by the toplogical Wess-Zumino (WZ) term,
\begin{eqnarray}
S_{\rm WZ}=\mu_{p}\int_{V_{p+1}} \left[C\wedge \mbox{Tr}
\left(e^{B+2\pi\alpha^\prime F}\right)\right]_{p+1}. \label{wzterm}
\end{eqnarray}
In above equations,
 the parameters
$\tau_p$ and $\mu_p$ are the tension and R-R charge of a $Dp$-brane.
which are actually equal because of the BPS-saturation property of a
$Dp$-brane,
\begin{eqnarray}
\tau_p=\mu_p=\frac{T_p}{\kappa_{\rm 10}}=\frac{1}{(2\pi)^p
{\alpha^{\prime}}^{(1+p)/2}},~~T_p=\sqrt{\pi}\left(2\pi\sqrt{\alpha^\prime}
\right)^{3-p};
\end{eqnarray}
The gravitational  coupling $\kappa_{\rm 10}$ is related to the
ten-dimensional Newtonian constant $G_{N(10)}$ derived from closed
string, $2\kappa^2_{\rm 10}=16\pi G_{N(10)}=(2\pi)^7
g_s^2\alpha^{\prime 4}$. $F_{\mu\nu}$ is the strength of gauge field
living on $Dp$-brane world-volume;
 $G_{\mu\nu}$, $B_{\mu\nu}$ and
 $C_{\mu_1\mu_2\cdots\mu_n}$ are the pull-backs of ten-dimensional
bulk metric $G_{MN}$, the antisymmetric NS-NS field $B_{(2)MN}$ and
R-R fields  $C_{M_1\cdots M_n}$ to the $p+1$-dimensional
world-volume,
\begin{eqnarray}
&& g_{\mu\nu}=G_{MN}\partial_\mu X^{M} \partial_\nu X^{N}, ~~
B_{\mu\nu}=B_{MN}\partial_\mu X^{M} \partial_\nu X^{N},\nonumber\\
&& C_{\mu_1\mu_2\cdots\mu_n}=C_{M_1M_2\cdots M_n}\partial_{\mu_1}
X^{M_1}\partial_{\mu_2} X^{M_2}\cdots \partial_{\mu_n} X^{M_n}, ~~
n\leq p+1;
\end{eqnarray}
Finally, the trace operation takes over certain representation of
gauge group.

Now let us see how the superconformal anomaly is covered in the
$D$-brane configuration. First, when the transverse space like
conifold or orbifold has singularity, the closed string states
consist of both the untwisted and twisted sectors \cite{poch3}. The
fractional $Dp$-branes frozen at the singular point of transverse
space are identical to the $D(p+2)$-branes wrapped on two-cycle
${\cal C}_2$ like $S^2$ and the singular point can be considered as
a vanishing two-cycle \footnote{ A roughly explanation is the
following \cite{di2}: the fixed point in orbifold $R^4/\Gamma$ for
$\Gamma=Z_2$ or $Z_2\times Z_2$ and the apex of  conifold  with base
$T^{1,1}=S^5/Z_2$ are singular limits of ALE spaces. Usually an ALE
space contains compact 2-cycles, which shrinks to zero size at the
fixed point of orbifold. Specifically, the McKay correspondence
theorem states that the simple root $\alpha_I$ of simply-laced Lie
algebra has one-to-one correspondence with a 2-cycle in any ALE
space and that especially the root corresponds to an irreducible
representation of $\Gamma$. On the other hand, the fractional brane,
as a boundary state of closed string theory, constitutes an
irreducible of $\Gamma$. Therefore, a connection between the
fractional brane and a vanishing 2-cycle can be established.}. The
fields corresponding to the twisted string states emitted by
fractional $Dp$-branes should locate at the singular point of target
space-time. Based on the viewpoint that a $Dp$-brane is considered
as a wrapped $D(p+2)$-brane, the twisted fields  decompose into two
parts, one part living on the $p+1$-dimensional world-volume of
fractional $Dp$-brane, and the other on the blow-up of the vanishing
two-cycle ${\cal C}_2$,
\begin{eqnarray}
V_{p+3}=V_{p+1}\times {\cal C}_2, ~~~ B_{(2)}=b (x)\omega_2,
~~~C_{(p+3)}=C_{(p+1)}\wedge \omega_2. \label{fracfield}
\end{eqnarray}
In above equation $\omega_2$ is the differential two-form defined on
the blow-up of the vanishing 2-cycle, the twisted scalar field
$b(x)$ and the pull-back $c_{\mu_1\cdots\mu_{p+1}}$ of R-R
antisymmetric tensor fields
 live on the $p+1$-dimensional
world-volume of $Dp$-branes. Substituting the fields in
(\ref{fracfield}) into the DBI action (\ref{dbiaction}) and WZ term
(\ref{wzterm}) and expanding them to the quadratic terms of the
gauge field strength, we obtain
\begin{eqnarray}
&& S_{\rm DBI}+S_{\rm WZ}\nonumber\\
&=&-\tau_{p+2} \int_{V_{p+1}} d^{p+1}x\, e^{(p-3)\phi/4}
\mbox{Tr}\sqrt{-\det\left(g_{\mu\nu}+ 2\pi\alpha^\prime e^{-\phi/2}
F_{\mu\nu}\right)}\int_{{\cal C}_2} B_{(2)}\nonumber\\
&&+\mu_{p+2}\left[\int_{V_{p+1}} C_{(p+1)}\int_{{\cal C}_{2}}\,
C_{(2)}+\frac{1}{2}(2\pi\alpha^\prime)^2 \int_{V_{p+1}}
C_{(p-3)}\wedge\mbox{Tr}\left(F\wedge F\right) \int_{{\cal
C}_2}\,C_{(2)}\right.\nonumber\\
&&+\left.\int_{V_{p+1}} C_{(p+1)}\int_{{\cal C}_2}\,
B_{(2)}+\frac{1}{2}(2\pi\alpha^\prime)^2 \int_{V_{p+1}}
C_{(p-3)}\wedge\mbox{Tr}\left(F\wedge F\right) \int_{{\cal
C}_2}\,B_{(2)} +\cdots\right]
\nonumber\\
&=& S_{\rm brane-bulk}+ S_{\rm gauge}, \label{expdbi0}
\end{eqnarray}
 where $S_{\rm brane-bulk}$ is the action  describing the
 interaction of fractional $Dp$-brane with bulk fields and $S_{\rm
 gauge}$ gives the gauge field action on the world-volume of
the $Dp$-brane,
 \begin{eqnarray}
 && S_{\rm brane-bulk} =
-\tau_p\int_{V_{p+1}} d^{p+1}x\, e^{(p-3)\phi/4}
\sqrt{-g}\,\frac{1}{(2\pi\sqrt{\alpha^\prime})^2} \int_{{\cal C}_2}
B_{(2)} \nonumber\\
&& +\mu_p\left[\int_{V_{p+1}} C_{(p+1)}\,
\frac{1}{(2\pi\sqrt{\alpha^\prime})^2} \int_{{\cal C}_2} C_{(2)}+
\int_{V_{p+1}} C_{(p+1)}\, \frac{1}{(2\pi\sqrt{\alpha^\prime})^2}
\int_{{\cal C}_2} B_{(2)}\right]+\cdots
;\nonumber\\
&& S_{\rm gauge} =-\alpha^\prime \tau_p \int_{V_{p+1}} d^{p+1}x\,
e^{(p-3)\phi/4} \sqrt{-g}
\left[-\frac{1}{4}\mbox{Tr}\left(F^{\lambda\rho}
F_{\lambda\rho}\right)\right]\,e^{-\phi} \int_{{\cal C}_2} B_{(2)} \nonumber\\
&& + \frac{1}{2}\alpha^\prime \mu_p \left[\int_{V_{p+1}}
\mbox{Tr}\left( F\wedge F\right)\wedge C_{(p-3)}\,\left(\int_{{\cal
C}_2} C_{(2)} +\int_{{\cal C}_2} B_{(2)}\right) \right]+\cdots,
\nonumber\\
 && \mu, \nu =0,\cdots, p; ~~~\alpha,\beta=0,\cdots,p, a,b;
 \label{expdbi}
\end{eqnarray}
where $a,b=1,2$  are the indices on the vanishing 2-cycle ${\cal
C}_2$. Note that in Eqs.\,(\ref{expdbi0}) and (\ref{expdbi})
 we choose $X^I$=Constant ($I=p+1,\cdots,9$) for the convenience
 of discussion, and hence the pull-back
$G_{IJ}$ to world volume vanishes. Recalling the relation between
the gauge couplings and string couplings,
\begin{eqnarray}
\frac{4i\pi}{g^2_{\rm YM}}+\frac{\theta^{\rm (CP)}}{2\pi}=i
e^{-\phi}+\frac{C_{(0)}}{2\pi},
\end{eqnarray}
and comparing Eq.\,(\ref{expdbi}) (in the case of $p=3$) with
Eqs.\,(\ref{claco})--(\ref{effcou}) we can see  that it is the
NS-NS- and R-R two-form fluxes $\int_{{\cal C}_2} B_{(2)}$,
$\int_{{\cal C}_2} C_{(2)}$ carried by fractional branes through the
shrunken 2-cycle that have created the running of gauge coupling and
the shift of $\theta$-angle and consequently lead to the internal
superconformal anomaly of a supersymmetric gauge theory. Therefore,
we conclude that the fractional branes frozen at the singular point
of background space-time is the origin of superconformal anomaly of
a supersymmetric gauge theory in brane configuration.

 This observation can be made more  explicitly and quantitatively
by the brane probe technique \cite{di2,cliff}.

A concrete example is a $D3$-brane configuration in type IIB
superstring theory. It consists of $N$ $D3$-branes and $M$
fractional $D3$-branes  in  a target space-time $M^4 \times {\cal
C}_6$. ${\cal C}_6$ is  a conifold with base $T^{1,1}=[SU(2)\times
SU(2)]/U(1)$, which is a Calabi-Yau threefold and has $SU(3)$
holonomy. The apex of ${\cal C}_6$ is  a singular point in
background space-time. Therefore, one can put not only a stack of
$N$ $D3$-branes moving freely in the transverse space, but also a
stack of $M$ fractional $D3$-branes fixed at the singularity. All
these $D3$-branes extend in  four-dimensional Minkowski space-time
$M^4$. Topologically $T^{1,1}\sim S^2\times S^3$, those fractional
$D3$-branes can be considered as $D5$-branes wrapped around the
vanished two-cycle $S^2$ at the apex of ${\cal C}_6$. The
supersymmetric gauge theory coming from this brane configuration is
just the ${\cal N}=1$ $SU(N+M)\times SU(N)$ supersymmetric gauge
theory with two flavors matter chiral superfields in the
bifundamental representations $(N+M,\overline{N})$ and
$(N,\overline{N+M})$. The superconformal anomaly multiplet listed in
Eqs.\,(\ref{chira2}), (\ref{scal2}) and (\ref{gammatrace2}) shows
that the anomaly coefficient is indeed proportional $M$, the number
of $D3$-fractional branes. This confirms that the fractional brane
is origin of the internal superconformal anomaly in brane
engineering.

\subsection{Modification on Brane Solution to Type Supergravity
and Symmetry Breaking by Fractional Brane}

\subsubsection{Deformation on $AdS_5\times T^{1,1}$ Background
by Fractional $D3$-branes}

Since in the strongly coupled type II superstring theory, a stack of
$Dp$-branes behaves as a $p$-brane solution to type supergravity,
the existence of fractional branes in brane configuration definitely
modifies the $p$-brane solution produced by bulk $Dp$-branes. In the
following we still take above example in type IIB superstring ---
the $N$ $D3$-branes and $M$ fractional $D3$-branes  in  the
 space-time  background $M^4 \times {\cal C}_6$ --- to expound how the
fractional $D3$-branes modify three-brane solutions of type IIB
supergravity and hence give rise to the symmetry breaking in the
geometric background described by three-brane solution.

We first consider the case without fractional $D3$-branes, i.e, the
case of $M=0$. The three-brane solution to type IIB supergravity
takes the following form,
\begin{eqnarray}
ds^2 &=& H^{-1/2}(r)\eta_{\mu\nu}dx^\mu dx^\nu+H^{1/2}(r)\left
(dr^2+
r^2 ds^2_{T^{1,1}}\right),\nonumber\\
F_{(5)}&=&{\cal F}_{(5)}+\widetilde{\cal F}_{(5)}, \nonumber\\
{\cal F}_{(5)} &=& \frac{1}{4}\pi \alpha^{\prime 2}N g^1 \wedge g^2
\wedge g^3 \wedge g^4 \wedge g^5=\frac{1}{2}\pi \alpha^{\prime 2}N
\omega^2\wedge \omega^3,\nonumber\\
 H(r) &=& 1+\frac{L^4}{r^4}, ~~ L^4=\frac{27\pi g_s
N(\alpha^\prime)^2}{4},\nonumber\\
ds^2_{T^{1,1}} &= &\frac{1}{9}
(\sigma^{\widetilde{3}}+\sigma^{\widehat{3}})^2+\frac{1}{6}
\sum_{{\widetilde{1}}=1}^2
(\sigma^{\widetilde{i}})^2+\frac{1}{6}\sum_{\widehat{i}=1}^2
(\sigma^{\widehat{i}})^2 \nonumber\\
&=& \frac{1}{9} (g^5)^2+\frac{1}{6}\sum_{m=1}^4 (g^m)^2, \nonumber\\
\omega_2 &=&\frac{1}{2}\left(g^1\wedge g^2+g^3\wedge
g^4\right)\nonumber\\
&=& \frac{1}{2}\left(\sin\theta_1 d\theta_1\wedge d\phi_1-
\sin\theta_2 d\theta_2\wedge d\phi_2\right), \nonumber\\
\omega_3 &=& g^5\wedge \omega_2,\nonumber\\
 \int_{S^2}\omega_2 &=& 4\pi, ~~~\int_{S^3}\omega_3=8\pi^2.
\label{nfsolu1}
\end{eqnarray}
In above equations, the basis $g^m$ are one-form basis on $T^{1,1}$,
\begin{eqnarray}
g^1 &\equiv&
\frac{\sigma^{\widetilde{1}}-\sigma^{\widehat{1}}}{\sqrt{2}},
~~~g^2\equiv\frac{\sigma^{\widetilde{2}}
-\sigma^{\widehat{2}}}{\sqrt{2}}, ~~~
g^3\equiv\frac{\sigma^{\widetilde{1}}
+\sigma^{\widehat{1}}}{\sqrt{2}}, \nonumber\\
g^4 &\equiv &
\frac{\sigma^{\widetilde{2}}+\sigma^{\widehat{2}}}{\sqrt{2}}, ~~~~
g^5\equiv\sigma^{\widetilde{3}}+\sigma^{\widehat{3}} ,
\end{eqnarray}
and
\begin{eqnarray}
\sigma^{\widetilde{1}}& {\equiv}&\sin\theta_1d\phi_1,
~~~\sigma^{\widehat{1}}{\equiv}\cos 2\beta\sin\theta_2d\phi_2-\sin
2\beta d\theta_2, \nonumber\\
 \sigma^{\widetilde{2}} &{\equiv}& d\theta_1,~~~\sigma^{\widehat{2}}{\equiv}\sin
2\beta\sin\theta_2d\phi_2+\cos 2\beta
d\theta_2,\nonumber\\
\sigma^{\widetilde{3}}&\equiv& \cos\theta_1d\phi_1, ~~~
\sigma^{\widehat{3}}\equiv 2d\beta+\cos\theta_2 d\phi_2 .
\end{eqnarray}
All other fields in this solution vanish. Near the horizon limit
$r\rightarrow 0$, Eq.\,(\ref{nfsolu1}) yields
\begin{eqnarray}
 ds_{10}^2 &=& \frac{r^2}{L^2}\eta_{\mu\nu} dx^\mu  dx^\nu
+\frac{L^2}{r^2} dr^2+L^2 ds^2_{T^{1,1}},\nonumber\\
 F_{(5)} &=&
{\cal F}_{(5)}+{}^\star{\cal F}_{(5)}\nonumber\\
&=& \frac{r^3}{g_s L^4} dx^0 \wedge dx^1 \wedge dx^2 \wedge dx^3
\wedge dr + \frac{L^4}{27 g_s} g^1\wedge g^2\wedge g^3\wedge
g^4\wedge g^5,\nonumber\\
 L^2 &=& \frac{3\sqrt{3\pi g_sN}\alpha^\prime}{2}.
 \label{at11}
\end{eqnarray}
It is exactly the $AdS_5\times T^{1,1}$ space-time background for
type IIB superstring. This leads to the correspondence between type
IIB supersting in $AdS_5\times T^{1,1}$ background and ${\cal N}=1$
$SU(N)\times SU(N)$ supersymmetric gauge theory with two flavors in
bifundamental representations $(N,\overline{N})$ and
$(\overline{N},N)$ at the fixed point of its renormalization group
flow.

When those $M$ $D3$ fractional branes are switched on at the apex of
the conifold, the resultant background to type IIB superstring  is
the celebrated solution to type IIB supergravity found by Klebanov
and Strassler \cite{klst}. This solution is composed of not only the
ten-dimensional space-time metric, the self-dual five-form field
strength $\widehat{F}_{(5)}$, but also the NS-NS- and R-R two-forms,
$B_{(2)}$ and $C_{(2)}$ living only on $T^{1,1}\sim S^2\times S^3$,
\begin{eqnarray}
ds_{10}^2 &=& h^{-1/2}(\tau) dx_{1,3}^2+ h^{1/2}(\tau) d s_6^2,\nonumber\\
\widehat{F}_{(5)}&=& {\cal
F}_{(5)}+\widetilde{F}_{(5)},~~~\widetilde{\cal F}_{(5)}=dx^0\wedge
dx^1\wedge dx^2\wedge dx^3\wedge dh^{-1},\nonumber\\
B_{(2)}&=&\frac{g_sM\alpha^\prime}{2}\left[ f(\tau)g^1\wedge
g^2+k(\tau)
g^3\wedge g^4\right], \nonumber\\
F_{(3)}&=&\frac{M\alpha^\prime}{2}\left\{g^3\wedge g^4\wedge
g^5+d\left[ F(\tau)\left(g^1\wedge g^3+g^2\wedge g^4\right) \right]
\right\}. \label{ksso}
\end{eqnarray}
In this solution $h(\tau)$ is the warp factor, $ds_6^2$ is the
metric of deformed conifold \cite{klst,rama},
\begin{eqnarray}
ds_6^2&=&\frac{1}{2} (12)^{1/3} K(\tau)\left\{\frac{1}{3[K(\tau)]^3}
\left[ d\tau^2+
(g^5)^2\right]\right.\nonumber\\
&&\left.+\sinh^2\left(\frac{\tau}{2}\right)\left[ (g^1)^2+(g^2)^2
\right]+\cosh^2\left(\frac{\tau}{2}\right)\left[ (g^3)^2+(g^4)^2
\right] \right\}, \nonumber\\
K(\tau) &=& \frac{(\sinh 2\tau-2\tau)^{1/3}}{2^{1/3}\sinh\tau},\nonumber\\
F(\tau) &=& \frac{\sinh\tau-\tau}{2\sinh\tau},\nonumber\\
f(\tau) &=&
\frac{\tau\coth\tau-1}{2\sinh\tau}(\cosh\tau-1),\nonumber\\
k(\tau) &=&
\frac{\tau\coth\tau-1}{2\sinh\tau}(\cosh\tau+1),\nonumber\\
h(\tau) &=&C (g_sM)^2\frac{2^{2/3}}{4}\int_\tau^\infty dx
\frac{x\coth x-1}{\sinh^2x}\left[\sinh (2 x)-2x \right]^{1/3},
\label{kssoe}
\end{eqnarray}
$C$ is the normalization constant. In the small $\tau$ limit,
$h(\tau\to 0)\rightarrow (g_sM)^2$, the background for
ten-dimensional type IIB supergravity is approximately the
$M^4\times$deformed conifold,
\begin{eqnarray}
&& \left.ds_{10}^2\right|_{\tau\to 0}\rightarrow (g_sM)^{-1}
\eta_{\mu\nu} dx^\mu dx^\nu+ g_sM
\left(\frac{1}{2}d\tau^2+d\Omega_3^2+\frac{1}{4}\tau^2
[(g^1)^2+(g^2)^2] \right),\nonumber\\
&& F(\tau\to 0)\rightarrow 0, ~~\left.F_{(3)}\right|_{\tau\to
0}\rightarrow M g^3\wedge g^4\wedge g^5,\nonumber\\
&&f(\tau\to 0)\rightarrow 0, ~~k(\tau\to 0)\rightarrow
0,~~~\left.B_{(2)}\right|_{\tau\to 0}\rightarrow 0. \label{ksso1}
\end{eqnarray}
Therefore, this solution is non-singular since the apex of the
conifold has been resolved by $F_{(3)}$ fluxes fractional brane
\cite{klst}, which prevents the three-cycle ($\sim S^3$) from
shrinking to a point.

In large-$\tau$ case, the radial coordinate $\tau$ is relevant to
the radial coordinate $r$ as the following,
\begin{eqnarray}
r^3\sim 12^{1/2}e^\tau ~~~ \mbox{at large}~\tau. \label{ksso2}
\end{eqnarray}
Consequently, the solution becomes the Klebanov-Tseytlin (K-T)
solution \cite{tseytlin},
\begin{eqnarray}
ds_{10}^2 &=& h^{-1/2}(r) dx_{1,3}^2+ h^{1/2}(r) d s_6^2,\nonumber\\
ds^2_6 &=& dr^2+r^2 ds^2_{T^{1,1}},\nonumber\\
 h(r)&=&\frac{27\pi (\alpha^\prime)^2}{4r^4}\left[ g_s
N+\frac{3}{2\pi}(g_s
M)^2\ln\left(\frac{r}{r_0}\right)+\frac{3}{8\pi} (g_sM)^2\right],
\nonumber\\
F_{(3)} &=& d C_{(2)}= c (r) \omega_3, ~~~B_{(2)}=b(r) \omega_2, \nonumber\\
{\cal F}_{(5)} &=& N_{\rm eff}(r) \omega_2
\wedge \omega_3, \nonumber\\
 b(r)&=& \frac{3g_s
 M\alpha^\prime}{2}\ln\left(\frac{r}{r_0}\right),~~
 c(r)=\frac{M\alpha^\prime}{2}, \nonumber\\
 N_{\rm eff}(r)&=& N+\frac{3}{2\pi} g_sM^2
 \ln\left(\frac{r}{r_0}\right),~~
 C_{(2)} = M\alpha^\prime \beta \omega_2~~\mbox{ (locally)}.
 \label{ksso3}
\end{eqnarray}
All other fields vanish. At large $\tau$ (or equivalently, large-$r$
limit) the solution shows clearly that there are
 $M$ units of R-R three-form fluxes passing through the 3-cycle $S^3$ of
$T^{1,1}$.

 On the field theory side, the ${\cal N}=1$ $SU(N+M)\times SU(N)$
 supersymmetric gauge theory  with two chiral superfields in
 the bifundamental representations
 $(N+M,\overline{N})$ and $(\overline{N+M},N)$ is no longer a
 superconformal field theory. The zero point of its beta function
 is removed due to the existence of  fractional
 branes.

 In above subsection the origin of
superconformal anomaly of the field theory  is identified as the
 fractional $D$-branes in  brane configuration. Therefore, we
see that the K-S solution (\ref{ksso}) furnishes a space-time
background for type IIB
 superstring (or supergravity) and it can be considered as a deformation
 on the three-brane solution (\ref{nfsolu1}) obtained from above $D3$-brane
 configuration without fractional $D3$-branes. In the following
 we analyze the symmetry breaking reflected in the space-time
 geometry described by the K-S solution so that an quantitative investigation
 on the gravity dual of the superconformal anomaly can proceed.

\subsubsection{Breaking of $U(1)$ Rotational Symmetry and Scale
Symmetry in Transverse Space by Fractional Brane}

Generally speaking, in a brane configuration, the scale symmetry and
chiral $R$-symmetry of a supersymmetric gauge theory living on the
world-volume of $Dp$-branes come from the geometrical scale- and
rotation transformation invariance  of transverse space,
\begin{eqnarray}
x^I\longrightarrow \mu e^{i\alpha}x^I, ~~p+1\leq I\leq 10.
\end{eqnarray}
The amount of supersymmetries in supersymmetric gauge theory depends
on how many supersymmetries are preserved by such a brane
configuration. Since a $D$-brane configuration is represented by a
brane solution to supergravity, one can thus equivalently analyze
the symmetry in the background geometry described by brane solution
to observe the effects of fractional branes.

  Based on
this argument, let us observe the manifestation of   the scale
symmetry and $U(1)\cong SO(2)$ rotation symmetry in the K-S
solution. First, in the case with no fractional branes ($M=0$). The
near-horizon limit of the three-brane solution  is $AdS_5\times
T^{1,1}$ listed in (\ref{at11}). This solution shows explicitly that
the scale symmetry is trivially present in the transverse space part
and the $U(1)$ symmetry is reflected in the invariance  under  the
following $\beta$-angle rotation,
\begin{eqnarray}
\beta\longrightarrow \beta+\alpha.
\end{eqnarray}

When  fractional $D3$-branes switch on at the apex of ${\cal C}_6$,
the solution
 (\ref{ksso}) shows that near the horizon $r\to 0$ the metric ceases to be
$AdS_5\times T^{1,1}$. This is because that not  only $h(r)$ and
$F_{(5)}$ get additional
 logarithmic  dependence,
\begin{eqnarray}
  h(r)&\sim& \frac{1}{r^4} \left[C_1+C_2\ln
 \frac{r}{r_0}\right], \nonumber\\
  \widehat{F}_5 &\sim& \left[N+C_3\ln \frac{r}{r_0}\right] [
 \omega_2\wedge\omega_3+{}^\star(\omega_2\wedge\omega_3)],
\label{logari}
 \end{eqnarray}
 but also the background fields $B_{(2)MN}$ and $C_{(2)MN}$
 have been generated by the fractional brane fluxes.
 The coefficients $C_i$  ($i=1,2,3$) in (\ref{logari}) can be
 extracted out from the solution (\ref{ksso3}).

 It was analyzed in Ref.\,\cite{kow} that it is the non-invariance
 of $C_{(2)}$ under the $\beta$-angle rotation that leads to
the chiral R-symmetry anomaly in ${\cal N}=1$ supersymmetric
$SU(N+M)\times SU(N)$ gauge theory. A similar analysis can show that
it is the non-invariance of $B_{(2)}$ under scale transformation in
transverse space that results in the scale anomaly in ${\cal N}=1$
supersymmetric $SU(N+M)\times SU(N)$ gauge theory. Under the scale
transformation $r\rightarrow \mu r$, the $B_{(2)}$ given in
Eq.\,(\ref{ksso}) transforms as
\begin{eqnarray}
B_{(2)}\longrightarrow
B_{(2)}+\frac{3g_sM\alpha^\prime}{2}\omega_2\ln \mu . \label{bfv}
\end{eqnarray}
The relation between string coupling and gauge couplings of the
${\cal N}=1$ supersymmetric $SU(N+M)\times SU(N)$ gauge theory
\cite{klst,klwi} gives
\begin{eqnarray}
\frac{1}{g_1^2}+\frac{1}{g_2^2}\sim e^{-\phi}\sim \frac{1}{g_s},
~~~\frac{1}{g_1^2}-\frac{1}{g_2^2}\sim e^{-\phi} \left[
\int_{S^2}B_{(2)}-\frac{1}{2}\right],
\end{eqnarray}
so we have
\begin{eqnarray}
\frac{1}{g_1^2}\sim \frac{1}{2g_s}\left[
\int_{S^2}B_{(2)}-\frac{1}{2}\right], ~~~\frac{1}{g_1^2}\sim
\frac{1}{2g_s}\left[-\int_{S^2}B_{(2)}+\frac{3}{2}\right].
\label{twogc}
\end{eqnarray}
Comparing  Eq.\,(\ref{twogc}) with Eq.\,(\ref{bfv}), one can see
that the variation of $B_{(2)}$ under scale transformation is
relevant to $\beta$-functions of the
 $ SU(N+M)\times SU(N)$ gauge couplings at the IR fixed point,
 \cite{klst},
\begin{eqnarray}
 \frac{d}{d(\ln \mu)}\frac{8\pi^2}{g_1^2(\mu)}\sim 3M, ~~~
\frac{d}{d(\ln \mu)}\frac{8\pi^2}{g_2^2(\mu)}\sim -3M.
\end{eqnarray}
Therefore, the  scale anomaly coefficients  of ${\cal N}=1$
$SU(N+M)\times SU(M)$ gauge theory given in Eq.\,(\ref{scal2}) is
reproduced.

\subsubsection{Breaking of Supersymmetry by Fractional Brane in  Three-brane
Solution Background }

 The supersymmetry breaking produced by fractional $D3$-branes
 can be detected by  checking the supersymmetry transformations
 of fermionic fields
$\Lambda$ and $\Psi_M$ of  type IIB supergravity in the space-time
background  described by K-S solutiopn (\ref{ksso}) and observing
the existence of Killing spinors. The field content of type IIB
supergravity consists of  the dilaton field $\phi$, the metric
$G_{MN}$ and  the second-rank antisymmetric tensor field $B_{(2)MN}$
from NS-NS sector and axion field $C_{(0)}$, the two-form potential
$C_{(2)MN}$ and  four-form potential $C_{(4)MNPQ}$ with self-dual
field strength from R-R sector. The fermionic fields are left-handed
complex Weyl gravitino  $\Psi_M$ and right-handed complex Weyl
dilatino ${\Lambda}$, $\Gamma^{11}\Psi_M=-\Psi_M$,
$\Gamma^{11}{\Lambda}={\Lambda}$. The supersymmetry transformation
laws for the fermionic fields of type IIB supergravity read
\cite{schw},
\begin{eqnarray}
\delta {\Lambda} &=& \frac{i}{\kappa_{\rm 10}}\Gamma^M
\epsilon^{\star} P_M -\frac{i}{24}\Gamma^{MNP}\epsilon
G_{(3)MNP}+\mbox{fermions~relevant
~terms}, \nonumber\\
\delta \Psi_M &=& \frac{1}{\kappa_{\rm
10}}\left(D_M-\frac{1}{2}iQ_M\right)\epsilon+\frac{i}{480}
\Gamma^{PQRST}\Gamma_M \epsilon \widehat{F}_{(5)PQRST} \nonumber\\
&& +\frac{1}{96} \left(\Gamma_M^{~~NPQ}G_{(3)NPQ}
-9\Gamma^{NP}G_{(3)MNP}\right)
\epsilon^{\star}+\mbox{fermions relevant terms},\nonumber\\
&& M,N,\cdots =0,1,\cdots, 9. \label{susytr1}
\end{eqnarray}
In above equations, the  supersymmetry transformation parameter
$\epsilon$ is a left-handed complex Weyl spinor,
$\Gamma^{11}\epsilon=-\epsilon$, and other quantities are listed as
the following,
\begin{eqnarray}
P_M &=& f^2\partial_M B, ~~ Q_M=f^2\mbox{Im}\left( B\partial_M
B^{\star}\right),~~ B\equiv \frac{1+i\tau}{1-i\tau}, \nonumber\\
 f^{-2} &=& 1-B B^{\star},~~\tau=C_{(0)}+ie^{-\phi},\nonumber\\
 G_{(3)MNP}&=& f\left(\widehat{F}_{(3)MNP}-B \widehat{F}_{(3)MNP}^\star\right), \nonumber\\
  \widehat{F}_{(3)MNP} &=&
3\partial_{[M}A_{(2)NP]},~~
A_{(2)MN}\equiv C_{(2)MN}+iB_{(2)MN},\nonumber\\
\widehat{F}_{(5)MNPQR}&=& F_{(5)MNPQR}-\frac{\kappa_{\rm
10}}{8}\times 10\, \mbox{Im} \left(A_{[(2)MN} \widehat{F}_{(3)PQR]}
\right),
\nonumber\\
F_{(5)MNPQR} &=& 5\partial_{[M} C_{(4)NPQR]},\nonumber\\
 D_M\epsilon &=& \partial_M \epsilon +\frac{1}{4}\omega_M^{~~AB}\Gamma_{AB}
 \epsilon.
\end{eqnarray}
To highlight the effects of fractional branes, we first consider the
supersymmetry transformation in the background without fractional
branes and work on the background furnished by three-brane solution
whose near-horizon limit is $AdS_5\times T^{1,1}$ (\ref{at11}). The
supersymmetry transformations for the fermionic fields in such a
background read:
\begin{eqnarray}
\delta \Lambda &=&0, \nonumber\\
\delta \Psi_M &=&\frac{1}{\kappa_{\rm
10}}\left(\partial_M+\frac{1}{4}
\omega_M^{~~AB}\Gamma^{AB}\right)\epsilon
+\frac{i}{480}\Gamma^{PQRST} {F}_{(5)PQRST}\Gamma_M\epsilon=0.
\label{susytr2}
\end{eqnarray}
The first equation $\delta \Lambda=0$  is trivially satisfied. We
only need  to substitute the background solution (\ref{at11}) into
$\delta \Psi_M =0$ and then solve it to observe whether there exist
non-trivial Killing spinors.

 First, the $AdS_5\times T^{1,1}$ metric (\ref{at11}) can be rewritten
 in terms of the velbein,
\begin{eqnarray}
ds_{\rm 10}^2
&=&({e}^a)^2+({e}^{\overline{r}})^2+\sum_{\widetilde{i}=1}^2
(V^{\widetilde{i}})^2+\sum_{\widehat{i}=1}^2 (V^{\widehat{i}})^2
+(V^{\widehat{3}})^2, \nonumber\\
{e}^a&=&\frac{r}{L}\delta^a_{~\mu} dx^\mu,
~~{e}^{\overline{r}}=\frac{L}{r}\delta^{\overline{r}}_{~r}dr,\nonumber\\
V^{\widetilde{1}} &=& \frac{1}{\sqrt{6}} L\sigma^{\widetilde{1}},
~~V^{\widetilde{2}}= \frac{1}{\sqrt{6}}
L\sigma^{\widetilde{2}},\nonumber\\
V^{\widehat{1}}&=&  \frac{1}{\sqrt{6}} L\sigma^{\widehat{1}},
~~V^{\widehat{2}}= \frac{1}{\sqrt{6}} L\sigma^{\widehat{2}},~~
V^{\widehat{3}}= \frac{1}{3}
L(\sigma^{\widetilde{3}}+\sigma^{\widehat{3}})
\end{eqnarray}
Using the structure equation and torsion-free condition,
$T^A=dE^A+\omega^A_{~B}\wedge E^B=0$, and taking into account
following identities on $T^{1,1}$,
\begin{eqnarray}
d\sigma^{\widetilde{i}}=-\frac{1}{2}
\epsilon_{\widetilde{i}\widetilde{j}\widetilde{k}}\sigma^{\widetilde{j}}\wedge
\sigma^{\widetilde{k}}, ~~~~
d\sigma^{\widehat{i}}=\frac{1}{2}\epsilon_{ijk}\sigma^{\widehat{j}}\wedge
\sigma^{\widehat{k}},
\end{eqnarray}
we derive spin connections such as
$\omega^a_{~\overline{r}}=-\omega_{~a}^{\overline{r}}=e^a/L$ etc..
Then the Killing spinor equation  (\ref{susytr2}) with the self-dual
five-form field given in (\ref{at11}) leads to \cite{rama},
\begin{eqnarray}
\epsilon &=& r^{\Gamma_\star/2}\left[  1+\frac{\Gamma_r}{2R^2} x^\mu
\Gamma_\mu \left( 1-\Gamma_\star\right)\right]\epsilon_0,\nonumber\\
\Gamma_{\star} &\equiv&
i\Gamma_{x_0}\Gamma_{x_1}\Gamma_{x_2}\Gamma_{x_3},~~~ \Gamma_\star^2
=1, \label{killing1}
\end{eqnarray}
where $\epsilon_0$ is an arbitrary constant spinor in ten dimensions
but constrained by
\begin{eqnarray}
\Gamma_{g_1g_2}\epsilon_0=\epsilon_0,
~~~\Gamma_{g_3g_4}\epsilon_0=-\epsilon_0. \label{cons1}
\end{eqnarray}
In above equations, for clarity we use the $AdS_5\times T^{1,1}$
space-time coordinates
 to label  the components of $\Gamma$-matrices. The Killing
spinor solution (\ref{killing1}) is consistent with the chirality
$\Gamma_{11}\epsilon=-\epsilon$. The constraint (\ref{cons1}) on
$\epsilon_0$ leads to
\begin{eqnarray}
\Gamma_{g_1g_2}\epsilon=\epsilon,
~~~\Gamma_{g_3g_4}\epsilon=-\epsilon \label{cons2}
\end{eqnarray}
because $\Gamma_{g_1g_2}$ and $\Gamma_{g_3g_4}$ commute with
$\Gamma_\star$ and $\Gamma_\mu$, $\mu=x_0,\cdots, x_3$.
Eq.\,(\ref{cons2}) means that the Killing spinor $\epsilon$ has only
eight independent components. This implies that  after the
spontaneous compactification on $T^{1,1}$, type IIB supergravity in
$AdS_5\times T^{1,1}$ background  gives rise to ${\cal N}=2$ $U(1)$
gauged $AdS_5$ supergravity coupled with $SU(2)\times SU(2)$
Yang-Mills fields and some Betti tensor multiplets due to
non-trivial topology of $T^{1,1}$ .

Furthermore, the Killing spinor (\ref{killing1}) is actually a
formally unified expression for the following two types of Killing
spinors \cite{rama,lpt},
\begin{eqnarray}
\epsilon_+&=& r^{1/2}\epsilon_{0+},
~~~\epsilon_-=r^{-1/2}\epsilon_{0-}+\frac{r^{1/2}}{L^2}\Gamma_r
x^\mu
\Gamma_{\mu}\epsilon_{0-},\nonumber\\
\epsilon_{0\pm} &=&\frac{1}{2}\left(1\pm
\Gamma_\star\right)\epsilon_0,
~~~\Gamma_\star\epsilon_{0\pm}=\pm\epsilon_{0\pm}.
 \label{killing2}
\end{eqnarray}
The above equations show that $\epsilon_+$ is independent of the
coordinates on $D3$-brane world-volume  and is a right-handed
eigenspinor of $\Gamma_\star$, i.e.,
$\Gamma_\star\epsilon_+=\epsilon_+$. It thus represents the ${\cal
N}=1$ Poinc\'{a}re supersymmetry in a four-dimensional
supersymmetric gauge theory. On the other hand, $\epsilon_-$ depends
on $x^\mu$ linearly and is not an eigenspinor of $\Gamma_\star$. So
it indicates the ${\cal N}=1$ conformal supersymmetry of
supersymmetric gauge theory in four dimensions.

 When we switch on $M$ fractional branes, the background is described
 by the K-S solution (\ref{ksso}). The above Killing
spinor equations from  the supersymmetric transformations on
dilatino and gravitino becomes
\begin{eqnarray}
\delta {\Lambda}
&=&-\frac{i}{24}\widehat{F}_{(3)MNP}\Gamma^{MNP}\epsilon
=0, \nonumber\\
\delta \Psi_M &=&\frac{1}{\kappa}\left(\partial_M+\frac{1}{4}\,
\omega_M^{~~AB}\Gamma^{AB}\right)\eta+\frac{i}{480}
\Gamma^{PQRST}\Gamma_M \epsilon \widehat{F}_{(5)PQRST} \nonumber\\
& +&\frac{1}{96} \left(\Gamma_M^{~~NPQ}\widehat{F}_{(3)NPQ}
-9\Gamma^{NP}\widehat{F}_{(3)MNP}\right) \epsilon^{\star}
 = 0. \label{susytr3}
\end{eqnarray}
The metric (\ref{ksso}) expressed  in terms of
$\sigma^{\widetilde{i}}$ and $\sigma^{\widehat{i}}$ takes the
following form \cite{rama},
\begin{eqnarray}
ds_{10}^2 &=& h^{-1/2}(\tau) \eta_{\mu\nu} dx^\mu dx^\nu +
h^{1/2}(\tau)\frac{1}{2}\mu^{4/3}
K(\tau)\left\{\frac{1}{3[K(\tau)]^3} \left[ d\tau^2+
(\sigma^{\widetilde{3}}+\sigma^{\widehat{3}} )^2\right]\right.\nonumber\\
&+&\left.\frac{\sinh^2\tau}{2\cosh\tau}\left[
(\sigma^{\widetilde{1}})^2+(\sigma^{\widetilde{2}})^2
\right]+\frac{\cosh\tau}{2}\left[
\left(\sigma^{\widehat{1}}+\frac{\sigma^1}{\cosh\tau} \right)^2
+\left(\sigma^{\widehat{2}}+\frac{\sigma^2}{\cosh\tau} \right)^2
\right]\right\},
\end{eqnarray}
where the vielbein one-forms read \cite{rama}
\begin{eqnarray}
{e}^a&=&h^{-1/4}(\tau)\delta^a_{~\mu} dx^\mu,
~~{e}^{\overline{\tau}}=\frac{\mu^{2/3}h^{1/4}(\tau)}{\sqrt{6}K(\tau)}d\tau
\delta^{\overline{\tau}}_{~\tau}d\tau,\nonumber\\
 V^{\widetilde{i}} &=&
\frac{\mu^{2/3}h^{1/4}(\tau)K^{1/2}(\tau)}{2}
\frac{\sinh\tau}{\sqrt{\cosh\tau}}\sigma^{\widetilde{i}}, ~~{\widetilde{i}}=1,2\nonumber\\
V^{\widehat{i}}&=&\frac{\mu^{2/3}h^{1/4}(\tau)K^{1/2}(\tau)}{2}\sqrt{\cosh\tau}
\left(\sigma^{\widehat{i}}+\frac{\sigma^{\widetilde{i}}}{\cosh\tau} \right),~~\widehat{i}=1,2\nonumber\\
V^{\widehat{3}}&=& \frac{\mu^{2/3}h^{1/4}(\tau)}{\sqrt{6}K(\tau)}
(\sigma^{\widetilde{3}}+\sigma^{\widehat{3}}).
\end{eqnarray}
 The same procedure as the case without fractional brane
 gives the following Killing spinor \cite{rama}
\begin{eqnarray}
\epsilon=h^{-1/8}(\tau) \exp\left(-\frac{\alpha}{2}\Gamma_{g_1g_2}
\right)\epsilon_0,
\end{eqnarray}
where $\alpha$ is defined by
\begin{eqnarray}
\sin\alpha\equiv -\frac{1}{\cosh\tau}, ~~~\cos\alpha \equiv
\frac{\sinh\tau}{\cosh\tau}.
\end{eqnarray}
$\epsilon_0$ is a constant spinor but suffers from following
constraints \cite{rama},
\begin{eqnarray}
\Gamma_\star\epsilon_0=-i\epsilon_0,
~~~\Gamma_{g_1g_2}\epsilon_0=-\Gamma_{g_3g_4}\epsilon_0, ~~\Gamma_{r
g_5} \epsilon_0=-i\epsilon_0. \label{kicon1}
\end{eqnarray}
These constraints on $\epsilon_0$  determine that the Killing spinor
$\epsilon$ should satisfy
\begin{eqnarray}
\Gamma_\star \epsilon=-i\epsilon,
~~~\Gamma_{g_1g_2}\epsilon=-\Gamma_{g_3g_4}\epsilon=
i\left(\cos\alpha+\sin\alpha\Gamma_{g_1g_3}\right)\epsilon,~~\Gamma_{rg_5}
\epsilon=-i\epsilon. \label{kicon2}
\end{eqnarray}
These three constraints imply that $\epsilon$ has only four
independent components. Specifically, now $\epsilon$ is independent
of the coordinate on the world-volume of $D3$-brane. This
four-component Killing spinor means the only existence of  ${\cal
N}=1$ Poin\'{a}re supersymmetry.  This indicates that  the conformal
supersymmetry in the dual four-dimensional supersymmetric gauge
theory disappear due to the presence of fractional $D3$-branes.

\section{Gravity  Dual of Superconformal anomaly as spontaneously breaking of
local  supersymmetry breaking in gauged $AdS_5$ supergravity and
consequent super-Higgs effect}

\subsection{Spontaneous Compactification of Type IIB Supergravity
and Symmetry Breaking in Gauged $AdS_5$ Supergravity}

The above analysis  shows that  the geometrical background
represented by K-S solution keeps less space-time symmetries in type
IIB supergravity than $AdS_5\times T^{1,1}$ does. We choose this
solution as a classical vacuum configuration for type IIB
supergravity and expand the theory around such a background. The
spontaneous of local symmetries in gauged $AdS_5$ supergravity will
occur and the consequent super-Higgs mechanism will take place.

To expound this physical phenomenon clearly, we first consider  the
case without $D3$-fractional branes and the vacuum configuration for
type IIB supergravity is $AdS_5\times T^{1,1}$ given in
(\ref{at11}). However, a  gravitational system has geometrical
meaning, expanding ten-dimensional type IIB supergravity around such
a vacuum configuration is actually a process of performing
spontaneous compactification of type IIB supergravity
\cite{freu,kim,mark} on the compact internal manifold $T^{1,1}$.
According to the general idea of the Kaluza-Klein (K-K)
compactification \cite{balov}, all the geometrical symmetries
reflected in the internal manifold including the supersymmetry it
preserves should emerge as local symmetries in the compactified
theory. For example, the isometric symmetry of internal manifold
becomes the gauge symmetry of compactified theory.  Therefore, for
type IIB supergravity in the $AdS_5\times T^{1,1}$ background,
 the isometry symmetry of $T^{1,1}$ is $SU(2)\times
SU(2)\times U(1)$ and it preserves eight supersymmetries, the
resultant theory after compactification should be ${\cal N}=2$
five-dimensional $U(1)$ gauged $AdS_5$ supergravity coupled with
${\cal N}=2$ $SU(2)\times SU(2)$ Yang-Mills vector multiplets and
several Betti tensor supermultiplets, whose presence attributes to
the nontrivial topology of $T^{1,1}$ \cite{ferrara}. The local
symmetries in ${\cal N}=2$ $U(1)$ gauged $AdS_5$ supergravity are
the ${\cal N}=2$ supersymmetry, $SO(2,4)$ space-time symmetry and
$U(1)$ gauge symmetry.

When fractional branes switch on, the fluxes carried by them deform
 $AdS_5 \times T^{1,1}$ space-time, and  the space-time
geometry is described by the K-S solution. In practical calculation
we use the UV limit of K-S solution --- the K-T solution
(\ref{ksso3}) \cite{tseytlin}).  The discussions in last section
show that the amount of the
 isometry symmetry of
the deformed $T^{1,1}$ and  supersymmetries it preserves are less
than that exploited from $T^{1,1}$. Therefore, the compactified
theory obtained from the compactification of type IIB supergravity
on  the deformed $T^{1,1}$, i.e., the five-dimensional gauged
supergravity, should possess less local symmetries than those
extracted from $T^{1,1}$. This means that some of local symmetries
break spontaneously since the breaking takes place on the classical
solution to type IIB supergravity.

A straightforward consequence of the spontaneous breaking of local
symmetry is the occurrence of Higgs mechanism.  Just like what
usually done in gauge theory, we reparametrize the field variable
and ``shift" the vacuum configuration described by the K-T solution
back to $AdS_5\times T^{1,1}$. The essence of this operation is
performing local symmetry transformation and transferring the
non-symmetric feature of the vacuum configuration to the classical
action. Then we expand type IIB supergravity around $AdS_5\times
T^{1,1}$ with newly
 defined field variables, the action of five-dimensional gauged supergravity
 should lose some of local symmetries and the graviton multiplet
in gauged $AdS_5$ supergravity should obtain mass by eating a
Goldstone multiplet relevant to  NS-NS- and R-R two-form fields in
the K-S solution. In this way, we reveal how the super-Higgs
mechanism due to the spontaneous breaking of local supersymmetry in
gauged  supergravity occurs.
 In fact, this phenomenon is well known in the Kaluza-Klein
supergravity \cite{duff}: when the internal manifold is deformed or
squashed, it keeps less symmetries  for the compactified theory than
the undeformed (or unsquashed) internal manifold. This is the
so-called ``space invader" scenario and
 can be naturally given an interpretation in terms of
 spontaneous breaking of local symmetry in K-K supergravity \cite{duff}.

Now we have realized  following physical phenomena. On the field
theory side the presence fractional $D3$-branes is the origin for
superconformal anomaly, while on gravity side, they deform the
$AdS_5\times T^{1,1}$ space-time geometry of type IIB supergravity
and hence lead to the spontaneous breaking of local symmetries and
the consequent super-Higgs mechanism in gauged $AdS_5$ supergravity.
These two physical phenomena have common origin in brane
configuration and thus should be dual to each other in the
gauge/gravity correspondence.

 Before proceeding to establish  the dual correspondence between
the superconformal anomaly in a  supersymmetric gauge theory and the
super-Higgs mechanism in gauged $AdS_5$ supergravity, we review
  the  compactification
 of type IIB supergravity on
$T^{1,1}$.

 The first step of performing
the compactification on $T^{1,1}$ is writing the metric in
(\ref{nfsolu1}) as the Kaluza-Klein metric form,
\begin{eqnarray}
ds^2&=& h^{-1/2}(r) \eta_{\mu\nu}dx^\mu dx^\nu+h^{1/2}(r)dr^2\nonumber\\
&+& h^{1/2}(r)
r^2\left\{\frac{1}{9}\left(g^5-2A\right)^2+\frac{1}{6}\left[\sum_{r=1}^{2}
\left(g^r-K^{ar}\widehat{W}^a\right)^2+\sum_{s=3}^4\left(g^s-L^{bs}
\widehat{\widetilde{W}}^b \right)^2\right]\right\}. \label{kkmetric}
\end{eqnarray}
In above equation, $K^{ar}$ and $L^{bs}$ are components of Killing
vectors $K^a$ and $L^b$ on $T^{1,1}$, they are generators for
$SU(2)\times SU(2)$ Yang-Mills gauge group in the resultant ${\cal
N}=2$ $U(1)$ gauged supergravity coupled with six Yang-Mills vector
supermultiplets,
\begin{eqnarray}
[K^{a_1},K^{a_2}]=if^{a_1a_2a_3}K^{a_3},
~~~[L^{b_1},L^{b_2}]=if^{b_1b_2b_3}L^{b_3};
\end{eqnarray}
$\widehat{W}^a=\widehat{W}^{a}_{~\alpha} dx^\alpha$ and
$\widehat{\widetilde{W}}^{b}=\widehat{\widetilde{W}}^{b}_{~\alpha}dx^\alpha$
are the corresponding six $SU(2)$ gauge fields in five-dimensional
space-time; $\widehat{A}=\widehat{A}_\alpha dx^\alpha$ is the $U(1)$
gauge field corresponding to the isometry symmetry $U(1)$ and it
constitutes a ${\cal N}=2$ supermultiplets with the graviton
$\widehat{h}_{\alpha\beta}$ and gravitini $\widehat{\psi}_\alpha^i$
($i=1,2$) in five-dimensional gauged supergravity. The isometry
symmetry and various other geometric symmetries manifested in the
internal manifold such as the supersymmetry it preserves all convert
into local symmetries of the compactified theory. To serve  our
purpose we concentrate only on the $U(1)$ symmetry since it is the
gauge symmetry that enters space-time supersymmetry  in  ${\cal
N}=2$ gauged $AdS_5$ supergravity.

Accompanying with the K-K metric (\ref{kkmetric}) the self-dual
five-form in Eq.\,(\ref{nfsolu1}) should also be modified to keep
its self-duality since the Hodge star operation is defined with
respect to the metric \cite{kow},
\begin{eqnarray}
\overline{F}_5 &=& d\overline{C}_4=\frac{1}{g_s}
\partial_r h^{-1}(r) dx^0\wedge dx^1\wedge dx^2\wedge dx^3
\wedge dr\nonumber\\
&&+\frac{L^4}{27}\left[\chi\wedge g^1 \wedge g^2 \wedge g^3 \wedge
g^4-dA\wedge g^5\wedge dg^5+\frac{3}{L}\left({}^{\star_5}
dA\right)\wedge dg^5 \right]. \label{kkfiveform}
\end{eqnarray}
Consequently, there exists locally
\begin{eqnarray}
 \overline{C}_4 &=& \frac{1}{g_s}
h^{-1}(r) dx^0\wedge dx^1\wedge
dx^2\wedge dx^3\nonumber\\
&&+\frac{2L^4}{27}\left[\beta g^1 \wedge g^2 \wedge g^3 \wedge
g^4-\frac{1}{2}A\wedge g^5\wedge
dg^5+\frac{3}{2r}h^{-1/4}(r)\left({}^{\star_5} dA\right)\wedge
dg^5\right], \label{kkfiveform2}
\end{eqnarray}
where $\chi=g^5-2A$, ${\star_5}$ is the
 five-dimensional Hodge dual defined with respect to
 $AdS_5$ metric
\begin{eqnarray}
ds_{{\rm AdS}_5}^2=\widehat{g}_{\alpha\beta}^{(0)}(x,r)dx^\alpha
dx^\beta=\frac{r^2}{L^2} \eta_{\mu\nu}dx^\mu dx^\nu+
\frac{L^2}{r^2}dr^2. \label{adsme}
\end{eqnarray}
  Obviously, the K-K metric (\ref{kkmetric}) and
$\overline{F}_5$ are gauge invariant under local $U(1)$ gauge
transformation in the gauged $AdS_5$ supergravity,
\begin{eqnarray}
\beta\rightarrow \beta+\Lambda, ~~~  A\rightarrow A+d\Lambda.
\end{eqnarray}
$\overline{F}_5$ satisfies the Bianchi identity $d\overline{F}_5=0$,
which determines $A$ is a massless vector fields in $AdS_5$ space,
i.e., $d{}^{\star_5}dA=0$.

Let us turn to the specifics of performing compactification on
$T^{1,1}$ \cite{ita}. Usually when performing compactification of a
certain $D$-dimensional supergravity on $AdS_{D-d}\times K^d$, one
should first linearize the classical equation of motion for field
function $\Phi (x,y)\equiv \Phi^{\{J\}[\lambda]}(x,y)$ in
$AdS_{D-d}\times K^d$ background. In the above, $K^d=G/H$ is certain
$d$-dimensional compact Einstein manifold, $x$ and $y$ are the
coordinates on $AdS_{D-d}$ and $K^d$, respectively; $\{J\}$ and
$[\lambda]$ label the representations of local Lorentz groups
$SO(2,D-d-1)$ and $SO(d)$ realized on field functions,respectively.
The next step is to expand the $y$-dependent part of field functions
in terms of $H$-harmonics on $K^d$, which are representations of the
group $G$ branched (or reduced) with respect to its subgroup $H$. If
the internal manifold is $K^d=S^d=SO(d+1)/SO(d)$, which is the
maximally symmetric Einstein manifold, this procedure works smoothly
since now $H$ is $SO(d)$ and it coincides with local Lorenz group of
the $d$-dimensional internal manifold. However, if the internal
manifold is a less symmetric one, then $H$ is usually a subgroup of
$SO(d)$. In this case, the field function representations of $SO(d)$
are usually reducible under the action of $H$. Therefore, only those
$H$-harmonics that are identical  to the $SO(d)$ field function
representations branched by $H$ can contribute to the expansion of
field functions on $K^d$.

The compactification of type IIB supergravity on $AdS_5\times
T^{1,1}$ is exactly like this. $T^{1,1}$ is the coset space
$G/H=[SU(2)\times SU(2)]/U_{\rm H}(1)$ and  the generator of $U_{\rm
H}(1)$ is the sum $T^3+\widehat{T}^3$ of two diagonal generators of
$SU(2)\times SU(2)$. The harmonics on $T^{1,1}$ are representations
of $SU(2)\times SU(2)$ labeled  by weights $\{\nu\}=(j,l)$,
\begin{eqnarray}
Y(y)\equiv \left(\left[Y^{(j,l,r)} (y)\right]^m\right).
\end{eqnarray}
In above equation, $m=1,\cdots, (2j+1)\times (2l+1)$ are the
representation space indices of $SU(2)\times SU(2)$ and $r$ is the
representation quantum number for the $U(1)$ group whose generator
is  $T_3-\widehat{T}_3$. This $U(1)$ group is actually the
$R$-symmetry group in the dual ${\cal N}=1$ supersymmetric gauge
theory, which on the gravity side originates from the isometry group
$SU(2)\times SU(2)\times U(1)$ of $T^{1,1}$. The representations
$(j,l)$ are reducible with respect to the subgroup $U_{\rm H}(1)$
and hence decompose into a direct sum of fragments labeled by the
$U_{\rm H}(1)$-charge $q_i$,
\begin{eqnarray}
\left[Y^{(j,l,r)} (y)\right]^m=\bigoplus_i\,\left[Y^{(j,l,r)}
(y)\right]^m_{~q_i}=\left(\begin{array}{c}
\left[Y^{(j,l,r)}(y)\right]^m_{~q_1}\\ \left[Y^{(j,l,r)}
(y)\right]^m_{~q_2}\\ \vdots\\
\left[Y^{(j,l,r)} (y)\right]^m_{~q_N}
\end{array} \right).
\end{eqnarray}
 The irreducible representations
$\left[Y^{(j,l,r)} (y)\right]^m_{~q_i}$ are called $U_{\rm
H}(1)$-harmonics on $T^{1,1}$.

 On the other
hand, since the field function $\Phi^{\{J\}[\lambda]}(x,y)$  on
$AdS_5\times T^{1,1}$  lies in certain representations of the local
Lorentz group $SO(2,4)\times SO(5)$,  here $\{J\}$, $[\lambda]$
denoting the representation wights for $SO(2,4)$ and $SO(5)$,
respectively. The $U_{\rm H}(1)$ is a subgroup of $SO(5)$ since the
representation of its generator can be naturally embedded into the
representation of  $SO(5)$ generators \cite{sast}. The
representations $[\lambda]$ of $SO(5)$ furnished by the
$y$-dependent part of $\Phi^{\{J\}[\lambda]} (x,y)$, which are
denoted as $\left[X^{[\lambda]}(y)\right]^n$, are also reducible
with respect to $U_{\rm H}(1)$,  $n$ being the representation
indices  of $SO(5)$ in field function space. The field function
representation space thus decomposes into a direct sum of
irreducible subspaces labeled by the $U_{\rm H}(1)$-charge $q_\xi$,
\begin{eqnarray}
 \left[X^{[\lambda]}(y)\right]^n
=\bigoplus_{\xi=1}^K\left[X^{[\lambda]}(y)\right]^n_{~q_\xi}
=\left(\begin{array}{c} \left[X^{[\lambda]}(y)\right]^n_{~q_1}\\
\left[X^{[\lambda]}
(y)\right]^n_{~q_2}\\ \vdots\\
\left[X^{[\lambda]} (y)\right]^n_{~q_K}
\end{array} \right).
\end{eqnarray}
 The irreducible representations
$\left[X^{[\lambda]}(y)\right]^n_{~q_\xi}$ in above equation are
usually called the $SO(5)$ harmonics. The field function naturally
admits an expansion in terms of the $SO(5)$ harmonics.
 Therefore,
 a $U_{\rm H}$-harmonics $Y$ can contribute to the expansion of a
field function on $T^{1,1}$ only when it
 is identical to (or contains)  certain $SO(5)$ harmonics $X$.
  A detailed analysis on how the $SU(2)\times SU(2)$ representations
branched with respect to $U_{\rm H}(1)$ appear in the decomposition
 of the field function representations of
$SO(5)$ with respect to $U_{\rm H}(1)$ is performed in
Ref.\,\cite{ita}. Once the expansion of  field function
$\Phi^{\{J\}[\lambda]} (x,y)$ in terms of those admissible $U_{\rm
H}(1)$-harmonics on $ T^{1,1}$ is known\cite{ita},
\begin{eqnarray}
\Phi^{\{J\}[\lambda]} (x,y) &=&\left(\Phi^{\{J\}}_{ab\cdots}
(x,y)\right)^{[\lambda]} =\bigoplus_{i=1}^N
\left(\Phi^{\{J\}}_{q_i}\right)^{[\lambda]}= \left(\begin{array}{c}
\Phi^{\{J\}}_{q_1}(x,y)\\  \Phi^{\{J\}}_{q_2}(x,y)\\ \vdots\\
\Phi^{\{J\}}_{q_N}(x,y)
\end{array} \right),
 \nonumber\\
 \Phi^{\{J\}}_{q_i}(x,y) &=& \sum_{(j,l)}\sum_m
\sum_r \Phi_{q_i m}^{\{J\}(j,l,r)}(x) \left[Y^{(j,l,r)}
(y)\right]^m_{~q_i}, \label{harmexp1}
\end{eqnarray}
 then we substitute it into the following linearized
 equations of motion of type IIB
supergravity in  $AdS_5 \times T^{1,1}$ background,
\begin{eqnarray}
\left( K_x^{\{J\}}+K_y^{[\lambda]}\right)\Phi^{\{J\}[\lambda]}
(x,y)=0. \label{lineareq}
\end{eqnarray}
In above equations, $a,b\cdots$ are the indices of the
$[\lambda]$-representation, $K_x^{\{J\}}$ and $K_y^{[\lambda]}$ are
the kinetic operators of type IIB supergravity on $AdS_5$ and
$T^{1,1}$, respectively.  Usually there are three types kinetic
operators $K_y^{[\lambda]}$ in supergravity: the Hodge-de Rahm- and
 Laplacian operators acting on the scalar, vector and antisymmetric
fields; the Dirac and Rarita-Schwinger operator acting on the
fermionic fields and the Lichnerowicz operator  on the symmetric
rank-two tensor field. All these three types of operators  are (or
are relevant to) certain Laplace-Beltrami operators on
$T^{1,1}=\left[SU(2)\times
 SU(2)\right]/U_{\rm H}(1)$. Thus the action of
 $K_y^{[\lambda]}$ on  $U_{\rm H}(1)$-harmonics gives \cite{ita}
 \begin{eqnarray}
K_y^{[\lambda]}\left[Y^{(j,l,r)}
(y)\right]^m_{~q_i}=M_{ik}^{(j,l,r)}\left[Y^{(j,l,r)}
(y)\right]^m_{~q_k}.
\end{eqnarray}
This equation together with the linearized equation of motion
(\ref{lineareq}) and $U_{\rm H}(1)$-harmonic expansion
(\ref{harmexp1}) of field functions shows that $M_{ij}^{(j,l,r)}$
are mass matrices for the K-K particle tower $\Phi_{q_k
m}^{\{J\}(j,l,r)}(x)$ in $AdS_5$ space,
 \begin{eqnarray}
\left(\delta_{ik}K_x^{\{J\}}+ M_{ik}^{(j,l,r)}\right)\Phi_{q_k
m}^{\{J\}(j,l,r)}(x)=0.
\end{eqnarray}
Therefore, the zero modes of the kinetic operator $K_y^{[\lambda]}$
will constitute field content of the compactified theory, which
turned out to be the ${\cal N}=2$ $U(1)$ gauged $AdS_5$ supergravity
coupled with $SU(2)\times SU(2)$ Yang-Mills vector supermultiplets
as well as some  Betti tensor multiplets \cite{ita}.

\subsection{${\cal N}=2$ $U(1)$ Gauged $AdS_5$ Supergravity
from Type IIB supergravity in $AdS_5 \times T^{1,1}$ Space-time
Background}

In this  subsection,  we show how the ${\cal N}=2$ $U(1)$ gauged
$AdS_5$ supergravity comes from type IIB supergravity in the $AdS_5
\times T^{1,1}$ vacuum configuration.

 The action of  type IIB supergravity is listed as the following
 \cite{gsw,schw},
\begin{eqnarray}
S_{\rm IIB}&=& S_{\rm NS-NS}+S_{\rm R-R}+S_{\rm CS}+S_{\rm F},
\nonumber\\
S_{\rm NS-NS} &=&\frac{1}{2\kappa_{\rm 10}^2}\int d^{10}x E
e^{-2\Phi} \left(R+4\partial_M\Phi\partial^M\Phi-\frac{1}{2\times
3!} H_{(3)MNP} H^{MNP}_{(3)}
\right),\nonumber\\
S_{\rm R-R}&=&-\frac{1}{2\kappa_{\rm 10}^2}\int d^{10}x E \left(
\frac{1}{2}\partial_M C_{(0)}\partial^M C_{(0)}+\frac{1}{2\times 3!}
\widetilde{F}_{(3)MNP}\widetilde{F}^{MNP}_{(3)}\right.\nonumber\\
&&\left.+\frac{1}{4\times
5!}\widehat{F}_{(5)MNPQR}\widehat{F}^{MNPQR}_{(5)}
 \right),\nonumber\\
S_{\rm CS} &=& -\frac{1}{4\kappa_{\rm 10}^2}\int C_{(4)} \wedge
F_{(3)} \wedge H_{(3)}\nonumber\\
&=&-\frac{1}{4\kappa_{\rm 10}^2}\int d^{10}x \frac{1}{4!\times
3!\times 3!} \epsilon^{M_1\cdots M_4N_1 \cdots N_3 P_1\cdots P_3}
C_{(4)M_1\cdots M_4}
F_{(3)N_1\cdots  N_3}H_{(3)P_1\cdots P_3},\nonumber\\
S_{\rm F}&=&\frac{1}{2\kappa_{\rm 10}^2}\int d^{10}x E \left[
-\frac{i}{2}\overline{\Lambda}\Gamma^M D_M {\Lambda} +
\frac{1}{4\times 5!}\overline{\Lambda}\Gamma^{MNPQR} {\Lambda}
\widehat{F}_{(5)MNPQR}\right.\nonumber\\
&& -\frac{i}{2}\overline{\Psi}_M\Gamma^{MNP}
 D_N\Psi_P-\frac{1}{8\times 5!}\overline{\Psi}_M \Gamma^{MNP}\left(
\Gamma^{UVWXY}\widehat{F}_{(5)UVWXY}\right)\Gamma_N\Psi_P\nonumber\\
&&\left.-\frac{i}{48}\kappa_{\rm
10}\overline{\Psi}_M\Gamma^{NPQ}\left({F}_{(3)NPQ}-iH_{(3)NPQ}\right)
\Gamma^M {\Lambda}
 + \mbox{four-fermion terms}\right], \label{type2b}
\end{eqnarray}
which is further supplemented with the self-dual constraint on the
five-form field strength ${}^\star \widetilde{F}_5=\widetilde{F}_5$.
In above equation
\begin{eqnarray}
\widetilde{F}_{(3)MNP}&\equiv&{F}_{(3)MNP}-C_{(0)}{H}_{(3)MNP}, \nonumber\\
\widehat{F}_{(5)MNPQR}&\equiv &{F}_{(5)MNPQR}-\frac{5}{4}\kappa_{\rm
10} C_{[(2)MN}{H}_{(3)PQR]} +\frac{5}{4} \kappa_{\rm 10}
B_{[(2)MN}{F}_{(3)PQR]}.
\end{eqnarray}

The arising of ${\cal N}=2$ $U(1)$ gauged supergravity from  type
IIB supergravity in the $AdS_5\times T^{1,1}$ background can be
observed from the classification of the particle spectrum under the
$AdS_5$ space-time supergroup $SU(2,2|1)$. The bosonic subgroup of
$SU(2,2|1)$ is $SO(2,4)\times U(1)$, of which  $SO(2,4)$ is actually
the isometry group of $AdS_5$ space-time. Further,
 $SO(2,4)$ has the subgroup $SO(2)\times SO(4)=SO(2)\times
SU(2)\times SU(2)$ and the $U(1)$ comes from the isometry group
$SU(2)\times SU(2)\times U(1)$ of $T^{1,1}$. This $U(1)$ is the
gauge group of the ${\cal N}=2$ gauged $AdS_5$ supergravity, while
on the dual field theory side it is the $R$-symmetry group for the
${\cal N}=1$ supersymmetric gauge theory in four dimensions. A
representation $D(E_0,j,l|r)$ of $SU(2,2|1)$ is labeled by the
$AdS_5$ energy $E_0$, the $SU(2)\times SU(2)$ quantum number
 $(j,l)$ and $U(1)$-charge $r$. According to
 the representation of $SU(2,2|1)$, the K-K spectrum of type IIB
supergravity on $AdS_5\times T^{1,1}$ can be classified into a
graviton multiplet,  four gravitino multiplets and four vector
multiplets and also some Betti multiplets. The massless gravitino
supermultiplet $(\widehat{h}_{\alpha\beta}, \widehat{\psi}_\alpha^i,
\widehat{A}_\alpha)$ constitutes the following representations of
$SU(2,2|1)$,
\begin{eqnarray}
D(4, 1,1|0)\oplus D(7/2,1,1/2|-1)\oplus D(7/2,1/2,1|1) \oplus
D(3,1/2,1/2|0).
\end{eqnarray}
These are the so-called  maximal shortening representations to
$SU(2,2|1)$ since the bound for a unitary irreducible representation
is saturated by this supermultiplet. That is, they satisfy
$E_0=2+j+l$, $r=2(j-l)/3$.

Further, the above graviton multiplet indeed arises from the
compactification of type IIB supergravity on $T^{1,1}$. First,
$\widehat{h}_{\alpha\beta}(x)$ is the massless mode in the scalar
$U_{\rm H}(1)$-harmonic expansion of $H_{\alpha\beta}(x,y)$, which
is the $AdS_5$ space-time component
 of ten-dimensional graviton $H_{MN}(x,y)$;
 $\widehat{A}_{\alpha}$ is the linear combination
 of two massless modes in the vector
$U_{\rm H}(1)$-harmonic expansion of $H_{\alpha a}$ and
$C_{(4)\alpha abc}$, which are the crossing  components on $AdS_5$
and $T^{1,1}$ of the graviton $H_{MN}(x,y)$ and the R-R four form
field $C_{(4)MNPQ}(x,y)$, respectively; $\widehat{\psi}_\alpha$
comes from the massless mode in the spinor $U_{\rm H}(1)$-harmonics
expansion of $\psi_\alpha$, which is the $AdS_5$  component of
ten-dimensional gravitino $\psi_M$ \cite{ita}. Based on above
compactification, the action of ${\cal N}=2$ $U(1)$ gauged $AdS_5$
supergravity  can be straightforwardly extracted out from  type IIB
supergravity in the $AdS_5\times T^{1,1}$ background,
\begin{eqnarray}
 S_{\rm gauged}=\frac{1}{\kappa^2_5} \int d^5x \widehat{e}
 \left[-\frac{1}{2} \widehat{R}-\frac{1}{2}
\overline{\widehat{\psi}}^i_\alpha\widehat{\gamma}^{\alpha\beta\gamma}
\widehat{D}_\beta\widehat{\psi}_{i\gamma}
-\frac{3l^2}{32}\widehat{F}^{\alpha\beta}\widehat{F}_{\alpha\beta}
+\cdots \right]. \label{fiveaction}
\end{eqnarray}
In above equation we  list only the quadratic sector of the
 action for graviton supermultiplet and omit
the terms involving other Yang-Mills vector  multiplets and Betti
multiplets. Eq.\,(\ref{fiveaction}) shows that the graviton
multiplet $(\widehat{A}_\alpha,
\widehat{\psi}_\alpha^i,\widehat{h}_{\alpha\beta})$ is massless.
 This graviton multiplet  should be
 dual to the current supermultiplet $(j_\mu ,
s_\mu, T_{\mu\nu})$ of the supersymmetric gauge field theory on the
$AdS_5$ boundary in terms of the holographic definition on
 $AdS/CFT$ correspondence \cite{witt1},.

This scenario will change in terms of the gauge/gravity
correspondence due to the superconformal anomaly of the
supersymmetric gauge theory.

\subsection{Dual of Chiral $R$-symmetry Anomaly $\partial_\mu j^\mu$}

We first review the dual description of the chiral $U_R(1)$ anomaly
of ${\cal N}=1$ $SU(N+M)\times SU(N)$ gauge theory   from type IIB
supergravity in the K-S solution background found in
Ref.\,\cite{kow}.

When fractional branes are absent, the K-K metric and the five-form
field shown in Eqs.\,(\ref{kkmetric}) and (\ref{kkfiveform}) are
$U_R(1)$ gauge invariant  after the compactification on $T^{1,1}$ is
performed. The graviton multiplet of gauged $AdS_5$ supergravity is
massless. When fractional branes switch on, both the K-K metric
(\ref{kkmetric}) and the five-form (\ref{kkfiveform}) get deformed,
but they still keep $U_R(1)$ gauge invariant. However, as shown in
Eq.\,(\ref{ksso}),  the field strength $F_{(3)}$ of R-R two-form
$C_{(2)}$ appears in the background solution and it depends on the
angle $\beta$ linearly,
\begin{eqnarray}
 F_{(3)}&=&\frac{M\alpha^\prime}{2}\omega_3=\frac{M\alpha^\prime}{2}
g^5\wedge \omega_2=\frac{M\alpha^\prime}{4} g^5\wedge
\left(g^1\wedge g^2+g^3\wedge g^4\right),\nonumber\\
g^5 &=& 2d\beta +\cos\theta_1 d\phi_1+\cos\theta_2 d\phi_2,
\end{eqnarray}
it is not invariant under the rotation of $\beta$-angle. Like what
usually done in dealing with the spontaneous  breaking of gauge
symmetry, one should consider the fluctuations around this $F_{(3)}$
and shift it as
\begin{eqnarray} \overline{F}_{(3)}
=\frac{M\alpha^\prime}{2} \left(g^5+2\partial_\alpha\theta
dx^\alpha\right) \wedge \omega_2.
\end{eqnarray}
by introducing a new field,
\begin{eqnarray}
 \theta\equiv F \int _{S^2} C_2.
 \label{fgold}
 \end{eqnarray}
In above equation $F$ is a field function in ten-dimensional
space-time. ${F}_{(3)}$ is obviously invariant under the combined
gauge transformation,
\begin{eqnarray}
 \beta\rightarrow \beta+\alpha, ~~~\theta\rightarrow
\theta-\alpha.
\end{eqnarray}
Further,  $\overline{F}_{(3)}$ can be  rewritten  in terms of the
gauge invariant quantity $\chi \equiv g^5-2{A}$ and  a newly defined
vector field ${W}_\alpha$,
\begin{eqnarray}
\overline{F}_3 &=& \frac{M\alpha^\prime}{2} \left(\chi+2 {W}_\alpha
dx^\alpha\right) \wedge \omega_2\nonumber\\
&\equiv& \overline{F}_3^{(0)}+M\alpha^\prime {W}_\alpha dx^\alpha
\wedge \omega_2; \nonumber\\
{W}_\alpha & \equiv & {A}_\alpha+\partial_\alpha\theta,~~~~
\overline{F}_3^{(0)} = \frac{M\alpha^\prime}{2}\chi .
 \label{rrthree}
\end{eqnarray}
The  vector field ${W}_\alpha$ and the $U(1)$ gauge field
${A}_\alpha$ have the same field strength ${F}_{\alpha\beta}$, but
${W}_\alpha$ has got the longitudinal component. Now we consider
type IIB supergravity in the symmetric vacuum configuration
furnished by the metric (\ref{kkmetric}), the self-dual five-form
(\ref{kkfiveform}) and this new R-R three-form field strength
$\overline{F}_3^{(0)}$ defined in (\ref{rrthree}). Taking into
account only  the Einstein-Hilbert- and the R-R three-form terms
 in the classical action (\ref{type2b}) of type IIB supergravity,
 one has \cite{kow}
\begin{eqnarray}
S_{\rm IIB}&=&\frac{1}{2\kappa_{10}^2}\int d^{10}x \left[E
R-\frac{1}{2}{F}_3\wedge {}^\star
{F}_3+\cdots\right]\nonumber\\
&=&\frac{1}{2\kappa_{10}^2}\int d^{10}x E\left[-\frac{1}{9}
h^{1/2}(r) r^2 F^{\alpha\beta}F_{\alpha\beta}- \left(
\frac{3M\alpha^\prime}{h^{1/2}(r)r^2}\right)^2 W^\alpha W_\alpha
+\cdots\right].
\end{eqnarray}
The above equation means that the $U_R(1)$ gauge field coming from
the compactification of  type IIB supergravity on $T^{1,1}$ has
gained a mass, whose value depends on the fluxes carried by
fractional branes \footnote{A rigorous analysis, which takes into
account  both the self-dual equation for five-form field strength
and the $(\alpha\chi)$-component in the Einstein equation of type
IIB supergravity, was performed in Ref.\,\cite{mik}. It showed that
there actually exists another vector field $K_\alpha$, which
originates from the fluctuation of the four-form potential $C_{(4)}$
and mixes with $W_\alpha$. If considering only transverse parts of
$W_\alpha$ and $K_\alpha$, one can diagonalize their coupled
equations by taking a linear combination $W^{(1)}_\alpha
=W_\alpha-54/[h(r) r^4] K_\alpha$, $W^{(2)}_\alpha =W_\alpha
+54/[h(r) r^4] K_\alpha$. The decoupled equations show that
$W^{(2)}_\alpha$ is actually massive in $AdS_5\times T^{1,1}$ limit.
It is $W^{(1)}_\alpha$ that acquires a mass through the spontaneous
breaking of the $U(1)$-symmetry. In particular, the mass of
$W_\alpha^{(1)}$ acquired through Higgs mechanism is  at the order
of $(M\alpha^\prime)^2$ rather than $M\alpha^\prime$.}. This
phenomenon  exactly means the occurrence of the Higgs mechanism:
when fractional branes are present, the classical solution breaks
the $U(1)$ gauge symmetry; after we modify the classical solution to
make it gauge invariant, the classical action of type IIB
supergravity around this symmetric vacuum configuration  loses gauge
symmetry and the $U(1)$ gauge field acquires a mass.

\subsection{ Dual of Scale Anomaly $\theta^{\mu}_{~\mu}$}

  To find the dual of scale anomaly, we first
analyze how the local symmetry  corresponding to the scale symmetry
on field theory side becomes broken in the K-S solution and how it
is related to the compactification of type IIB supergravity on the
deformed $T^{1,1}$. Obviously, the  local symmetry that corresponds
to the scale symmetry on field theory side should have nothing to do
with the isometry symmetry of internal manifold $T^{1,1}$, and it
would rather be related  to the diffeomorphism symmetry of the
$AdS_5$ space. The argument for this statement is motivated from
Ref.\,\cite{imbi}, where it was shown that near the $AdS_5$ boundary
the bulk diffeomorphism symmetry preserving the form of $AdS_5$
metric decomposes into a combination of the four-dimensional
diffeomorphism symmetry and Weyl symmetry. According to the
$AdS/CFT$ correspondence at supergravity level, the four-dimensional
diffeomorphism- and Weyl symmetries are equivalent to the
conservation and tracelessness of the energy-momentum tensor of the
four-dimensional supersymmetric gauge theory \cite{chch1}.
 This suggests that we should observe how the diffeomorphism
symmetry of  $AdS_5$ space is spoiled by  fractional branes to look
for the dual of scale anomaly.

When the fractional branes are absent,  the near-horizon limit of
this solution is $AdS_5\times T^{1,1}$. Let us observe the
diffeomorphism symmetry of type IIB supergravity (\ref{type2b}) in
this background. Since we care about how the breaking of
diffeormorphism symmetry affects the graviton, so only the
Einstein-Hilbert- and the self-dual five-form terms in the action
(\ref{type2b}) are taken into account
  \footnote {It is necessary to
have this five-form term present since it produces the $AdS_5$
space-time background.},
\begin{eqnarray}
S_{\rm IIB} =\frac{1}{2\kappa_{\rm 10}^2}\int d^{10}x E
\left(R-\frac{1}{4\times 5!}{F}_{(5)MNPQR}{F}^{~~MNPQR}_{(5)}
 \right)+\cdots.
\end{eqnarray}
Using the expansion $G_{MN}=G^{(0)}_{MN}+h_{MN}$,  we obtain the
quadratic action for the graviton in the background with only
non-vanishing $G^{(0)}_{MN}$ and $F^{(0)}_{(5)MNPQR}$,
\begin{eqnarray}
S_{\rm Q.G.}&=&\frac{1}{2\kappa_{10}^2}\int d^{10}x
\sqrt{-G^{(0)}}\left[-\frac{1}{4} \nabla^P h^{MN}  \nabla_P
h_{MN}+\frac{1}{2}\nabla^P h^{MN} \nabla_M h_{PN}\right.\nonumber\\
&&+\frac{1}{4}\nabla^M h\left( \nabla_M h-\frac{1}{2}
\nabla^N h_{MN}\right)\nonumber\\
&& + \left( R^{(0)}-\frac{1}{4\times
5!}F_{(5)}^{(0)MNPQR}F_{(5)MNPQR}^{(0)}
\right)\frac{1}{4}\left( \frac{1}{2}h^2-h^{ST}h_{ST} \right)\nonumber\\
 &&-\frac{1}{4\times 5!}\left(10 h^{M^\prime [M}h^{NN^\prime}
 F_{(5)MNPQR}^{(0)}F_{(5)M^\prime
N^\prime}^{(0)~~~~~PQR]}\right.\nonumber\\
&& \left.\left.-\frac{5}{2}h\,h^{M^\prime[M}
 F_{(5)MNPQR}^{(0)}F_{(5)M^\prime
}^{(0)~~~NPQR]} +5 h^{M^\prime
M^{\prime\prime}}h_{M^{\prime\prime}}^{~~~[M}
F_{(5)MNPQR}^{(0)}F_{(5)M^\prime}^{(0)~~NPQR}\right) \right]
\nonumber\\
&=&\frac{1}{2k_{10}^2}\int d^{10}x \sqrt{-G^{(0)}}\left[-\frac{1}{4}
\nabla^P h^{MN}  \nabla_P
h_{MN}+\frac{1}{2}\nabla^P h^{MN} \nabla_M h_{PN}\right.\nonumber\\
&&+\frac{1}{4}\nabla^M h\left( \nabla_M h-\frac{1}{2} \nabla^N
h_{MN}\right)
+\frac{1}{2}R_{MPNQ}h^{MN}h^{PQ}-\frac{1}{2}h^{MN}h_{N}^{~P} R_{PM}\nonumber\\
&&\left. +\frac{27}{32}\frac{\pi \alpha^{\prime 2}N}{h^2(r) r^5}
\left(h^{MN}h_{MN}- \frac{1}{2}h^2 \right) \right]\nonumber\\
&=&\frac{1}{2k_{10}^2}\int d^{10}x \sqrt{-G^{(0)}}\left[-\frac{1}{4}
\nabla^P h^{MN}  \nabla_P
h_{MN}+\frac{1}{2} \nabla_M h^{MN}  \nabla^P h_{PN}\right.\nonumber\\
&&\left.+\frac{1}{4}\nabla^M h\left( \nabla_M h-\frac{1}{2} \nabla^N
h_{MN}\right)-\frac{9}{32}\frac{\pi \alpha^{\prime 2}N}{h^2(r) r^5}
\left(h^{MN}h_{MN}+ \frac{1}{2}h^2 \right) \right].
\label{qudraction}
\end{eqnarray}
 We then perform the compactification
 on $T^{1,1}$ with  the quadratic action (\ref{qudraction}).
 Taking into account the background dependent and covariant gauge-fixing
  conditions $
D^\xi h_{\xi\alpha}=D^\xi h_{(\xi\varepsilon)}=0$, one can  expand
$h_{MN}(x,y)$ in terms of the $U_{\rm H}(1)$-harmonics on $T^{1,1}$
\cite{ita},
\begin{eqnarray}
h_{MN}(x,y) & = &\left( h_{\alpha\beta} (x,y), h_{\alpha \xi}(x,y),
h_{\xi\varepsilon}(x,y) \right),\nonumber\\
h_{\alpha\beta} (x,y) &=& \sum_{j,l,r}
\widehat{h}_{\alpha\beta}^{(j,l,r)} (x) Y^{(j,l,r)}_0(y),
\nonumber\\
h_{\alpha \xi} (x,y) &=& \sum_{j,l,r} \widehat{A}^{(j,l,r)}_{\alpha}
(x) Y^{(j,l,r)}_\xi (y)=\sum_{i=1}^5\sum_{j,l,r_i}
\widehat{A}^{(j,l,r_i)}_{\alpha} (x) Y^{(j,l,r_i)}_{q_i} (y) ,
\nonumber\\
h_{(\xi\varepsilon)} (x,y) &\equiv&
h_{\xi\varepsilon}-\frac{1}{5}g_{\xi\varepsilon}h^\tau_{~\tau}=
\sum_{j,l,r} \widehat{\varphi}^{(j,l,r)} (x)
Y^{(j,l,r)}_{(\xi\varepsilon)}(y)\nonumber\\
&=& \sum_{i=1}^{10}\sum_{j,l,r_i}\widehat{\varphi}^{(j,l,r_i)}
(x)Y^{(j,l,r_i)}_{q_i} (y),\nonumber\\
h^\xi_{~\xi} (x,y) &=& \sum_{j,l,r} \widehat{\pi}^{(j,l,r)} (x)
Y^{(j,l,r)}_0 (y), \label{harmexpan2}
\end{eqnarray}
where $D_\xi$ is the $SO(5)$ covariant derivative defined on
$T^{1,1}$.

The eigenvalues of mass matrices of the K-K modes and their
classifications under the supersymmetry group $SU(2,2|1)$ in the
$AdS_5$ space-time are listed  in Ref.\,\cite{ita}.
 The zero modes will constitute the ${\cal N}=2$ $U(1)$ gauged
 $AdS_5$ supergravity coupled with Yang-Mills vector multiplets
 and some Betti
 multiplets.  Substituting the expansions (\ref{harmexpan2}) into
 (\ref{qudraction}) and considering only  zero modes, we get
   the quadratic action for the massless graviton
 of ${\cal N}=2$ $U(1)$ gauged $AdS_5$ supergravity,
\begin{eqnarray}
S_{\rm q.g.} &=&\frac{1}{2k_{5}^2}\int d^{5}x
\sqrt{-\widehat{g}^{(0)}}\left[-\frac{1}{4} \widehat{\nabla}^\gamma
\widehat{h}^{\alpha\beta} \widehat{\nabla}_\gamma
\widehat{h}_{\alpha\beta}+\frac{1}{2}\widehat{\nabla}^\gamma
\widehat{h}^{\alpha\beta} \widehat{\nabla}_\alpha
\widehat{h}_{\gamma\beta}\right.\nonumber\\
&&\left.+\frac{1}{4}\widehat{\nabla}^\alpha \widehat{h}\left(
\widehat{\nabla}_\alpha \widehat{h}-\frac{1}{2}
\widehat{\nabla}^\beta \widehat{h}_{\alpha\beta}\right)-\frac{1}{l
^2} \left(\widehat{h}^{\alpha\beta}\widehat{h}_{\alpha\beta}+
\frac{1}{2}\widehat{h}^2 \right)+\cdots \right]. \label{afac}
\end{eqnarray}

 In the following we analyze the symmetry of  above
graviton action in $AdS_5$  space-time background. First, the
graviton in the $AdS_5$ background is the fluctuation around
space-time metric,
\begin{eqnarray}
ds^2_5=\widehat{g}_{\alpha\beta} dx^\alpha
dx^\beta=\left[\widehat{g}_{\alpha\beta}^{(0)}
+\widehat{h}_{\alpha\beta}(x)\right]
dx^\alpha dx^\beta. \label{conback}
\end{eqnarray}
The general coordinate transformation invariance determines the
following infinitesimal gauge symmetry for the graviton in the
$AdS_5$ space-time background,
\begin{eqnarray}
\delta \widehat{h}_{\alpha\beta}=\widehat{\nabla}^{(0)}_\alpha
\widehat{\xi}_\beta +\widehat{\nabla}^{(0)}_\beta
\widehat{\xi}_\alpha, \label{diff}
\end{eqnarray}
where the covariant derivative $\nabla^{(0)}_\alpha$ is defined with
respect to the $AdS_5$ metric $g^{(0)}_{\alpha\beta}$. To verify the
masslessness of the graviton $\widehat{h}_{\alpha\beta}$  in $AdS_5$
space, we observe the equation of motion derived from the action
(\ref{afac}) \cite{higu},
\begin{eqnarray}
E_{\alpha\beta} &\equiv & \frac{1}{2}\left( \widehat{\nabla}^\gamma
\widehat{\nabla}_\gamma
\widehat{h}_{\alpha\beta}-\widehat{\nabla}_\alpha\widehat{\nabla}_\gamma
\widehat{h}^\gamma_{~\beta}-\widehat{\nabla}_\beta\widehat{\nabla}_\gamma
\widehat{h}^\lambda_{~\alpha}+
\widehat{\nabla}_\alpha\nabla_\beta \widehat{h}\right)\nonumber\\
&&+\frac{1}{2}\widehat{g}^{(0)}_{\alpha\beta}
\left(\widehat{\nabla}_\gamma\widehat{\nabla}_\delta
\widehat{h}^{\gamma\delta}- \widehat{\nabla}^\gamma
\widehat{\nabla}_\gamma \widehat{h}\right)-\frac{2}{l^2} \left(
\widehat{h}_{\alpha\beta}+\frac{1}{2}\widehat{g}^{(0)}_{\alpha\beta}
\widehat{h} \right) =0. \label{eomgr}
\end{eqnarray}
 $\widehat{h}_{\alpha\beta}$
contain non-physical modes due to the local symmetry (\ref{diff})
and the physical ones should be the traceless and divergence-free
(i.e., transverse and traceless) part of
$\widehat{h}_{\alpha\beta}$. Eq.\,(\ref{eomgr}) gives the identity,
$\widehat{\nabla}^{(0)}_\beta E^{\alpha\beta}=0$. This equation
together  gauge-fixing  condition $\widehat{\nabla}_\alpha^{(0)}
\widehat{h}^{\alpha\beta}=0$ gives exactly the right physical
degrees of freedom for the massless  graviton in the $AdS_5$ space.

 When  fractional $D3$-branes switch on, the $AdS_5\times T^{1,1}$
 background get deformed by R-R- and NS-NS three-form
 fluxes carried by fractional branes. The isometric
  symmetries of the deformed $AdS_5$ and $T^{1,1}$ reduce.
   As explained above, when we
  consider type IIB supergravity in
  this background, the compactification on deformed
  $T^{1,1}$ occurs, and the isometric
  symmetries of the deformed $AdS_5$ and $T^{1,1}$
  convert into the space-time- and gauge symmetries for the compactified
   theory. Therefore, less symmetries appear in the compactified theory
   than  the case of type IIB supergravity
   in the $AdS_5\times T^{1,1}$ background. This phenomenon can be
   considered as the spontaneous breaking of local symmetry in
  the gauged $AdS_5$ supergravity \cite{duff}.
    The broken symmetry should  be the symmetry corresponding to
  conformal supersymmetry on field they side,
  and the symmetry corresponding to Poincar\'{e} supersymmetry should still
  persist. Further, the Higgs effect should take place
  since those broken symmetries are local ones
in the gauged $AdS_5$
  supergravity. In the following we demostrate  how the
Higgs mechanism  corresponding to the spontaneous breaking of local
symmetry $SO(2,4)$ to $SO(1,4)$ happens.

We start from type IIB supergravity  in ten dimensions and consider
the Einstein-Hilbert-, self-dual five-form $F_{(5)}$ and NS-NS
two-form $B_{(2)}$ terms listed in Eq.\,(\ref{type2b}) \footnote{For
the simplicity of discussion we do not consider the R-R three-form
sector.},
\begin{eqnarray}
S_{\rm IIB}=\frac{1}{2\kappa_{\rm 10}^2}\int d^{10}x \overline{E}
\left(\overline{R}-\frac{1}{2\times 3!} H_{(3)MNP} H^{MNP}_{(3)}
-\frac{1}{4\times 5!}{F}_{(5)MNPQR}{F}^{MNPQR}_{(5)}
 \right). \label{t2bs}
\end{eqnarray}
The Einstein equation derived from the above classical action is
\begin{eqnarray}
\overline{R}_{MN}=\frac{1}{96}F_{(5)MPQRS}F_{(5)N}^{~~~~PQRS}+\frac{1}{4}\left(
H_{(3)MPQ} H_{(3)N}^{~~~~PQ}-\frac{1}{12}G_{MN} H_{(3)PQR}
H^{(3)PQR} \right).
\end{eqnarray}
This equation of motion shows clearly that without the three-form
field $H_{(3)PQR}$, the self-dual five-form flux should lead to a
three-brane solution whose near-horizon limit is $AdS_5\times
T^{1,1}$ if the flat limit is $M^4\times {\cal C}_6$ \cite{balov},
while the presence of three-form flux  deforms the $ASdS_5\times
T^{1,1}$ and leads to a less symmetric vacuum configuration for type
IIB supergravity. According to the general idea of the Higgs
mechanism, we should make the vacuum configuration symmetric by
performing a gauge transformation which re-parametrises the field
variable. This can be done by shifting the Ricci curvature and
absorbing the three-form flux contribution into the shifted Ricci
curvature. Thus we define
\begin{eqnarray}
R\equiv \overline{R}-\frac{1}{2\times 3!} H_{(3)MNP}
H^{~~MNP}_{(3)}.
\end{eqnarray}
Consequently, the above Einstein equation becomes  the one with only
the five-form flux as the matter source,
\begin{eqnarray}
R_{MN}&\equiv & \overline{R}_{MN}-\frac{1}{4}\left( H_{(3)MPQ}
H_{(3)N}^{~~~~PQ}-\frac{1}{12}G_{MN} H_{(3)PQR} H^{PQR}_{(3)}
\right)\nonumber\\
& = &\frac{1}{96}F_{(5)MPQRS}F_{(5)N}^{~~~~PQRS}. \label{ree}
\end{eqnarray}
It will give the three-brane  solution whose near horizon limit is
$AdS_5\times T^{1,1}$. Eq.\,(\ref{ree}) also means a shift of the
Riemannian curvature,
\begin{eqnarray}
R_{KMLN}&=&\overline{R}_{KMLN}-\frac{1}{4}\left[ H_{(3)KMQ}
H_{(3)LN}^{~~~~~~Q}\right.\nonumber\\
 &&\left.-\frac{1}{12\times
9}\left(G_{KL}G_{MN}-G_{KN}G_{ML}\right)H_{(3)PQR}
H^{PQR}_{(3)}\right].
\end{eqnarray}

In the following we expand the gravitational field around this
``shifted" $AdS_5 \times T^{1,1}$ vacuum configuration.  Just like
revealing the Higgs phenomenon in gauge theory, $h_{MN}$ must
undergo a gauge transformation, which  makes $h_{MN}$ pick up a
longitudinal component and  become massive. That is, we have the
following operation on background metrics and graviton fields,
\begin{eqnarray}
G_{MN}=\overline{G}_{MN}^{(0)}+{h}_{MN}={G}_{MN}^{(0)}+\overline{h}_{MN},
\label{reme}
\end{eqnarray}
where ${G}_{MN}^{(0)}$ and $\overline{G}_{MN}^{(0)}$ denote the
symmetric and deformed $AdS_5\times T^{1,1}$ metrics, respectively,
$\overline{h}_{MN}$ and ${h}_{MN}$ are  graviton excitations around
these space-time background metrics, respectively.

The above discussion is a qualitative analysis in ten dimensions. In
the following we go to five-dimensions. Using the explicit form of
the K-T solution, we can find the explicit conversion between the
undeformed and deformed $AdS_5$ geometry.   The deformed $AdS_5$
background  in the K-T solution (\ref{ksso3}) is
\begin{eqnarray}
ds_{\rm d-AdS_5}^2&=&\widehat{\overline{g}}_{\alpha\beta}^{(0)}
dx^\alpha dx^\beta  \approx
\frac{r^2}{L^2}\left[1-A(r)\right]\eta_{\mu\nu}
dx^\mu dx^\nu +\frac{L^2}{r^2}\left[1+A(r)\right] dr^2, \nonumber\\
A(r)&=&\frac{3}{4\pi}\frac{M^2}{N}g_s\left(\frac{1}{4}+ \ln
\frac{r}{r_0}\right), \label{comf}
\end{eqnarray}
The logarithmic dependence on the radial coordinate in the deformed
$AdS_5$ space-time background means that the $SO(2,4)$ isometry
symmetry of $AdS_5$ space breaks to $SO(1,4)$.

 The $(\alpha\beta)$ component of Eqs.\,(\ref{ree})
 and the  explicit form of NS-NS two-form
$B_{(2)}$  given in (\ref{ksso3}) determine that the deformed
$AdS_5$ vacuum background
$\widehat{\overline{g}}_{\alpha\beta}^{(0)}$ can be restored back to
the $AdS_5$ metric $\widehat{g}^{(0)}_{\alpha\beta}$ by the
diffeomorphism transformation,
\begin{eqnarray}
\widehat{\overline{g}}_{\alpha\beta}^{(0)}=
\widehat{g}^{(0)}_{\alpha\beta}-\widehat{\nabla}^{(0)}_\alpha
\widehat{B}_\beta -\widehat{\nabla}^{(0)}_\beta \widehat{B}_\alpha .
\label{eme}
\end{eqnarray}
In above equation $\widehat{B}_\alpha $ is a newly defined vector
field relevant to
 the NS-NS two-form field $B_{(2)}$ in the K-S solution,
\begin{eqnarray}
\widehat{B}_\alpha = \partial_\alpha\left[ G\int _{S^2}
B_{(2)}\right], \label{sagoldb}
\end{eqnarray}
where $G$, as $F$ in (\ref{fgold}), is a function in ten dimensions.
The restored space-time background metric
$\widehat{g}^{(0)}_{\alpha\beta}$ now becomes invariant under the
following diffeomorphism transformation in the $AdS_5$ space-time,
\begin{eqnarray}
\delta x^\alpha=-\widehat{\xi}^\alpha, ~~~\delta
\widehat{B}_\alpha=-\widehat{\xi}_\alpha.
\end{eqnarray}
Therefore, the $AdS_5$ space-time background is re-gained and the
full $SO(2,4)$ space-time symmetry is recovered. A straightforward
observation for this symmetry restoration is that
$\widehat{B}_\alpha$ provides a compensation to the non-symmetric
part since $\widehat{B}_\alpha$ also has $\ln (r/r_0)$ dependence.

 Eqs.\,(\ref{reme}) and (\ref{eme}) determine  that  in the
 restored $AdS_5$ space-time background the graviton $\widehat{\overline{h}}_{\alpha\beta}$
 is related to the graviton $\widehat{h}_{\alpha\beta}$ in the
 deformed $AdS_5$ space-time as the
 the following,
\begin{eqnarray}
\widehat{h}_{\alpha\beta}=
\widehat{\overline{h}}_{\alpha\beta}+\widehat{\nabla}^{(0)}_\alpha
\widehat{B}_\beta +\widehat{\nabla}^{(0)}_\beta \widehat{B}_\alpha .
\label{sgold}
\end{eqnarray}
Hence it picks up a longitudinal component. This shows
 that $\widehat{B}_\alpha$ plays the role of a Goldstone vector
field in the $AdS_5$ space \cite{porra}.

Based on above analysis, we expand the classical  action
(\ref{t2bs}) to the second order of graviton ${h}_{\alpha\beta}$
around the symmetric $AdS_5$ background. Furthermore, as in  usual
gauge theory, all of other fields such as $B_{(2)}$ must be replaced
by their $SO(2,4)$ transformed versions if they belong to certain
non-trivial representations of $SO(2,4)$. The quadratic action of
type IIB supergravity in the deformed $AdS_5\times T^{1,1}$
background can be recast into the one around the symmetric
$AdS_5\times T^{1,1}$ vacuum configuration,
\begin{eqnarray}
\overline{S}_{\rm Q.G.} &=&\frac{1}{2k_{10}^2}\int d^{10}x
\sqrt{-G^{(0)}}\left\{-\frac{1}{4} \nabla^P \overline{h}^{MN}
\nabla_P \overline{h}_{MN}+\frac{1}{2}\nabla^P \overline{h}^{MN}
\nabla_M
\overline{h}_{PN}\right.\nonumber\\
&+&\frac{1}{4}\nabla^M \overline{h}\left( \nabla_M
\overline{h}-\frac{1}{2} \nabla^N \overline{h}_{MN}\right)
+\frac{27}{32}\frac{\pi \alpha^{\prime 2}N}{\overline{h}^2(r) r^5}
\left(\overline{h}^{MN}\overline{h}_{MN}- \frac{1}{2}h^2 \right)\nonumber\\
&+& \left.\frac{1}{2}\left[R^{(0)}_{MPNQ}-\frac{1}{90}
\left(G^{(0)}_{MN}G^{(0)}_{PQ}-G^{(0)}_{MQ}G^{(0)}_{NP}
\right)\frac{1}{2\times 3!} H^{(0)}_{(3)RST} H^{(0)RST}_{(3)}\right]
\overline{h}^{MN}\overline{h}^{PQ}\right\}\nonumber\\
&=&\frac{1}{2k_{10}^2}\int d^{10}x \sqrt{-G^{(0)}}\left[-\frac{1}{4}
\nabla^P \overline{h}^{MN}  \nabla_P \overline{h}_{MN}+\frac{1}{2}
\nabla_M \overline{h}^{MN} \nabla^P
\overline{h}_{PN}\right.\nonumber\\
&&+\frac{1}{4}\nabla^M \overline{h}\left( \nabla_M
\overline{h}-\frac{1}{2} \nabla^N
\overline{h}_{MN}\right)-\frac{9}{32}\frac{\pi \alpha^{\prime
2}N}{h^2(r) r^5}
\left(\overline{h}^{MN}\overline{h}_{MN}+ \frac{1}{2}\overline{h}^2 \right)\nonumber\\
&&\left.-\frac{1}{4}\times \frac{1}{2\times 45}\left(\frac{3g_s
M\alpha^\prime}{4} \right)^2 \frac{72}{r^6h^{3/2}(r)} \left(
\overline{h}^{MN} \overline{h}_{MN}-\overline{h}^2\right) \right].
\end{eqnarray}
The last term indicates the arising of the celebrated Pauli-Fierz
mass term for the graviton in $AdS_5$ space \cite{pauli}.

Performing compactification on $T^{1,1}$ and taking into account
only zero modes in the K-K spectrum, we obtain the massive $AdS_5$
graviton due to the partially spontaneous breaking of local
diffeomorphism symmetry in $AdS_5$ space-time,
\begin{eqnarray}
\overline{S}_{\rm q.g.} &=&\frac{1}{2k_{5}^2}\int d^{5}x
\sqrt{-\widehat{g}^{(0)}}\left[-\frac{1}{4} \widehat{\nabla}^\gamma
\widehat{\overline{h}^{\alpha\beta}} \widehat{\nabla}_\gamma
\widehat{\overline{h}}_{\alpha\beta}+\frac{1}{2}\widehat{\nabla}^\gamma
\widehat{\overline{h}}^{\alpha\beta} \widehat{\nabla}_\alpha
\widehat{\overline{h}}_{\gamma\beta}\right.\nonumber\\
&&+\frac{1}{4}\widehat{\nabla}^\alpha \widehat{\overline{h}}\left(
\widehat{\nabla}_\alpha \widehat{\overline{h}}-\frac{1}{2}
\widehat{\nabla}^\beta
\widehat{\overline{h}}_{\alpha\beta}\right)-\frac{1}{L^2}
\left(\widehat{\overline{h}}^{\alpha\beta}\widehat{\overline{h}}_{\alpha\beta}+
\frac{1}{2}\widehat{\overline{h}}^2
\right)\nonumber\\
&&\left.-\frac{1}{4}m^2\left(\widehat{\overline{h}}^{\alpha\beta}
\widehat{\overline{h}}_{\alpha\beta}-\widehat{\overline{h}}^2\right)
\right]. \label{afac11}
\end{eqnarray}
The mass $m$ can be explicitly evaluated through integrating over
the internal manifold $T^{1,1}$.

\subsection{Dual of $\gamma$-trace Anomaly $\gamma_\mu s^\mu$ of Supersymmetry Current}

In this section we investigate the gravity dual of $\gamma$-trace
anomaly of supersymmetry current $s_\mu$. As discussed in Sect.\,V,
the breaking or preservation of supersymmetry by the  K-S solution
can be detected by observing the Killing spinor equation deduced
from  supersymmetry transformations on graviton $\Psi_M$ and
dilatino ${\Lambda}$ in the space-time background described by the
K-S solution. The Killing spinor equations in $AdS_5\times T^{1,1}$
background is listed in Eq.\,(\ref{susytr2}),
\begin{eqnarray}
\kappa_{\rm 10} \delta {\Lambda} &=& 0 \nonumber\\
\kappa_{\rm 10} \delta {\Psi}_M &=&\left(\partial_M+\frac{1}{4}
\omega_M^{~~AB}\Gamma^{AB}\right)\epsilon +\frac{i}{480}\kappa_{\rm
10} \Gamma^{PQRST} {F}_{(5)PQRST}\Gamma_M\epsilon=0. \label{susytr5}
\end{eqnarray}
On the other hand, Eq.\,(\ref{susytr3}) gives the Killing spinor
equation in the deformed $AdS_5\times T^{1,1}$ background,
\begin{eqnarray}
\kappa_{\rm 10}\,\delta {\Lambda} &=& -\frac{i}{24}\kappa_{\rm
10}\left({F}_{(3)MNP}+i{H}_{(3)MNP} \right)\Gamma^{MNP} \epsilon
=0, \nonumber\\
\kappa_{\rm 10}\, \delta \Psi_M &=&\left(\partial_M+\frac{1}{4}\,
\omega_M^{~~AB}\Gamma^{AB}\right)\epsilon \nonumber\\
&+&\frac{i}{480} \kappa_{\rm 10}\Gamma^{PQRST}\Gamma_M
\epsilon\left( \overline{F}_{(5)PQRST}-\frac{5}{4}\kappa_{\rm 10}
C_{[(2)PQ}{H}_{(3)RST]}
+\frac{5}{4}\kappa_{\rm 10} B_{[(2)PQ}{F}_{(3)RST]}\right)\nonumber\\
&+&\frac{1}{96}\kappa_{\rm 10}
\left[\Gamma_M^{~~NPQ}\left({F}_{(3)NPQ}+i{H}_{(3)NPQ} \right)
-9\Gamma^{NP}\left({F}_{(3)MNP}+i{H}_{(3)MNP}
\right)\right]\epsilon^\star \nonumber\\
 &=& 0. \label{susytr4}
\end{eqnarray}
 Now we employ the same idea as
looking for the dual of scale anomaly to find the dual description
to $\gamma$-trace anomaly of supersymmetry current. That is, we
should define new ${\Lambda}$ and ${\Psi}_M$ and make the Killing
spinor equation (\ref{susytr3}) (or (\ref{susytr4})) in deformed
$AdS_5\times T^{1,1}$ background recover to the form of
Eq.\,(\ref{susytr2}) (or (\ref{susytr5})).

A straightforward comparison between Eq.\,(\ref{susytr5}) and
Eq.\,(\ref{susytr4}) implies that that we should introduce a complex
right-handed Weyl spinor $\Upsilon$ and define
\begin{eqnarray}
{\Lambda}^\prime &\equiv& {\Lambda}
-4i\Upsilon, \nonumber\\
{\Psi}^\prime_M &\equiv&\Psi_M-\Gamma_M \Upsilon.
\label{fermionshift}
\end{eqnarray}
Then we use the identities of $\Gamma$-matrix in ten dimensions,
\begin{eqnarray}
\left\{\Gamma_M,\Gamma_N \right\}&=& 2 g_{MN},\nonumber\\
\Gamma_{M_1M_2\cdots M_n}&=&\Gamma_{[M_1}\Gamma_{M_2}\cdots
\Gamma_{M_n]}=\frac{1}{n!}\sum_P
(-1)^{\delta_P}\Gamma_{a_{P(1)}}\Gamma_{a_{P(2)}}\cdots
\Gamma_{a_{P(n)}},\nonumber\\
\Gamma_{M_1M_2\cdots M_n N}&=& \Gamma_{M_1M_2\cdots M_n} \Gamma_N -
n
\Gamma_{[M_1M_2\cdots M_{n-1}} G_{M_n] N},\nonumber\\
\Gamma_{NM_1M_2\cdots M_n }&=& \Gamma_N \Gamma_{M_1M_2\cdots M_n} -
nG_{N[M_1 } \Gamma_{M_1M_2\cdots M_{n}]} ,
\end{eqnarray}
and assign the supersymmetry transformation of $\Upsilon$ as the
following,
\begin{eqnarray}
\delta \Upsilon = -\frac{1}{96}\left({F}_{(3)MNP}+i{H}_{(3)MNP}
\right)\Gamma^{MNP} \epsilon .
 \label{susytrans6}
\end{eqnarray}
In this way,  the supersymmetry transformation (\ref{susytr5}) will
be reproduced from (\ref{susytr4}) (up to the linear terms in
fermionic fields and
 to the first order in the gravitational coupling $\kappa_{\rm 10}$). Later we shall see
 that after the
 compactification on $T^{1,1}$, $\Upsilon$ will lead to the Goldstone fermion
 . This Goldstone fermionic field should arise when
 the NS-NS- and R-R three-form fluxes breaks one-half of local
 supersymmetries. Indeed, we can see that
 the supersymmetric transformation for the compactified $\Upsilon$ is
 proportional to the transformation parameter with the
 proportionality  coefficient given by the three-form fluxes
 passing through $S^3$.
 This is a typical feature of the Goldstone fermion \cite{wezu}.

To show how the super-Higges effect takes place,  we consider the
quadratic part of the
 gravitino action of type IIB supergravity,
\begin{eqnarray}
S_{\rm gravitino}&=&\frac{1}{2k_{\rm 10}^2} \int d^{10}x\,E\left[
-\frac{i}{2}\overline{\Psi}_M\Gamma^{MNP}
 D_N\Psi_P\right.\nonumber\\
 &&\left.-\frac{1}{8\times 5!}\overline{\Psi}_M \Gamma^{MNP}\left(
\Gamma^{UVWXY}{F}_{(5)UVWXY}\right)\Gamma_N\Psi_P+\cdots \right].
\label{qttb}
\end{eqnarray}
When the fractional D3-branes are absent, the space-time background
is $AdS_5 \times T^{1,1}$ with the self-dual five-form field
strength,
\begin{eqnarray}
F_{x_0 x_1 x_2 x_3 r}=\frac{r^3}{g_sL^4}, ~~~F_{g_1g_2 g_3g_4
g_5}=\frac{L^4}{27 g_s}.
\end{eqnarray}

We put above fermionic action in the $AdS_5 \times T^{1,1}$
background. As stated above, the compactification on $T^{1,1}$ will
take place. The compactification of fermionic fields in type IIB
supergravity was processed as the following \cite{kim}. First, the
explicit representation for Dirac matrices $\Gamma^A$ in ten
dimensions should be specified  \cite{ita,kim},
\begin{eqnarray}
\Gamma^A &=& E^A_{~~M}\Gamma^M, ~~\Gamma^a = \widehat{\gamma}^a
\otimes 1_4 \otimes \sigma^1, ~~\Gamma^m= 1_4 \otimes \tau^m \otimes
(-\sigma^2),\nonumber\\
M&=& (\alpha,\xi),~~ A=(a,m),~~  a = 0,\cdots,4, ~~m =5,\cdots,9,
\label{gmat1}
\end{eqnarray}
\begin{eqnarray}
&& \left\{\Gamma^A,\Gamma^B \right\}=2\eta^{AB},
~~\left\{\widehat{\gamma}^a,\widehat{\gamma}^b
\right\}=2\eta^{ab},~~\left\{\tau^m,\tau^n
\right\}=2\delta^{mn},\nonumber\\
&& \widehat{\gamma}^0=-\widehat{\gamma}_0=i\sigma^1 \otimes 1_2=i
\left(\begin{array}{cc} 0 & 1_2 \\ 1_2 & 0 \end{array}\right), ~
\widehat{\gamma}^i=\widehat{\gamma}_i=\sigma^2 \otimes \sigma^i
=\left(\begin{array}{cc} 0 & i\sigma^i
\\ -i\sigma^i & 0 \end{array}\right),\nonumber\\
&& \widehat{\gamma}^4=\widehat{\gamma}_4=
i\widehat{\gamma}^0\widehat{\gamma}^1\widehat{\gamma}^2\widehat{\gamma}^3=
\sigma^3\otimes 1_2=\left(\begin{array}{cc}  1_2 & 0 \\ 0 &-1_2
\end{array}\right), ~~~i=1,2,3;\nonumber\\
&& \tau^5=\tau_5=i\widehat{\gamma}_0,
~~\tau^6=\tau_6=\widehat{\gamma}_1,~~\tau^7=\tau_7=\widehat{\gamma}_2,
~~\tau^8=\tau_8=\widehat{\gamma}_3,~~\tau^9=\tau_9=\widehat{\gamma}_4.
\label{gmat2}
\end{eqnarray}
In above equations, $\gamma^a$ and $\tau^m$ denote the Dirac
matrices in $AdS_5$ space and $T^{1,1}$, respectively.
 The above  choice on $\Gamma$-matrices determine an
 explicit form for the four-dimensional $\gamma_5$-analogue in ten dimensions,
\begin{eqnarray}
\Gamma^{11}=\Gamma_{11}=\Gamma^0 \Gamma^1\cdots \Gamma^9=1_4\otimes
1_4\otimes \sigma_3=\left(\begin{array}{cc} 1_{16} & 0 \\ 0 &
-1_{16}
\end{array}\right).
\end{eqnarray}
Consequently, the left-handed gravitino $\Psi_M$, the supersymmetry
transformation parameter $\epsilon$ and the right-handed dilatino
${\Lambda}$ decompose as the following,
\begin{eqnarray}
\Psi_\alpha (x,y) &=& {\psi} _\alpha (x,y)
\otimes \left(\begin{array}{c} 1\\
0\end{array}\right)= \left(\begin{array}{c} {\psi} _\alpha (x,y)
\\
0\end{array}\right),\nonumber\\
\Psi_\xi (x,y) &=&{\psi} _\xi (x,y)
\otimes \left(\begin{array}{c} 1\\
0\end{array}\right)= \left(\begin{array}{c} {\psi}_\xi (x,y)
\\
0\end{array}\right), \nonumber\\
\Lambda (x,y) &=& {\lambda}(x,y)\otimes \left(\begin{array}{c} 0\\
1\end{array}\right)=\left(\begin{array}{c} 0\\
{\lambda} (x,y)\end{array}\right), \nonumber\\
\epsilon (x,y)
&=&  \widetilde{\epsilon} (x,y) \otimes \left(\begin{array}{c} 1\\
0\end{array}\right)=\left(\begin{array}{c} \widetilde{\epsilon} (x,y)\\
0\end{array}\right),\nonumber\\
\Upsilon (x,y) &=& \widetilde{\chi}(x,y) \otimes \left(\begin{array}{c} 0\\
1\end{array}\right)=\left(\begin{array}{c} 0\\
\widetilde{\chi} (x,y)\end{array}\right),
 \label{mwspinor}
\end{eqnarray}
where $x^\alpha$, $y^\xi$ are the coordinates on $AdS_5$ and
$T^{1,1}$, respectively, and ${\psi}_\alpha$, ${\psi}_\xi$,
$\widetilde{\lambda}$ and $\widetilde{\eta}$ are 16-component Weyl
spinor fields in ten dimensions. Further, we expand these field
functions and the local supersymmetry transformation parameter in
terms of the four-component $SO(5)$ spinor harmonics
$\Xi^{(j,l,r)}(y)$, and they can be further decomposed into the
one-dimensional $U_{\rm H}(1)$-harmonics $Y^{(j,l,r)}_q$ on
$T^{1,1}$ \cite{ita},
\begin{eqnarray}
{\psi}_{\alpha} (x,y) &=&\sum_{j,l,r}
\widehat{\psi}_{\alpha}^{(j,l,r)}
(x)\Xi^{(j,l,r)}(y)=\sum_{j,l,r}\left(\begin{array}{c}
\widehat{\psi}^{(j,l,r-1)}_{\alpha} (x) Y^{(j,l,r-1)}_{0} (y)\\
\widehat{\psi}^{(j,l,r+1)}_{\alpha} (x) Y^{(j,l,r+1)}_{0} (y)\\
\widehat{\psi}^{(j,l,r)}_{\alpha} (x) Y^{(j,l,r)}_{-1} (y)\\
\widehat{\psi}^{(j,l,r)}_{\alpha} (x) Y^{(j,l,r)}_{+1} (y)
 \end{array}
\right),
 \nonumber\\
{\lambda} (x,y) &=&\sum_{j,l,r} \widehat{\lambda}^{(j,l,r)}
(x)\Xi^{(j,l,r)}(y)=\sum_{j,l,r}\left(\begin{array}{c}
\widehat{\lambda}^{(j,l,r-1)} (x) Y^{(j,l,r-1)}_{0} (y)\\
\widehat{\lambda}^{(j,l,r+1)} (x) Y^{(j,l,r+1)}_{0} (y)\\
\widehat{\lambda}^{(j,l,r)} (x) Y^{(j,l,r)}_{-1} (y)\\
\widehat{\lambda}^{(j,l,r)}(x) Y^{(j,l,r)}_{+1} (y)
 \end{array}
\right),\nonumber\\
\widetilde{\epsilon} (x,y) &=&\sum_{j,l,r}
\widehat{\epsilon}^{(j,l,r)}
(x)\Xi^{(j,l,r)}(y)=\sum_{j,l,r}\left(\begin{array}{c}
\widehat{\epsilon}^{(j,l,r-1)} (x) Y^{(j,l,r-1)}_{0} (y)\\
\widehat{\epsilon}^{(j,l,r+1)} (x) Y^{(j,l,r+1)}_{0} (y)\\
\widehat{\epsilon}^{(j,l,r)} (x) Y^{(j,l,r)}_{-1} (y)\\
\widehat{\epsilon}^{(j,l,r)} (x) Y^{(j,l,r)}_{+1} (y)
 \end{array}
\right),
\nonumber\\
\widetilde{\chi} (x,y) &=& \sum_{j,l,r} \widehat{\chi}^{(j,l,r)}
(x)\Xi^{(j,l,r)}(y)=\sum_{j,l,r}\left(\begin{array}{c}
\widehat{\chi}^{(j,l,r-1)} (x) Y^{(j,l,r-1)}_{0} (y)\\
\widehat{\chi}^{(j,l,r+1)}(x) Y^{(j,l,r+1)}_{0} (y)\\
\widehat{\chi}^{(j,l,r)}(x) Y^{(j,l,r)}_{-1} (y)\\
\widehat{\chi}^{(j,l,r)} (x) Y^{(j,l,r)}_{+1} (y)
 \end{array}\right). \label{spexpan}
\end{eqnarray}
In Eq.\,(\ref{spexpan}) $\widehat{\psi}_\alpha^{(j,l,r)} (x)$,
$\widehat{\lambda}^{(j,l,r)} (x)$, $\widehat{\epsilon}^{(j,l,r)}
(x)$ and $\widehat{\chi}^{(j,l,r)} (x)$ are  the four-component
spinor fields on $AdS_5$; $\Xi^{(j,l,r)} (y)$ are the four-component
$SO(5)$ spinor harmonics on $T^{1,1}$, and $Y^{(j,l,r)}_{q}(y)$ are
the $U_{\rm H}(1)$-harmonics with $U_{\rm H}(1)$-charge $q$.
Substituting the chiral decomposition (\ref{mwspinor})
 and further the $U_{\rm H}(1)$-harmonic expansions for $\Psi_\alpha$
 and ${\Lambda}$ into their linearized equations of motion
 and  further taking into account only zero modes of the corresponding   kinetic operators on
 $T^{1,1}$, one can obtain the following quadratic action for the gravitino
 of the ${\cal N}=2$ $U(1)$
 gauged $AdS_5$ supergravity \cite{ita},
\begin{eqnarray}
S_{\rm gravitino}&=&\frac{1}{2\kappa_{\rm 5}^2} \int d^{5}x\,e\left[
-\frac{i}{2}\overline{\widehat{\psi}}^i_\alpha\widehat{\gamma}^{\alpha\beta\gamma}
 \widehat{D}_\beta\widehat{\psi}_{i\gamma} + \frac{3}{4 L}
  \overline{\widehat{\psi}}_\alpha^i \widehat{\gamma}^{\alpha\beta}
\widehat{\psi}_{i\beta}
 +\cdots\right].
 \label{maslac}
\end{eqnarray}
Eq.\,(\ref{maslac}) is the action for the massless gravitino of
${\cal N}=2$ $U(1)$  gauged  $AdS_5$ supergravity, and the second
mass-like term is required by supersymmetry to accompany the
cosmological term in $AdS_5$ space \cite{send}; $\psi_\alpha^i$ is
the $SU(2)$ symplectic Majorana spinor and the indices $i=1,2$ label
two gravitini  of $R$-charges $r=\pm 1$, which arise automatically
when the compactification on $T^{1,1}$ is performed \cite{ita}.

Now we come to the case with fractional branes. The $AdS_5\times
T^{1,1}$ background is deformed by the three-form fluxes carried by
fractional branes. The gravitino action (\ref{qttb}) can be
expressed in terms of either $\Psi_M$, ${\Lambda}$ in the deformed
$AdS_5 \times T^{1,1}$ background or the shifted  fields
$\Psi_M^\prime$ and ${\Lambda}^\prime$ in   the $AdS_5 \times
T^{1,1}$ background. We focus on  the second  term of (\ref{qttb})
and re-write it in terms of the shifted  fermionic fields shown
(\ref{fermionshift}). Up to the leading order of $\kappa_{\rm 10}$,
we obtain
\begin{eqnarray}
&& \overline{S}_{\rm gravitino}=\frac{1}{2\kappa_{\rm 10}^2} \int
d^{10}x\,\overline{E}\left[ -\frac{1}{8\times 5!}\overline{\Psi}_M
\Gamma^{MNP}\left(
\Gamma^{UVWXY}\overline{\widetilde{F}}_{(5)UVWXY}\right)\Gamma_N\Psi_P
+\cdots\right]\nonumber\\
&=& \frac{1}{2\kappa_{\rm 10}^2} \int
d^{10}x\,{E}\left(-\frac{1}{8\times 5!} \right)\left(
\overline{\Psi}^\prime_M + \overline{\Upsilon}\Gamma_M
\right)\Gamma^{MNP}
\left(\Gamma^{(5)}\cdot\widetilde{F}_{(5)}\right)\Gamma_N \left(
{\Psi}_P^\prime- \Gamma_P {\Upsilon} \right) \nonumber\\
&=& -\frac{1}{2\kappa_{\rm 10}^2}\frac{1}{ 5!} \int
d^{10}x\,{E}\left[\frac{1}{4}\overline{\Psi}_\alpha^\prime
\Gamma^{\alpha\beta}
\left(\Gamma^{\alpha_1\cdots\alpha_5}\widetilde{F}_{\alpha_1\cdots\alpha_5}
-\Gamma^{\alpha_1\cdots\alpha_5}\widetilde{F}_{\alpha_1\cdots\alpha_5}
\right)\Psi_\beta^\prime \right.\nonumber\\
&&+\overline{\Psi}_\alpha^\prime
\Gamma^{\alpha}\Gamma^{\xi}\left(\Gamma^{\alpha_1\cdots\alpha_5}
\widetilde{F}_{\alpha_1\cdots\alpha_5}
-\Gamma^{\xi_1\cdots\xi_5}\widetilde{F}_{\xi_1\cdots\xi_5}
\right)\Psi_\xi^\prime\nonumber\\
&&+ \overline{\Upsilon}\Gamma^\alpha
\left(\Gamma^{\alpha_1\cdots\alpha_5}\widetilde{F}_{\alpha_1\cdots\alpha_5}
-\Gamma^{\alpha_1\cdots\alpha_5}\widetilde{F}_{\alpha_1\cdots\alpha_5}
\right)\Psi_\alpha^\prime-\overline{\Psi}^\prime_\alpha\Gamma^\alpha
\left(\Gamma^{\alpha_1\cdots\alpha_5}\widetilde{F}_{\alpha_1\cdots\alpha_5}
+\Gamma^{\xi_1\cdots\xi_5}\widetilde{F}_{\xi_1\cdots\xi_5}
\right)\Upsilon\nonumber\\
&&- \overline{\Psi}_\xi^\prime\Gamma^\xi
\left(\Gamma^{\alpha_1\cdots\alpha_5}\widetilde{F}_{\alpha_1\cdots\alpha_5}
+\Gamma^{\xi_1\cdots\xi_5}\widetilde{F}_{\xi_1\cdots\xi_5}
\right)\Upsilon+\overline{\Upsilon}\Gamma^\xi
\left(\Gamma^{\alpha_1\cdots\alpha_5}\widetilde{F}_{\alpha_1\cdots\alpha_5}
-\Gamma^{\xi_1\cdots\xi_5}\widetilde{F}_{\xi_1\cdots\xi_5}
\right)\Psi_\xi^\prime\nonumber\\
&&-2\times5 \times 5\overline{\Upsilon}\,\Gamma^\alpha
\left(\Gamma^{\alpha_1\cdots\alpha_5}\widetilde{F}_{\alpha_1\cdots\alpha_5}
+\Gamma^{\xi_1\cdots\xi_5}\widetilde{F}_{\xi_1\cdots\xi_5}
\right)\Upsilon\nonumber\\
&&\left.-\frac{1}{4}\,\overline{\Psi}_\xi^\prime
\Gamma^{\xi\varepsilon}\left(\Gamma^{\alpha_1\cdots\alpha_5}
\widetilde{F}_{\alpha_1\cdots\alpha_5}
-\Gamma^{\xi_1\cdots\xi_5}\widetilde{F}_{\xi_1\cdots\xi_5}
\right)\Psi_\varepsilon^\prime \right], \label{shiftaction}
\end{eqnarray}
where we have decomposed the ten-dimensional components  of the
fields into those on $AdS_5$ and $T^{1,1}$ and  used the following
identities,
\begin{eqnarray}
\left[\Gamma^\alpha,\Gamma^{\alpha_1\cdots\alpha_5}
\right]\widetilde{F}_{\alpha_1\cdots\alpha_5}&=&\left(5 g^{\alpha
[\alpha_1}\Gamma^{\alpha_2\cdots\alpha_5]}-5
\Gamma^{[\alpha_1\cdots\alpha_4}g^{\alpha_5]\alpha}
\right)\widetilde{F}_{\alpha_1\cdots\alpha_5} \nonumber\\
&=&
5g^{\alpha\alpha_1}\Gamma_{\alpha_1}-5g^{\alpha\alpha_1}\Gamma_{\alpha_1}=0,
\nonumber\\
\left[\Gamma^\xi,\Gamma^{\xi_1\cdots\xi_5}
\right]\widetilde{F}_{\xi_1\cdots\xi_5}
&=&\left[\Gamma^{\xi\varepsilon},\Gamma^{\xi_1\cdots\xi_5}
\right]\widetilde{F}_{\xi_1\cdots\xi_5}
=\left[\Gamma^{\alpha\beta},\Gamma^{\alpha_1\cdots\alpha_5}
\right]\widetilde{F}_{\alpha_1\cdots\alpha_5}=0.
\end{eqnarray}
Further, according to the explicit representations of
$\Gamma$-matrices
 listed in (\ref{gmat1}) and (\ref{gmat2}),  we have
\begin{eqnarray}
&&\Gamma^{\alpha_1\cdots\alpha_5}=-i\epsilon^{\alpha_1\cdots\alpha_5}1_4\otimes
1_4\otimes \sigma_1,~~~
\Gamma^{\xi_1\cdots\xi_5}=-\epsilon^{\xi_1\cdots\xi_5}1_4\otimes
1_4\otimes \sigma_2,\nonumber\\
&&
\Gamma^{\alpha_1\cdots\alpha_5}\widetilde{F}_{\alpha_1\cdots\alpha_5}
\pm
\Gamma^{\xi_1\cdots\xi_5}\widetilde{F}_{\xi_1\cdots\xi_5}\nonumber\\
&=&-i 1_4\otimes 1_4\otimes \left( \begin{array}{cc} 0 &
\epsilon^{\alpha_1\cdots\alpha_5}\widetilde{F}_{\alpha_1\cdots\alpha_5}\mp
\epsilon^{\xi_1\cdots\xi_5}\widetilde{F}_{\xi_1\cdots\xi_5}
\\
\epsilon^{\alpha_1\cdots\alpha_5}\widetilde{F}_{\alpha_1\cdots\alpha_5}\pm
\epsilon^{\xi_1\cdots\xi_5}\widetilde{F}_{\xi_1\cdots\xi_5} & 0
\end{array} \right).
\label{gacal}
\end{eqnarray}
Substituting (\ref{gacal}) and the Weyl spinor representations
(\ref{mwspinor}) into the action (\ref{shiftaction}), we can reduce
$\overline{S}_{\rm gravitino}$ to the following,
\begin{eqnarray}
\overline{S}_{\rm gravitino}&=&\frac{1}{2\kappa_{\rm 10}^2}\frac{i}{
5!} \int d^{10}x\,{E}\left[\left(\frac{1}{4}\overline{\psi}_\alpha
(1_4\otimes\gamma^{\alpha\beta}){\psi}_\beta-2\overline{\psi}_\alpha
(1_4\otimes\gamma^\alpha) {\chi}+5 \overline{\chi}
{\chi} \right)\right.\nonumber\\
&&\times \left.
\left(\epsilon^{\alpha_1\cdots\alpha_5}\widetilde{F}_{\alpha_1\cdots\alpha_5}
-\epsilon^{\xi_1\cdots\xi_5}\widetilde{F}_{\xi_1\cdots\xi_5}
\right)\right]\nonumber\\
&=&-\frac{i}{2\kappa_{\rm 10}^2} \int d^{10}x\,\sqrt{-G_{\rm
10}}\,\left[\frac{3}{L}\,\frac{1}{g_s}\left(1+\frac{3}{2\pi}\frac{g_s
M^2}{N}\ln\frac{r}{r_0}\right)\right.\nonumber\\
&&\left.\times \left(\frac{1}{4}\overline{\psi}_\alpha
(1_4\otimes\gamma^{\alpha\beta}){\psi}_\beta-2\overline{\psi}_\alpha
(1_4\otimes\gamma^\alpha){\chi}+5 \overline{\chi}{\chi}
\right)\right]. \label{superhi}
\end{eqnarray}
In above derivation we have used
\begin{eqnarray}
&& \frac{1}{5!}\left(\epsilon^{\alpha_1\cdots\alpha_5}
\widetilde{F}_{\alpha_1\cdots\alpha_5} -\epsilon^{\xi_1\cdots\xi_5}
\widetilde{F}_{\xi_1\cdots\xi_5}\right)\nonumber\\
&=&-\sqrt{-g_{AdS_5}}\,F_{0123r}-\sqrt{g_{
T^{1,1}}}\,F_{g^1g^2g^3g^4g^5} \nonumber\\
&=&-3 h^{-5/4}(r)\frac{27}{4}\frac{\pi \alpha^{\prime 2}N_{\rm
eff}}{r^5}\nonumber\\
&=& -\frac{3}{L}\,\frac{1}{g_s}\left(1+\frac{3}{2\pi}\frac{g_s
M^2}{N}\ln\frac{r}{r_0} \right).
\end{eqnarray}

Eq.\,(\ref{superhi}) shows the sign of super-Higgs effects in the
gauged $AdS_5$ supergravity \cite{wezu}.  Finally, substituting the
$U_{\rm H}(1)$-harmonic expansions (\ref{spexpan})
 for ${\psi}_\alpha$ and
${\chi}$ into (\ref{superhi}) and integrating over the internal
manifold $T^{1,1}$, we obtain the five-dimensional gauged
supergravity from the compactification on $T^{1,1}$. If only the
contributions from zero-modes are taken into account,
Eq.\,(\ref{superhi}) yields the following terms of the
five-dimensional gauged supergravity,
\begin{eqnarray}
\overline{S}_{\rm gravitino} &=&-\frac{i}{2\kappa_{\rm 5}^2} \int
d^{5}x\,\sqrt{-\widehat{g}^{(0)}}\,\left(\frac{3}{4L}\,\frac{1}{g_s}+m\right)
\left(-\overline{\widehat{\psi}}^i_\alpha
\widehat{\gamma}^{\alpha\beta}\widehat{\psi}_{i\beta}
-8\overline{\widehat{\psi}}^i_\alpha \widehat{\gamma}^\alpha
\widehat{\chi}_i+20 \overline{\widehat{\chi}}^i\widehat{\chi}_i
\right). \label{superhi2}
\end{eqnarray}
Finally,  making a shift
\begin{eqnarray}
\widehat{\psi}^{i\prime}_\alpha
=\widehat{\psi}^i_\alpha-\widehat{\gamma}_\alpha \widehat{\chi}^i,
\end{eqnarray}
in above equation, we obtain an  elegant result,
\begin{eqnarray}
\overline{S}_{\rm gravitino} &=&-\frac{i}{2\kappa_{\rm 5}^2} \int
d^{5}x\,\sqrt{-\widehat{g}^{(0)}}\,\left(\frac{3}{4L}\,\frac{1}{g_s}+m\right)
\overline{\widehat{\psi}}_\alpha^{i\prime}
\widehat{\gamma}^{\alpha\beta}\widehat{\psi}_\beta^{i\prime}.
\label{superhi3}
\end{eqnarray}
 In above action, the first term with  coefficient
proportional to $1/(Lg_s)$ is the term accompanying cosmological
constant term, which is  required by the supersymmetry  in $AdS_5$
space \cite{send}; The second one is the mass term for the gravitino
of  five-dimensional gauged supergravity. This mass is generated by
the super-Higgs mechanism and is proportional to fluxes $M$ carried
by the fractional branes, which can be seen clearly from
Eq.\,(\ref{superhi}).

\subsection{${\cal N}=1$ Goldstone Chiral Supermultiplet on $AdS_5$ Boundary}

We have shown  that  the superconformal anomaly multiplet of an
${\cal N}=1$ $SU(N+M)\times SU(N)$  supersymmetric gauge theory in
four dimensions is dual to the spontaneous breaking of local
supersymmetry from ${\cal N}=2$ to ${\cal N}=1$ and the consequent
super-Higgs mechanism in ${\cal N}=2$ $U(1)$ gauged $AdS_5$
supergravity, through which the ${\cal N}=2$ graviton supermultiplet
($\widehat{h}_{\alpha\beta}$,$\widehat{\psi}^i_\alpha$,$\widehat{A}_\alpha$)
becomes massive. A crucial ingredient in implementing this
super-Higgs mechanism is the Goldstone fields $\theta$, $B_\alpha$
and $\Upsilon$ in ten dimensions, which are defined in
Eqs.\,(\ref{fgold}), (\ref{sagoldb}) and (\ref{fermionshift}) and
(\ref{susytrans6}). Since the superconformal anomaly in an ${\cal
N}=1$ four-dimensional supersymmetric gauge theory constitutes an
${\cal N}=1$ chiral supermultiplet \cite{grisaru,afgj}. Therefore,
the gauge/gravity dual  means that $\theta$, $B_\alpha$ and
$\Upsilon$ should constitute a supermultiplet. Further, according to
the holographic version on $AdS/CFT$ correspondence at the
supergravity level \cite{gkp,witt1}, $\theta$, $B_\alpha$ and
$\Upsilon$  after the compactification on $T^{1,1}$ should form an
${\cal N}=2$ Goldstone supermultiplet in five dimensions and its
$AdS_5$ boundary value should be an ${\cal N}=1$ chiral
supermultiplet in four dimensions. This fact can be verified as the
following. First, we make use of the supersymmetric transformations
for $B_{(2)MN}$ and $F_{(3)MNP}$ in type IIB supergravity
\cite{schw},
\begin{eqnarray}
\delta G_{(3)MNP}&=& 3\partial_{[M}\delta
A_{(2)NP]}\nonumber\\
&=& \delta
\left[F_{(3)MNP}+iH_{(3)MNP}\right]\nonumber\\
&=& 3\partial_{[M}\delta
C_{(2)NP]}+3i \partial_{[M}\delta B_{(2)NP]}\nonumber\\
&=& 3\partial_{[M}\overline{\epsilon}\Gamma_{NP]}{\Lambda}
-12i\partial_{[M}\overline{\epsilon}^\star \Gamma_N \Psi_{P]}.
\end{eqnarray}
 This implies that locally there should exist
\begin{eqnarray}
 \delta A_{(2)NP]} &=& \delta
C_{(2)NP}+i \delta B_{(2)NP}\nonumber\\
&=& \overline{\epsilon}\Gamma_{NP}{\Lambda}
-2i\overline{\epsilon}^\star \left(\Gamma_N
\Psi_{P}-\Gamma_P\Psi_N\right).
\end{eqnarray}
 When we restore the deformed $AdS_5\times T^{1,1}$ back to the symmetric
 $AdS_5\times T^{1,1}$
 vacuum background,
   the fermionic fields $\Psi_M$ and ${\Lambda}$
   are reparametrized as shown  in (\ref{fermionshift}),
   the above supersymmetric transformation becomes
\begin{eqnarray}
 \delta {A}_{(2)NP}^\prime &\equiv& \delta {A}_{(2)NP}+  \delta {A}_{(2)NP}^{(\Upsilon)}
 \nonumber\\
 &=&
\delta{C}^\prime_{(2)NP}+i \delta {B}_{(2)NP}^\prime \nonumber\\
&=& \overline{\epsilon}\Gamma_{NP}\left(\Lambda-4i\Upsilon\right)
-2i\overline{\epsilon}^\star \left[\Gamma_N
\left(\Psi_{P}-\Gamma_P\Upsilon\right)
-\Gamma_P\left(\Psi_N-\Gamma_P\Upsilon\right)\right]\nonumber\\
&=&\overline{\epsilon}\Gamma_{NP}{\Lambda}
-2i\overline{\epsilon}^\star \left(\Gamma_N
\Psi_{P}-\Gamma_P\Psi_N\right)-4i \left(
\overline{\epsilon}-\overline{\epsilon}^\star\right)
\Gamma_{NP}\Upsilon. \label{shiftst}
\end{eqnarray}
where we use ${A}_{(2)NP}^{(\Upsilon)}$ to represent the parts of
$B_{(2)MN}$ and $C_{(2)MN}$ whose supersymmetric transformations
depend only on $\Upsilon$ due to the shift on fermionic fields.
Eq.\,(\ref{shiftst}) yields the following supersymmetric
transformation involving the Goldstone spinor $\Upsilon$,
\begin{eqnarray}
\delta {A}_{(2)NP}^{(\Upsilon)}=-4i \left(
\overline{\epsilon}-\overline{\epsilon}^\star\right)
\Gamma_{NP}\Upsilon =4 \left(\mbox{Im}\,\overline{\epsilon}\right)
\,\Gamma_{NP}\Upsilon. \label{goldsusy1}
\end{eqnarray}

We turn to  the supersymmetric transformation of $\Upsilon$ given in
(\ref{susytrans6}),
\begin{eqnarray}
\delta\Upsilon &=&-\frac{1}{96}\times 3\,\partial_{[M}
A^{(\Upsilon)}_{(2)NP]}\Gamma^{MNP}\epsilon \nonumber\\
&=&-\frac{1}{96}\times 3\left(\partial_{[M}
C^{(\Upsilon)}_{(2)NP]}+i\partial_{[M}B^{(\Upsilon)}_{(2)NP]}
\right)\Gamma^{MNP}\epsilon\nonumber\\
&=&-\frac{1}{32}\Gamma^M\left(  \partial_M
A^{(\Upsilon)}_{(2)NP}\right)\Gamma^{NP}\epsilon, \label{goldsusy2}
\end{eqnarray}
where the following identity among $\Gamma$-matrix in ten dimensions
is used,
\begin{eqnarray}
\Gamma^{MNP}=\Gamma^M\Gamma^{NP}-
\left(G^{MN}\Gamma^P-G^{MP}\Gamma^N \right).
\end{eqnarray}
Eqs.\,(\ref{goldsusy1}) and (\ref{goldsusy2}) show that
$A^{(\Upsilon)}_{(2)NP}$ and $\Upsilon$ constitute a supermultiplet
in ten dimensions since their supersymmetric transformations form a
closed algebra.

Further, from the definitions on Goldstone bosons given in
(\ref{fgold}) and (\ref{sgold}),
\begin{eqnarray}
\theta = F\int _{S^2} C_{(2)}, ~~~B_\alpha=\partial_\alpha \left[
G\int_{S^2} B_{(2)}\right]\equiv \partial_\alpha \omega,
\end{eqnarray}
we choose $F=G$ \footnote{This choice is always possible since in
(\ref{fgold}) and (\ref{sgold}) $F$ and $G$ are introduced as
arbitrary field functions in ten-dimensions. Actually, the
requirement that the Goldstone fields should constitute a
supermultiplet imposes this choice.} and define a new complex scalar
field in ten dimensions,
\begin{eqnarray}
\varphi \equiv \theta +i \omega  =F\left[\int _{S^2} \left(
C_{(2)}+i B_{(2)}\right)\right]. \label{comsca}
\end{eqnarray}
Since in the K-S solution both $C_{(2)}$ and $B_{(2)}$ are
proportional to $\omega_2=\left( g^1\wedge g^2+g^3\wedge g^4
\right)/2$, the complex scalar field $\varphi$ defined in
(\ref{comsca}) actually contains two complex (four real) scalar
fields,
\begin{eqnarray}
\varphi^1 &\sim & \int_{S^2}\left( C_{(2)g_1g_2}+i
B_{(2)g_1g_2}\right),
\nonumber\\
\varphi^2 &\sim & \int_{S^2}\left( C_{(2)g_3g_4}+i
B_{(2)g_3g_4}\right).
\end{eqnarray}
Consequently, the supersymmetric transformations (\ref{goldsusy1})
and (\ref{goldsusy2}) for the Goldstone multiplet
$(A^{(\Upsilon)}_{(2)NP},\Upsilon)$ reduce to the following forms,
\begin{eqnarray}
\delta \varphi^1 &=& 4\left(\mbox{Im}\,\overline{\epsilon}\right) \,
\Gamma_{g_1g_2}\widetilde{\Upsilon}, \nonumber\\
\delta \varphi^2 &=& 4\left(\mbox{Im}\,\overline{\epsilon}\right) \,
\Gamma_{g_3g_4}\widetilde{\Upsilon}, \nonumber\\
\delta \widetilde{\Upsilon} &=& -\frac{1}{32}\Gamma^M \left[
\left(\partial_M \varphi^1\right)\Gamma_{g_1g_2}+ \left(\partial_M
\varphi^2\right)\Gamma_{g_3g_4}\right]{\epsilon}, \nonumber\\
\widetilde{\Upsilon} &\equiv &\int_{S^2}\Upsilon . \label{goldsusy3}
\end{eqnarray}
Further, recall that  $\Upsilon$ and its supersymmetric
transformation (\ref{susytrans6}) are introduced to counter the
supersymmetric transformation in the K-S solution background so that
the Killing spinor equation (\ref{susytr4})  can be restored back to
the Killing spinor equation (\ref{susytr5}) in  the $AdS_5\times
T^{1,1}$ background. Therefore, the supersymmetry transformation
parameter $\epsilon$ in (\ref{goldsusy1})-(\ref{goldsusy3}) should
satisfy the constraint equations (\ref{kicon2}). Since we use the UV
(large-$\tau$) limit of the K-S solution (i.e., the K-T solution),
Eq.\,(\ref{kicon2}) yields
\begin{eqnarray}
\Gamma_{g_1g_2}\epsilon=-\Gamma_{g_3g_4}\epsilon=\lim_{\tau\to\infty}\,
i\left(\frac{\sinh\,\tau}{\cosh\,\tau}-\frac{1}{\cosh\,\tau}
\Gamma_{g_1g_3}\right)\epsilon=i\epsilon. \label{goldsusy4}
\end{eqnarray}
So finally we get an elegant form for the supersymmetric
transformation of the Goldstone multiplet without resorting to the
explicit forms of $\Gamma_{g_m}$ ($m=1,2,3,4)$,
\begin{eqnarray}
\delta \varphi^1 &=&
-4i\,\left(\mbox{Im}\,\overline{\epsilon}\right)
\,\widetilde{\Upsilon}, \nonumber\\
\delta \varphi^2 &=& 4i\,\left(\mbox{Im}\,\overline{\epsilon}\right)
\,
\widetilde{\Upsilon}, \nonumber\\
\delta \widetilde{\Upsilon} &=& -\frac{i}{32}\Gamma^M
\left(\partial_M \varphi^1- \partial_M \varphi^2\right){\epsilon}.
\label{goldsusy5}
\end{eqnarray}
Substituting the ten-dimensional complex right-handed Weyl spinor
$\Upsilon$ and the supersymmetry transformation parameter $\epsilon$
listed in (\ref{mwspinor}) into (\ref{goldsusy5}) and using
\begin{eqnarray}
\overline{\epsilon}={\epsilon}^\dagger \Gamma_0=
\left(\widetilde{\epsilon}^\dagger  (x,y), 0\right) \,
\gamma^0\otimes 1_4\otimes\left( \begin{array}{cc} 0 & 1 \\ 1 & 0
\end{array} \right) =\left(0, \overline{\widetilde{\epsilon}} (x,y)
\right)
\end{eqnarray}
 we obtain
\begin{eqnarray}
\delta \varphi^1 (x,y) &=& -4i
\left(\mbox{Im}\,\overline{\widetilde{\epsilon}}\right) \,
{\widetilde{\chi}}(x,y), \nonumber\\
\delta \varphi^2 (x,y) &=& 4i
\left(\mbox{Im}\,\overline{\widetilde{\epsilon}}\right) \,
\widetilde{\chi}(x,y), \nonumber\\
\delta \widetilde{\chi}(x,y) &=& -\frac{i}{32} \left[\gamma^\alpha
\partial^x_\alpha \left(\varphi^1-\varphi^2\right)
-i\tau^\xi\partial^y_\xi
\left(\varphi^1-\varphi^2\right)\right]\widetilde{\epsilon}(x,y).
\label{goldsusy6}
\end{eqnarray}
The following task is performing compactification on $T^{1,1}$ and
reducing above supersymmetric transformations to five dimensions.
That is, we expand $\varphi^p(x,y)$ ($p=1,2$),
$\widetilde{\chi}(x,y)$ and $\widetilde{\epsilon}(x,y)$ in terms of
scalar- and spinor $U_H(1)$ harmonics as shown in (\ref{spexpan}),
and then insert these expansions into Eq.\,(\ref{goldsusy6}). Taking
into account only zero modes in the expansions  and  comparing both
sides of Eq.\,(\ref{goldsusy6}), we obtain
\begin{eqnarray}
\delta \widehat{\chi}^i &=&-\frac{i}{32}f^{ij}_p
\widehat{\gamma}^\alpha
\partial_\alpha
\varphi^p \widehat{\eta}_j,\nonumber\\
\delta  \widehat{\varphi}^p &=& 4i
f^p_{ij}\overline{\widehat{\eta}}^i \widehat{\chi}^j.
 \label{bulksusy}
\end{eqnarray}
In above equation, $f^{ij}_p\equiv\delta^{ij} N^{j p}$ ($j$ is not
summed), $i,j=1,2$ are the indices labeling fermions with opposite
$U_R(1)$-charges, which arise naturally in performing the
compactification of fermionic fields on $T^{1,1}$; $N^{jp}$ are
defined as the following:
\begin{eqnarray}
N^{11}=N^{21}=1,~~~ N^{12}=N^{22}=-1. \label{bulksusy2}
\end{eqnarray}
 $f^p_{ij}$ is related to
$f_p^{ij}$ in the following way:
\begin{eqnarray}
f^p_{ij}\equiv \epsilon^{pq}\delta_{ik}\delta_{jl}f^{kl}_q,
~~~~\epsilon^{12}=-\epsilon^{21}=1.
 \label{bulksusy3}
\end{eqnarray}
Eq.\,(\ref{bulksusy}) together with (\ref{bulksusy2}) and
(\ref{bulksusy3}) is exactly the supersymmetry transformation for
${\cal N}=2$ hypermultiplet in five-dimensions \cite{gst,sie}.

We should further take the hypermultipelt $(\phi^p,\chi^i)$ to the
boundary of $AdS_5$ space. In principle, we can take a similar
procedure as in initiated in Ref.\,\cite{witt1} and carried out
explicitly in Ref.\,\cite{nish,bala}, where on-shell fields in
gauged $AdS_{d+1}$ supergravity can reduce to off-shell fields of
$d$-dimensional conformal supergravity on $AdS_{d+1}$ boundary.
However, here the
 equations of  motion for $\varphi^p$ and $\chi^{i}$ are not clear.
 We naively require their behaviors near the $AdS_5$ boundary according to the
 physical reasonableness.
A careful analysis
 shows that near the $AdS_5$ boundary Eq.\,(\ref{bulksusy}) may convert into
 the supersymmetry transformation  for  ${\cal N}=1$ chiral
supermultiplet in four dimensions. This means that  as expected from
the holographic version on gauge/gravity correspondence, the ${\cal
N}=2$ Goldstone supermultiplet $(\phi^p,\chi^{i})$ in the $AdS_5$
space can constitute a chiral supermultiplet on $AdS_5$ boundary,
which corresponds  exactly to the superconformal anomaly multiplet
of the ${\cal N}=1$ supersymmetric gauge theory in four dimensions.

\section{Summary and Discussion}

We  have highlighted that there exist two types of superconformal
anomalies in a four-dimensional classically conformal invariant
supersymmetric gauge theory and they have distinct gravitational
correspondences in the gauge/gravity dual. One type of
superconformal anomaly arises when the supersymmetric gauge theory
couples with an external conformal supergravity background and its
existence depends on the non-trivial topology of the field
configuration of the external conformal supergravity background,
which is  usually called  the external superconformal anomaly
\cite{grisaru,afgj}. The other one originates purely from the
dynamics of  the supersymmetric gauge theory itself and its anomaly
coefficient is proportional to the beta function of the theory. It
has nothing to do with the external field  background and is usually
called the internal superconformal anomaly \cite{grisaru,afgj}.

 We have emphasized that the external
superconformal anomaly can be partially calculated from the
$AdS/CFT$ correspondence at the supergravity level, i.e., the
holographic version of the gauge/gravity dual. The AdS/CFT
correspondence conjecture at supergravity level, after the
compactification on the compact five-dimensional Einstein manifold,
predicts that there should exist a holographic correspondence
between the gauged $AdS_5$ supergravity and the superconformal gauge
theory on the boundary of $AdS_5$ space. The off-shell conformal
supergravity in four dimensions plays the role of an intermedium: on
one hand, it comes from the on-shell five-dimensional gauged
supergravity around its $AdS_5$ vacuum configuration and  the
superconformal symmetry of the conformal supergravity is relevant to
boundary decomposition of the $AdS_5$ supersymmetry; on the other
hand, the four-dimensional conformal supergravity furnishes an
external field background for the four-dimensional classically
conformal  invariant supersymmetric gauge theory and hence the
various conservation laws in the supersymmetric gauge theory can be
reflected in the preservation of the local symmetries in conformal
supergravity. These relations provide the justifications why the
external superconformal anomaly can be evaluated from the gauged
$AdS_5$ supergravity. Of course, since the holography between the
gauged $AdS_5$ supergravity and four-dimensional supersymmetric
gauge theory is only the lowest-order approximation to the $AdS/CFT$
correspondence, so only partial results of the external
superconformal anomaly can be reproduced. One must go to the
$AdS/CFT$ correspondence at string level to get the whole external
superconformal  anomaly.

We have found the dual correspondence of the  internal
superconformal anomaly. We first figure out that the internal
superconformal anomaly reflects  in the quantum effective action as
the running of gauge coupling and the shift of the CP-violation
parameter as well as the shift their superpartner in superspace.
Further, we make use of the two distinct features of D-branes in the
weakly- and strongly coupled type II superstring theory. In a weakly
coupled case, a $D$-brane behaves as a geometric and dynamical
object from which a supersymmetric gauge theory can be extracted
out; While in the strongly coupled case, the D-brane arises as a
solution to the type II supergravity, the low-energy effective
theory of type II superstring theory. Based on this fact, we realize
that the fractional branes  in non-perturbative string theory are
the source of the internal superconformal  anomaly. Then we go to
the strongly coupled side of type II superstring theory and have
observed that the same fractional branes deform the brane solution.
This brane solution  is obtained in the absence of fractional branes
and its near-horizon limit is the product of the $AdS_5$ space and a
compact five-dimensional Einstein manifold $X^5$. Choosing the brane
solution as a vacuum configuration of type II supergravity, we get a
five-dimensional gauged supergravity since the compactification on
the compact Einstein manifold will occur and its isometry symmetry
will become the gauge symmetry of the gauged supergravity. The
deformation on $AdS_5\times X^5$ makes the space-time background
less symmetric. Therefore,  we obtain a gauged $AdS_5$ supergravity
with spontaneously broken gauge symmetry and consequently the
super-Higgs mechanism will take place. Since the origin of this
super-Higgs mechanism is exactly the fraction branes that lead to
the internal superconformal anomaly. We thus claim the dual of
internal superconformal anomaly of a four-dimensional supersymmetric
gauge theory is the super-Higgs mechanism in the gauged $AdS_5$
supergravity.

Anomaly is a typical quantum phenomenon in gauge field theory.
Therefore, a clear understanding   to the gravitational
correspondences of these two types of superconformal anomaly will be
significant for the comprehension on the gauge/garvity duality.

\bigskip

 \acknowledgments
I would like to thank Professor M. Chaichian and Professor R. Jackiw
for continuous encouragement. During doing this project I have been
benefited from discussions, conversations and communications with O.
Aharony, A. Buchel, E. Bergshoeff, C. Burgess, C.T. Chan, B. de Wit,
M. Grisaru, M. G\"{u}naydin, M. Henningson,  A. Kobakhidze, M.
Nishmura, C. N\'{u}\~{n}ez and Y. Yang and hence acknowledge here. I
would also like to thank my colleagues at the string theory group of
Taiwan for comments when I gave several seminars relevant to  this
topic. Last but not the least, I would like to thank Professors S.
Bulman-Flemming, D. Vaughan and E. Wang in the Department of
Mathematics of the Wilfrid Laurier University for various helps and
encouragements. This work is partially supported by the National
Research Council of Taiwan through NCTS under the grant NSC
94-2119-M-002-001 and the Faculty of Science, Wilfrid Laurier
University.

\end{document}